\documentclass[12pt,a4paper,oneside]{report}

\usepackage{xcolor}
\usepackage[unicode]{hyperref}
\PassOptionsToPackage{unicode}{hyperref}
\PassOptionsToPackage{naturalnames}{hyperref}

\usepackage{graphicx}
\usepackage{lscape}
\usepackage[bf]{caption}
\usepackage{amssymb}
\usepackage{amsmath}
\usepackage{float}
\usepackage{fancyhdr}
\usepackage{longtable}
\usepackage{graphicx}
\usepackage{url}
\usepackage{amsmath}
\usepackage{natbib}
\usepackage{sidecap}
\usepackage{rotating}
\usepackage[nottoc]{tocbibind}
\usepackage{breakcites}
\usepackage{nomencl}
\usepackage{sectsty}
\usepackage{sidecap}

\fancypagestyle{plain}{%
   \fancyhead{} %
}

\chapterfont{\color{cyan}}  % sets colour of chapters
\newcounter{chapcntr}
\newcommand*\toccolor{%
    \ifcase\value{chapcntr}%
         \color{cyan}%----- 0 --
    \or  \color{cyan}%---- 1 --
    \or  \color{cyan}%--- 2 --
    \or  \color{cyan}%---- 3 --
    \or \color{cyan}%-- default
    \or \color{cyan}%-- default
        \or \color{cyan}%-- default
            \or \color{cyan}%-- default
            \or \color{cyan}%-- default
                        \or \color{cyan}%-- default
                                    \or \color{cyan}%-- default
    \or \color{cyan}%-- default
    \else \color{cyan}
    \fi}

\usepackage{tocloft}

\newcommand{\be}{\begin{equation}}
\newcommand{\ee}{\end{equation}}
\newcommand{\ba}{\begin{eqnarray}}
\newcommand{\ea}{\end{eqnarray}}

\newcommand{\arcsec}{\mbox{\ensuremath{^{\prime\prime}}}}

\newcommand{\ppt}[2]					% 2nd partial der.
	{\frac{\partial^{2}\! #1}{\partial#2^{2}}}
\newcommand{\ppm}[3]						% 2nd partial mixed
	{\frac{\partial^{2}\! #1}{\partial#2\partial#3}}	
\newcommand{\grumpy}				% draws a grumpy face
{\mbox{\begin{picture}(20,20)(0,7)
\put(10,10){\circle{15}}
\put(12.5,12.5){\circle*{2}}
\put(7.5,12.5){\circle*{2}}
\put(6,1){\shortstack[l]{$\mbox{}^{\frown}$}}
\end{picture}}}

\newcommand{\happy}				% draws a happy face
{\mbox{\begin{picture}(20,20)(0,7)
\put(10,10){\circle{15}}
\put(12.5,12.5){\circle*{2}}
\put(7.5,12.5){\circle*{2}}
\put(6,1){\shortstack[l]{$\mbox{}^{\smile}$}}
\end{picture}}}

\begin{document}

%
%       --- title.tex ---
%
% Title page for my thesis...

\begin{titlepage}

\begin{center}

\vspace*{2cm}

\huge{\textbf{The spatial, spectral and polarization properties of solar flare X-ray sources}}

\vspace{1.2cm}
\large{\textbf{Natasha Louise Scarlet Jeffrey, M.Sci.}}
\vspace{1.2cm}

\small Astronomy and Astrophysics Group\\
School of Physics and Astronomy\\
Kelvin Building\\
University of Glasgow\\
Glasgow, G12 8QQ\\
Scotland, U.K.
\setlength{\belowcaptionskip}{0.cm}
\vspace{1.cm}
   \begin{center}
   \centering
     \includegraphics[width=0.5\textwidth]{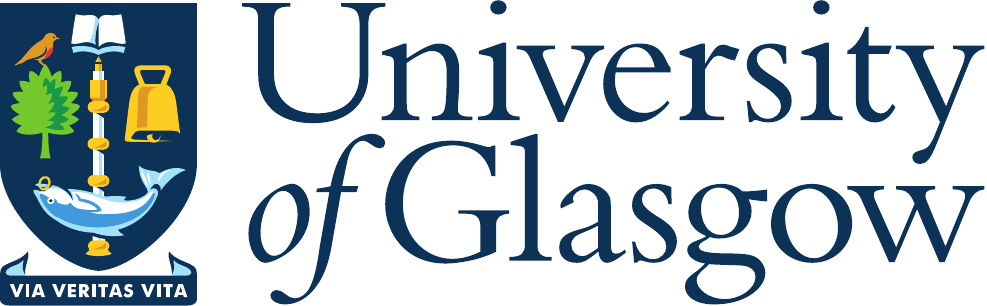}
   \end{center}

\large        Presented for the degree of\\
        Doctor of Philosophy\\
        The University of Glasgow\\
        March 2014

\end{center}
\setlength{\belowcaptionskip}{0.3cm}
\end{titlepage}

%%
%% End of file...
%%

\cleardoublepage
\thispagestyle{empty}

\newpage\markboth{}{}

\vspace*{4cm}
\begin{flushright}
\parbox{130mm}{
\hrulefill

This thesis is my own composition except where indicated in
the text. No part of this thesis has been submitted elsewhere for any other degree
or qualification.
\vspace*{1cm}

{\bf Copyright \copyright ~2014 by Natasha Jeffrey}
\vspace*{0.4cm}

17th March 2014

\hrulefill  }
\end{flushright}

\thispagestyle{empty}

\newpage\markboth{}{}

\vspace*{8cm}
\emph{For my parents, James and Catherine Jeffrey.}

\thispagestyle{empty}
%\addcontentsline{toc}{chapter}{Abstract}

\chapter*{Abstract}
X-rays are a valuable diagnostic tool for the study of high energy accelerated electrons. Bremsstrahlung X-rays produced by, and directly related to, high energy electrons accelerated during a flare, provide a powerful diagnostic tool for determining both the properties of the accelerated electron distribution, and of the flaring coronal and chromospheric plasmas. This thesis is specifically concerned with the study of {\it spatial, spectral and polarization} properties of solar flare X-ray sources via both modelling and X-ray observations using the Ramaty High Energy Solar Spectroscopic Imager ({\it RHESSI}). Firstly, a new 
model is presented, accounting for finite temperature, pitch angle scattering and initial pitch angle injection. This is developed to accurately infer the properties of the acceleration region from the observations of dense coronal X-ray sources. Moreover, examining how the spatial properties of dense coronal X-ray sources change in time, interesting trends in length, width, position, number density and thermal pressure are found and the possible causes for such changes are discussed. Further analysis of data in combination with the modelling of X-ray transport in the photosphere, allows changes in X-ray source positions and sizes due to the X-ray albedo effect to be deduced. Finally, it is shown, for the first time, how the presence of a photospheric X-ray albedo component produces a spatially resolvable polarization pattern across a hard X-ray (HXR) source. It is demonstrated how changes in the degree and direction of polarization across a single HXR source can be used to determine the anisotropy of the radiating electron distribution.

\cleardoublepage
%\phantomsections

\pagenumbering{roman}

\renewcommand{\cfttoctitlefont}{\bf\Huge\color{cyan}}
\renewcommand{\cftloftitlefont}{\bf\Huge\color{cyan}}
\renewcommand{\cftlottitlefont}{\bf\Huge\color{cyan}}

\tableofcontents

\cleardoublepage
%\addcontentsline{toc}{chapter}{\listtablename}
\listoftables
\cleardoublepage
%\addcontentsline{toc}{chapter}{\listfigurename}
\listoffigures
\cleardoublepage

\thispagestyle{empty}

\chapter*{Preface}

\addcontentsline{toc}{chapter}{Preface}

Chapter \ref{ref:Chapter1} provides a brief introduction to the topics and theory required for the following chapters: the interactions of electrons and ions in a plasma, the emission mechanisms required to create solar flare X-rays, the interactions of solar flare X-rays in the photosphere (the albedo effect) and our current understanding of solar flare X-ray observations, using instruments such as Ramaty High Energy Solar Spectroscopic Imager ({\it RHESSI\,}).

Chapters \ref{ref:Chapter2} and \ref{ref:Chapter3} examine an interesting flare type with strong coronal X-ray emission from a dense loop, with little or no emission from the chromosphere. Observations of these events with instruments such as {\it RHESSI\,} have enabled the detailed study of their structure, revealing that amongst other interesting trends, the spatial parameter parallel to the guiding field increases with X-ray energy. This variation has been discussed in the context of a beam of non-thermal electrons in a one-dimensional cold target model, and the results used to constrain both the physical extent of, and density within, an electron acceleration region believed to be situated within the coronal loop itself. In Chapter \ref{ref:Chapter2}, the investigation is extended to a physically realistic model of electron transport that takes into account the finite temperature of the ambient plasma, the initial pitch angle distribution of the accelerated electrons, and the effects of collisional pitch angle scattering. The implications of the results when determining parameters such as number density and acceleration region length from observation are discussed. In Chapter \ref{ref:Chapter3}, the observational analysis of such flare types is further advanced, and the spatial and spectral properties of three dense coronal X-ray loops are studied temporally before, during, and after the peak X-ray emission. Using observations from {\it RHESSI\,}, the temporal changes in emitting X-ray length, width, volume, position, number density and thermal pressure are deduced. Collectively, the observations also show for the first time three temporal phases given by peaks in temperature, X-ray emission, and thermal pressure, with the minimum volume coinciding with the X-ray peak. The possible explanations for the observed trends are discussed.

Chapters \ref{ref:Chapter5} and \ref{ref:Chapter6} examine solar flare X-ray albedo, an effect produced by the Compton backscattering of solar flare produced X-rays in the photosphere. This is studied via Monte Carlo simulations of X-rays in the photosphere. Chapter \ref{ref:Chapter5} investigates quantitatively for the first time the resulting positions and sizes of solar flare hard X-ray chromospheric sources due to the presence of an albedo component, for various chromospheric X-ray source sizes, spectral indices and directivities. It is shown how the albedo effect can alter the true source positions and substantially increase the measured source sizes; this is greater for flatter primary X-ray spectra, stronger downward anisotropy, and for sources closer to the solar disk centre, between the peak albedo energies of 20 and 50 keV. Chapter \ref{ref:Chapter5} demonstrates how the albedo component should be taken into account when X-ray footpoint positions, footpoint motions and source sizes are observed and analysed by instruments such as {\it RHESSI\,}. In Chapter \ref{ref:Chapter6}, this study is extended to investigate the polarization of solar flare chromospheric X-ray sources, by investigating how the presence of an X-ray albedo component produces a variation in the spatial distribution of polarization across a single X-ray source. From this, polarization maps for each of the modelled electron distributions are calculated at various heliocentric angles from the solar centre to the solar limb. The investigation shows how Compton scattering produces a distinct polarization variation across the albedo patch at peak albedo energies of 20-50 keV. It discusses how spatially resolved hard X-ray polarization measurements from future X-ray polarimeters could provide important information about the directivity and energetics of the radiating electron distribution, using both the degree and direction of polarization.

Chapter \ref{ref:Chapter7} provides conclusions, discussion and some final remarks regarding the thesis as a whole, in the context of current solar flare understanding and possible future missions. Unless indicated, $CGS$ units are used throughout the thesis. 

\thispagestyle{empty}

\chapter*{Acknowledgements}

\addcontentsline{toc}{chapter}{Acknowledgements}

\thispagestyle{empty}

It must be mentioned that each chapter (Chapters \ref{ref:Chapter2}, \ref{ref:Chapter3}, \ref{ref:Chapter5} and \ref{ref:Chapter6}) is published and hence I wish to thank my publication co-authors: Drs. Eduard P. Kontar, Nicolas H. Bian and A. Gordon Emslie, and highlight their contribution to the work residing within this thesis. I would also like to thank Dr. Brian Dennis and the {\it RHESSI} team at NASA GSFC for their help with {\it RHESSI\,} imaging and spectroscopy during my short stay at Goddard.
\\\\
However, I wish to solely acknowledge and express my sincerest gratitude to Dr. Eduard Kontar, for his invaluable help and insightful guidance during my postgraduate study and undergraduate summer projects.

\pagebreak

\pagenumbering{arabic}
\chapter{Introduction}
\label{ref:Chapter1}

\newcommand{\hilight}[1]{\colorbox{green}{#1}}
\newcommand{\hili}[1]{\colorbox{red}{#1}}

\section{The Sun, its atmosphere and solar flares}\label{intro_intro}
Our star, the Sun is a G2 main sequence star. It has a mass, radius, luminosity and effective surface temperature of M$_{\bigodot}=1.99\times10^{33}$ g, R$_{\bigodot}=6.96\times10^{10}$ cm, $\mathcal{L}_{\bigodot}=3.84\times10^{33}$ erg s$^{-1}$ and T$_{\bigodot}=5778$ K respectively \citep[e.g.,][]{2004suin.book.....S}, with an estimated age of 4.6 Gyr \citep{2011MNRAS.418.1217H}. The solar atmosphere, which extends into the solar wind, is the largest continuous structure in the solar system, permeating the entire heliosphere. The solar magnetic field governs the evolution of the solar corona and hence it is widely believed to be responsible for transient phenomenon such as solar flares. Solar flares are uninterestingly defined as a ``rapid, sudden brightening in the solar atmosphere", yet they are responsible for the largest release of energy in our solar system, which can be greater than $10^{32}$ erg. Most solar flares occur within active regions on the Sun; regions where the solar magnetic field is particularly strong. The physics associated with the production of, and processes throughout, a solar flare is immense; in order to fully understand the entire flare mechanism, large scale processes describing the evolution of the magnetic field within an entire active region must be coupled with the small scale processes describing the interactions of high energy particles accelerated during the flare. This thesis is concerned with the latter.

\begin{figure}
\centering
\includegraphics[width=14cm]{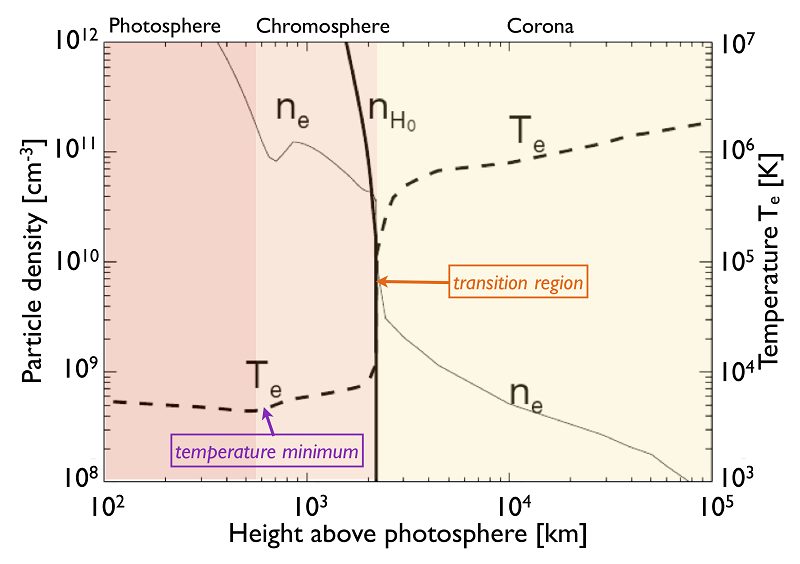}
\caption[The changing number density and temperature structure of the solar atmosphere.]{Original figure taken from \cite{2004psci.book.....A} and then adapted. The figure shows how electron number density $n_{e}$, hydrogen number density $n_{H_{0}}$ and electron temperature $T_{e}$ change with height above the solar photosphere. The photosphere, chromosphere, corona, temperature minimum region and transition region are noted on the figure.}
\label{temp_num}
\end{figure}

The solar atmosphere is a continuous structure with many layers of varying temperature and number density. A semi-empirical model of the solar atmosphere is shown in Figure \ref{temp_num}. It is usual to split the solar atmosphere into three layers defined as the: photosphere, chromosphere, and the corona, which eventually extends into, and is renamed, the solar wind at roughly $3R_{\bigodot}$, filling the entire heliosphere. The photosphere is the optical `surface' of the Sun; the point at which the solar atmosphere becomes opaque to optical wavelengths. The temperature $T$ and number density $n$ of the photosphere fall with increasing height, with $T$ falling from $\sim6000$ K to $\sim4000$ K at the highest point of the photosphere, known as the \textit{temperature minimum} region. Hydrogen number densities within the photosphere are of the order $10^{17}$ cm$^{-3}$, falling to around $10^{15}$ cm$^{-3}$ at the temperature minimum region \citep{2008ApJS..175..229A, Vernazzaetal1981}. Within hydrogen number densities of the order $10^{17}$ cm$^{-3}$, high energy X-rays can interact with free or bound electrons, and a significant proportion of this thesis is dedicated to studying these interactions (Chapters \ref{ref:Chapter5} and \ref{ref:Chapter6}). After the temperature minimum region, there is a $\sim2000$ km layer known as the chromosphere, where the temperature of the solar atmosphere begins to rise, reaching $\sim2\times10^{4}$ K at the top. At the top of chromosphere, hydrogen number densities have fallen to around $10^{11}$ cm$^{-3}$ (Figure \ref{temp_num}). The higher hydrogen number densities deeper within the chromosphere collisionally stop high energy electrons transported to the chromosphere during a solar flare, producing bremsstrahlung X-rays. At the top of the chromosphere lies the {\it transition region}. Here, there is sudden two magnitude increase in temperature and decrease in number density over a very small height of around $100$ km. After the transition region, there is the final and largest layer of the solar atmosphere; the corona. The lower corona is a low $\beta$ plasma where the thermal pressure is much less than that the magnetic pressure, of the order $\sim10^{-2}$. However $\beta$ can vary dramatically with coronal height and solar activity \cite[e.g., models by][]{2001SoPh..203...71G}. However, in general the corona is magnetically dominated and highly conductive. At quiet Sun times, the corona has a high temperature of $\sim1-2$ MK and hence can be observed at X-ray energies. The high temperature of the corona is indicated by the presence of lines from highly ionised elements such as iron (Fe) and calcium (Ca) in the coronal emission spectrum. The method of heating the corona to such high temperatures is still not properly understood and is an outstanding problem in astrophysics \cite[e.g.][]{2012RSPTA.370.3217P}. The energy release process that causes the onset of a solar flare is believed to occur within the corona, where the temperature of the plasma in the vicinity of the region of energy release can be tens of mega Kelvin. The number density of the quiet corona is low; $\sim10^{8}-10^{9}$ cm$^{-3}$ or less. During a solar flare, regions of the corona can have a number density as high as $10^{11}$ cm$^{-3}$, possibly from heated material moving into the corona from the denser chromosphere below; this is known as chromospheric evaporation \citep[cf.,][]{1980ApJ...239..725D,1983SoPh...86...67A}. As in the chromosphere, high coronal densities are important for the interaction of particles, mainly electrons, via Coulomb collisions with the background plasma, and the emission of X-rays. This is particularly important in Chapters \ref{ref:Chapter2} and \ref{ref:Chapter3} of this thesis.

It is widely believed that the onset of a solar flare is caused by the release of stored magnetic energy in the corona, due to reconnecting magnetic fields \cite[cf.,][]{2000mare.book.....P}. During a flare, coronal plasma in the vicinity of the energy release region is heated to temperatures greater than $10$ MK. Particles, primarily electrons, but also protons and heavier ions, are accelerated to high energies greater than $\sim20$ keV and often up to MeV and even GeV energies, out of the background thermal plasma. The acceleration of a large number of particles during a solar flare requires an efficient acceleration mechanism. This is a topic of ongoing debate within the solar physics community. Popular candidates are: DC electric field acceleration, stochastic acceleration (second order Fermi acceleration) and shock acceleration (first order Fermi acceleration) \cite[see][as a recent review of such mechanisms]{2011SSRv..159..107H}. The energy released during a solar flare propagates into the lower layers of the corona, transition region and chromosphere, either in the form of precipitating high energy electrons, protons and heavier ions, or by thermal conduction, due to the even steeper temperature gradient created between the corona and chromosphere during a flare. The chromosphere and transition region react to this heating; dense, heated chromospheric material bound by the magnetic field has to expand up into the corona, causing the chromospheric evaporation mentioned in the previous paragraph.
\begin{figure}
  \begin{minipage}[c]{0.55\textwidth}
   \includegraphics[width=\textwidth]{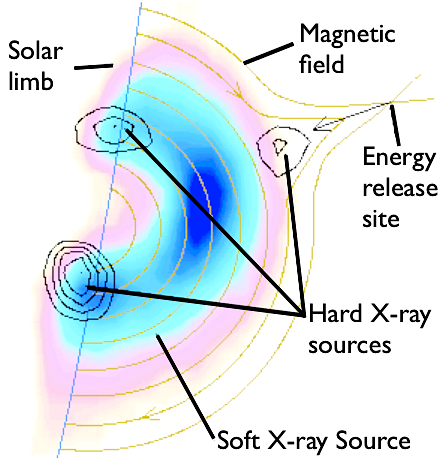}
  \end{minipage}\hfill
  \begin{minipage}[c]{0.40\textwidth}
\caption[{\it Yohkoh} soft and hard X-ray images of a flare from the 13th January 1992.]{X-ray image of a flare (13th January 1992) using Soft and Hard X-ray Telescopes (SXR and HXR) on-board {\it Yohkoh}. HXR contours are overlaid onto the SXR loop. The positions of X-ray sources are discussed in Section \ref{intro_imaging}. This image is taken and adapted from \url{http://hesperia.gsfc.nasa.gov/hessi/images/fd-close.gif}.}
\label{sfm_with_xray}
  \end{minipage}
\end{figure}

During a solar flare, radiation is emitted across the entire electromagnetic spectrum from radio to X-rays and even gamma rays for the largest flares; from the corona to the photosphere. Hard X-rays (HXRs) with energies greater than $\sim10$ keV are produced collisionally by the electrostatic interactions of electrons with background particles in both the corona and chromosphere, mainly as free-free bremsstrahlung emission. Soft X-rays (SXRs) in the range of $\sim0.1-10$ keV are also produced as bremsstrahlung but mainly from particles interacting within a high temperature plasma.  Gamma-rays, if present, above around $300$ keV can also be produced by the interaction of protons, heavier ions and flare produced neutrons. For example gamma-rays can be emitted from the photosphere by the interactions of neutrons combining with neutral hydrogen to form deuterium \citep[e.g.,][]{2009RAA.....9...11C}.

Solar flare sizes are classified by their soft X-ray flux; specifically by the 1-8~\AA flux measured by the Geostationary Orbiting Environmental Satellites ({\it GOES\,}) at 1 AU. The flare classifications are A, B, C, M and X with an X-class flare being the largest. The flux of each class increases by an order of magnitude. The flux of an X-class flare is equal to or greater than $10^{-4}$ W m$^{-2}$, while the flux of a smaller M-class flare is of the order $10^{-5}$ W m$^{-2}$. For classes A to M, the numbers 1 to 10 also denote the strength of the flare, that is, a M10 flare has a higher flux than a M5 flare. There is no limit on the numbers for an X-class flare \cite[e.g.,][]{2011SSRv..159...19F}.

X-rays, and even more so, gamma-rays if present, only represent a small proportion of the total flare radiative output \citep{2004GeoRL..3110802W,2006JGRA..11110S14W,2011A&A...530A..84K}, with the majority of the emission actually coming from larger wavelength emissions of extreme ultraviolet, ultraviolet and visible light. However, the chromosphere and corona are optically thin at high X-ray and gamma-ray energies, and studying their temporal, energetic, spatial and polarization properties can provide a direct link not only to the accelerated electrons, protons and ions responsible for their production, but also the conditions in the corona or chromosphere during a flare; the main topics of study within this thesis. Therefore, the rest of this chapter will discuss the observation and analysis of solar flare X-rays, starting with a brief review of the particle interactions and emission mechanisms required for the production of solar flare X-rays in the solar atmosphere.

\section{Electron and ion interactions the solar atmosphere}
\subsection{Coulomb collisions}
\begin{figure*}
\includegraphics[width=15cm]{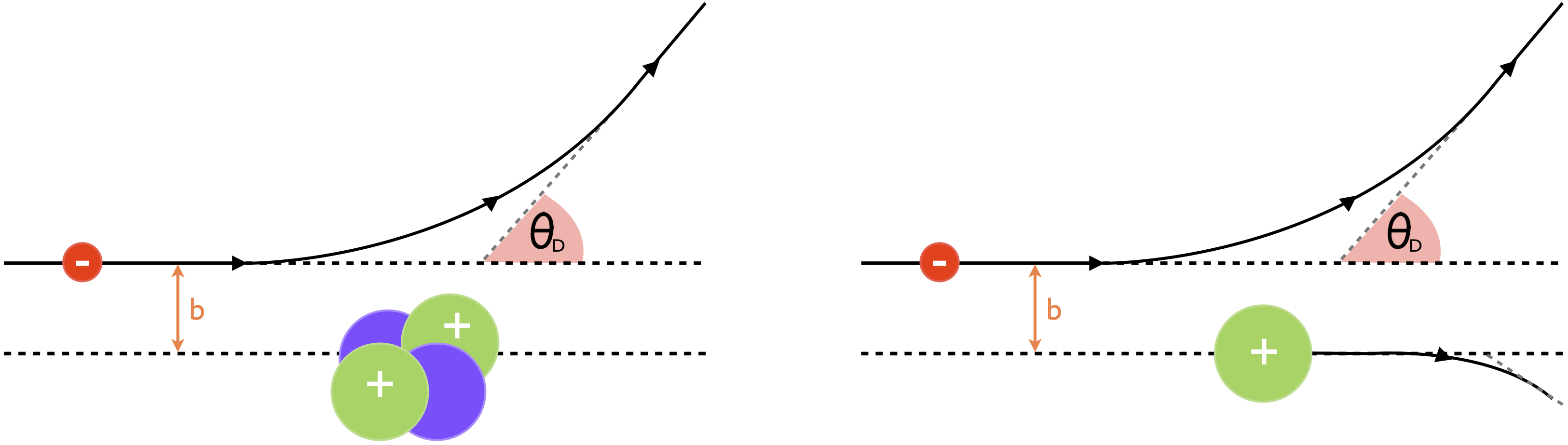}
\caption[Diagram of a Coulomb collision between an electron and an ion.]{{\it Left:} Electron deflected by a heavy ion in a Lorentz collisional model. {\it Right:} In general, both particles are deflected during a Coulomb collision, and momentum and energy are transferred.}
\label{cc}
\end{figure*}

In a fully or partially ionised plasma such as the solar corona or chromosphere, electrons and ions will interact by the Coulomb electrostatic force, via `Coulomb collisions'. When an electron passes close to an ion or another electron, it is deflected by some angle $\theta_{D}$ due to the Coulomb electric field of the ion. This is shown in Figure \ref{cc}.  In the simplest model describing Coulomb collisions, an electron moves through a background plasma of heavy, stationary ions. This is known as a \textit{Lorentz model}. The background electrons required for neutrality in the plasma are neglected, since the Lorentz model assumes that the ion atomic number $Z$ is large, meaning that the electron-ion collisions (e-i) have a dominant effect over the electron-electron (e-e) collisions. The cross section $\sigma_{R}$ for the small angle scatter of a moving electron due to the Coulomb field of a heavy, stationary ion can be given by the Rutherford formula \cite[cf.,][]{1981phki.book.....L}:
\begin{equation}\label{ruth_cs}
\sigma_{R}=\frac{4\pi Ze^{2}}{m_{e}^{2}v_{e}^{4}}\int_{b_{min}}^{b^{max}}\frac{db}{b}
\end{equation}
where $e$ [esu] is the charge of an electron, $m_{e}$ [g] is the mass of the electron and $v_{e}$ [cm s$^{-1}$] is the total electron speed. The encounter is characterised by $b$ [cm], the impact parameter; the expected closest distance of approach between the electron and ion, had the electron not been deflected during their encounter (as shown in Figure \ref{cc}). The integral
\begin{equation}\label{clog}
\int_{b_{min}}^{b^{max}}\frac{db}{b}=\ln\frac{\lambda_{D}}{b_{min}}=\ln\Lambda
\end{equation}
is defined as the Coulomb logarithm $\ln\Lambda$. The role of the Coulomb logarithm is to take into account the total effect of all deflections with different impact parameters $b$ ranging from a minimum $b_{min}$ to $b_{max}=\lambda_{D}$, where $\lambda_{D}$ is the Debye length. Within a flaring solar corona, where the plasma is fully ionised, values of $\ln\Lambda\sim20$ are often used. An electron undergoing many deflections within a field of ions will experience a \textit{collisional drag force}, causing an electron to lose momentum in its direction of travel, and this is transferred to the components of momentum perpendicular to the direction of travel. The time $\tau_{0}$ (or equivalently the frequency $\nu_{0}$) it takes for all particle momentum to be lost in the direction of travel, is given by,
\begin{equation}\label{lorentz_nu_cgs}
\centering
\tau_{0}^{-1}=\nu_{0}=n_{i}v_{e}\sigma=\frac{4\pi n_{e}Ze^{4}\ln\Lambda}{m_{e}^{2}v_{e}^{3}}=\frac{\Gamma}{v^{3}},\;\;\;\;\;\;\;\;\;{\rm where}\;\;\;\;\;\;\;\;\;\Gamma=\frac{4\pi n_{e}Ze^{4}\ln\Lambda}{m_{e}^{2}}
\end{equation}
and $n_{e},n_{i}$ are the electron and ion number densities [cm$^{-3}$]. This is known as the \textit{Lorentz collisional time or frequency}. In a Lorentz model, where the background particles are stationary, there is no {\it exchange of energy} between the electrons and the ions.

In solar flare conditions, collisions are not fully described by the Lorentz model, where there will be e-i, e-e and i-i collisions. Electrons and ions will be in motion and {\it exchange energy} during an interaction.  In this case, there are two main timescales to consider:
\begin{enumerate}
\item{the momentum loss time $\tau^{p}$, and}
\item{the energy exchange time $\tau^{E}$.}
\end{enumerate}
Assuming, during a solar flare, the background distribution of particles are Maxwellian in form, and in thermal equilibrium, then 1. describes the time it takes for a particle distribution to isotropize in angle with the thermal background, and 2. describes the time required for a particle distribution to form an energy equilibrium with the thermal background. Each timescale is slightly different depending on the particle species involved in the collision.
The timescales for {\it energy exchange} can be related by,
\begin{equation}\label{timescales}
\tau_{ee}^{E}:\tau_{ii}^{E}:\tau_{ei}^{E}\sim1:\left(\frac{m_i}{m_e}\right)^{1/2}:\left(\frac{m_{i}}{m_{e}}\right)
\end{equation}
showing that the quickest equilibrium time for, and hence the largest change in energy occurs during, an $e-e$ interaction \citep{1981phki.book.....L} . For $e-e$ collisions, the energy loss is given by,
\begin{equation}\label{high_energy_loss}
\frac{dE}{dt}=-\frac{E}{\tau_{ee}^{E}}=-\frac{2E}{\tau_{0}}=-\frac{2E\Gamma}{v_{e}^{3}}=-\frac{Kn_{e}}{E}v_{e}
\end{equation}
where $K=\Gamma m_{e}^{2}/2n_{e}$. Equation \ref{high_energy_loss} is often used to describe collisions in solar flare physics vis a {\it collisional thick target model} \cite[e.g.,][]{1971SoPh...18..489B,1972SvA....16..273S}. The loss in electron energy over a distance along $z$ from an initial energy $E_{0}$ is then found to be,
\begin{equation}\label{energy_loss_2}
E^{2}=E_{0}^{2}-2K\int_{0}^{z}n(z^{'})dz^{'}\;\;\;\; {\rm where\;the\;column\;density\;\;}\;N(z)=\int_{0}^{z}n(z^{'})dz^{'}
\end{equation}
Assuming the density of the target is constant such that $N(z)=n_{0}z$, then a typical chromospheric density of $n_{0}=10^{13}$ cm$^{-3}$ would collisionally stop a 30 keV electron over a distance $\Delta z\sim0.5\arcsec$. Within a target density of $n_{0}=10^{11}$ cm$^{-3}$, which is the number density of a dense corona (Chapters \ref{ref:Chapter2} and \ref{ref:Chapter3}), then a 30 keV electron would lose all energy over a distance $\Delta z\sim47\arcsec$. If the target is not fully ionised \citep[e.g.,][]{1978ApJ...224..241E} then the value of the Coulomb logarithm is decreased. Often values of $\ln\Lambda\sim7$ are used in the chromosphere to account for the presence of atoms in cooler chromospheric regions. Equations \ref{high_energy_loss} and \ref{energy_loss_2} only describe the energy loss of an electron in the high energy limit, that is when $E>>E_{th}$, where $E_{th}$ is the average thermal energy of the background plasma. The energy variation of electrons close to the average thermal energy of a background plasma is discussed in Chapter \ref{ref:Chapter2}.

\section{Solar flare X-rays: bremsstrahlung}
During a Coulomb collision, on average only a very small fraction of the energy lost by an accelerated electron is radiated as a photon. The radiation that is emitted is termed bremsstrahlung and means ``braking radiation". Although other emission mechanisms may contribute in the corona, such as free-bound emission \citep[e.g.,][]{1970MNRAS.151..141C,2010A&A...515C...1B} from the recombination of an ion and electron for example, overall bremsstrahlung is the most important emission mechanism for the production of X-rays during a solar flare \citep{1967SvA....11..258K} and is produced by both electron-ion and electron-electron Coulomb collisions \citep{1975SoPh...45..453H,2007ApJ...670..857K}, in the solar corona and chromosphere. Below $\sim300$ keV, the bulk of solar flare bremsstrahlung emission comes from electron-ion interactions.

\subsection{Bremsstrahlung produced by a single accelerated electron}
In the simplest situation, where a single electron is moving at a non-relativistic velocity, the total power $P_{rad}$ [erg s$^{-1}$] radiated by the accelerated electron is given by Larmor's formula
\begin{equation}\label{Larmor_cgs}
P_{rad}=\left(\frac{dE}{dt}\right)_{rad}=\frac{2e^{2}|{\dot{\bf v}}|^{2}}{3c^{3}}
\end{equation}
where $e$ [esu] is the electron charge, $\dot{\bf v}$ is the electron acceleration [cm s$^{-2}$] and $c$ [cm s$^{-1}$] is the speed of light. Larmor's formula gives the radiation loss rate in the frame of the electron. The total energy per unit frequency $\frac{dE}{d\omega}$ emitted the entire time a single charge is accelerated can be found via Parseval's theorem \citep[cf.,][]{1981cup..book.....L},
\begin{equation}\label{longair}
\int_{-\infty}^{\infty}\frac{dE}{dt}dt
=\int_{-\infty}^{\infty}\frac{2e^{2}}{3c^{3}}|{\bf \dot{v}(\omega)}|^{2}d\omega
=2\int_{0}^{\infty}\frac{2e^{2}}{3c^{3}}|{\bf \dot{v}(\omega)}|^{2}d\omega
\end{equation}
giving,
\begin{equation}\label{I_2}
I(\omega)=\frac{4e^{2}}{3c^{3}}|{\bf \dot{v}(\omega)}|^{2},
\end{equation}
the frequency spectrum of a single accelerated charge.

\subsection{Bremsstrahlung X-rays from a solar flare}
In solar flare physics, the general form of the total angle-averaged X-ray distribution $I$ [photons cm$^{-2}$ s$^{-1}$ keV$^{-1}$] produced by an electron flux density [electrons cm$^{-2}$ s$^{-1}$ keV$^{-1}$] undergoing Coulomb collisions in the corona or chromosphere is given by,
\begin{equation}\label{I_1}
I(\epsilon)=\frac{1}{4\pi R^{2}}\int_{\epsilon}^{\infty}\int_{V}n({\rm\textbf{r}})F(E,{\rm\textbf{r}})\sigma(\epsilon,E)dEd^{3}{\rm\textbf{r}},
\end{equation}
where $R=1$ AU is the Sun-Earth distance, $\epsilon$ [keV] is the photon energy, $V$ [cm$^{3}$] is the emitting volume, $n$ [cm$^{-3}$] is the number density of the emitting region, ${\rm\textbf{r}}$ is the position on the Sun and $\sigma$ [cm$^{2}$] is the angle-averaged bremsstrahlung cross section.
Often for clarity or simplicity the electron-ion (e-i) bremsstrahlung cross section is approximated by the Kramers formula $\sigma=Q_{K}$ or can be better estimated by the non relativistic Bethe-Heitler cross section $\sigma=Q_{BH}$. Both are given by
\begin{equation}\label{K_cross}
Q_{K}(\epsilon,E)=\frac{Z^{2}8\alpha}{m_{e}c^{2}r_{0}^{2}\epsilon E} ,\,\,\,\,\,\,\,\,\, Q_{BH}(\epsilon,E)=Q_{K}\ln\left(\frac{1+\sqrt{1-\epsilon/E}}{1-\sqrt{1-\epsilon/E}}\right),
\end{equation}
where $\alpha\sim1/137$ is the fine structure constant and $r_{0}=2.82\times10^{-13}$ cm is the classical electron radius \citep[cf.][]{2011SSRv..159..301K}. Equation \ref{I_1} is an inverse problem; from observation the ultimate goal is to deduce the electron flux distribution $F$ from a measured photon flux $I$. The form of the inferred electron distribution is dependent upon the form of the bremsstrahlung cross section. Both the Kramers and non-relativistic Bethe-Heitler forms are often used as they allow the analytical deduction of $F(E,{\rm\textbf{r}})$ \cite[e.g.,][]{1971SoPh...18..489B,1977ApJS...35..419R,2002SoPh..210..373B}. However, \cite{1997A&A...326..417H} noted that relativistic changes to the e-i bremsstrahlung cross section should be taken into account above even $\sim30$ keV, and found an analytical form up to semi-relativistic energies. The full form of the angle-averaged e-i bremsstrahlung cross section is shown in \cite{1959RvMP...31..920K}, formula 3BN, with a more useable form for numerical simulation given by \cite{1997A&A...326..417H}.
Measuring the X-ray photon spectrum alone without any spatial information, implies that Equation \ref{I_1} can be spatially integrated to give,
\begin{equation}\label{I_2a}
I(\epsilon)=\frac{1}{4\pi R^{2}}\int_{\epsilon}^{\infty}\left[\bar{n}V\bar{F(E)}\right]\sigma(\epsilon,E)dE
\end{equation}
where $\left[\bar{n}V\bar{F(E)}\right]$ is known as the mean electron flux spectrum \citep{2003ApJ...595L.115B}.
As well as being dependent upon the X-ray energy $\epsilon$, initial electron energy $E$ and the atomic number of the target $Z$, the bremsstrahlung cross section $\sigma$ is also angular and polarization dependent. The angular dependent
\begin{figure}
 % \begin{minipage}[c]{0.60\textwidth}
    \includegraphics[width=7cm]{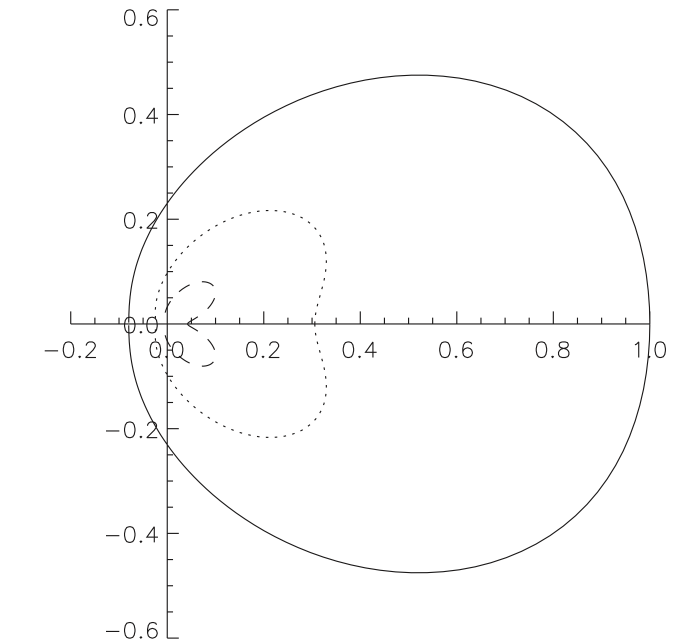}
    \hspace{2 cm}
    \includegraphics[width=5cm]{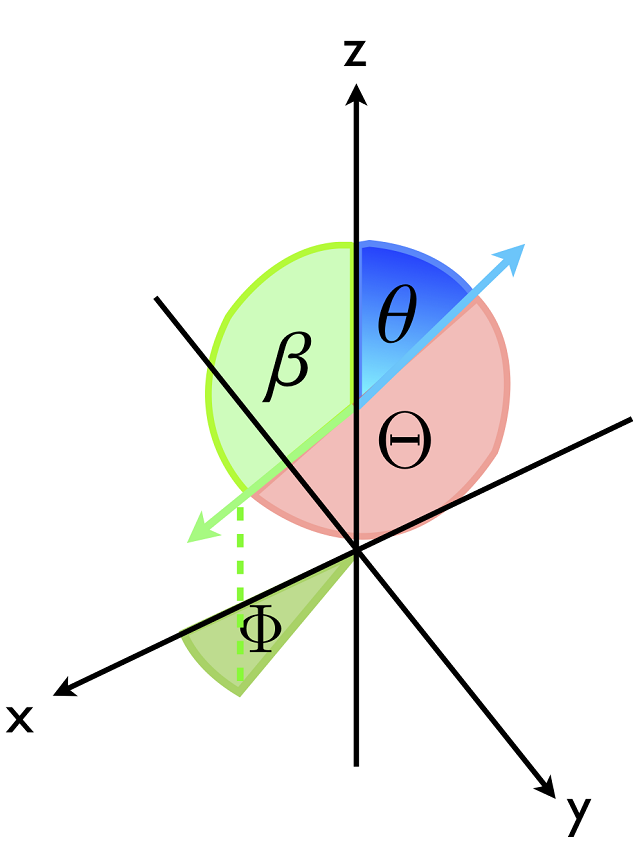}
  %\end{minipage}\hfill
  %\begin{minipage}[c]{0.40\textwidth}
\caption[A polar diagram of the angular dependent e-i bremsstrahlung cross section.]{{\it Left:} Figure taken from \cite{Massoneetal2004}. Angle dependent e-i bremsstrahlung cross section for a 100 keV electron and the emission of a 30 keV (solid), 50 keV (dotted) and 80 keV (dashed) photon. The radial distance gives the size of the cross section while the angle from the x-axis is the angle between the photon emission and the incoming electron. {\it Right:} Diagram showing the X-ray emission angle $\theta$, the electron angle to the guiding field $\beta$, the electron azimuthal angle $\phi$ and the angle between $\theta$ and $\beta$, $\Theta$.}
\label{massone_figures}
%\end{minipage}
\end{figure}
e-i cross section summed over all polarization states is given by \cite{GlucksternHull1953}.  A polar diagram showing the form of the polarization integrated angular dependent e-i bremsstrahlung cross section is shown in Figure \ref{massone_figures} (left) and is taken from \cite{Massoneetal2004}. It is plotted for a $100$ keV electron emitting either a $30$, $50$ or $80$ keV photon. This figure shows the e-i cross section is larger for lower energy photons. It is more likely a low energy photon will be emitted during an interaction and the direction of emission is more likely to peak away from the direction of the incoming electron as the emitted photon energy increases. Using an angle dependent e-i bremsstrahlung cross section $\sigma$, the angular and energy dependent photon flux distribution $I(\epsilon,\theta)$ can be given by,
\begin{equation}\label{I_3}
I(\epsilon,\theta)\propto\int_{E=\epsilon}^{\infty}\int_{\phi=0}^{2\pi}\int_{\beta=0}^{\pi}F(E,\beta)\sigma(E,\epsilon,\beta,\theta,\phi)\sin\beta d\beta d\phi dE,
\end{equation}
where $\theta$ is the photon emission angle, $\beta$ is the electron angle to the guiding field and $\phi$ is the electron azimuthal angle in the plane perpendicular to the guiding field. Each angle is related by,
\begin{equation}
\cos\Theta=\cos\theta\cos\beta+\sin\theta\sin\beta\cos\Phi.
\label{cosTheta_intro}
\end{equation}
This is further described in Chapter \ref{ref:Chapter6}, and each angle can be seen pictorially in Figure \ref{massone_figures} (right).
Depending upon the disk location (viewing angle) of the X-ray source and the electron anisotropy, \cite{Massoneetal2004} found that using the angle-averaged, instead of the angle-dependent e-i bremsstrahlung cross section can cause significant changes to the inferred electron flux distribution, particularly above 50 keV, leading to suspect inferred mean electron spectra and total injected energies.
\cite{GlucksternHull1953} gives the polarization dependent parallel and perpendicular components of the angular dependent e-i bremsstrahlung cross section. These are used in Chapter \ref{ref:Chapter6} of this thesis and hence are further discussed there. The total polarization and angular dependent e-i bremsstrahlung cross section $\sigma$ is then the sum of the components of the cross section parallel $\sigma_{||}$ and perpendicular $\sigma_{\perp}$ to the plane of X-ray emission
\begin{equation}\label{Q_par_perp}
\sigma=\sigma_{||}+\sigma_{\perp}.
\end{equation}
Importantly, the angular distribution of the X-ray and hence electron distribution is positively correlated with the X-ray polarization. This will be discussed further in Section \ref{intro_pol} and Chapter \ref{ref:Chapter6}.

\subsection{Electron-ion versus electron-electron bremsstrahlung}\label{eeb}
As mentioned, bremsstrahlung X-rays can be produced by both electron-ion and electron-electron Coulomb collisions. In the previous section, only e-i bremsstrahlung was considered, as most solar flare problems only need to account for the electron-ion collisions. Below $\sim300$ keV the e-e bremsstrahlung cross section decreases rapidly and the emission is negligible compared to that of e-i bremsstrahlung \citep{1975SoPh...45..453H,2007ApJ...670..857K}. However, \cite{2007ApJ...670..857K} found that the presence of e-e bremsstrahlung should not be ignored above $\sim300$ keV. For a given X-ray distribution, the presence of an e-e bremsstrahlung component requires a steeper electron spectrum at higher energies.
Unlike e-i bremsstrahlung, e-e bremsstrahlung cannot produce X-rays of all energies up to the energy of the emitting electron. The maximum e-e bremsstrahlung energy is bounded by the angle between the direction of the incoming electron and the emitted X-ray. In general the bremsstrahlung cross section should be a combination of both e-i and e-e interactions \citep{1975SoPh...45..453H,1998SoPh..178..341H,2007ApJ...670..857K}, given by
\begin{equation}\label{total_brem_cs}
\sigma(\epsilon,E)=Z^{2}\sigma_{ei}(\epsilon,E)+Z\sigma_{ee}(\epsilon,E),
\end{equation}
where $Z$ is the effective atomic number of a plasma or quasi-neutral target. In Chapter \ref{ref:Chapter6}, only the angular and polarization dependent e-i bremsstrahlung cross section is used, since the majority of the work in Chapter \ref{ref:Chapter6} studies X-rays in the range of 20-50 keV, where the emission due to e-e interactions is negligible.

\subsection{Thermal bremsstrahlung}\label{tb}
Thermal bremsstrahlung is the term given to the production of bremsstrahlung X-rays by a distribution of electrons in thermal equilibrium. Often the spectrum of lower energy X-rays (below $\sim30$ keV) during a solar flare has an exponential form, representative of the emission from a thermal distribution of particles, \cite[e.g., first suggested via observation by][]{1966JGR....71.3611C}. Although realistically, the flaring region will have a temperature distribution, it is often useful to fit this part of the X-ray spectrum with a single isothermal function (see Figure \ref{x_gam_ray_spect}), in order to obtain an average temperature $T$ [K] and emission measure $EM=n^{2}V$ [cm$^{-3}$], where $V({\rm\textbf{r}})$ [cm$^{3}$] is the volume of the emitting flare region. The electron flux density $F(E,{\rm\textbf{r}})$ of such a distribution and the resulting photon flux distribution $I(\epsilon,{\rm\textbf{r}})$ can be given by,

\begin{equation}\label{F_th}
F(E,{\rm\textbf{r}})=\frac{2^{3/2}}{(\pi m_{e}^{1/2})}\frac{n({\rm\textbf{r}})E}{(k_{B}T^{3/2}({\rm\textbf{r}}))}\exp\left(\frac{-E}{k_{B}T({\rm\textbf{r}})}\right)\;\;\rightarrow\;\;I(\epsilon,{\rm\textbf{r}})\propto\frac{n^{2}({\rm\textbf{r}})V({\rm\textbf{r}})}{\epsilon T^{1/2}({\rm\textbf{r}})}\exp\left(\frac{-\epsilon}{k_{B}T({\rm\textbf{r}})}\right)
\end{equation}
where \textbf{r} is the position on the Sun.

\subsection{Non-thermal bremsstrahlung}\label{ntb}
Often in X-ray solar flare physics, the higher energies of the X-ray distribution have a form that can be approximated by either a single or double power law (see Figure \ref{x_gam_ray_spect}) \cite[e.g.,][]{1968JGR....73..434C,1981ApJ...251L.109L,1985SoPh..100..465D}. Hence, this means that the parent electron energy distribution can also be approximated by a power law,
\begin{equation}\label{ntb1}
F(E)\propto E^{-\delta}\;\;\;\;\;\rightarrow\;\;\;\;\;I(E)\propto \epsilon^{-\gamma}.
\end{equation}
In a collisional thick target model, the electrons lose all of their kinetic energy in the target region. In general, the spectral index of the target electron spectrum differs from the injected electron spectral index by $\delta_{T}\sim\delta-2$ and the spectral index of the resulting X-ray distribution is given by,  $\gamma_{thick}=\delta+1$ \cite[e.g.,][]{1971SoPh...18..489B}. The chromosphere and also the corona, depending on its density, can act as a thick target during a solar flare. In a thin target model, electrons do not lose all of their energy as they move through a thin target region and the resulting spectral index of the photon distribution is given by $\gamma_{thin}=\delta-1$. A low density corona may act as a thin target. The above approximations are for non-relativistic e-i interactions. The relationship between the spectral index of the electron distribution $\delta$ and the spectral index of the X-ray distribution $\gamma$ flattens if the X-ray emission is due to relativistic e-i interactions or e-e collisions. For example, e-e interactions in a thick target model produce an X-ray spectrum of the form $\delta\sim\gamma$ \citep{1989A&A...218..330H,2007ApJ...670..857K}.

\section{Solar flare X-rays: photon interaction processes}
In a dense plasma or neutral atmosphere X-ray photons can interact with free or bound electrons by Compton scattering.
The corona and chromosphere are mainly optically thin to X-ray and gamma-ray energies. However, this is not true in the high densities of the photosphere ($\sim10^{17}$ cm$^{-3}$), as noted in Section \ref{intro_intro}. Compton scattering in the photosphere is the cause of photospheric albedo which is discussed in Section \ref{intro_albedo} of this chapter and in Chapters \ref{ref:Chapter5} and \ref{ref:Chapter6} of this thesis. In the absence of energy exchange in the low energy limit below $\sim1$ keV, Compton scattering can be described by, and is equivalent to Thomson scattering.

\subsection{Thomson scattering}
\begin{figure}
\centering
\hspace{-25pt}
\includegraphics[width=7cm]{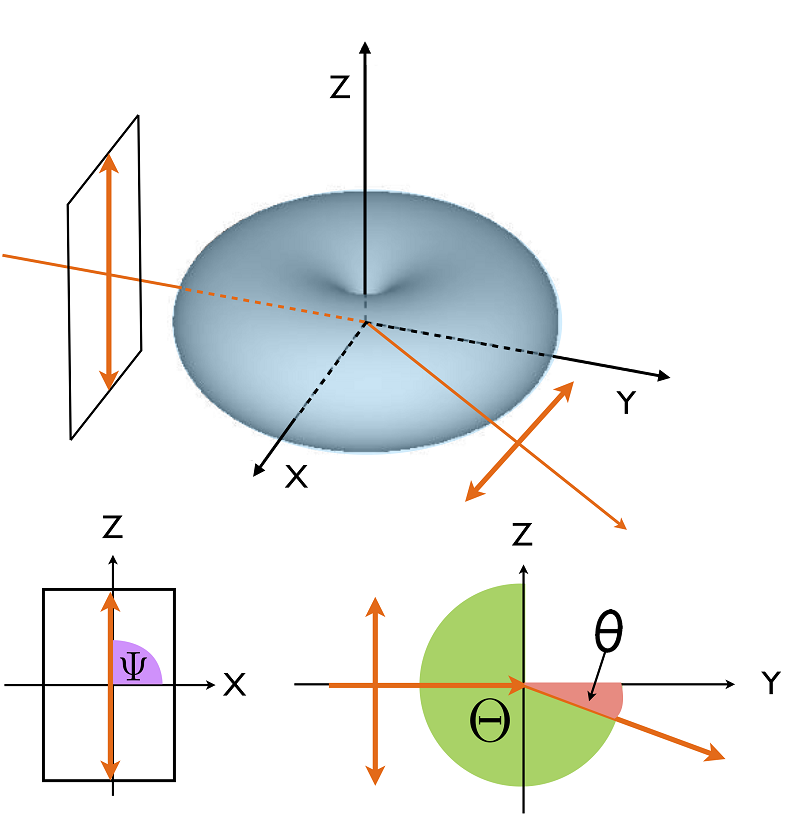}
\includegraphics[width=8cm]{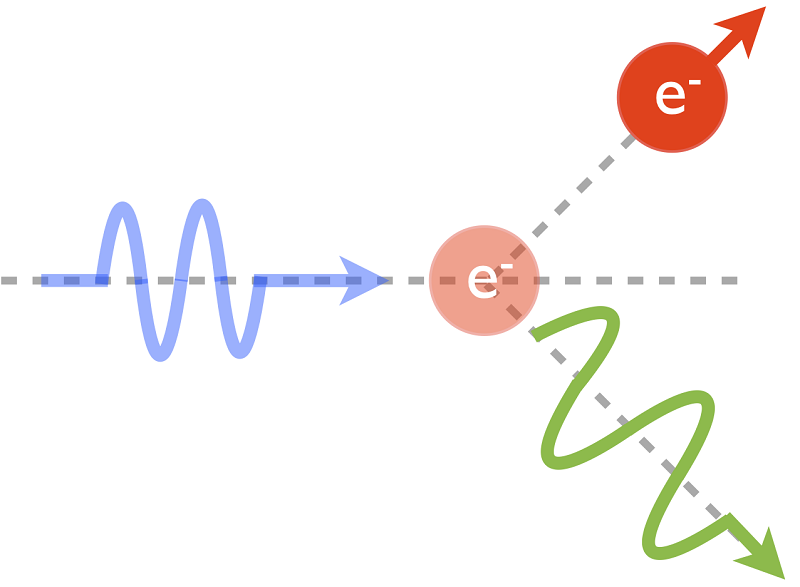}
\caption[Diagrams of Thomson and Compton scattering]{{\it Left:} Cartoon showing a polarized incident plane wave interacting with an electron, causing it to oscillate and re-radiate. The original figure was taken from \url{http://www.exul.ru/education/1/Note3b.pdf} and then adapted for this thesis. Each of the angles $\theta$ (angle between the incident and scattered radiation), $\Theta$ (angle between the direction of electron acceleration and the propagation direction of the outgoing radiation) and $\Psi$ (the direction of the incoming polarization measured from the x-axis) are shown. {\it Right:} diagram of the Compton interaction between a photon and an electron.}
\label{thom_scatter}
\end{figure}

Thomson scattering describes the interaction of an incident plane wave with an electron, and is a purely classical process \cite[cf.,][]{1962clel.book.....J,1981cup..book.....L}. The radiation field of the incident wave causes the electron to oscillate, with the direction of oscillation dependent upon the polarization of the incident wave. The oscillating electron is an accelerated charge and subsequently radiates its own radiation in a different direction; the scattered wave. For low energy interactions described by Thomson scattering, the wavelength of the incident wave is equal to wavelength of the scattered wave. The energy per unit time, $dP$ scattered into a solid angle $d\Omega$ or equivalently radiated by the electron is given by Equation \ref{dpdomega} (left) and the time-averaged energy flux of the incident electromagnetic wave (time-averaged Poynting flux) $\langle U \rangle$, [erg s$^{-1}$ cm$^{-2}$], is given by Equation \ref{dpdomega} (right),
\begin{equation}\label{dpdomega}
\frac{dP}{d\Omega}=\frac{e^{4}\mathcal{E}^{2}}{8\pi m_{e}^{2}c^{3}}\sin^{2}\Theta,\;\;\;\;\;\;\;\langle U\rangle=\frac{c}{8\pi}\mathcal{E}^{2}
\end{equation}
where $\mathcal{E}$ is the oscillating electric field of the scattered electromagnetic wave and $\Theta$ is the {\it angle between the direction of electron acceleration and the propagation direction of the outgoing radiation, not the scattering angle}.
The differential Thomson scattering cross section is the ratio of these two quantities, giving,
\begin{equation}\label{thom_dcs_pol}
\frac{d\sigma_{thom}}{d\Omega}=\left(\frac{e^{2}}{m_{e}c^{2}}\right)^{2}\sin^{2}\Theta.
\end{equation}
Equation \ref{thom_dcs_pol} describes completely polarized radiation. The angle $\Theta$ is related to the polar scattering angle $\theta$ and the direction of the incoming polarization $\Psi$ by $\cos\Theta=\sin\theta\cos\Psi$. Each of these angles is shown in Figure \ref{thom_scatter} (left).
Rearranging gives $\sin^{2}\Theta=1-\sin^{2}\theta\cos^{2}\Psi$. For unpolarized radiation, the average polarization angle is required and hence this gives $\sin^{2}\Theta=1-\sin^{2}\theta\langle\cos^{2}\Psi\rangle=1-\sin^{2}\theta/2=\left(1+\cos\theta\right)/2$. The Thomson scattering differential cross section for unpolarized incident radiation is then given by,
\begin{equation}\label{thom_dcs_unpol}
\frac{d\sigma_{thom}}{d\Omega}=\left(\frac{e^{2}}{m_{e}c^{2}}\right)^{2}\frac{1}{2}\left(1+\cos^{2}\theta\right)
\end{equation}
where $\theta$ is the scattering angle. Integrating either Equation \ref{thom_dcs_pol} or \ref{thom_dcs_unpol} over solid angle $d\Omega=2\pi\sin\theta d\theta$ gives,
\begin{equation}\label{thom_cs}
\sigma_{thom}=\left(\frac{e^{2}}{m_{e}c^{2}}\right)^{2}\frac{2\pi}{2}\int_{0}^{\pi}\left(1+\cos^{2}\theta\right)\sin\theta d\theta
=\frac{8\pi}{3}\left(\frac{e^{2}}{m_{e}c^{2}}\right)^{2},
\end{equation}
which is the total Thomson scattering cross section $\sigma_{thom}$. There is no energy exchange during a Thomson scattering since the electron does not recoil. Figure \ref{thom_scatter} (left) depicts the Thomson scattering of completely polarized radiation. Figure \ref{cs_graphs} plots the total Thomson cross section against energy and the unpolarized differential Thomson cross section against scattering angle (Equation \ref{thom_dcs_unpol}).

\subsection{Compton scattering}
In general, when a photon scatters from an electron, there is energy exchange and the energy of the outgoing photon is decreased. Arthur Compton's original result \citep{1923PhRv...21..483C} was derived from experiment and the formula was given in terms of a shift in photon wavelength, $\Delta\lambda$. This can be found easily by studying the kinematics of the collision, assuming that the incident radiation acts as a particle, that is a photon. The resulting wavelength $\lambda$ or energy $\epsilon$ of the outgoing photon can be found from,
\begin{equation}\label{cs_1}
\Delta\lambda=\lambda-\lambda_{0}=\frac{h}{m_{e}c}\left(1-\cos\theta\right),\;\;\;\;\;\;
\frac{\epsilon}{\epsilon_{0}}=\frac{1}{1+\frac{\epsilon}{m_{e}c^{2}}\left(1-\cos\theta\right)}.
\end{equation}
$\lambda_{0}$ and $\epsilon_{0}$ are the incoming photon wavelength and energy respectively, $h$ is the Planck constant and $\theta$ is the scattering angle of the outgoing photon relative to the direction of incoming photon. Equation \ref{cs_1} (left hand side) is easily converted to photon energy (right hand side) by $\epsilon=\frac{hc}{\lambda}$.
\begin{figure}
  \begin{minipage}[c]{0.55\textwidth}
    \includegraphics[width=\textwidth]{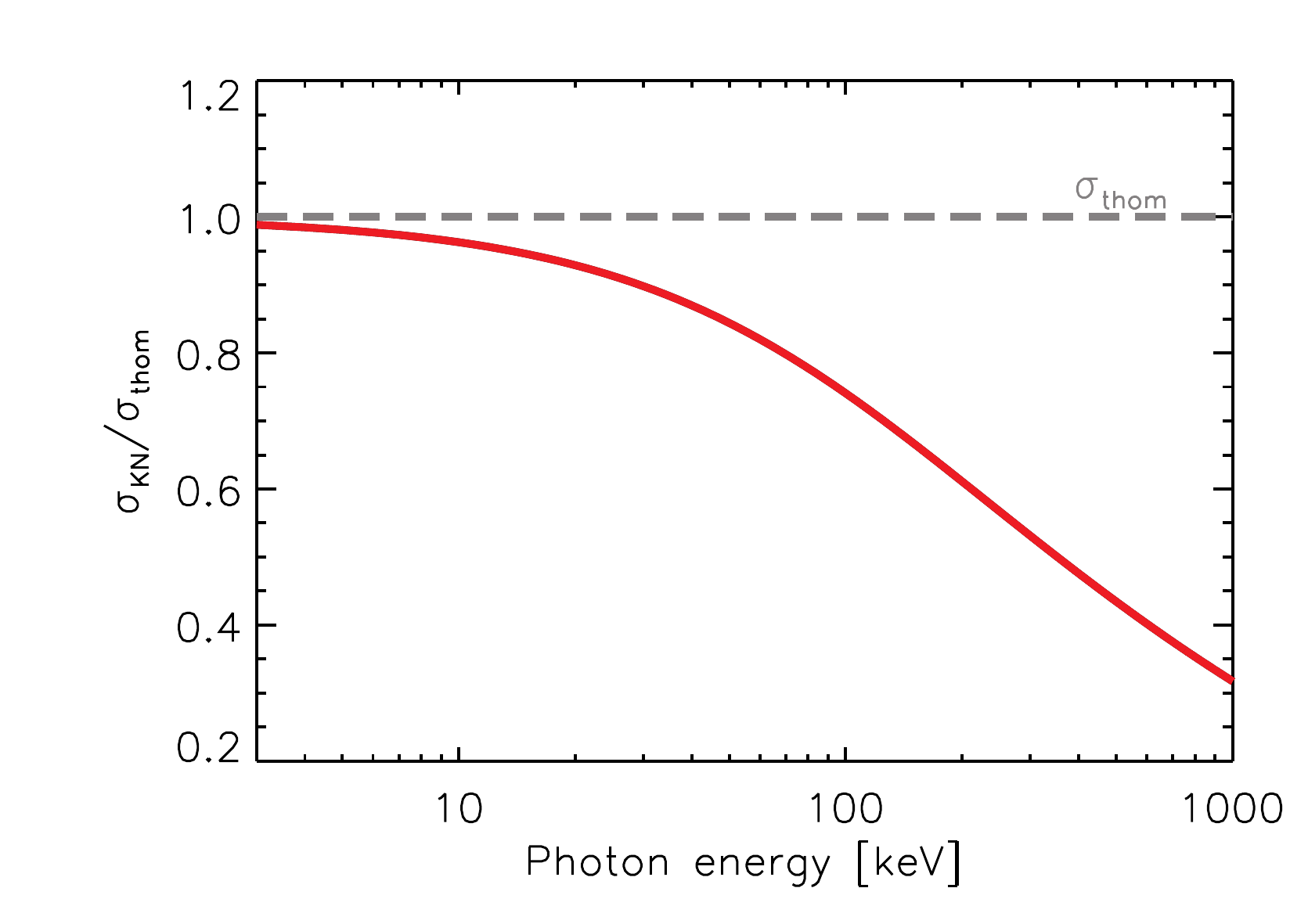}
     \includegraphics[width=\textwidth]{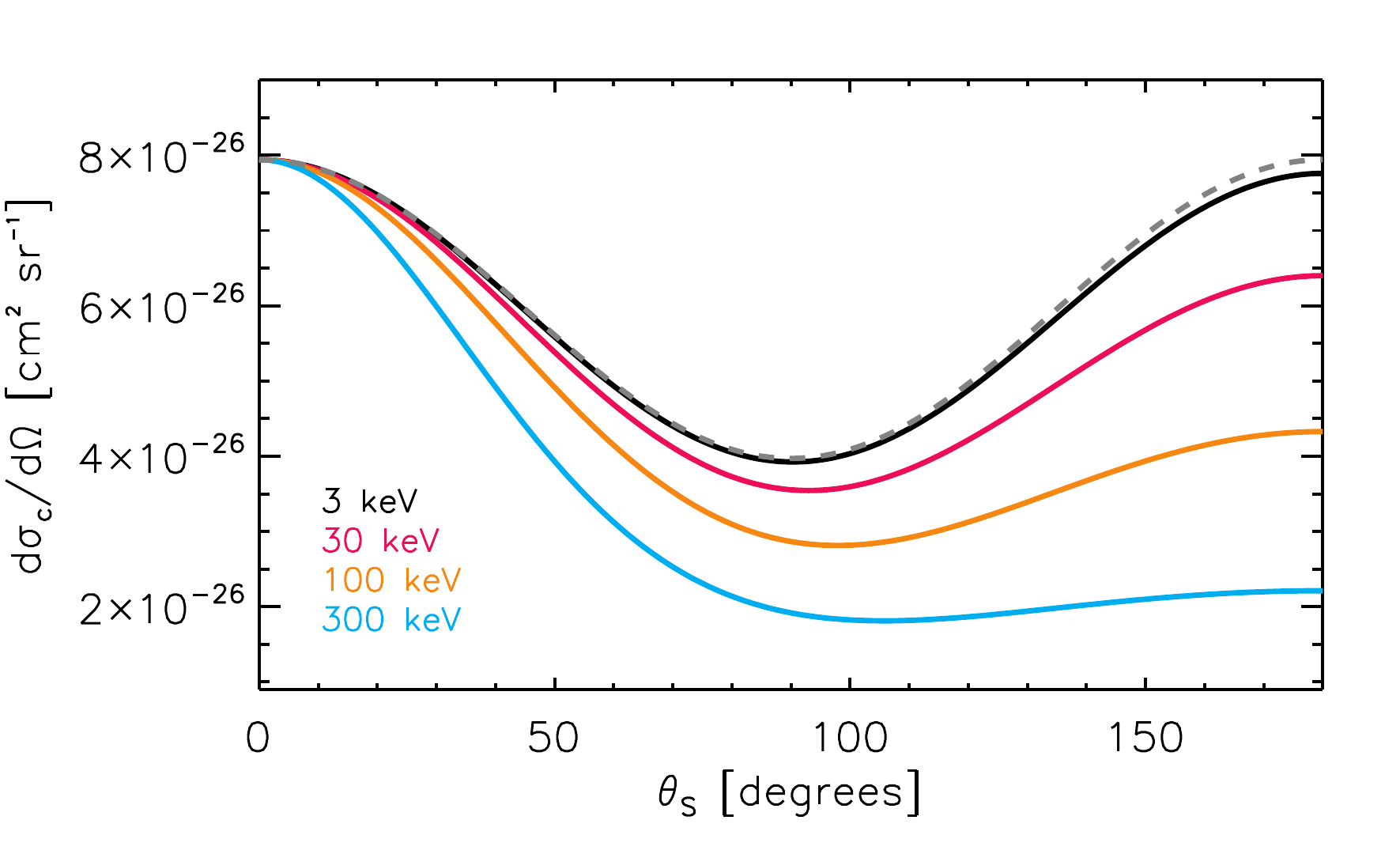}
\includegraphics[width=\textwidth]{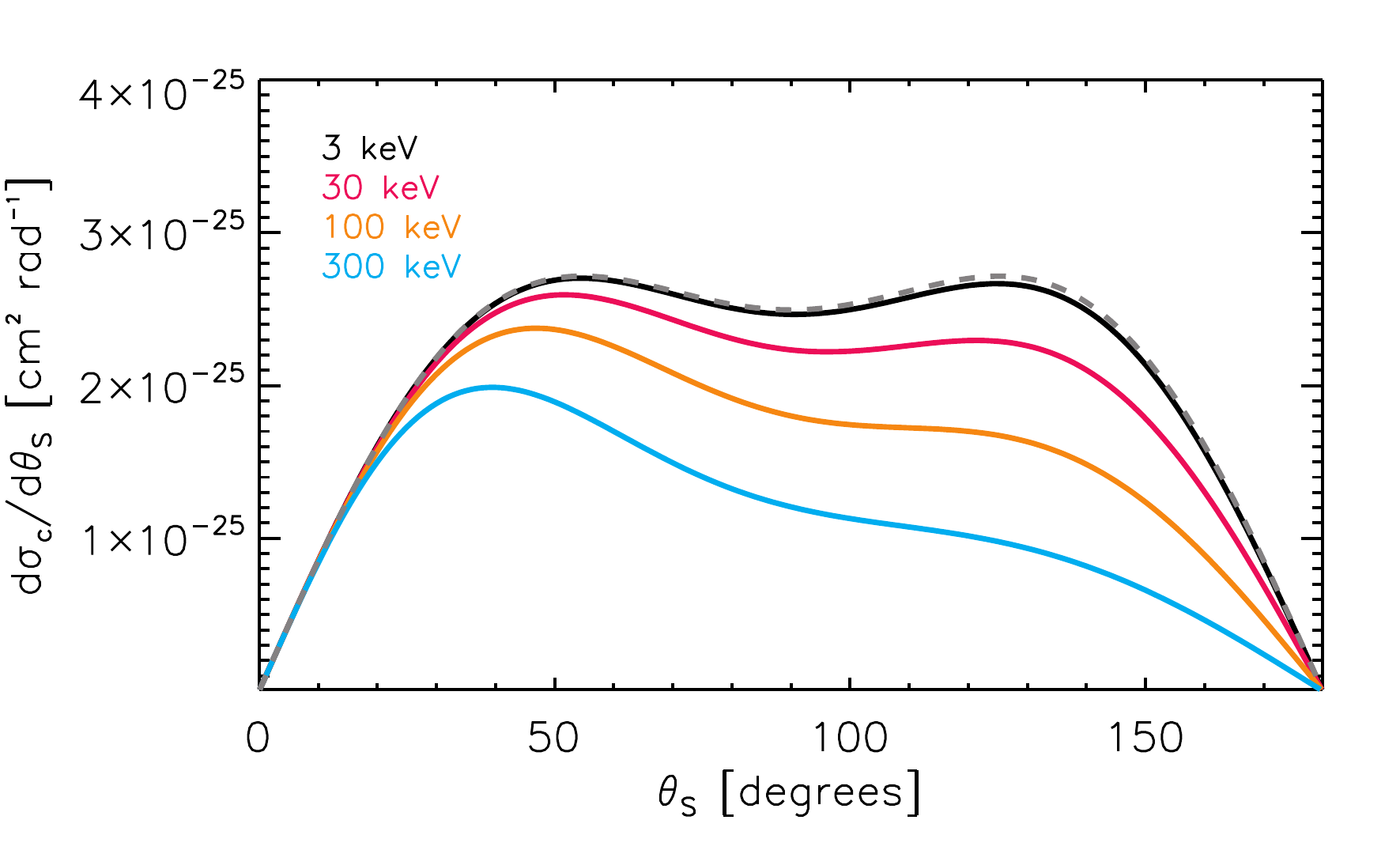}
  \end{minipage}\hfill
  \begin{minipage}[c]{0.35\textwidth}
\caption[The full and differential Thomson and Compton scattering cross sections plotted against X-ray energy and scattering angle.]{{\it Top:} A comparison of the KN $\sigma_{c}$ (red solid) and Thomson $\sigma_{thom}$ (grey dashed) scattering cross sections in units of $\sigma_{thom}$. Both cross sections only match at low energies less than $\sim1$ keV and hence the KN scattering cross section should be used to describe the Compton interaction in X-ray solar flare studies. {\it Middle:} The differential KN Compton scattering cross section $d\sigma_{c}/d\Omega$ versus scattering angle $\theta_{S}$. {\it Bottom} the azimuthal angle-averaged differential KN scattering cross section $d\sigma_{c}/d\theta_{S}$ versus scattering angle $\theta_{S}$. The case for Thomson scattering is shown by a grey dashed line in each case.}
\label{cs_graphs}
\end{minipage}
\end{figure}
The original formulation of Thomson scattering only described light as a plane wave, it did not take into account its quantum particle properties as a photon, and hence it can not account for the energy exchange between the incoming photon and the electron, which recoils during the interaction, gaining energy at the loss of the photon. A simple diagram of the Compton interaction is shown in Figure \ref{thom_scatter} (right). In order to find the Compton scattering cross section $\sigma_{c}$, modifications have to be made to the Thomson scattering cross section $\sigma_{thom}$ in order to account for the change in energy. From an approximate quantum mechanical derivation, the Compton scattering differential cross section is found to be,\\
\begin{equation}
\frac{d\sigma_{c}}{d\Omega}=\frac{d\sigma_{thom}}{d\Omega}\left(\frac{\epsilon}{\epsilon_{0}}\right)^{2}.
\end{equation}\\
A proper derivation of the differential Compton scattering cross section, fully taking into account both quantum and relativistic effects is performed in quantum electrodynamics. This gives the Klein-Nishina (KN) Compton differential scattering cross section \citep{KleinNishina1929},\\
\begin{equation}\label{comp_cs_intro}
\frac{d\sigma_{c}}{d\Omega}=\frac{1}{2}r_{0}^{2}\left(\frac{\epsilon}{\epsilon_{0}}\right)^{2}{\bigg\lgroup}\frac{\epsilon}{\epsilon_{0}}+
\frac{\epsilon_{0}}{\epsilon}-\sin^{2}\theta_{S}{\bigg\lgroup}1-Q\cos2\phi_{S}
 -U\sin2\phi_{S}{\bigg\rgroup}{\bigg\rgroup},
\end{equation}\\
where $\phi_{S}$ is the azimuthal scattering angle and $Q$ and $U$ are linear Stokes parameters used to describe linear polarization. This is discussed in detail in Chapter \ref{ref:Chapter6}. The total KN cross section against energy and the differential KN cross section against scattering angle for the completely unpolarized case (that is setting Q and U to zero in Equation \ref{comp_cs_intro}) are plotted in Figures \ref{cs_graphs}. From Figure \ref{cs_graphs}, it can be seen that Compton cross section deviates greatly from the constant Thomson cross section at high energies, decreasing due to the fact that a high energy photon is less influenced by an electron. In Figure \ref{cs_graphs}, both $d\sigma_{c}/d\Omega$ and $d\sigma_{c}/d\theta_{S}=(d\sigma_{c}/d\Omega) 2\pi\sin\theta_{S}$ are plotted. At low energies, the unpolarized $d\sigma_{c}/d\Omega$ matches that of the Thomson case and is symmetrical, with the smallest value occurring at a scattering angle of $90^{\circ}$, where the scattering angle is measured from the direction of the incident photon. As the incident photon energy increases, the scattered radiation becomes more and more forward beamed, it is scattered at a smaller angle and there is a smaller change in photon energy. Removing the azimuthal dependency and plotting $d\sigma_{c}/d\theta_{S}$ shows that the photons are more likely to be scattered between $50^{\circ}$ and $130^{\circ}$ at low energies, and the importance of this is discussed in Chapter \ref{ref:Chapter5}. At higher energies, the maximum scattering angle falls to a lower $\theta_{S}$ due to the forward beaming.

\section{Solar flare X-rays: observations}
In this section, the main X-ray observables during a solar flare: the X-ray temporal evolution, the X-ray spectrum, the X-ray source location and spatial properties, and finally the X-ray polarization will be discussed. The space-borne satellite, {\it The Ramaty High Energy Solar Spectroscopic Imager (RHESSI\,)} \citep{2002SoPh..210....3L} is currently used for high resolution imaging spectroscopy of solar flare X-rays from $3$ keV. {\it RHESSI} is discussed in Section \ref{intro_rhessi}.

\subsection{X-ray temporal evolution of a solar flare}\label{intro_temp}
\begin{figure}\centering
\includegraphics[width=8cm]{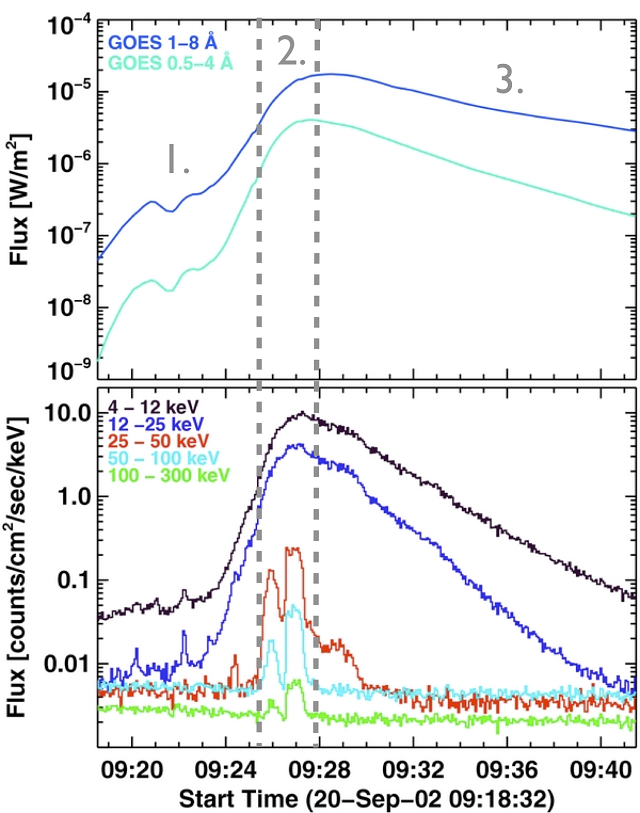}
\caption[{\it GOES} and {\it RHESSI} lightcurves for a flare that occurred on the 20th September 2002.]{Figure taken from \cite{2011ApJ...733...37F} showing both {\it GOES} and {\it RHESSI} lightcurves during a solar flare occurring on the 20th September 2002. {\it Stages 1, 2 and 3} are labelled on the figure and described in \ref{intro_temp}.}
\label{intro_lc}
\end{figure}

The duration of a solar flare can usually be separated into three stages: the rise or precursor stage ({\it stage 1}), the impulsive stage ({\it stage 2}) and lastly, the decay stage ({\it stage 3}).  {\it RHESSI} and {\it Geostationary Operational Earth Satellites (GOES)} light curves showing the typical temporal evolution of soft X-rays (SXRs) $\leq10$ keV and hard X-rays (HXRs) $\geq10$ keV are shown in Figure \ref{intro_lc}. Each stage is labelled on the figure. During {\it stage 1}, there is usually a slow, gradual increase in SXRs and lower energy HXRs ($\sim 1-20$ keV for the flare shown in Figure \ref{intro_lc}), where the coronal plasma is being heated to tens of mega-Kelvin.  During {\it stage 2}, there is usually a sudden, fast increase in HXRs above 20 keV, lasting for only $\sim$ 1 or 2 minutes, where a large number of electrons are accelerated to high non-thermal energies. The SXR and lower energy HXR emission usually peaks after the impulsive HXR emission, and then starts to gradually decrease. This denotes {\it stage 3}. The overall time and the length of each stage is individual for each flare; for example, the SXR emission may take hours to decrease during {\it stage 3}, while for other flares it decays over a much quicker period.
\subsection{The X-ray and gamma-ray solar flare energy spectrum}\label{intro_spect}
\begin{figure}\centering
\includegraphics[width=15cm]{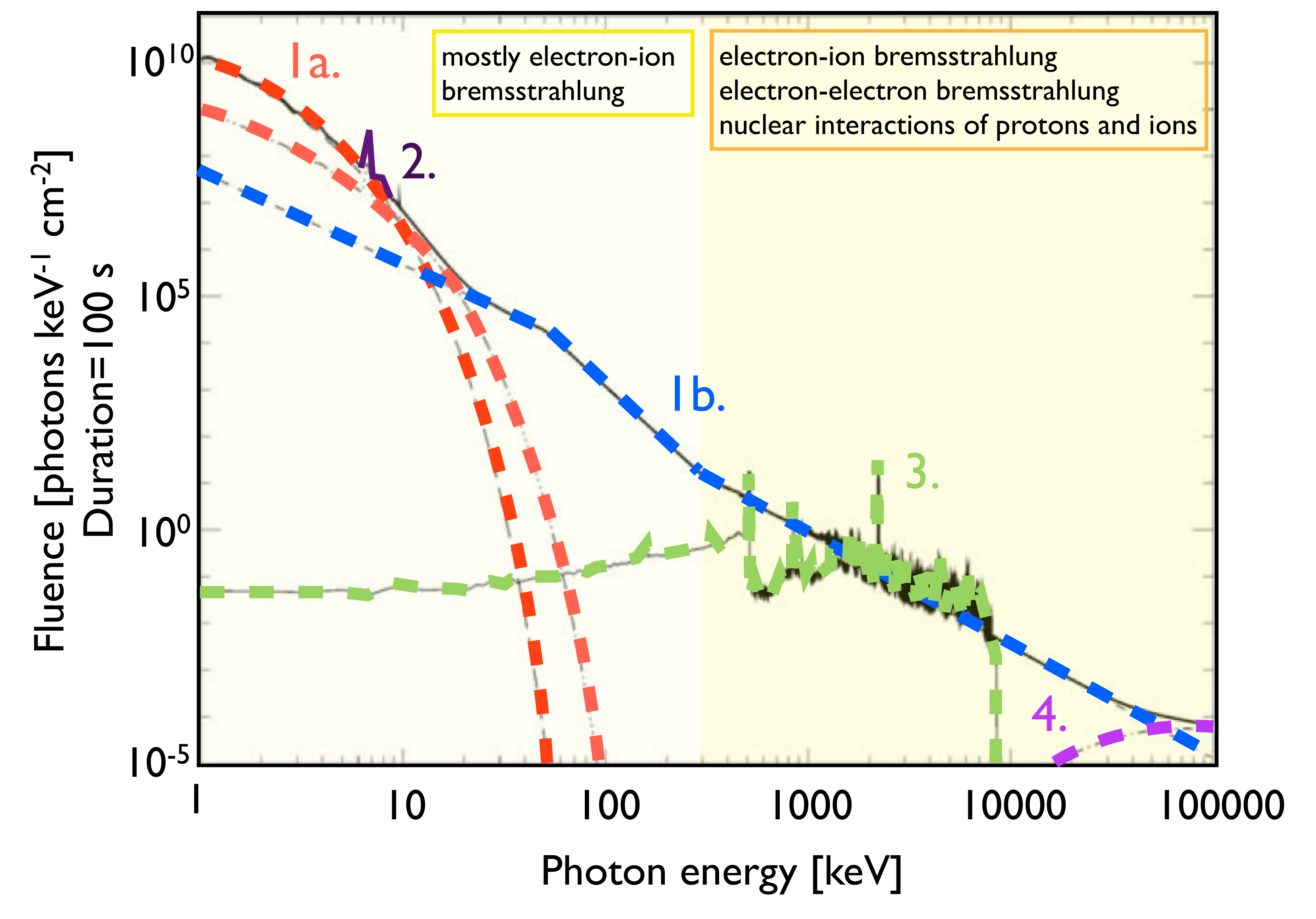}
\caption[An example X-ray and gamma ray solar flare spectrum.]{This figure was taken and then adapted from \cite{2002SoPh..210....3L}. A typical composite X-ray and gamma-ray spectrum from a solar flare. The X-ray spectrum is dominated by both thermal (1a. red) and non-thermal (1b. blue) bremsstrahlung emission. Prominent line emissions at 6.7 keV and 8.1 keV can often be observed (2. purple) due to highly ionised Fe and Ni in the corona. Above $\sim500$ keV, the gamma-ray spectrum is more complex, with continuum emission from bremsstrahlung, line emissions from nuclear interactions (3. green) and emission from pion decay (4. violet). Each process in described in \ref{intro_spect}.}
\label{x_gam_ray_spect}
\end{figure}
A general example of an expected solar flare X-ray and gamma-ray spectrum is shown in Figure \ref{x_gam_ray_spect}. The continuum emission in the spectrum from 1 keV onwards to 100 MeV, is predominantly bremsstrahlung emission produced by mostly e-i Coulomb collisions below $\sim400$ keV (see Section \ref{eeb}) and both e-i and e-e Coulomb collisions at higher energies. The spectrum is usually exponential in form at lower energies below $\sim30$ keV, suggesting the emission comes from collisions within a hot, thermal plasma. Spectral fits often suggest temperatures of $18-30$ MK and the use of imaging spectroscopy with instruments such as {\it RHESSI} shows that the majority of the thermal emission originates from the corona, possibly close to point of energy release during the flare. In this range, instruments like {\it RHESSI} can often see two line emissions: one at 6.7 keV due to highly ionised iron (Fe) and one at 8.1 keV due to highly ionised Fe and nickel (Ni) in the solar corona. Both the peaks and widths of these lines are highly dependent on the temperature and iron abundance of the corona and act as a useful diagnostic tool \citep{2006ApJ...647.1480P,2012ApJ...748...52P}. At higher energies $\geq 25$ keV, a non-thermal bremsstrahlung spectrum can be fitted by either one or two power laws, suggesting the emission comes from high energy particles accelerated out of the background thermal distribution. In general, but not always (see the following sections), the bulk of the HXR emission comes from the chromosphere. Line emissions in the gamma-ray range above $\sim500$ keV are mostly produced by nuclear interactions of accelerated protons and heavier ions. At 511 keV and 2.223 MeV, two clear emission lines can be seen; 511 keV is the electron-positron annihilation line and 2.223 MeV is the neutron capture line. This is the capture of neutrons by hydrogen in the photosphere. In an extremely rare case, if the spectrum can be seen up to 100 MeV, gamma-ray emission may be produced from pion decay \citep[e.g.,][]{1979ApJS...40..487R,2009RAA.....9...11C,2011SSRv..159..167V}.

\newpage
\subsection{X-ray imaging of a solar flare}\label{intro_imaging}
\subsubsection{The locations of X-ray sources}
Typically, there are X-ray sources located in both the chromosphere and corona during a solar flare. In a standard flare model, it is expected that the bulk of the HXR emission will be produced in the chromosphere, which is dense enough to stop high energy electrons accelerated in the corona. For the majority of flares, chromospheric HXR sources or {\it HXR footpoints}, as they are known, are observed, usually at X-ray energies greater than $\sim25$ keV. HXR footpoints generally sit at the bottom of the legs of a loop, formed by the reconnecting magnetic field (see Figure \ref{sfm_with_xray}), and hence for many flares they come in pairs at a given time and energy; one footpoint at the end of each loop leg. In the context of other observations, the HXR footpoints usually straddle the `magnetic inversion line', an imaginary line in the photosphere that separates regions of opposite vertical magnetic polarity in active regions. The majority of flare X-ray source morphologies are {\it footpoint dominated} \citep[e.g.,][]{1981ApJ...246L.155H, 1982SoPh...78..107A, 1982SoPh...81..137D, 1994PhDT.......335S,1996AdSpR..17...67S}. Most footpoint dominated flares also produce coronal SXR and HXR emission which can be a mixture of thermal, thin target and thick target emissions depending on the properties of the corona.  Usually this emission is observed up until $\sim30$ keV, with the majority emanating from the loop-top region (again see Figure \ref{sfm_with_xray}), possibly close to the energy release site. Many flares of this type were observed with instruments such as {\it Yohkoh} and now routinely with {\it RHESSI}. An excellent example of such a flare morphology is shown in Figure \ref{xray_sources} (top left).

\begin{figure*}[t]
\centering
\hspace{-30pt}
\includegraphics[width=8cm]{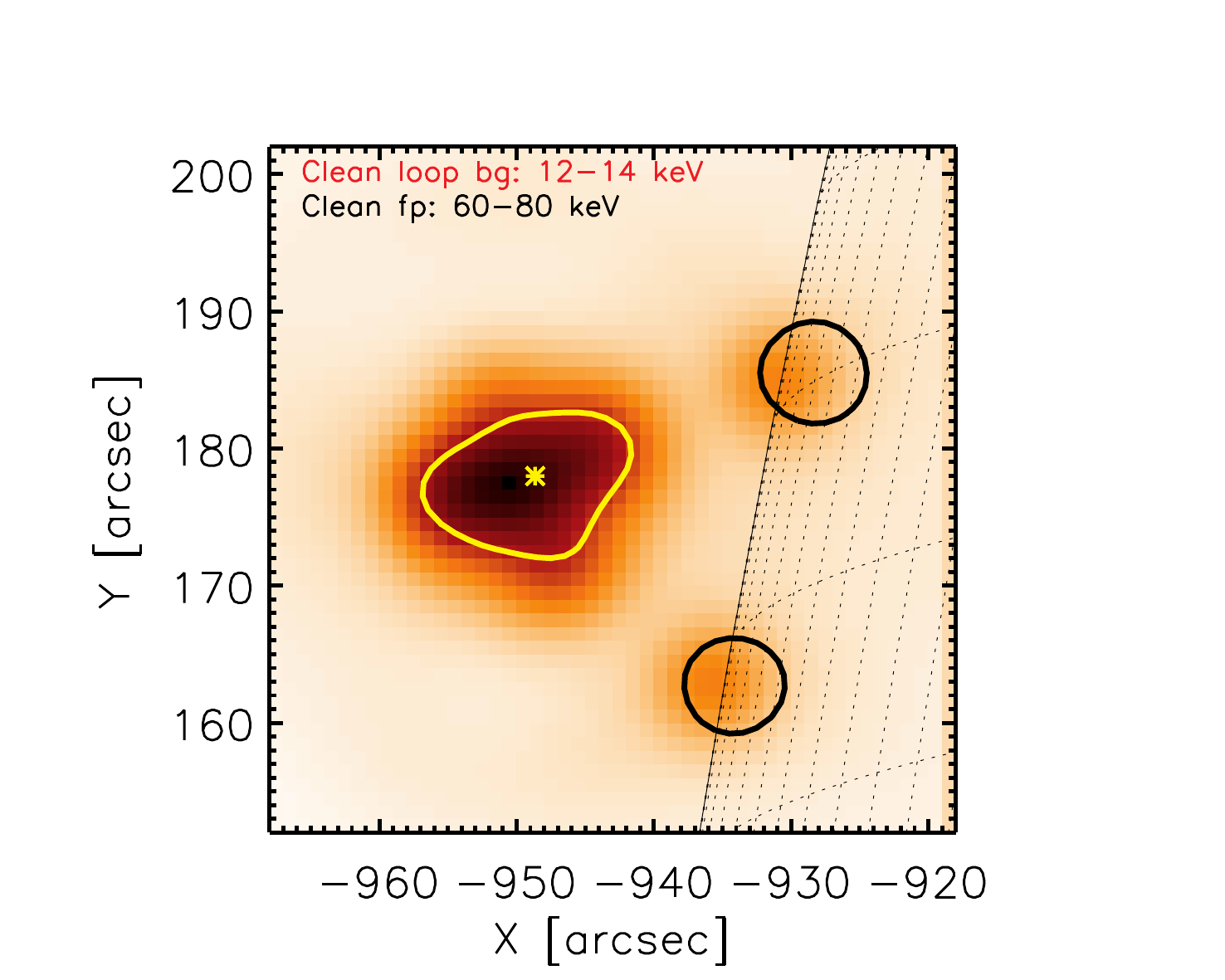}
\hspace{-50pt}
\includegraphics[width=7 cm]{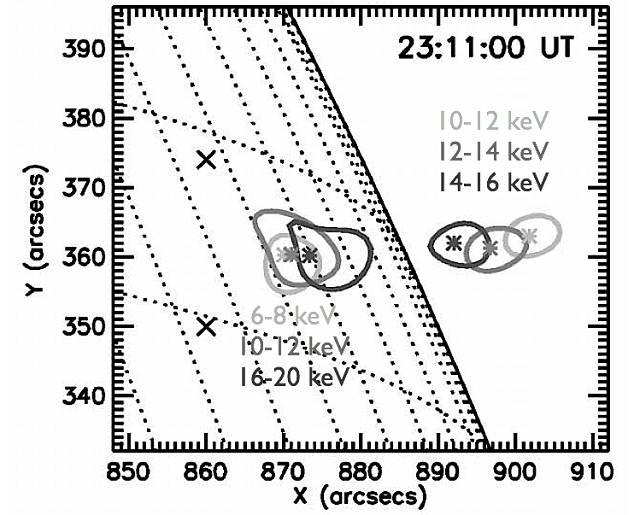}
\hspace{-30pt}
\end{figure*}
\begin{figure*}[h]
\centering
\hspace{-40pt}
\includegraphics[width=8cm]{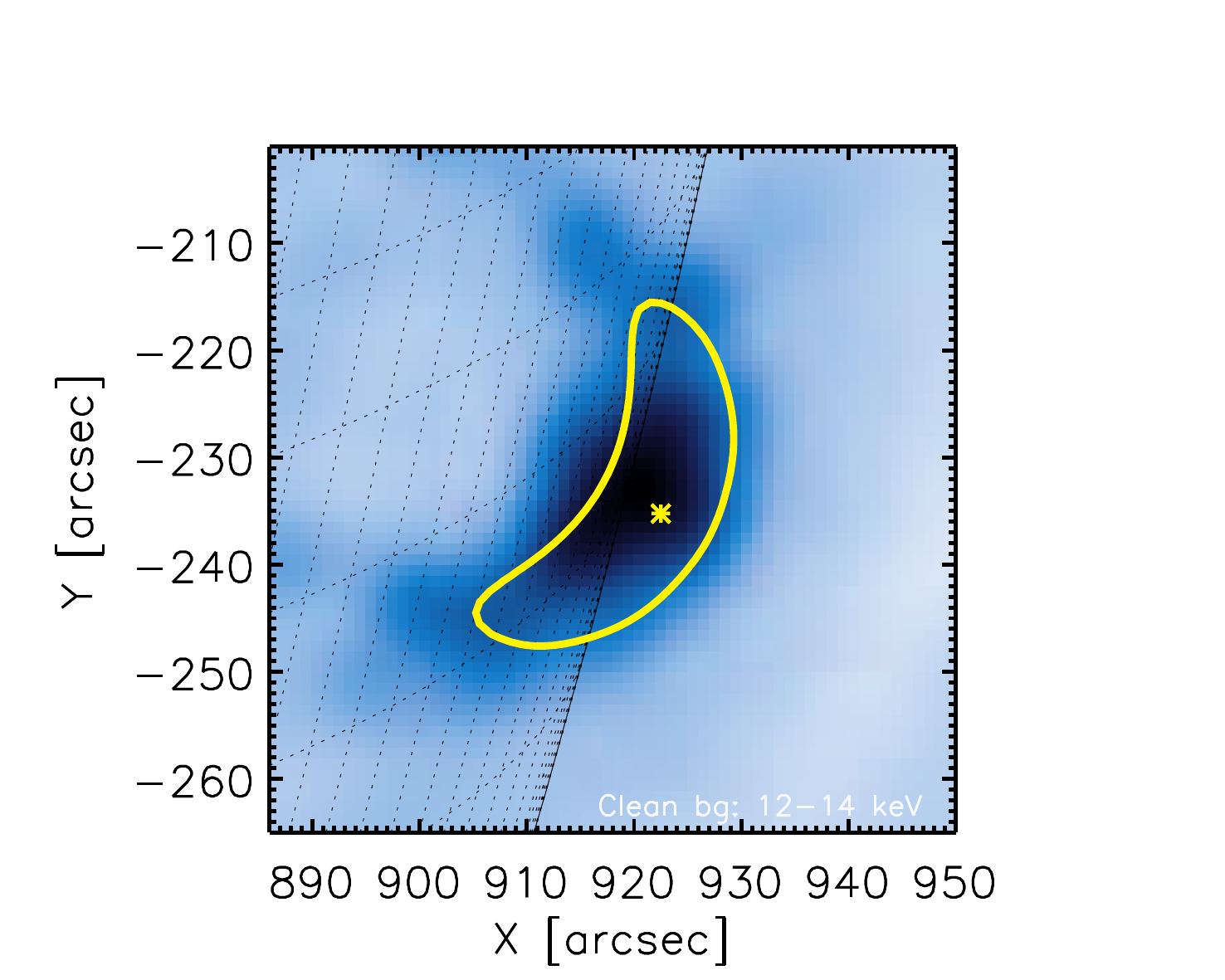}
\hspace{-10pt}
\includegraphics[width=5.5 cm]{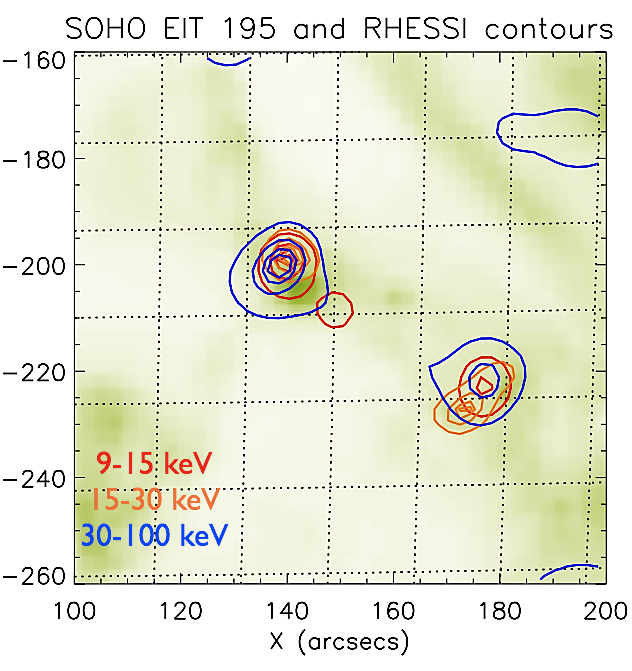}
\hspace{-40pt}
\caption[Four different examples of solar flare X-ray source morphologies.]{Four different flare X-ray source morphologies. {\it Top left:} `standard' flare with strong HXR footpoint emission and lower energy coronal emission. {\it Top right:} as top left with the observation of an additional above the loop-top X-ray source (figure taken from \cite{2003ApJ...596L.251S}). {\it Bottom left:} a coronal thick target source with very little HXR footpoint emission. {\it Bottom right:} a rare event with only the observation of HXR footpoints (figure taken from \cite{2011ApJ...731L..19F}). The energy bands of each X-ray source are displayed on the figure. Legends have been added to the top right and bottom right figures.}
\label{xray_sources}
\end{figure*}
For some flares, a HXR source above the lower energy X-ray loop-top can also be observed \citep[e.g.,][]{1994Natur.371..495M}. Although it is speculated that this HXR emission comes from the site of energy release itself, and may offer evidence for the formation of a current sheet in the corona during magnetic reconnection, it is only observed on rare occasions. One interesting case in particular, investigated by \cite{2003ApJ...596L.251S} with follow-on studies by \cite{2004ApJ...612..546S,2008ApJ...676..704L} is shown in Figure \ref{xray_sources} (top right). For this event both a loop top and an above the loop top source can be observed. While the height of the loop top source at a given energy decreased before the peak in X-ray emission, the above the loop-top source showed the opposite trend and moved away from the Sun at a high inferred velocity of $\sim300$ km s$^{-1}$. Sometimes, the above the loop-top source can be observed at very high energies, greater than that of the X-ray source within the loop \citep[e.g.,][]{1994Natur.371..495M,2011ApJ...737...48I,2014ApJ...780..107K}. However, the observation of an above the loop-top source is rare; this may be due to the fact they do not exist for every flare morphology or simply because instruments such as {\it RHESSI} have a limited dynamic range ($\sim10:1$) and are not sensitive to the low intensity emission from a low density corona.

Another type of flare morphology that is particularly important for this thesis, are flares where the majority of the X-ray emission comes from not the chromosphere, but the corona. Often this type of coronal X-ray source is interpreted as a {\it thick target coronal source} \citep[e.g.,][]{1995SoPh..158..283W, 2004ApJ...603L.117V}, where a high coronal density of the order $1\times10^{11}$ [cm$^{-3}$] allows the corona to stop electrons up to $30$ keV or so. Chapters \ref{ref:Chapter2} and \ref{ref:Chapter3} are dedicated to studying this type of event, and their properties will be further explained in these chapters, and in the following section. An example of this type of flare morphology is shown in Figure \ref{xray_sources} (bottom left).

In a very rare case \cite{2011ApJ...731L..19F} observed and analysed an interesting event with {\it only} non-thermal HXR footpoint emission. It was found that electrons were accelerated up to $100$ keV or so, but the measured temperature during the flare did not exceed 6.1 MK. An example of this event is shown in Figure \ref{xray_sources} (bottom right).
\subsubsection{Observations of X-ray source spatial properties}\label{fp_loop}
\begin{figure*}
\centering
\vspace{20pt}
\includegraphics[width=15cm]{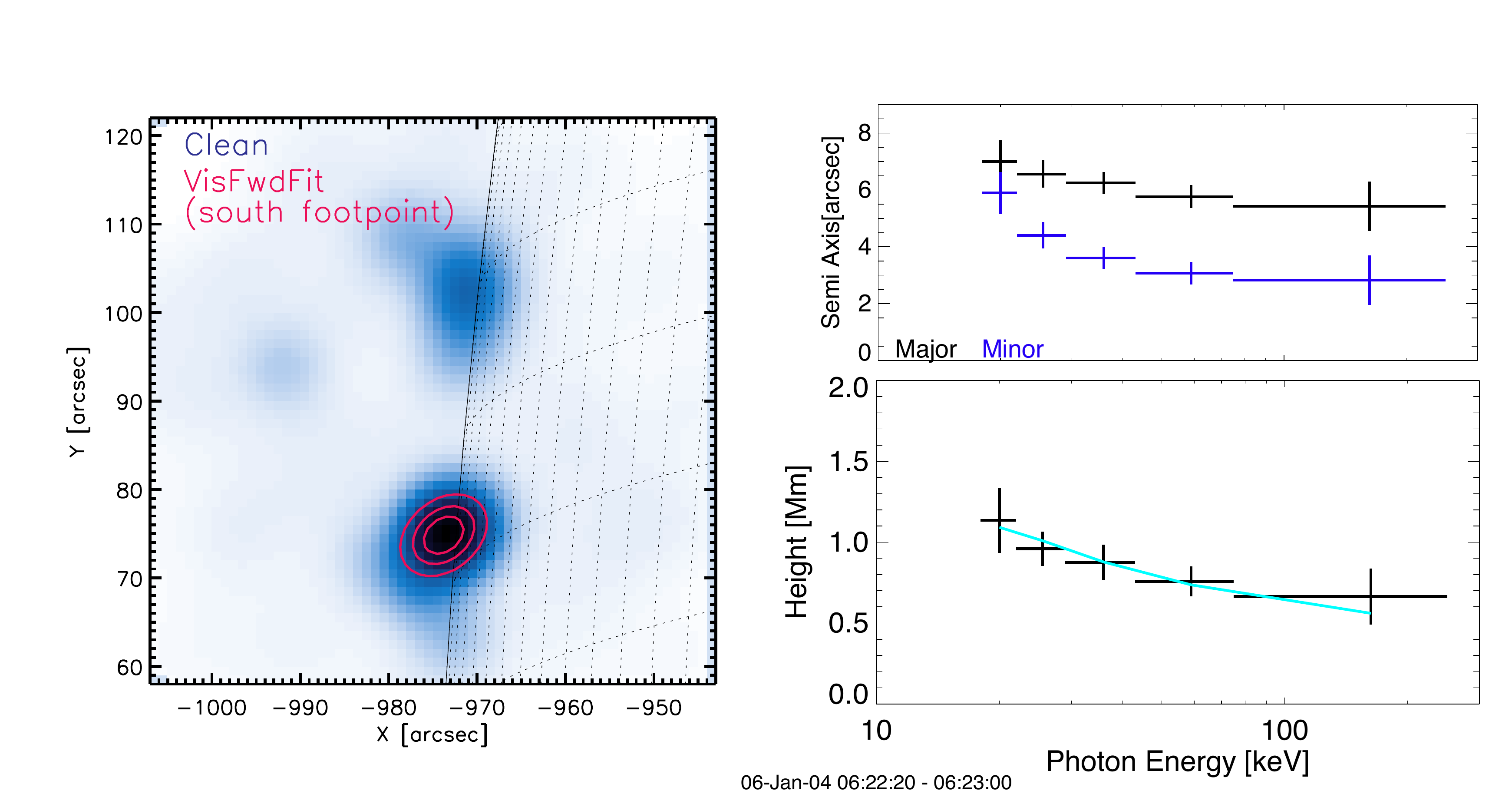}
\includegraphics[width=15cm]{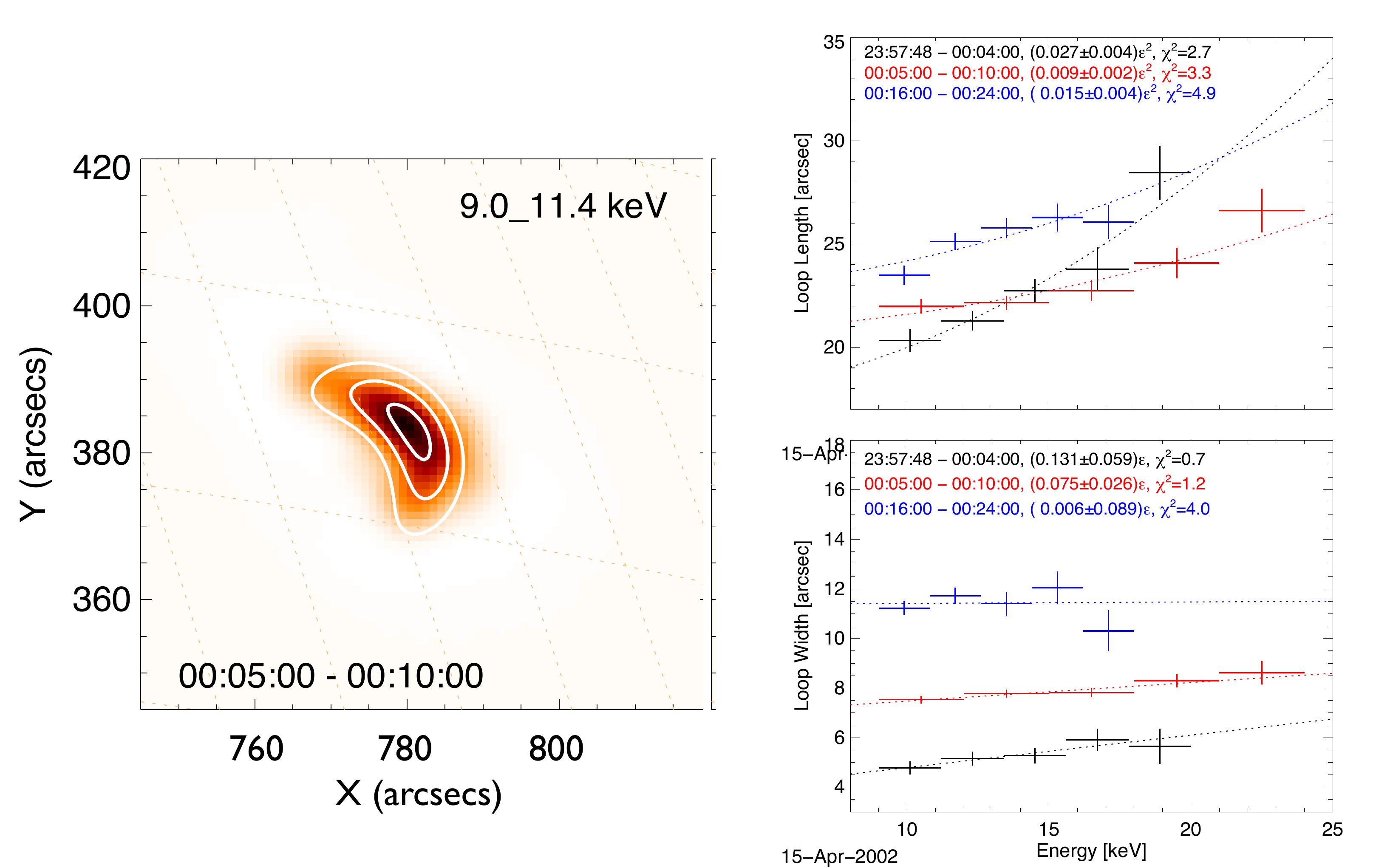}
\caption[Changes in X-ray spatial parameters with energy, for both a chromospheric HXR footpoint (top) and coronal X-ray source (bottom).]{Changes in X-ray spatial parameters with energy. {\it Top:} Changes in radial height and semi-major and semi-minor axes for a chromospheric HXR footpoint. {\it Bottom:} Changes in loop length and loop width for a coronal X-ray source. The bottom figure is taken from \cite{2011ApJ...730L..22K}. See Section \ref{fp_loop} for details.}
\label{footpointsandloop}
\end{figure*}

Often HXR footpoints in the chromosphere exhibit a circular or elliptical shape.  \citet*{2010ApJ...717..250K}, performed a study examining the changing spatial properties of HXR sources with X-ray energy. This study used a forward fitting method known as Visibility Forward Fitting (Vis FwdFit) that `fits' simple shapes to the X-ray visibilities ({\it RHESSI} is a non-direct imager and creates an image from Fourier components in $uv$ space; this is discussed further, along with the imaging algorithm of Vis FwdFit in Section \ref{rhessi_imaging}), allowing the source spatial properties to be studied.  This study built upon previous spatial studies with {\it RHESSI}, specifically \cite{2008A&A...489L..57K} where the Vis FwdFit method was employed and simple circular Gaussians were fitted to the HXR footpoint sources visibilities. \citet*{2010ApJ...717..250K} advanced the study by using an elliptical source that allowed the radial height and the shape of the HXR footpoint to be analysed; giving both the semi-major and semi-minor axes of the ellipse and hence the HXR footpoint. For this study, a flare located at the solar limb was chosen, where the changes in radial distance could be interpreted as height changes and the semi-major and semi-minor axes give the horizontal and vertical extents of the HXR source respectively. It was found that all three measured spatial properties decreased with increasing X-ray energy, and properties of the chromosphere could be deduced from the {\it RHESSI} observations. An image of the HXR footpoints for this event and a Vis FwdFit ellipse fitted to the bright southern footpoint is shown in Figure \ref{footpointsandloop}. The graphs for radial height, horizontal size and vertical size are also shown in Figure \ref{footpointsandloop}. By analysing the changes in both radial height and vertical extent of the HXR source together, it was found that the number density structure of the target chromosphere could not be explained by a simple uniform density structure but was much better fitted by a multi-threaded number density structure (see \citet*{2010ApJ...717..250K} for details); this would not have been deduced from the radial height measurements alone. The observation of a decreasing HXR source horizontal size with X-ray energy, and therefore with height in the chromosphere, suggests that the magnetic field guiding the electrons through the chromosphere is converging. From the conservation of magnetic flux $B_{1}A_{1}=B_{2}A_{2}$ where $B$ and $A$ are the magnetic field strength and area perpendicular to the field respectively, the magnetic field strength $B$ must increase as the cross sectional area $A$ decreases at lower heights in the chromosphere as described by \cite{2008A&A...489L..57K}.
\\
The observation, simulation and analysis of thick target coronal X-ray source spatial properties, of the type shown in Figure \ref{xray_sources} (bottom left), are discussed and then further examined in Chapters \ref{ref:Chapter2} and \ref{ref:Chapter3}. As is discussed in Chapter \ref{ref:Chapter3}, the changing lengths and widths of these loops with energy have also been previously analysed with the Vis FwdFit imaging algorithm. Many studies have found \citep{2008ApJ...673..576X, 2011ApJ...730L..22K, 2012ApJ...755...32G,2013ApJ...766...28G} that the source length parallel to the guiding field grows with energy, indicative of a collisional thick target model in a high density corona of the order $10^{11}$ cm$^{-3}$. The width changes of these loops perpendicular to the guiding field are particularly interesting. \cite{2011ApJ...730L..22K} performed a detailed study of one flare examining how the loop width increased proportionally with X-ray energy. However, unlike increases in coronal loop length, changes in width are more difficult to explain since the electrons are bound to the guiding field and classical cross field transport should be negligible. \cite{2011ApJ...730L..22K} and \cite{2011A&A...535A..18B} inferred that the width increase could be due to the presence of magnetic turbulence within the loop. An example of one such coronal X-ray source displaying these trends in length and width, found from the Vis FwdFit algorithm, is shown in Figure \ref{footpointsandloop}. Again, these types of study indicate the usefulness of observing changes in the spatial properties of X-ray sources, particularly in determining the properties of the chromosphere and corona during a flare.

\subsection{Solar flare X-ray and gamma ray polarization}\label{intro_pol}
\begin{figure}\centering
\includegraphics[width=12cm]{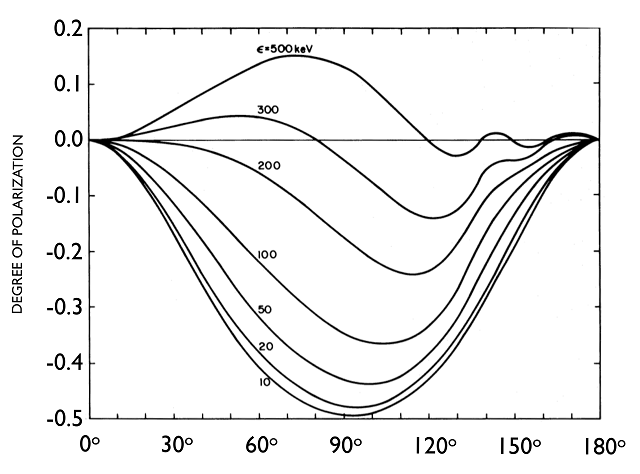}
\caption[The degree of polarization plotted against X-ray emission angle for different energies.]{Figure taken from \cite{BaiRamaty1978} showing the degree of polarization against X-ray emission angle for different energies. The sign determines the polarization angle, see Section \ref{intro_pol}.}
\label{bai_ramaty_3}
\end{figure}

\begin{figure}
\includegraphics[width=15cm]{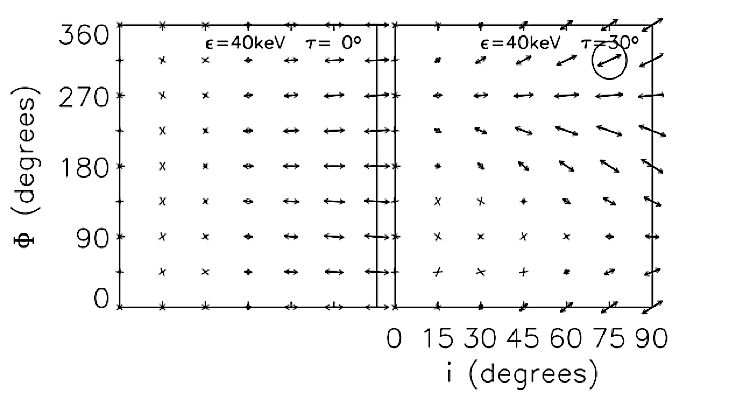}
\caption[The azimuthal X-ray emission angle plotted against the polar X-ray emission angle, showing the changing polarization angle with loop tilt.]{Figure taken from \cite{Emslieetal2008}. The azimuthal X-ray emission angle $\Phi$ against polar X-ray emission angle (heliocentric angle) $i$ for an X-ray energy of 40 keV, for two different loop tilt angles $\tau$ of $0^{\circ}$ and $30^{\circ}$. The arrow direction from the horizontal line gives the polarization angle $\Psi$.}
\label{emslie_pol}
\end{figure}

The pitch angle distribution of solar flare electrons in both the corona and chromosphere should be determinable through X-ray and gamma ray linear polarization measurements. The bremsstrahlung X-rays emitted from a highly beamed distribution of electrons for instance, should be highly polarized. In general the level of polarization detected is dependent on the photon energy, the level of beaming, the location of the X-ray source on the solar disk and whether Coulomb collisions have isotropised the electrons as they are transported along the guiding field. This has been extensively modelled \citep[e.g.][]{ElwertHaug1970,Brown1972, Haug1972, BaiRamaty1978, LeachPetrosian1983} for both thin and thick target scenarios. \cite{1980ApJ...237.1015E} demonstrated that the emission from a purely thermal source should have some level of polarization. In a simple model, the degree of polarization should increase with viewing angle. This means, that for the same distribution of electrons producing X-rays during a flare, the degree of polarization will be highest for sources viewed at the solar limb, at angles perpendicular to the guiding field. Figure \ref{bai_ramaty_3} taken from \cite{BaiRamaty1978} plots, for a given simulation model using an electron distribution with pitch angles uniformly distributed in a cone with a half opening angle of $30^{\circ}$, the resulting degree of polarization versus solar heliocentric angle. For this model, the highest degree of polarization for low energies (10-20 keV) is at a viewing angle of $90^{\circ}$, equivalent to viewing the flare at the solar limb. The plot also shows how the sign of the degree of polarization changes from negative to positive at high energies of $\sim300$ keV. This indicates that the preferred direction of polarization, the polarization angle, changes from being aligned to the local radial direction (the direction along the line connecting the centre of the X-ray source with the solar centre) to the perpendicular to radial direction. This is discussed further in Chapter \ref{ref:Chapter6}.
Measuring the spatially integrated polarization angle provides information about the overall geometry of the flaring loop and hence the orientation of the magnetic field \citep{Emslieetal2008}. If the magnetic field of the loop is tilted away form the local solar vertical direction, then the polarization angle will not lie along the radial direction. The work of \cite{Emslieetal2008} grew from the non-radial polarization angle observations from \cite{McConnelletal2003} and the suggestions of \cite{2003ApJ...595L..81S}. Figure \ref{emslie_pol} taken from \cite{Emslieetal2008} plots the azimuthal X-ray emission angle on the solar disk $\Phi$ against viewing angle $i$ for two different loop tilts $\tau=0^{\circ},30^{\circ}$ for an X-ray energy of $\epsilon=40$ keV. The direction of each arrow gives the angle of polarization. For a loop tilt of $0^{\circ}$, the polarization angle is always equal to $0^{\circ}$ and the arrows lie along the horizontal radial direction. However for the case of a loop tilt of $30^{\circ}$, the polarization angle can have values of $\Psi\ne0^{\circ}$ depending on the X-ray source position on the solar disk. It will also be shown in Chapter \ref{ref:Chapter6} that the direction of the polarization angle parallel or perpendicular to the radial direction is dependent upon the highest energies in the electron distribution and the degree of beaming.
The {\it RHESSI} satellite, which is discussed in detail in Section \ref{intro_rhessi} has limited polarization capabilities. Rare measurements for seven flares \citep{2006SoPh..239..149S} showed a range of polarization degrees from $0-60\%$ using an X-ray energy range of $100-350$ keV. \cite{Boggsetal2006} used {\it RHESSI} gamma-ray observations to measure the polarization of two flares. They found that the degree of polarization at these higher energies was consistent with a beamed distribution of electrons. However, a number of other observations with {\it RHESSI} over the last decade suggest that the emitted X-ray distribution, and hence the radiating electron distribution, is close to being isotropic, which should produce a very low level of polarization. One such method that has been used to determine this isotropy is that of X-ray albedo \citep[e.g.,][]{KontarBrown2006,Kasparovaetal2007,2013SoPh..284..405D}.
\begin{figure}
\centering
\includegraphics[width=7.6cm]{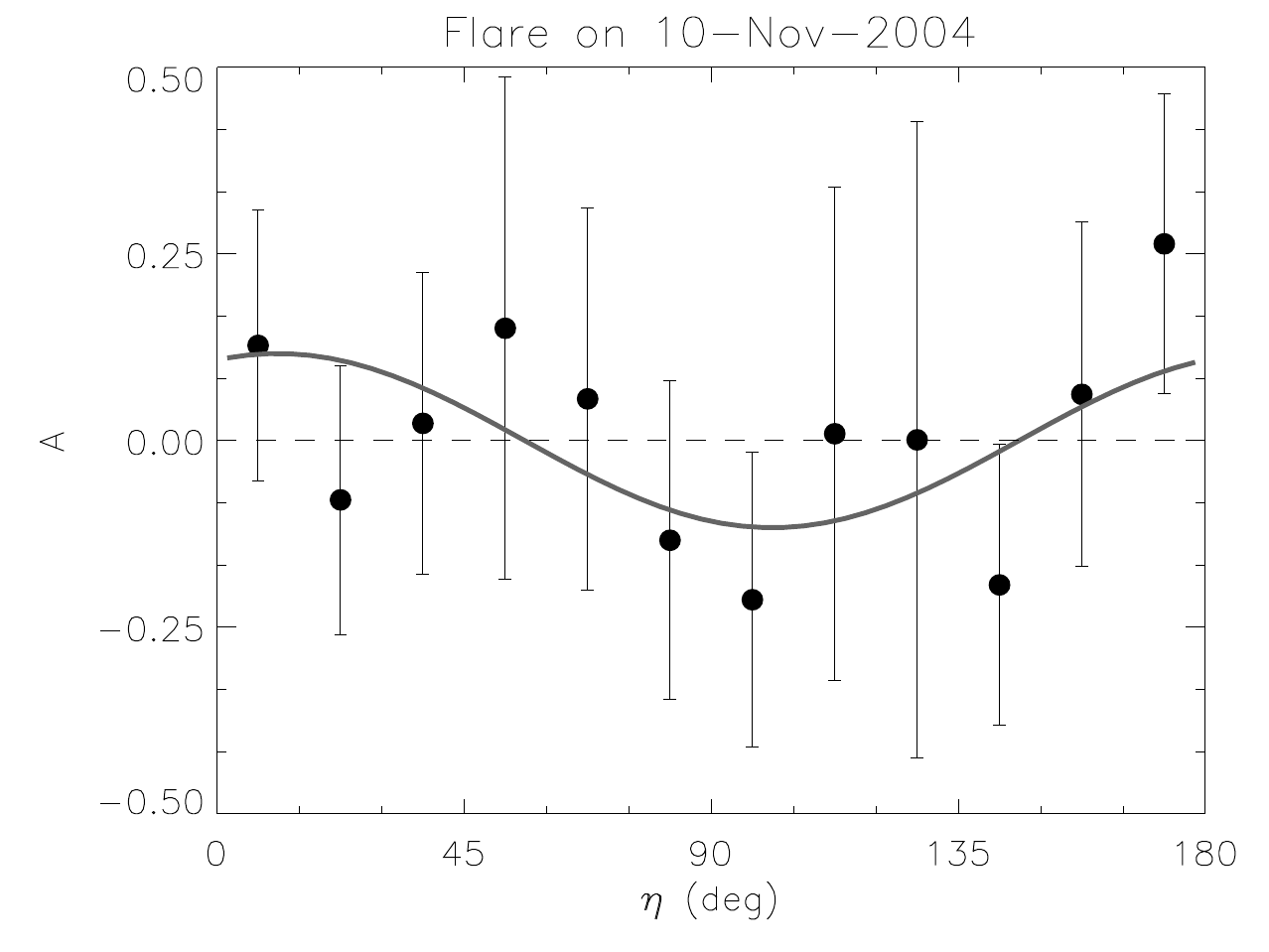}
\includegraphics[width=7.6cm]{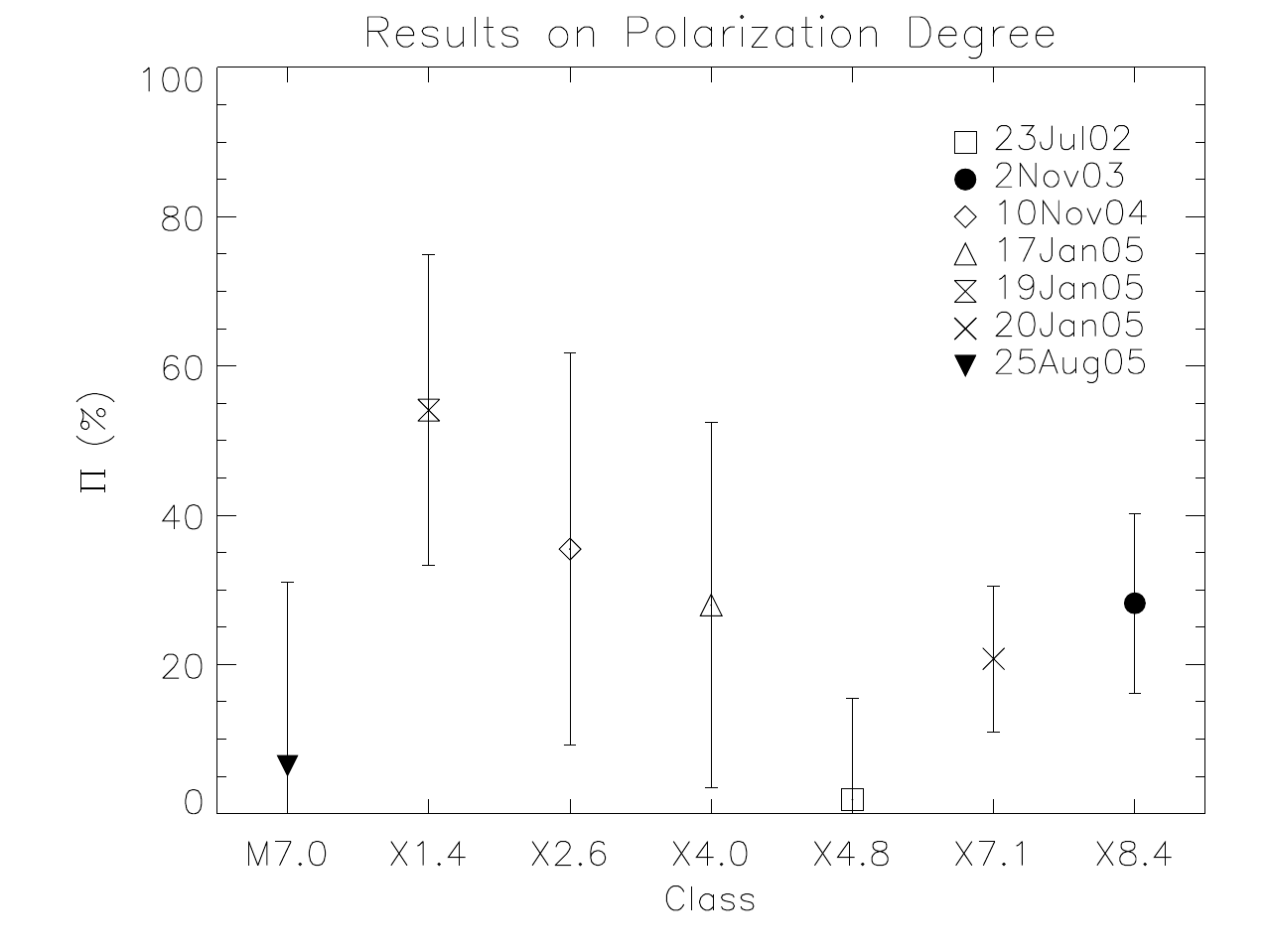}
\caption[Solar flare polarization measurements from {\it RHESSI}.]{Figure taken from \cite{2006SoPh..239..149S}. {\it Left:} Solar flare polarization measurements from {\it RHESSI} for one event. {\it Right:} the measured degree of polarization for all seven events.}
\end{figure}
\subsection{X-rays from the photosphere and albedo emission}\label{intro_albedo1}
The corona and chromosphere are optically thin at X-ray and gamma-ray energies. Therefore X-rays emitted at these wavelengths from chromosphere and corona can travel down into the denser layers of the solar atmosphere until they reach the much higher densities of the photosphere. For X-ray photons, the photosphere acts as either an absorber or `mirror', scattering a proportion of the X-rays back into the chromosphere and hence out into interplanetary space. The scattered X-ray flux depends on many factors: the energy of the X-rays, the heliocentric angle of the X-ray emitter in the chromosphere, the height of the X-ray emitter above the photosphere and the proportion of the X-rays emitted towards the photosphere. X-rays scattered from the photosphere are known as the albedo X-rays and the scattering mechanism is Compton scattering. The study of albedo is a major part of this thesis. The first studies of the X-ray photospheric albedo component were examined analytically by \cite{Tomblin1972} and through simulation by \cite{Santangeloetal1973}. A more comprehensive study was provided by \cite{BaiRamaty1978}. The peak of the X-ray albedo flux appears at around 20-50 keV due to: (1) lower energy X-rays below 10 keV are more likely to be photoelectrically absorbed than scattered and (2) higher energy electrons, particularly above 100 keV, are lost within the depths of the photosphere. This produces the well-known albedo reflectivity curve, the ratio of reflected flux to directly emitted primary flux which is shown in the bottom plot of Figure \ref{intro_albedo}. The scattered albedo flux also varies with X-ray emission angle or equivalently heliocentric angle on the solar disk; X-ray sources at the solar centre have the greatest proportion of albedo photons, while X-ray sources located at the solar limb have the smallest proportion, since the fraction of reflected photons seen by an observer is smaller at the solar limb ($90^{\circ}$) than at the solar centre ($0^{\circ}$). The properties inferred from an observed X-ray source will be tainted if it is assumed that the X-ray source consists of only directly emitted X-rays from the chromosphere. The albedo component will change the spectral, spatial and polarization properties of the observed X-ray emission. Some of the spatial and polarization changes are analysed for the first time in Chapters \ref{ref:Chapter5} and \ref{ref:Chapter6} of this thesis and are published in \cite{KontarJeffrey2010} and \cite{2013ApJ...766...75J}. The known spectral and polarization changes due to photospheric albedo are now briefly reviewed.
\begin{figure}
\centering
\includegraphics[width=15cm]{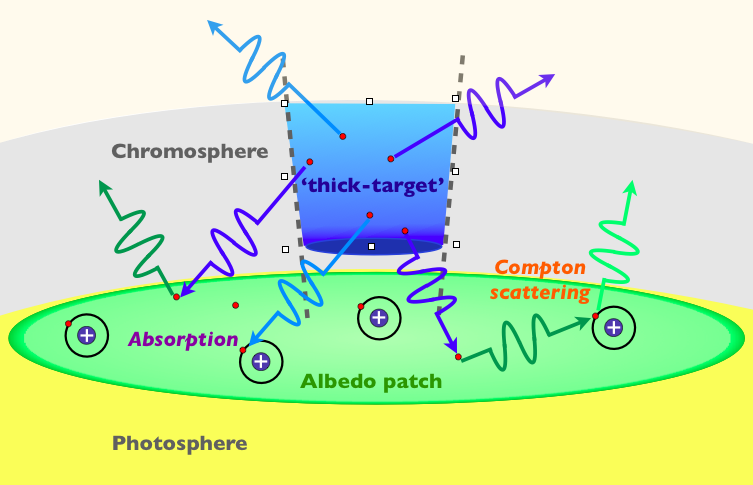}
\caption[A cartoon of solar flare X-ray interactions in the photosphere, after being emitted from the chromosphere.]{Cartoon showing the X-rays emitted directly from electrons stopped collisionally in the dense chromospheric thick target. Many of the X-rays are directly emitted away from the solar surface. These are known as the primary X-rays, while some travel down into the photosphere. Here they interact by photoelectric absorption or Compton scattering. Those X-rays Compton backscattered out of the photosphere are known as albedo X-rays from the albedo patch in the photosphere.}
\end{figure}
\subsubsection{Spectral changes and the determination of the electron distribution}
\begin{figure}
\centering
\includegraphics[width=15cm]{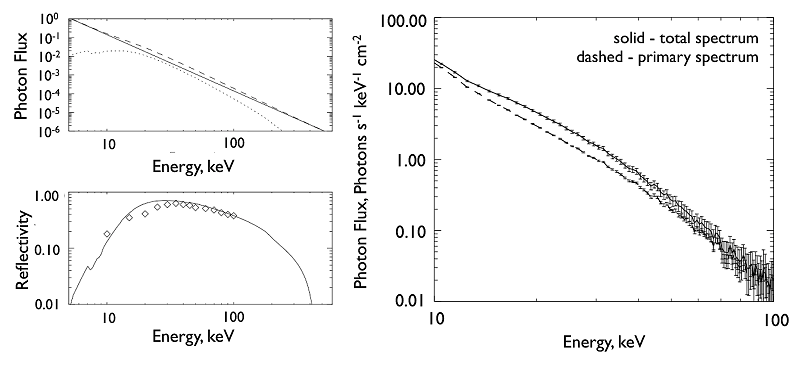}
\caption[X-ray albedo reflectivity and an X-ray spectrum with and without an albedo contribution, calculated using a Green's function. The figure also shows a real flare X-ray spectrum observed with {\it RHESSI} before and after albedo correction.]{Figures taken from \cite{Kontaretal2006}. {\it Left:} a photon spectrum calculated from Green's functions and an initial primary photon distribution of $I_{P}(\epsilon)\sim\epsilon^{-3}$, where the black solid line represents the primary spectrum, the black dotted line the albedo spectrum and the black dashed line the total spectrum. The bottom figure plots the reflectivity (flux down/flux up). The black diamonds plot the reflectivity from \cite{BaiRamaty1978}. {\it Right:} The spectrum of a flare that occurred on the 17th September 2002. The observed spectrum is corrected by accounting for the Compton scattered albedo component with the solid and dashed lines showing the spectrum before and after the correction respectively.}
\label{intro_albedo}
\end{figure}

An albedo component produces a `bump' in the photon spectrum which peaks at around 30-50 keV, due to the peak albedo flux at these energies. Although the presence of an albedo component from an X-ray spectrum should always be accounted for, it should be noted that other effects can produce similar flattening at lower energies in the spectrum (10-50 keV) such as return current, count pile-up in the {\it RHESSI} detectors \citep[cf.][]{2011SSRv..159..107H} and even wave-particle interactions \cite{2012A&A...539A..43K}. The size of the `bump' is dependent on the factors mentioned in the previous section and changes in the photon spectrum have been extensively studied through the use of simulations with a known primary X-ray spectrum \cite[e.g.,][]{BaiRamaty1978}. However, this approach is not so useful for actual observations with instruments such as {\it RHESSI}, where the properties of the initial primary X-ray distribution are unknown and have to be disentangled from the albedo spectrum. In \cite{Kontaretal2006}, a Green's function approach was used to try and remove the albedo component from the unknown primary X-ray spectrum by adapting \cite{MagdziarzZdziarski1995}. The primary distribution $I_{P}$ can be found using,
\begin{equation}\label{I_greens}
I(\epsilon)=I_{P}(\epsilon)+I_{S}(\epsilon)=I_{P}(\epsilon)+\int_{\epsilon}^{\epsilon_{max}}I_{P}G(\mu,\epsilon,\epsilon_{0})d\epsilon_{0}
\end{equation}
where $I$ is the total distribution and $I_{S}$ is the scattered albedo distribution, which is the integral over energy of the primary distribution $I_{P}$ and a Green's function $G(\mu,\epsilon,\epsilon_{0})$, where $\mu,\epsilon$ and $\epsilon_{0}$ are the cosine of heliocentric angle, scattered X-ray energy and incoming X-ray energy respectively. Writing the problem discretely, gives,
\begin{equation}\label{I_greens_2}
I(\epsilon_{i})=\left(1_{ij}+\alpha G_{ij}\right)I_{P}(\epsilon_{j}).
\end{equation}
where $G_{ij}$ is the Green's matrix, $1_{ij}$ is a diagonal matrix with values of 1 and $\alpha$ is the anisotropy of the X-ray distribution \citep{Kontaretal2006}. The best solution of $I_{P}$ can be found numerically and is incorporated into the {\it RHESSI} OSPEX software via the detector response matrix (as discussed in Section \ref{intro_rhessi}). Figure \ref{intro_albedo} shows a RHESSI photon spectrum before and after the correction. The presence of an albedo component should produce the `bump' discussed above in the observed spectrum between $\sim10-100$ keV, which can be seen in the black solid line in Figure \ref{intro_albedo}. The albedo corrected spectrum is shown by the black dashed line. Determining the mean electron spectrum from the observed X-ray distribution often produces a strange minimum at $\sim40$ keV. However, properly accounting for an X-ray albedo component can remove this strange feature \citep{Kontaretal2006} and clearly indicates the importance of accounting for the albedo component when interpreting information from an observed X-ray spectrum.
\subsubsection{Changes in total polarization}
\begin{figure}
  \begin{minipage}[c]{0.55\textwidth}
    \includegraphics[width=\textwidth]{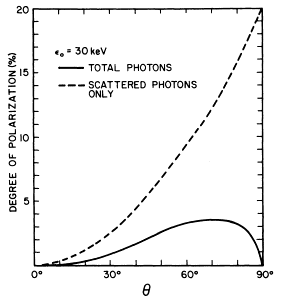}
  \end{minipage}\hfill
  \begin{minipage}[c]{0.40\textwidth}
\caption[The varying magnitude of polarization for a completely isotropic source viewed at different locations on the solar disk due to the presence of a photospheric backscattered albedo component.]{Figure taken from \cite{BaiRamaty1978} showing how the degree of polarization changes for a completely isotropic source viewed at different locations on the solar disk due to the presence of a photospheric backscattered albedo component.}
\label{br_pol}
\end{minipage}
\end{figure}

The albedo component will change the measured polarization of an X-ray source. For the case of a completely isotropic non-thermal X-ray distribution, the total degree of polarization will be zero for an X-ray source viewed at all solar heliocentric angles. However the degree of polarization for the scattered albedo component will {\it increase} with emission angle, as the observer views more and more X-rays scattered at angles closer to $90^{\circ}$. The polarization of the scattered contribution is hence $0\%$ for an X-ray source at the solar centre and becomes $\sim20\%$ for an isotropic distribution at the solar limb. This is shown in Figure \ref{br_pol} taken from \cite{BaiRamaty1978}. However, this will only produce a degree of polarization of $\sim4\%$ at the solar limb due to the much higher flux of the primary X-ray component at large heliocentric angles. Larger changes in polarization are possible for more anisotropic X-ray distributions. In contrast, the backscattered albedo X-rays generally act to reduce the polarization of an anisotropic source \citep{Henoux1975, LangerPetrosian1977,BaiRamaty1978}. This is discussed in greater detail in Chapter \ref{ref:Chapter6}, where the changes in spatially resolvable polarization due to albedo are discussed for the first time in solar physics.
\subsubsection{Determining the electron anisotropy using albedo}
\begin{figure}
  %\begin{minipage}[c]{0.60\textwidth}
    \includegraphics[width=7 cm]{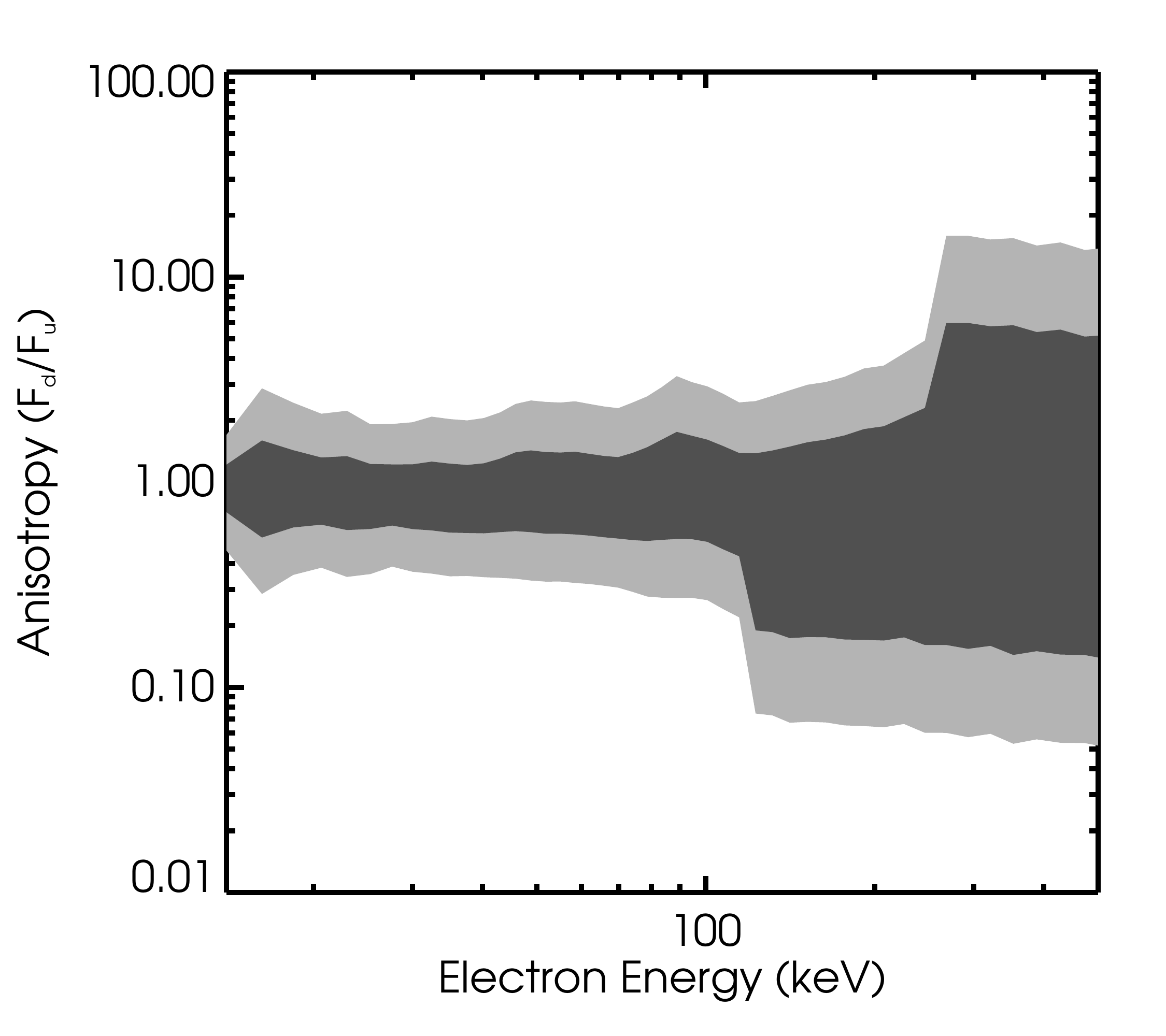}
    \includegraphics[width=8 cm]{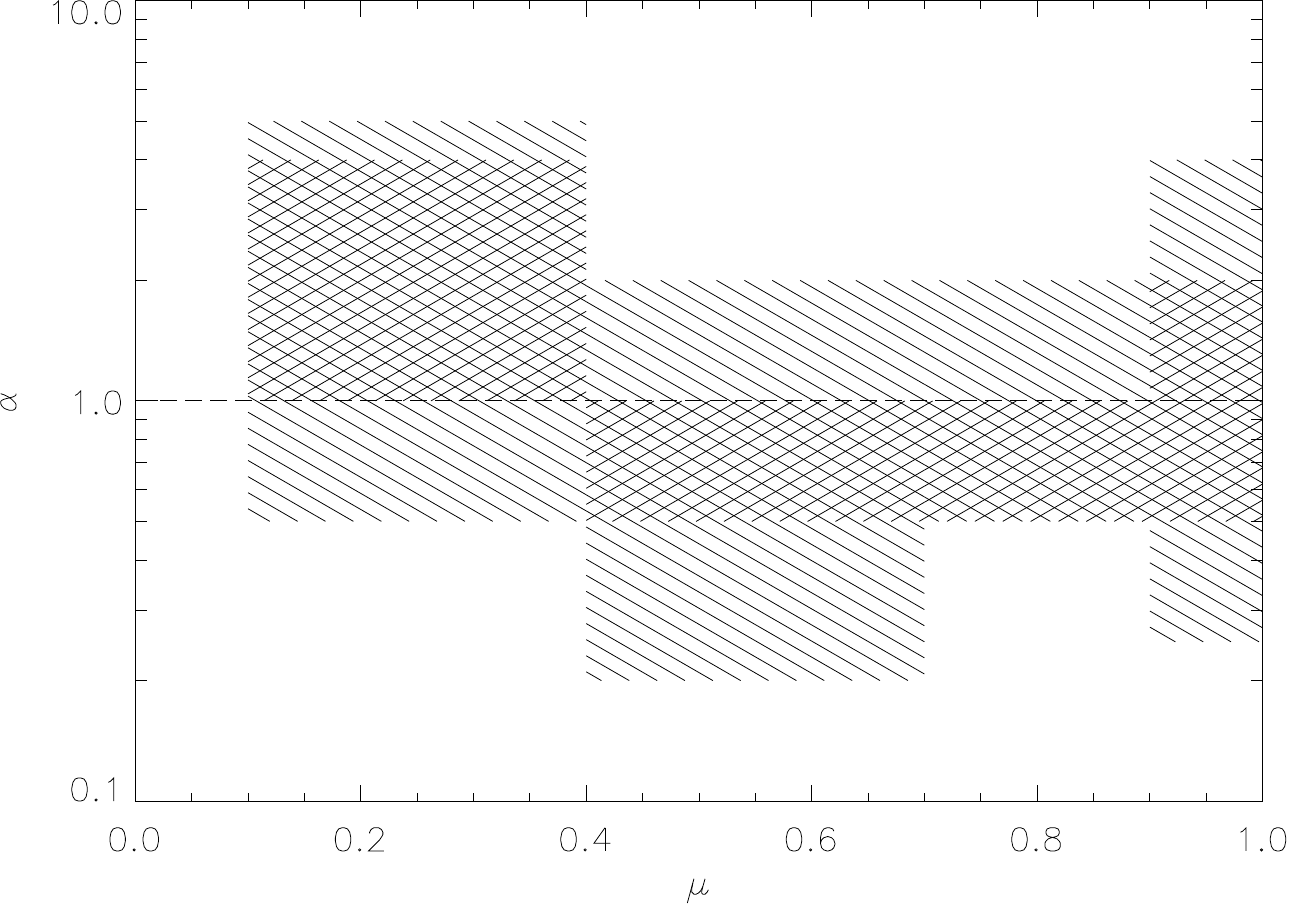}
  %\end{minipage}\hfill
  %\begin{minipage}[c]{0.35\textwidth}
\caption[Measured X-ray anisotropy from {\it RHESSI} observations using two different methods that take advantage of the X-ray albedo component.]{{\it Left:} Figure taken from \cite{2013SoPh..284..405D} showing the anisotropy of the electron spectrum (Flux down /Flux up) for one flare where dark grey represents a 1-sigma and light grey a 3-sigma, confidence interval. {\it Right:} Figure taken from \cite{Kasparovaetal2007} showing determined directivity $\alpha=$Flux down/Flux up for different heliocentric angles $\mu$ for an energy range of $15-20$ keV.}
\label{ed_figures}
%\end{minipage}
\end{figure}

\cite{KontarBrown2006} used a technique to effectively separate the `upward' and `downward albedo' components of X-ray flux that contribute to the measured X-ray spectrum and hence electron spectrum. Determining the X-ray albedo flux allows the anisotropy of a single flare to be measured, using a single instrument such as {\it RHESSI}. The results of this study determined that both flares observed had extremely isotropic distributions, a result that is not expected if the distribution of electrons is beamed; an assumption often made in a standard flare collisional thick target model. Follow on studies by \cite{2013SoPh..284..405D} performed a larger examination of eight events and again found a lack of anisotropy for each event below $\sim150$ keV. An example of one of the flares examined in \cite{2013SoPh..284..405D} is shown in Figure \ref{ed_figures}. A centre-to-limb statistical survey of 398 flare spectral indices in three different energy bands was performed by \cite{Kasparovaetal2007}. They found that there was a clear change in spectral index in the low energy $15-20$ keV band, with the spectral index increasing towards the limb. This is consistent with the presence of an albedo component flattening the X-ray spectrum at low energies. Further, \cite{Kasparovaetal2007} determined the directivity of the X-ray emission for a number of their flares, using the albedo Green's function method of \cite{Kontaretal2006}. From this the ratio of downward to upward flux was found to lie anywhere between $0.2$ and $5$ in the $15-20$ keV band (see Figure \ref{ed_figures} (right)). Hence this study gave no clear conclusion regarding the X-ray anisotropy, that is, the results are consistent with the predictions of a beamed downward distribution of electrons {\it and} with an isotropic distribution.
\section{Current X-ray telescopes and X-ray imaging methods}\label{intro_rhessi}
The X-ray observations shown in Chapter \ref{ref:Chapter3} of this thesis and some of the simulation work shown in Chapters \ref{ref:Chapter2} and \ref{ref:Chapter5} are concerned with the current solar X-ray and gamma ray imaging spectroscopy performed by the {\it Ramaty High Energy Solar Spectroscopic Imager, RHESSI} \citep{2002SoPh..210....3L} . Hence, here in Chapter \ref{ref:Chapter1}, a brief review of the instrument is given.
\subsection{RHESSI: instrument overview}
\begin{figure}
\includegraphics[width=15cm]{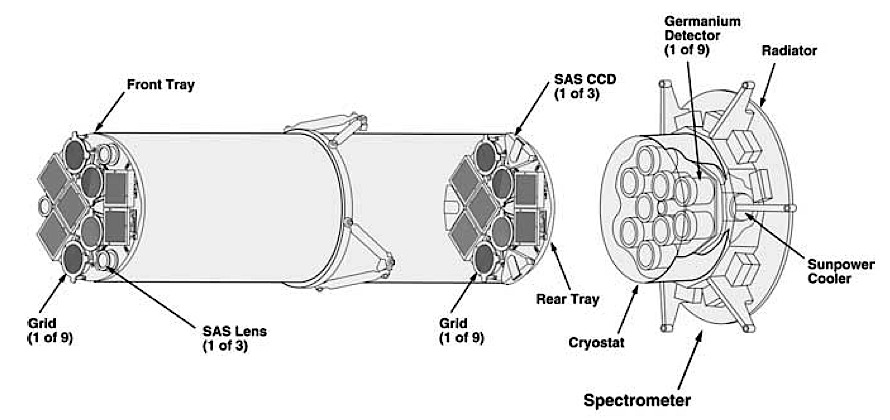}
\caption[Diagram of the {\it RHESSI} grids and detectors.]{Diagram of {\it RHESSI} grids and detectors. This image was taken from \cite{2002SoPh..210...61H}.}
\label{fig:rhessi_1}
\end{figure}

{\it RHESSI} is a NASA-led mission, launched in February 2002. For more than a decade, it has provided unparalleled hard X-ray observations of the Sun and solar flares in particular. {\it RHESSI} observes the full disk of the Sun from a low Earth orbit over the energies of $3$ keV to $17$ MeV. {\it RHESSI} performs imaging spectroscopy of X-rays and gamma-rays and it was designed specifically to study particle acceleration and energy release in solar flares. The {\it RHESSI} instrument consists of a spectrometer; nine cooled Germanium detectors placed at the rear of the spacecraft, with additional imaging apparatus consisting of nine pairs of widely spaced grids at a distance of $1.5$ m called rotating modulation collimators (RMCs) placed in front of each detector. The instrument setup is shown in Figure \ref{fig:rhessi_1}.
\subsection{RHESSI imaging}\label{rhessi_imaging}
\begin{figure}
\hspace{-15pt}
\includegraphics[width=16.2cm]{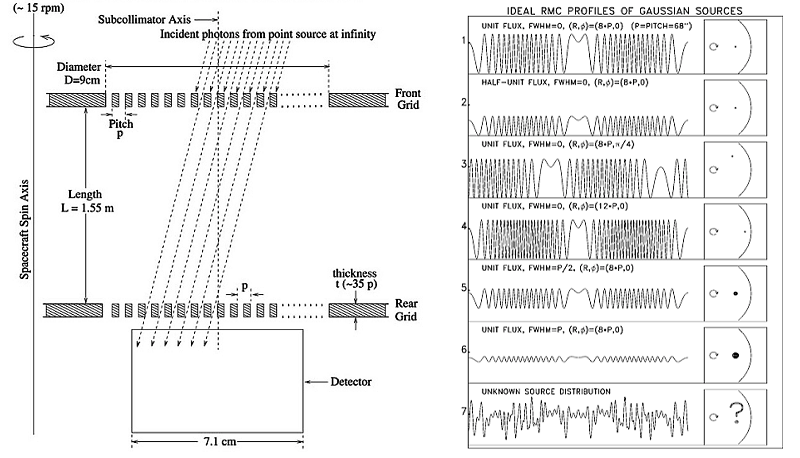}
\caption[An example of a photon entering a {\it RHESSI} RMC and {\it RHESSI} time modulation curves.]{{\it Left:} Diagram showing how photons entering a {\it RHESSI} RMC at a given time, during a spacecraft rotation, are either blocked by the slats of a grid or travel towards the detector, depending on their incident position. {\it Right:} An example of {\it RHESSI} time modulation curves for different simulated sources sitting on the solar disk. Both figures were taken from \cite{2002SoPh..210...61H}.}
\label{fig:rhessi_2}
\end{figure}

\begin{figure}
  \begin{minipage}[c]{0.60\textwidth}
    \includegraphics[width=\textwidth]{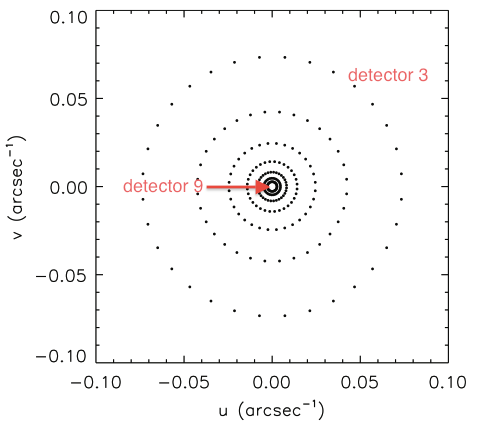}
  \end{minipage}\hfill
  \begin{minipage}[c]{0.30\textwidth}
\caption[Diagram of the {\it RHESSI} $uv$ plane.]{Diagram of the {\it RHESSI} $uv$ plane. Detector 9 with the largest angular resolution produces the smallest circle. Only detectors 3-9 are shown in this diagram. Image taken and adapted from \cite{2009ApJ...703.2004M}.}
\label{fig:rhessi_uv_plane}
\end{minipage}
\end{figure}

Due to their high energies, X-rays above $1$ keV are difficult to image directly, although astrophysical missions such as {\it The Nuclear Spectroscopic Telescope Array} NuSTAR \citep{2013ApJ...770..103H} and recent balloons missions: {\it The Focusing Optics X-ray Solar Imager} FOXSI \citep[e.g.][]{2011SPIE.8147E...4K} and {\it The High Energy Replicated Optics to Explore the Sun} HEROES \citep[e.g.][]{2013SPD....44...76C} are pioneering the technology and techniques required for future direct X-ray imaging in solar physics. {\it RHESSI} on the other hand, creates X-ray images via a non-direct Fourier technique using its RMCs. {\it RHESSI} rotates continuously around its axis pointing towards the Sun. As the spacecraft rotates, the incoming X-ray signal passes through the slits and slats of its nine RMCs, with the slits and slats either impeding or allowing the X-rays path to the detectors. This produces a time modulated signal, that is dependent upon the position and size of the X-ray source \cite[cf.,][]{2002SoPh..210...61H}. Figure \ref{fig:rhessi_2} (left) shows a diagram of the X-rays passing through the front and rear grids of one RMC and onto a detector while Figure \ref{fig:rhessi_2} (right) shows examples of different modulation patterns, and how the pattern varies with X-ray source position and size. The time modulated signal can be stacked per roll bin (fraction of a spacecraft rotation) over so many spacecraft rotations, creating {\it X-ray visibilities}. The X-ray visibilities $V$ are the two-dimensional Fourier components of the X-ray source in $uv$ space, given by,
\begin{equation}\label{rhessi_vis}
V(u,v;\epsilon)=\int_{x}\int_{y}I(x,y;\epsilon)e^{2\pi i(xu+yv)}dxdy.
\end{equation}
The inverse Fourier transform of Equation \ref{rhessi_vis} gives the X-ray image $I$ at a given energy $\epsilon$ in the real $xy$ plane. The X-ray visibilities are represented in $uv$ space by circles of constant radius, with each circle representing the angular resolution of RMC 1-9. This is shown for RMCs 3-9 in Figure \ref{fig:rhessi_uv_plane}. The grids of RMC 1 have the finest spatial resolution of 2.26 arc second and the grids of RMC 9 have the coarsest spatial resolution of 183.2 arc second, where the spatial resolution of each RMC increases by factor $\sqrt{3}$, with increasing number from 1 to 9. The creation of an image from Equation \ref{rhessi_vis} or directly from the time modulated signal is an inverse problem and performing a two-dimensional Fourier transform produces a Back projection image \cite{1986JOSAA...3.2167M}. However this method produces a poor image quality with side lobes and hence a number of different imaging algorithms have been created, or adapted from radio astronomy in order to solve the problem and improve the image, such as CLEAN \citep{1974A&AS...15..417H, 2002SoPh..210...61H}, Pixon \citep{1993PASP..105..630P,1996ApJ...466..585M} and forward fitting algorithms such as Visibility Forward Fitting (Vis FwdFit) \citep{2002SoPh..210...61H,2007SoPh..240..241S}. The CLEAN, Pixon and Vis FwdFit imaging algorithms are used in this thesis and hence are briefly discussed:
\begin{enumerate}
\item{CLEAN - this algorithm assumes that the X-ray image consists of a superposition of many X-ray point sources. The process of CLEAN-ing is an iterative process that continuously searches for the highest intensity pixel in the image. At each iteration, once the highest pixel is found, a chosen proportion of the highest intensity, convolved with the Point Spread Function (PSF) of the instrument, is centred at the highest pixel and subtracted from the image. This process usually repeats for a chosen number of iterations or until the peak flux in the image is negative. The resulting final image is a CLEANed map consisting of the positions and amplitudes of each chosen pixel at each iteration, convolved with the PSF.}
\item{Pixon - this algorithm wants to construct the simplest model for the image that is consistent with the data \citep{2002SoPh..210...61H}. Pixon uses different size pixels or `pixons' together to try and reproduce the X-ray modulation patterns and aims to use the least number of pixons to achieve this and reconstruct the image.}
\item{Vis FwdFit - imaging methods such as CLEAN or Pixon try to find a solution to Equation \ref{rhessi_vis} and often errors in the reconstructed images occur due to problems such as finite coverage in Fourier space. This can make estimating X-ray source spatial parameters difficult using CLEAN and Pixon \citep*{2010ApJ...717..250K}. Therefore, in certain situations, such as estimating X-ray source spatial parameters, it may be more helpful to use forward fitting methods. One such method is Vis FwdFit. This algorithm works by taking one of three simple shapes: a circular Gaussian, an elliptical Gaussian or a curved elliptical Gaussian, and matches the chosen shape with the X-ray visibilities given by Equation \ref{rhessi_vis}. Unlike other imaging algorithms, if a good comparison between the chosen model and the actual X-ray visibilities is achieved then the spatial properties of the X-ray source can be determined through the moments of the chosen Gaussian model: the position from the first moment and the spatial extent from the second moment. Another major advantage of this type of algorithm is the estimation of errors for each X-ray source spatial parameter, which can be found by propagating the error associated with the difference between the model and actual X-ray visibilities. The major drawback of this type of forward fitting method is the lack of source shapes, and hence only a limited number of X-ray sources that match well with one of the three simple Gaussian shapes, should ever by analysed using this method. However, the use of Vis FwdFit has, in the last couple of years, allowed the spatial properties of both chromospheric and coronal sources to be estimated, which is much harder with algorithms such as CLEAN and Pixon.}
\end{enumerate}
Vis FwdFit is discussed further in Chapters \ref{ref:Chapter2}, \ref{ref:Chapter3} and \ref{ref:Chapter5} of this thesis. An example of two CLEANed images is shown in Figure \ref{footpointsandloop}: one for HXR chromospheric footpoints at the limb and another of an X-ray coronal source, also at the limb. Over-plotted are Vis FwdFIt contours; an elliptical Gaussian was fitted to a HXR footpoint and a curved elliptical Gaussian was fitted to the coronal source. The parameters found from each fit are also plotted in the figure.
\subsection{RHESSI spectroscopy and polarimetry}
The nine cooled ($<75$ K) Germanium detectors make up the {\it RHESSI} spectrometer. The spectrometer has an energy resolution of $\le$ $1$ keV at $3$ keV and this increases to $\sim5$ keV at $5$ MeV \citep{2002SoPh..210....3L}. Each Ge detector consists of two segments: front and rear. The front segments can absorb photons up to around $\sim250$ keV and the rear segments up to $\sim17$ MeV. The photons that reach the detectors and cause a small current are registered as counts. Therefore, {\it RHESSI} produces a {\it count rate spectrum} and this is related to the {\it X-ray photon spectrum} via,
\begin{equation}\label{crs}
\textit{\textbf{C=B+SRM\;I}}
\end{equation}
where $\textit{\textbf{C}}$ is the count rate spectrum, $\textit{\textbf{B}}$ is the background, $\textit{\textbf{SRM}}$ is called the spectrometer or detector response matrix and $\textit{\textbf{I}}$ is the photon rate spectrum. {\it RHESSI} is an unshielded spacecraft and hence any background counts due to high energy cosmic rays and trapped high energy electrons and protons in the Earth's Van Allen belt, for example must be removed before a detailed spectroscopy of a flare is performed. The detector response matrix $\textit{\textbf{SRM}}$ accounts for the many effects that could modify the input count spectrum such as photons being Compton scattered in multiple detectors or detector radiation damage. If an effect only changes the efficiency of the instrument to detect a photon at its proper energy, then it contributes to the diagonal elements of the detector response matrix. If an effect changes the value of photon energy from its true value, then the effect contributes to the off-diagonal elements of the matrix \citep[cf.][]{2002SoPh..210...33S}. Each detector has two shutters or `attenuators'. The job of the attenuators is to stop the detectors saturating at high X-ray count rates, say, during a large solar flare. There are two different shutters: a thin and a thick shutter, and there are three attenuation states or combinations: A0 - no shutters, A1- thin shutter only and A3 - thin and thick shutter. The spectral analysis of a solar flare is performed in software called OSPEX \citep{2002SoPh..210..165S}, where different functions can be fitted to the data. {\it RHESSI} spectroscopy is performed in Chapter \ref{ref:Chapter3} and the functions fitted to the data are discussed there, for the flares analysed.

{\it RHESSI} is also capable of measuring the linear polarization of solar flare HXRs \citep{McConnelletal2002}. For this purpose from $20-100$ keV, {\it RHESSI} has a beryllium (Be) scatterer. {\it RHESSI} polarimetry is achievable since, as described above, the Ge detectors consist of two parts: a front and a rear segment. Photons in the range of $20-100$ keV should not be able to reach the rear segments and hence any detection of these low energy photons in the Ge rear segments must be due to firstly Compton scattering in the Be scatterer. The magnitude and direction of polarization is then found by measuring the scattered photon count rates in each rear segment of the detector. As mentioned in Section \ref{intro_pol}, {\it RHESSI} is also capable of measuring the polarization of higher energies above $100$ keV. For this, the same method is used, as described, except that the possible scattering between different Ge detectors is utilised at higher energies, instead of scattering from the Be scatterer. The simulated polarization measurements described in Chapter \ref{ref:Chapter6} are not currently possible with {\it RHESSI} or other dedicated astrophysical polarimeters, and hence Chapter \ref{ref:Chapter6} looks towards future instruments with imaging polarimetry capabilities.

\chapter{The variation of solar flare coronal X-ray source sizes with energy}
\label{ref:Chapter2}

\normalsize{\it This work can be found in the publication \cite{2014ApJ...787...86J}}

\section{Introduction to the chapter}\label{intro}
Chapter \ref{ref:Chapter1} discussed how, during a solar flare, the surrounding plasma is heated to tens of mega-Kelvin and electrons are accelerated to deka-keV energies and beyond. In a simple model, electrons travel through a tenuous corona and deposit energy into a dense chromospheric `thick target' via Coulomb collisions, with only a small fraction ($\sim 10^{-5}$) of the energy emitted as bremsstrahlung hard X-rays, mostly at the dense chromospheric footpoints.  Hard X-rays emitted from the corona are usually interpreted as predominantly thermal bremsstrahlung from a hot coronal plasma or as a combination of thermal and thin-target emissions.

Over the last decade, the Ramaty High Energy Solar Spectroscopic Imager \citep[{\em RHESSI};][]{2002SoPh..210....3L}
has provided unprecedented imaging spectroscopy observations of both chromospheric and coronal X-ray sources
\citep[for recent reviews of this topic see][]{2011SSRv..159..107H,2011SSRv..159..301K}.
Chapter \ref{ref:Chapter1} discussed how the design of the {\em RHESSI} instrument is such that spatial information is fundamentally encoded as two-dimensional
Fourier transforms, or {\it visibilities}.  The subsequent development of sophisticated and reliable visibility-based image reconstruction algorithms, such as visibility forward
fitting \citep{2002SoPh..210...61H,2007SoPh..240..241S}
and \textit{uv\_smooth} \citep{2009ApJ...703.2004M}, coupled with the use of {\it electron visibilities}, which are spectral inversions of the count visibility data provided by {\em RHESSI } \citep{2007ApJ...665..846P}, have
allowed the quantitative analysis of solar hard X-ray sources in both photon and electron space.

{\em RHESSI} observations have revealed the morphology details of flares with high plasma density \citep[e.g.,][]{1980ApJ...238L..43M,1981A&A....97..210C,1994ApJ...424..444F}, in which the bulk of the hard X-rays come from the corona, with only very weak or no footpoint emission from the chromosphere \citep[e.g.,][]{2004ApJ...603L.117V,2004ApJ...612..546S,2007ApJ...666.1256B,2008ApJ...673..576X,2013ApJ...769L..11L}.
The behaviour of the source extent with energy is not consistent with a thermal source characterised by a temperature distribution with a peak at the loop-apex, since for such a source the source size should decrease with energy.  Rather, the X-ray source extent {\it grows} with energy \citep{2008ApJ...673..576X}, indicative of a non-thermal model in which the propagation distance increases with energy. Apparently, the density within the coronal region in such sources is high enough to stop electrons prior to reaching the chromosphere; the source is a coronal ``thick target''.

Studying these events is particularly valuable since: (1) the coronal X-ray component and hence acceleration region can be studied without contamination from an intense chromospheric source; and (2) such sources exhibit trends in source extent with energy \citep{2008ApJ...673..576X, 2011ApJ...730L..22K, 2012ApJ...755...32G,2013ApJ...766...28G}
and time \citep[and Chapter \ref{ref:Chapter3}]{2013ApJ...766...75J}, which can be used to study particle acceleration and transport processes \citep[e.g.,][]{2012SoPh..277..299G,2013SoPh..284..489G}.  Further, unlike footpoint-dominated solar flares \citep[e.g.,][]{1982SoPh...78..107A,1982SoPh...81..137D,1995PASJ...47..355T,1996AdSpR..17...67S,1999ApJ...527..945P,2003ApJ...595L.107E,2004A&A...415..377M,2007A&A...461..315T,2011A&A...533L...2B,2011ApJ...731L..19F,2013ApJ...777...33C},
the HXR spectra of such ``coronal thick target sources'' tends to be softer than, and the sources higher than, chromospheric sources, which generally reduces the albedo contribution to X-ray images \citep[and see Chapter \ref{ref:Chapter5}]{KontarJeffrey2010}, making the interpretation of the spectro-spatial structure of such sources more straightforward.

Observations of compact coronal non-thermal hard X-ray sources typically show that the extent of the source parallel to the guiding magnetic field increases approximately quadratically with photon energy. Since the collisional stopping distance of an electron in a plasma also increases quadratically with particle energy, \cite{2008ApJ...673..576X} explained this behaviour in terms of an extended acceleration region, from which accelerated electrons emerge and subsequently undergo collisional transport in a background medium of uniform density.  As shown by \citet{2008AIPC.1039....3E}, application of such a model allows parameters such as the number density $n$ of the region and the specific electron acceleration rate $\eta$ (electrons~s$^{-1}$ per ambient electron) to be estimated.

However, the simple one-dimensional cold target approximation used by these authors is not completely realistic, for two main reasons. Firstly, it assumes that the injected electron trajectories are completely aligned with the guiding magnetic field, and it does not take into account pitch angle scattering (collisional or otherwise) of the accelerated electrons in the target. Secondly, it neglects effects associated with the finite temperature of the ambient medium; electrons with energies comparable to the thermal energy of the plasma $\sim k_BT$, where $k_{B}$ is the Boltzmann constant and $T$ is the temperature, are just as likely to gain
as lose energy during a collision, unlike the monotonic energy loss experienced by suprathermal electrons interacting with a cold plasma \citep[e.g.,][]{1978ApJ...224..241E}.  Even for electrons that do lose energy, they do so at a rate that is not the same as in a cold target, so that a quadratic behaviour of source extent with energy is not necessarily expected.

\cite{2003ApJ...595L.119E} and \cite{2005A&A...438.1107G} investigated analytically the effects of a finite target temperature, and both found that the associated velocity diffusion cannot be neglected when interpreting the results of flare hard X-ray spectra. \cite{2003ApJ...595L.119E} found that, because of the reduced energy losses suffered by accelerated electrons in  warm target, the inferred energy content of the injected electron distribution was significantly reduced. Indeed, he showed that allowance for this effect obviated the need to introduce a low-energy cutoff in the electron distribution. \cite{2005A&A...438.1107G} found that changes occurring close to the thermal energy of the plasma meant that many flare X-ray spectra may not be well fitted by a simple isothermal-plus-power-law model as discussed in Chapter \ref{ref:Chapter1}.

The motivation of Chapter \ref{ref:Chapter2} is to incorporate the effects of pitch angle scattering and finite target temperature in models of the variation of X-ray source size with electron energy.  How the inclusion of each of these processes changes the behaviour of the variation of source extent with electron energy and the estimation of parameters such as number density $n$ and acceleration region length $L_{0}$, is investigated. The conclusion (Section \ref{ref:chap2_con}) also briefly discusses how inferred parameters such as the filling factor $f$ and specific electron acceleration rate $\eta$ will change.

\section{Electron collisional transport in a cold plasma}\label{cold_theory}

Firstly electron transport within a uniform cold target is briefly reviewed. Here the electron energy $E>>k_BT$ where $k_B$ is Boltzmann's constant and $T$ is the target temperature.
The variation of energy $E$ [erg] with position $z$ [cm] in such a model is given by \citep[cf.][]{Brown1972,1978ApJ...224..241E}

\begin{equation}\label{eq:et}
E(E_{0},z)=\sqrt{E_{0}^{2}-2KN(z)}
=\sqrt{E_{0}^{2}-2Kn \, |z-z_{0}|} \,\,\, ,
\end{equation}
where $z_{0}$ is the (single) point of injection, $K=2\pi e^{4} \ln\Lambda$ (where $e$ [esu] is the electron charge and $\ln\Lambda$ the Coulomb logarithm), and $N$ and $n$ are the column density [cm$^{-2}$] along the trajectory and ambient number density [cm$^{-3}$], respectively.

This expression allows the stopping position $L_{S}$ of an electron of initial energy $E_{0}$ within a plasma of density $n$ to be found \citep[cf.][]{2002SoPh..210..373B}, viz.

\begin{equation}\label{eq: sd2}
L_{S}=z_{0}+\frac{E_{0}^{2}}{2Kn} \,\,\, .
\end{equation}

Using Equation~(\ref{eq:et}) and the one-dimensional continuity equation, the form of the electron spectrum $F(E,z)$ [electrons s$^{-1}$ cm$^{-2}$ erg$^{-1}$] as a function of position $z$ in the target can also be obtained:

\begin{equation}\label{eq: fex}
F(E,z)=F_{0}(E_0) \, \frac{dE_{0}}{dE} = F_{0}(E_0) \, \frac{E}{E_0}
= F_{0}(E_0[E,z]) \, E(E^{2}+2Kn|z-z_{0}|)^{-\frac{1}{2}} \,\,\, .
\end{equation}

Setting $z_{0}=0$ for simplicity and assuming a power-law injection spectrum $F_{0}(E_0) \propto E_{0}^{-\delta}$ gives,
\begin{equation}\label{eq:fex1}
F(E,z)=E(E^{2}+2Kn|z|)^{-\frac{(\delta+1)}{2}} \,\,\,,
\end{equation}
and an expression for the source extent as the square root of the variance $var(E)$ can be derived

\begin{equation}\label{eq: varfex1}
var(E)=\frac{\int_{0}^{\infty} z^2 \, F(E,z) \, dz}
{\int_{0}^{\infty} F(E,z) \, dz}=
\frac{\int_{0}^{\infty} z^2 \, (E^{2}+2Kn|z|)^{-(\delta+1)/2} \, dz}
{\int_{0}^{\infty} (E^{2}+2Kn|z|)^{-(\delta+1)/2} \, dz} \,\,\, ,
\end{equation}
where the symmetry about $z=0$ has been used.  Evaluating the integrals gives

\begin{equation}\label{eq:stdfex}
std(E)=\sqrt{var(E)}
=\frac{1}{2Kn} \, \sqrt{\frac{8}{(\delta-3)(\delta-5)}} \, E^{2} \,\,\,
\end{equation}
where $std(E)$ is the standard deviation. The spatial extent at a given energy $E$ depends on the spectral index $\delta$; for $\delta=7$ the form of the stopping distance is obtained, $std(E)=L_{s}$ given by Equation~(\ref{eq: sd2}). It should be noted that Equation (\ref{eq: varfex1}), and hence the spatial extent defined by Equation~(\ref{eq:stdfex}), is applicable only for $\delta >5$; for $\delta \leq 5$, the integral on the numerator diverges at the upper limit. This is related to the fact that the collisional stopping length is an increasing function of energy $\propto E^2$, so that large energies give the largest contribution to the integral for $\delta \leq 5$.  This issue can be formally avoided by imposing an upper energy cut-off $E_{max}$ to $F_0(E_0)$, so that the upper limit in the integral (\ref{eq: varfex1}) is finite, given by $E_{max}^2/2Kn$.

Equations \ref{eq: varfex1} and \ref{eq:stdfex} assumed the initial electron distribution was injected as a point source at $z=z_{0}=0$. However, if the initial electron distribution is injected over a finite region, with the injected flux profile having the form of (say) a Gaussian distribution with standard deviation $d$, then the equation  for $F(E,z)$ becomes \citep[see, e.g.,][]{2014ApJ...780..176K}
\begin{figure*}
\label{fig: fex_num_g_2}
\centering
\vspace{-20pt}
\includegraphics[width=15.5cm]{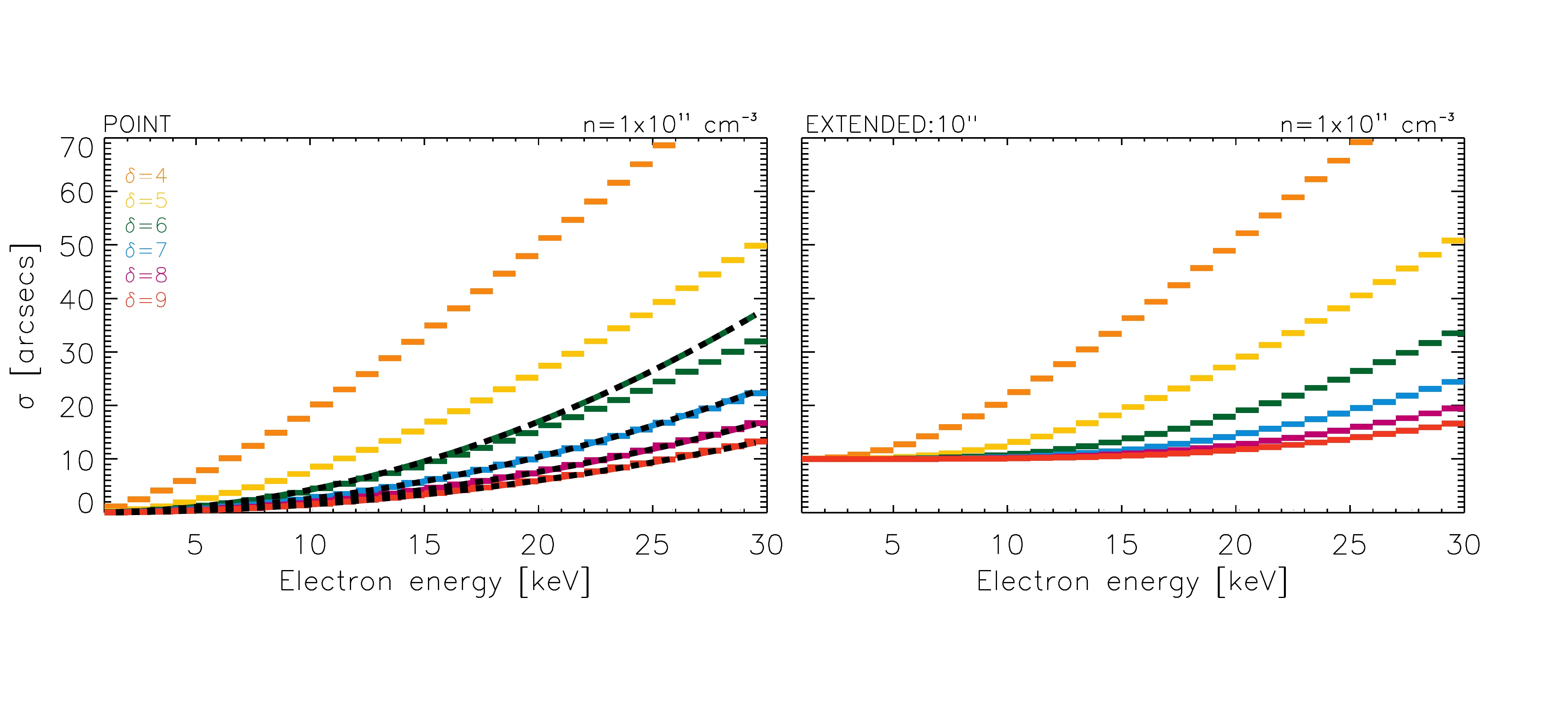}
\includegraphics[width=15.5cm]{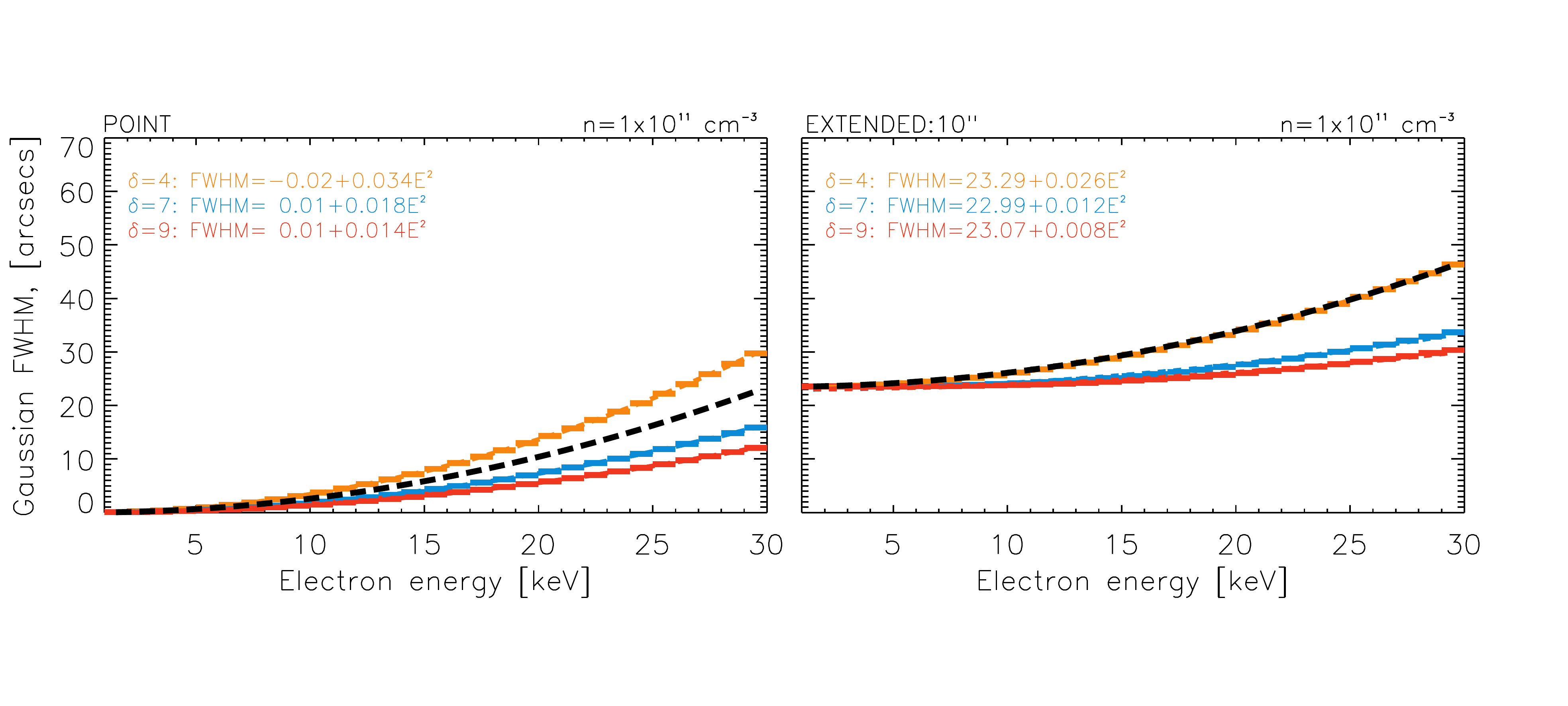}
\caption[The standard deviation and FWHM plotted against electron energy for a point source and a source of Gaussian standard deviation $d=10\arcsec$.]{{\it Top panels}: the standard deviation $\sigma$ calculated for a point source (left) and a source of Gaussian standard deviation $d=10\arcsec$ (right), using the moment-based Equation~(\ref{eq: varfex1}) and the distribution of electron flux with energy and position given by Equation~(\ref{eq:fexgauss}), for a target density $n=1\times10^{11}$~cm$^{-3}$. For the point source, the curves calculated using Equation~(\ref{eq:stdfex}) for $\delta=6-9$ and a maximum injected energy of $30$~keV are over-plotted as dashed lines of the same colour. {\it Bottom panels}: Gaussian FWHM calculated by fitting Gaussian curves to $F(E,z)$ for a point source (left) and a $10''$ source (right). Equation~(\ref{L}) is fitted to each curve:  the corresponding values of $L_0$ and $\alpha$ are shown on each panel. The curve FWHM $=2 \, \sqrt{2 \ln 2} \, d + E^2/2Kn$ \citep[black dashed curve; used by previous authors - e.g., ][]{2011ApJ...730L..22K} is overplotted.}
\label{fex_good}
\end{figure*}
\begin{equation}F(E,z)\label{eq:fexgauss}
\sim \frac{1}{d\sqrt{2\pi}} \, \int_{-\infty}^{\infty} E \, \left(E^{2}+2Kn \, \vert z-z^{'}\vert \right)^{-(\delta+1)/2}
\exp{\left(-\frac{{z^{'}}^2}{2d^{2}}\right)} \, dz^{'} \,\,\, .
\end{equation}\\
For this case, the evaluation of $F(E,z)$ and the corresponding standard deviation $std(E)$ cannot be evaluated analytically, and hence is calculated numerically. Figure~\ref{fex_good} (top) shows the numerical results for $std(E)$ for $\delta=4,5,6,7,8$ and $9$ using the initial source sizes of $d=0\arcsec\;\mbox{and}\;10\arcsec$ and a number density $n=1\times10^{11}$~cm$^{-3}$. For the $d=0\arcsec$ case, and for cases with $\delta > 5$ (cf. Equation~(\ref{eq:stdfex})), the $std(E)$ results calculated from the point-injection case~(Equation~(\ref{eq:stdfex})) are over-plotted for comparison and match well with the numerically calculated curves as expected.

The form of the spatially resolved spectrum $F(z)$ at a given energy $E$ at distances further away from the peak, where $F(E,z) \sim 0.15\max(F(E))$ (cf. Equation~(\ref{eq:fexgauss})) is not well determined by the {\em RHESSI} observations. Since {\em RHESSI} data is created as two-dimensional spatial Fourier transforms or X-ray visibilities (Chapter \ref{ref:Chapter1}, Section \ref{intro_rhessi}), the source sizes in practice are determined by fitting a Gaussian-like shape to the observed visibilities. Due to this indirect imaging approach and the finite dynamic range of the instrument, the brightest part of the image is the most reliable. Thus, $std(E)$ is calculated not through a moment-based approach, but rather through a shape-based analysis that focuses  on the high-intensity ``core'' of the spatial distribution of flux at a given electron energy $E$.

Therefore, Gaussian curves are fitted to $F(E,z)$ in order to determine the Gaussian standard deviation $std_{G}(E)$ at each energy. This allows the calculation of the Gaussian Full Width at Half Maximum FWHM$=2 \, \sqrt{2 \ln 2 \,var_{G}(E)}$.
The curves for $\delta= $4, 7, and 9 are plotted in the bottom panels of Figure~\ref{fex_good}. In general, and as expected, the $std_{G}(E)$ values deduced from the shape of the core of the $F(E,z)$ profile are smaller than the $std_{G}(E)$ values deduced from the moment-based analysis.

Each curve in Figure~\ref{fex_good} (bottom) was then fitted with an equation of the form

\begin{equation}
\mbox{FWHM}(E)=L_{0}+\alpha E^{2}
\label{L}
\end{equation}
and the values of $L_{0}$ and $\alpha$ are shown on each plot. In the bottom panels of Figure~\ref{fex_good},  FWHM$=2 \, \sqrt{2 \ln 2} \, d + E^2/2Kn$ is also over-plotted for comparison, since this simple approximation (basically the stopping distance approximation) has been
used \citep[e.g.,][]{2011ApJ...730L..22K,2012A&A...543A..53G} to infer information from observations; it is given by the black dashed curve.

From Figure~\ref{fex_good}, two main points are noted:

\begin{enumerate}
\item{For a given energy $E$, $std(E)$ decreases with increasing spectral index $\delta$.  This is because as $\delta$ increases there are a lower proportion of higher energy electrons in the overall electron distribution.  The lower energy electrons that are representative of steeper spectra travel a smaller distance through the plasma.}
\item{For a given spectral index $\delta$, the value of the quadratic coefficient $\alpha$ decreases somewhat with source size $d$. This is because of the increased contribution of the acceleration region to the overall source extent; the ``propagation'' region is to a large extent contained within the acceleration region itself.}
\end{enumerate}
Observationally, $L_{0}$ has been used to infer the size of the acceleration region, while $\alpha\propto 1/n$ has allowed the
number density of the propagation region to be inferred. This is assumed to be the same as the density of the acceleration region. Using the simplest one-dimensional cold plasma approximation ($\alpha=1/2Kn$), $n$ can be inferred easily. However, from Figure~\ref{fex_good}, it can be seen that, in general $\alpha=\mathcal{B}/2Kn$, where the value of the dimensionless number $\mathcal{B}$ and hence the number density $n$, depends upon the properties of both the acceleration region and the electron distribution.

Further, Equations~(\ref{eq:stdfex}) and~(\ref{eq:fexgauss}) do not account for three important processes expected to occur within a real flaring coronal plasma:
\begin{enumerate}
\item {a finite range of pitch angles in the injected pitch angle distribution,}
\item{any form of pitch angle scattering (collisional or non-collisional) within the target, and}
\item{the finite temperature of the plasma through which the electrons travel.}
\end{enumerate}
All of these physically important effects impact the form of $E(E_0,z)$, the variation of electron energy with position in the source, and incorporating them will thus change the resulting forms of $std(E)$ and FWHM$(E)$, in a manner which will now be investigated in the following sections of this chapter.

\section{Electron transport in a hot plasma with collisional pitch angle scattering}

\subsection{The Fokker-Planck Equation and coefficients}

In order to describe the transport of electrons through a coronal plasma of finite temperature $T$, accounting for collisional pitch angle scattering, a Fokker-Planck type equation can be used, as was briefly mentioned in Chapter \ref{ref:Chapter1}. For the purposes required in this investigation, a three-dimensional form from e.g., \cite{1981phki.book.....L,1986CoPhR...4..183K} in spherical coordinates is used. Assuming azimuthal symmetry and adding a source term for electrons $S$, this is given by

\begin{equation}\label{eq:fp}
\frac{df(v,z,\beta,t)}{dt}=\frac{\partial f}{\partial t}+v\cos\beta \frac{\partial f}{\partial z}=-\frac{1}{v^{2}} \, \frac{\partial}{\partial v}
\left(v^{2} \, J_{v}\right) - \frac{1}{v\sin\beta}\frac{\partial}{\partial \beta}
\left(\sin\beta \, J_{\beta}\right) +S(v,z,\beta,t),
\end{equation}
where $f(v,z,\beta,t)$ is the phase-space distribution function [electrons~cm$^{-3}$~{[cm~s$^{-1}$]}$^{-3}$], $v$ [cm~s$^{-1}$] is the total particle speed, $\beta$ is the particle pitch angle to the guiding magnetic-field (along the direction $z$ [cm]), $t$ is time [s] and $J_{v}$ and $J_{\beta}$ are given by

\begin{equation}\label{eq:SS}
J_{v}=-D_{vv} \, \frac{\partial f}{\partial v} + F_{v} \, f
 \,\,\, , \qquad J_{\beta}=-D_{\beta\beta} \, \frac{1}{v} \, \frac{\partial f}{\partial \beta} \,\,\, .
\end{equation}
Here $D_{vv}$ and $D_{\beta\beta}$ are the velocity and pitch angle diffusion terms while $F_{v}$ is the velocity collisional friction term. These three terms are respectively given by

\begin{eqnarray}\label{eq:dcvv}
D_{vv}=\frac{\Gamma}{2v} \, \left(\frac{{\rm erf}(u)}{u^{2}
}-\frac{{\rm erf}^{'}(u)}{u}\right) &\equiv & \frac{\Gamma}{v} \, G(u)
\cr
D_{\beta\beta}=\frac{\Gamma}{4v}\left( \left[2-\frac{1}{u^{2}}\right]{\rm erf}(u)
+ \frac{{\rm erf}^{'}(u)}{u} \right) &\equiv & \frac{\Gamma}{2v}\biggl ( {\rm erf}(u)-G(u) \biggr )
\cr
F_{v}=-\frac{\Gamma}{v^{2}}\left({\rm erf}(u)-u \, {\rm erf}^{'}(u)\right)
&\equiv & -\frac{2 \, \Gamma}{v^{2}} \, u^{2} \, G(u) \,\,\, ,
\end{eqnarray}
where the dimensionless velocity $u=v/(\sqrt{2} \, v_{th})$, $v_{th}=\sqrt{k_{B}T/m_{e}}$ where $k_{B}$ is the Boltzmann constant [erg K$^{-1}$], $T$ is the temperature of the background plasma [K] and $m_{e}$ is the electron rest mass [g], $\Gamma=4\pi e^{4} \ln\Lambda \, n /m_{e}^{2}$ where $e$ [esu] is the electron charge, $n$ is the number density [cm$^{-3}$] and $\ln\Lambda$ is the Coulomb logarithm. ${\rm erf}(u)$ is the error function and $G(u)$ is the Chandrasekhar function,

\begin{equation}\label{eq:gcha}
{\rm erf}(u)\equiv (2/\sqrt{\pi})\int\limits_{0}^{u}\exp(-t^2) \, dt
\,\,\,\,\,\,\,{\rm and}\,\,\,\,\,\,\,
G(u)=\frac{{\rm erf}(u)-u \, {\rm erf}^{'}(u)}{2u^{2}} \,\,\, .
\end{equation}
Substituting into the Fokker-Planck equation~(\ref{eq:fp}) gives

\begin{eqnarray}\label{eq:fpv}
\frac{d f(v,z,\beta,t)}{d t}
&= & \frac{\Gamma}{2v^{2}} \left \{ \frac{\partial}{\partial v}\left(2 \, v \, G(u) \, \frac{\partial f(v,z,\beta,t)}{\partial v}
+4 \, G(u) \, u^{2} \, f(v,z,\beta,t)\right) + \right . \cr
&+ & \left . \frac{1}{v\sin\beta} \, \frac{\partial}{\partial \beta}\left(\sin\beta \, \biggl [ {\rm erf}(u) -G(u) \biggr ] \, \frac{\partial f(v,z,\beta,t)}{\partial \beta} \right) \right \}+S(v,z,\beta,t). \,\,\,\,\,\,\,\,\,\,\,\,\,\,\,\,\,\,
\end{eqnarray}
\\
Current imaging spectroscopy X-ray observations with instruments such as {\em RHESSI} have a time resolution the order of several seconds; it takes a full spacecraft rotation period $\sim$4 s to yield a reliable image, which is much longer than the timescale for transport of deka-keV electrons ($v \sim 10^{10}$~cm~s$^{-1}$) along the typical length of a coronal loop ($\sim 10^9$~cm).  Therefore, it is appropriate to consider the time-independent case. It is also convenient to convert from the variable $\beta$ to the variable $\mu=\cos{\beta}$, giving

\begin{eqnarray}\label{eq:fpvxmu}
\mu \, v \, \frac{\partial f(v,z,\mu)}{\partial z}
&= & \frac{\Gamma}{2v^{2}} \left \{ \frac{\partial}{\partial v}\left(2 \, v \, G(u) \, \frac{\partial f(v,z,\mu)}{\partial v}
+4 \, u^{2} \, G(u) \, f(v,z,\mu)\right) + \right . \cr
&+ & \left . \frac{1}{v} \, \frac{\partial}{\partial \mu}\left( (1-\mu^2) \, \biggl [ {\rm erf}(u) -G(u) \biggr ] \, \frac{\partial f(v,z,\mu)}{\partial \mu} \right) \right \}+S(v,z,\mu). \,\,\,\,\,\,\,\,\,\,\,\,\,\,\,\,\,\,
\end{eqnarray}
\\
It is assumed that the source term $S(v,z,\mu)$ is separable in $v,\;\mu$ and $z$, with the spatial variation assumed to have a Gaussian form:

\begin{equation}\label{eq:source_v}
S(v,z,\mu)=f_{0}(v) \, \frac{1}{\sqrt{2\pi d^{2}}}\exp{\left(-\frac{z^{2}}{2d^{2}}\right)} \, H(\mu) \,\,\, ,
\end{equation}
where $f_{0}(v)$ and $H(\mu)$ are the initial velocity and pitch angle distribution functions. Equation~(\ref{eq:fpvxmu}) describes the evolution of an injected electron distribution through a non-evolving finite temperature background Maxwellian distribution.

\subsection{Steady-state solution}
For a background plasma with a finite temperature $T$, the input velocity distribution will evolve to a thermal distribution of the form

\begin{equation}\label{eq:fv_the}
f(v)\sim\exp{\left(-\frac{m_{e}v^{2}}{2k_{B}T}\right)} \,\,\, ,
\end{equation}
leading to an average kinetic energy of
\begin{equation}\label{eq:fv_kT}
\left\langle\frac{m_{e}v^{2}}{2}\right\rangle
=\frac{\int_{0}^{\infty}\frac{m_{e}v^{2}}{2}f(v)d^{3}v}{\int_{0}^{\infty}f(v)d^{3}v}
=\frac{3}{2} \, k_{B}T \,\,\, .
\end{equation}

\subsection{High velocity limit}
In the high electron velocity limit $u\gg1$, one finds  ${\rm erf} (u)\rightarrow 1$ and $G(u)\rightarrow 1/2u^2=(v_{th}/v)^{2}$. In this limit Equation~(\ref{eq:fpvxmu}) becomes

\begin{eqnarray}\label{eq:fpvxmu_he}
\mu \, v \, \frac{\partial f(v,z,\mu)}{\partial z}
&= & \frac{\Gamma}{v^2} \left \{ \frac{\partial}{\partial v} \left( \frac{v_{th}^2}{v} \frac{\partial f(v,z,\mu)}{\partial v} \,
+ \, f(v,\mu,z)\right) + \right . \cr
&+ & \left . \frac{1}{2v}
\frac{\partial}{\partial \mu} \left( (1-\mu^2) \, \frac{\partial f(v,z,\mu)}{\partial \mu} \right) \right \}  + S(v,z,\mu)\,\,\, .
\end{eqnarray}

\subsection{Cold plasma limit}
If the temperature of the plasma is also small compared to the typical particle energies, then it can be formally taken that $T=0$ (that is, $v_{th}=0$).  Equation (\ref{eq:fpvxmu_he}) then becomes

\begin{equation}\label{eq:fpvxmu_noT}
\mu \, \frac{\partial f(v,z,\mu)}{\partial z}
= \frac{\Gamma}{v^3} \, \left \{ \frac{\partial f(v,z,\mu)}{\partial v} + \frac{1}{2 v} \, \frac{\partial}{\partial \mu} \left( (1-\mu^2) \, \frac{\partial f(v,z,\mu)}{\partial \mu} \right) \right \}  + S(v,z,\mu)\,\,\, ,
\end{equation}
which is the transport equation for a cold plasma with azimuthal symmetry, a more familiar form often used in solar physics
\citep[e.g.,][]{1981SvA....25..215K}.

\subsection{Conversion to the electron flux distribution}
The electron flux spectrum $F(E,z,\mu)$ [electrons~cm$^{-2}$~s$^{-1}$~keV$^{-1}$] as a function of field-aligned coordinate $z$ [cm], energy $E$ [keV] and pitch angle cosine $\mu$ is related to the three-dimensional phase-space distribution function $f(v,z,\mu)$ by

\begin{equation}\label{con_fF1}
v \, f(v,z,\mu) \, d^{3}v = v \, f(v,z,\mu) \, v^2 \, dv = F(E,z,\mu) \, dE \,\,\, ,
\end{equation}
so that
\begin{equation}\label{con_fF}
f(v,z,\mu)=\frac{dE}{dv}\frac{1}{v^{3}}\,F(E,z,\mu)= \frac{m_{e}}{v^{2}} \, F(E,z,\mu) = \frac{m_{e}^{2}}{2E} \, F(E,z,\mu) \,\,\, .
\end{equation}
\\
Using this relation, the Fokker-Planck equation~(\ref{eq:fpvxmu}) can be re-written in terms of electron energy $E$
and the electron flux distribution $F(E,z,\mu)$, which is a more useful form for comparison with observations.  The result is

\begin{eqnarray}\label{eq: fp_e}
\mu \, \frac{\partial F}{\partial z} &= & \Gamma m_{e}^2 \left \{ \frac{\partial}{\partial E} \left[ G (u[E] ) \, \frac{\partial F}{\partial E} + \frac{G (u[E] )}{E} \, \left ( \frac{E}{k_B T}-1 \right ) \, F \right] + \right . \cr
&+ & \left . \frac{1}{8E^2} \, \frac{\partial}{\partial \mu} \left [ (1-\mu^{2}) \biggl ( {\rm erf} (u[E] ) - G (u[E] ) \biggr ) \, \frac{\partial F}{\partial \mu} \right ] \right \} + S_{F}(E,z,\mu),\,\,\,\,
\end{eqnarray}
\\
where $u(E)=\sqrt{E/k_B T}$ is used.
The solar corona also contains elements other than hydrogen, and for an element with atomic number $Z$, the Coulomb energy loss scales as $Z^2$ \citep[e.g.,][]{1978ApJ...224..241E}, and these additional elements are accounted for by adopting an effective atomic number $Z_{eff}=\sum_{i} n_{i} Z_{i}^{2}/\sum_i n_i$. Defining $\Gamma_{eff}=\Gamma Z_{eff} m_{e}^{2}$ and $G(u)=G\left(\sqrt{\frac{E}{k_{B}T}}\right)$, Equation \ref{eq: fp_e} becomes

\begin{eqnarray}\label{eq: fp_e_zefftext}
\mu \, \frac{\partial F}{\partial z} &= &
\Gamma_{eff} \left \{ \frac{\partial}{\partial E} \left [ G \left (\sqrt{\frac{E}{k_B T}} \, \right ) \, \frac{\partial F}{\partial E} +\frac{1}{E} \, \left ( \frac{E}{k_B T} - 1 \right ) \, G \left (\sqrt{\frac{E}{k_B T}} \, \right ) \, F \right ] \right .
+ \cr
&+ & \left . \frac{1}{8E^2} \, \frac{\partial}{\partial \mu} \left [ (1-\mu^{2}) \, \left ( {\rm erf} \left ( \sqrt{\frac{E}{k_B T}} \, \right ) -G\left (\sqrt{\frac{E}{k_B T}} \, \right ) \right ) \frac{\partial F}{\partial \mu} \right ]\right \} 
+ \cr
&+ &
+ S_{F}(E,z,\mu)),
\end{eqnarray}
The source term again consists of three separable functions for $E,\,\mu$ and $z$:
\begin{equation}\label{eq:source}
S_{F}(E,z,\mu)) = F_{0}(E) \; \frac{1}{\sqrt{2\pi d^{2}}} \, \exp{\left(-\frac{z^{2}}{2d^{2}}\right)} \, H(\mu) \,\,\, ,
\end{equation}
where $F_0(E)\propto E_{0}^{-\delta}$ and $H(\mu)$ describe the forms of the initial energy spectrum and pitch angle distribution, respectively.

\subsection{Derivation of the stochastic differential equations}
For use in the simulations, Equation \ref{eq: fp_e_zefftext} must be converted to a set of stochastic differential equations (SDE) for the field-aligned coordinate $z$ [cm], energy $E$ [keV] and pitch angle cosine $\mu$. If the source term is ignored, focusing on electron transport, Equation~(\ref{eq: fp_e_zefftext}) can be rewritten in the form

\begin{equation}\label{eq: fp_eABin}
\mu \, \frac{\partial F}{\partial z}
=\frac{\partial}{\partial E}\biggl ( A_E(E) \, F \biggr )
+ \frac{\partial^{2}}{\partial E^{2}} \biggl ( D_{EE}(E) \, F \biggr )
+\frac{\partial}{\partial \mu} \biggl ( A_\mu(E,\mu) \, F \biggr )
+\frac{\partial^2}{\partial \mu^2} \biggl ( D_{\mu\mu}(E,\mu) \, F \biggr )  \,\,\, ,
\end{equation}
where the coefficients are given by
\begin{align}\label{eq:A-D}
A_E(E) &=&  \frac{\Gamma_{eff}}{2E} \left [ {\rm erf} \left (\sqrt{\frac{E}{k_B T}} \, \right ) -2 \sqrt{\frac{E}{k_B T}} \, {\rm erf}^{'} \left (\sqrt{\frac{E}{k_B T}} \, \right ) \right ] \,\,\, && \cr
&\equiv& \frac{\Gamma_{eff}}{2E} \, g_{th} \left (\sqrt{\frac{E}{k_B T}} \right )\,\,\, &;& \cr
D_{EE}(E)\equiv\frac{1}{2} \, B_E^2(E) &=&   \, \Gamma_{eff} \, G \left (\sqrt{\frac{E}{k_B T}} \, \right )  \,\,\, &;&
\cr
A_\mu(E,\mu) &=& \frac{\mu \, \Gamma_{eff}}{4E^2} \left [ {\rm erf} \left ( \sqrt{\frac{E}{k_B T}} \, \right ) -G \left ( \sqrt{\frac{E}{k_B T}} \, \right ) \right ] \,\,\, &;&
\cr
D_{\mu\mu}(E,\mu)\equiv\frac{1}{2} \, B_\mu^2(E,\mu) &=&
 \frac{(1-\mu^{2}) \, \Gamma_{eff}}{8 E^2} \, \left [ {\rm erf}\left (\sqrt{\frac{E}{k_B T}} \, \right ) - G \left (\sqrt{\frac{E}{k_B T}} \, \right ) \right ] \,\,\, &.&
\end{align}

This general form of the Fokker-Planck equation is equivalent to the following stochastic differential equations (SDE) for $E$ and $\mu$ in the It\^o form \cite[cf.][]{1991A&A...251..693M,1994hsmp.book.....G}

\begin{equation}\label{eq:sto_Emu}
{dE}= - \, A_E \, ds + B_E \, \, dW_{E}
 \,\,\, ; \qquad {d\mu}= - \, A_\mu \,ds  + B_\mu \, \, dW_{\mu} \,\,\, ,
\end{equation}
where the independent Wiener processes $W_{\mu}$ and $W_{E}$ are stochastic processes with independent increments. These two equations suggest the numerical stepping algorithm

\begin{equation}\label{eq:sto_x}
z_{j+1}=z_{j}+\mu_{j} \, \Delta s \,\,\, ;
\end{equation}
\begin{equation}\label{eq:sto_E}
E_{j+1}=E_{j}-\frac{\Gamma_{eff}}{2E_{j}} \, g_{th}(u_j) \, \Delta s
+\sqrt{2 \, \Gamma_{eff} \, G(u_{j}) \, \Delta s} \,\, W_{E} \,\,\, ; \,\,\,
\end{equation}
\begin{equation}\label{eq:sto_mu}
\mu_{j+1}=\mu_{j}-\frac{\Gamma_{eff} \biggl ( {\rm erf}(u_{j})-G(u_{j}) \biggr ) } {4 E_{j}^{2}} \,\, \mu_{j} \, \Delta s
+\sqrt{\frac{ (1-\mu_{j}^{2}) \, \Gamma_{eff} \, \biggl ( {\rm erf}(u_{j})-G(u_{j}) \biggr ) } {4 E_{j}^{2}} \, \Delta s} \, \, W_{\mu} \,\,\, ,
\end{equation}
where $u_j=\sqrt{E_j/k_B T}$ and $W_E$ and $W_{\mu}$ are drawn at random from the Gaussian distribution $N(0,1)$ such that $\langle W_{\mu}\rangle=\langle W_{E}\rangle  =0$, $\langle W^2_{\mu}\rangle=\langle W^2_{E}\rangle= 1$. Equations~(\ref{eq:sto_x}) through~(\ref{eq:sto_mu}) are
the form of the SDEs used in the numerical simulations, which must be amended for low energies. This will be discussed in Section \ref{low-energy-limit}.

It should be noted that a root mean square (rms) atomic number of $Z_{eff}=1$ is taken for simplicity (that is, a pure hydrogen target), but the equation for a general $Z_{eff}$ is provided as it may prove useful in other studies.

\begin{sidewaysfigure*}[ht]
\centering
\includegraphics[width=23cm]{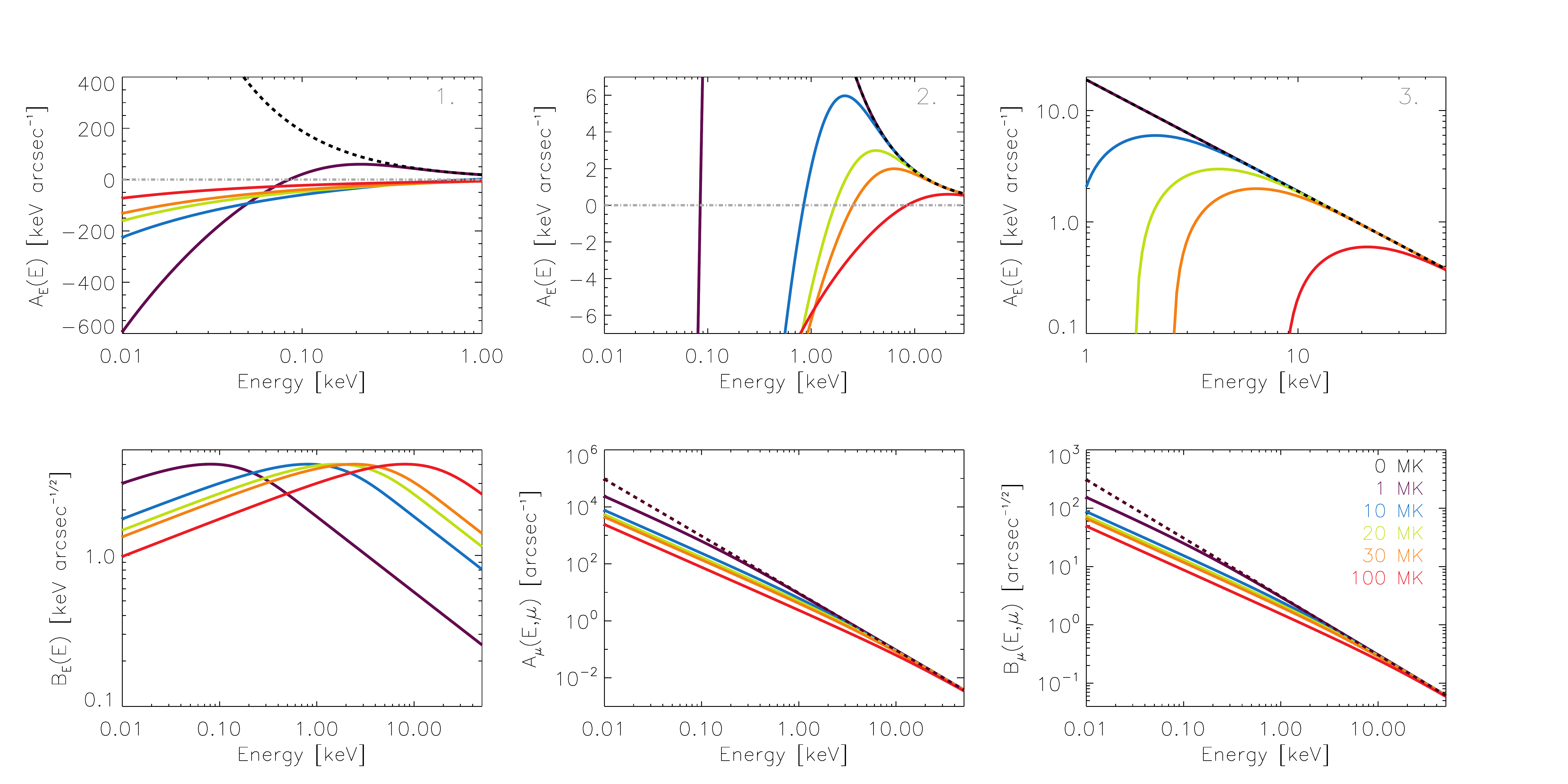}
\caption[Plots of the energy $A_E$, $B_{E}$ and pitch angle $A_\mu(E,\mu=1)$, $B_{\mu}(E,\mu=0)$ coefficients against
electron energy $E$ for different plasma temperatures from $T=0 - 100$~MK.]{Plots of the energy $A_E$, $B_{E}$ and pitch angle $A_\mu(E,\mu=1)$, $B_{\mu}(E,\mu=0)$ coefficients against
electron energy $E$ for different plasma temperatures from $T=0 - 100$~MK. The colours corresponding to each plasma temperature are shown on the bottom right plot. The $A_E$ coefficient is plotted three times (top row) so that all the features in different energy ranges can be seen clearly.}
\label{fig:check_ABCD}
\end{sidewaysfigure*}

The coefficients $A_E$, $A_\mu$, $B_E (=\sqrt{2D_{EE}})$ and $B_\mu (=\sqrt{2D_{\mu\mu}})$ are plotted against energy $E$ in Figure~\ref{fig:check_ABCD}, for a number of different plasma temperatures $T$ ranging from $T = 0$ (cold plasma) to $T=100$~MK. For ease of presentation, the $A_\mu$ and $B_\mu$ terms are shown as a function of $E$ for a fixed value of $\mu$ ($\mu=1$ for $A_\mu$ and $\mu=0$ for $B_\mu$). Below an energy $E_c\simeq k_BT$ the coefficient $A_E$ becomes negative; that is, electrons on average {\it gain} energy; the value of $E_c$ for which $A_E=0$ increases linearly with the ambient
\clearpage
temperature. In order that these features can be seen clearly, the coefficient $A_E$ is plotted (top row of Figure~\ref{fig:check_ABCD}) over three different energy ranges: two (below 1~keV, and below 30~keV) plotted on linear $y$-axes and 1-50~keV plotted on a logarithmic $y$-axis.  Further, the stochastic term $B_{E}$ peaks at $\simeq k_B T$. Therefore, in a warm plasma, electrons with $E \sim k_B T$ are more likely to gain energy, both secularly and through diffusion, rather than to lose it.

\begin{figure*}
\centering
\includegraphics[scale=0.8]{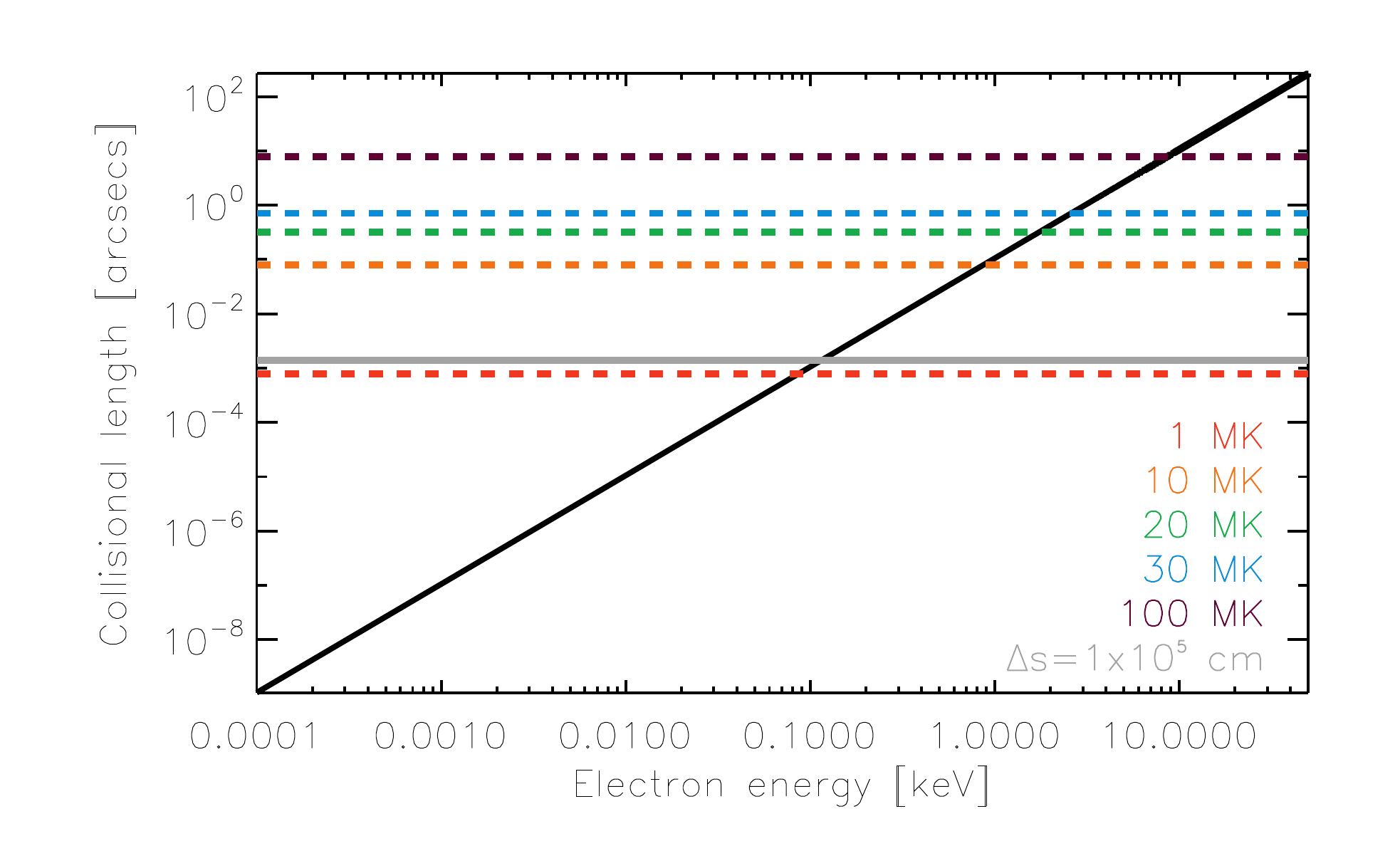}
\caption[Electron collisional length versus electron energy in a cold plasma (black) with the thermal collisional lengths over-plotted for $T=1$, $10$, $20$, $30$ and $100$ MK.]{Electron collisional length versus electron energy in a cold plasma (black) with the thermal collisional lengths over-plotted for $T=1$, $10$, $20$, $30$ and $100$ MK. In the simulations, sensible flaring plasma temperatures of $T=10$, $20$ and $30$ MK are used and hence a stochastic distance step is chosen to be $\Delta s=1\times10^{5}$ cm, which is at least one order of magnitude smaller than the thermal collisional length for $T=10$ MK.}
\label{plot_tau}
\end{figure*}

To get reliable results from the simulations, an appropriate value of the length step $\Delta s$ (Equations~(\ref{eq:sto_x}) through~(\ref{eq:sto_mu})) must be chosen. This was chosen by calculating the thermal collision length (mean free path) $\lambda_c(E)$ and ensuring that $\Delta s$ was much smaller than $\lambda_c$ for all $E$ of interest. The thermal collisional length is given by $\lambda_{c}=v\tau_{c}$, where $\tau_{c}$ is the thermal collisional time as discussed in Chapter \ref{ref:Chapter1}.  The mean-free path $\lambda_c$ for a 1~keV electron in a cold target of density $n=1 \times 10^{11}$~cm$^{-3}$ is approximately $10^6$~cm; the mean-free paths in warm targets are even greater.  For all simulations, a length step $\Delta s=1\times10^{5}$~cm is used, much smaller than the mean free path in all cases and this is shown in Figure \ref{plot_tau}.

\subsection{The low-energy limit}\label{low-energy-limit}

As the plots in Figure~\ref{fig:check_ABCD} show, $A_E$, $A_\mu$ and $B_{\mu}$ diverge as $E\rightarrow 0$. Therefore, following \citet{2009JCoPh.228.1391L} and \citet{2010ITPS...38.2394C}, for low energies $E$ the finite difference Equation~(\ref{eq:sto_E}) is replaced with an analytic expression for the energy evolution. To obtain this expression, the functions ${\rm erf}(u)$ and ${\rm erf'}(u)$ for small $u$ are expanded in a MacLaurin series, so that the coefficients $A_E$ and $B_E$ become

\begin{equation}\label{low_e_A}
A_E = \frac{\Gamma_{eff}}{2E} \left ( {\rm erf}(u)-2u \, {\rm erf}^{'}(u) \right )
\simeq - \, \frac{\Gamma_{eff}}{\sqrt{\pi}E} \, u \,\,\, ,
\end{equation}

\begin{equation}\label{low_e_B}
B_E = \sqrt{2 \, \Gamma_{eff} \, G (u)}
\simeq \sqrt{\frac{4 \, \Gamma_{eff}}{3 \, \sqrt{\pi}} \, u} \,\,\, ,
\end{equation}
and in the low-energy limit, $E\rightarrow 0$, the energy equation becomes

\begin{equation}\label{eq:low_de_sde}
\frac{dE}{ds} \simeq \frac{\Gamma_{eff}}{E} \, \sqrt{\frac{E}{\pi k_{B} T}} + \left ( \frac{4 \, \Gamma_{eff}}{3} \, \sqrt{\frac{E}{\pi k_B T}} \, \right )^{1/2} W^{E} \,\,\, .
\end{equation}
For low values of $E$, the second (stochastic) term can be neglected in comparison with the first (secular) term to give

\begin{equation}\label{eq:low_de_sde_A}
\frac{dE}{ds}\simeq \frac{\Gamma_{eff}}{\sqrt{\pi k_{B} T}} \, \frac{1}{\sqrt{E}} \,\,\, ,
\end{equation}
which can be integrated analytically, giving

\begin{equation}\label{eq:low_e_sde_A_2}
E=\left[E_{0}^{3/2}+\frac{3 \, \Gamma_{eff}}{2\sqrt{\pi k_{B} T}} \, (s-s_0)\right]^{2/3} \,\,\, .
\end{equation}
Equation (\ref{eq:low_e_sde_A_2}) was used for energies below

\begin{equation}
E_{low} = \left[\frac{3\Gamma_{eff}}{2\sqrt{\pi k_{B} T}} \, \Delta s  \right]^{2/3}\,\,\, ,
\end{equation}
thus guaranteeing that $E \ge 0$ everywhere. To avoid divergence, the pitch angle cosine $\mu$ for energies $E\leq E_{low}$ was sampled from a uniform distribution between $-1$ and $1$. Since electrons with $E_{low}<<\bar{E}$, where $\bar{E}$ is the average thermal energy of the plasma, are likely to be part of the background thermal distribution, then it is sensible to draw their pitch angle from an isotropic distribution.
\\\\
In the cold plasma limit $T\rightarrow0$, the stochastic equation for $E$ becomes

\begin{equation}\label{eq:cold_E}
E_{j+1}=E_{j}-\frac{\Gamma_{eff}}{2 E_{j}} \, \Delta s \,\,\, ,
\end{equation}
which can be solved to give the usual cold target result

\begin{equation}
E_{j+1}=\sqrt{E^2_{j} - 2 K n \, \Delta s} \,\,\, ,
\end{equation}
where $K=2\Gamma_{eff}/n$. In this limit, the pitch angle behaviour is given by

\begin{equation}\label{eq:cold_mu}
\mu_{j+1}=\mu_{j}-\frac{\Gamma_{eff}}{4 E_{j}^{2}} \, \mu_{j} \, \Delta s + \sqrt{\frac{\Gamma_{eff}}{4 E_{j}^2} \, (1-\mu^{2}) \, \Delta s} \,\, W_{\mu}\,\,\, .
\end{equation}

\section{Simulations}

The aim of the simulations is to determine how collisional pitch angle scattering and the finite temperature of the target plasma affect the transport of electrons through the plasma compared to the one-dimensional cold target result, and hence to determine how the observed length of a hard X-ray source varies with electron energy in a more realistic physical scenario.  The simulations use the stochastic equations for $z$, $E$, and $\mu$ given by Equations~(\ref{eq:sto_x}) through~(\ref{eq:sto_mu}) with initial conditions for each injected electron provided by sampling the source term $S(E,z,\mu)$ -- see Equation (\ref{eq:source}). The simulations model the evolution of an injected distribution of electrons, moving either within a cold plasma or a plasma of finite temperature, they do not account for the evolution of the background plasma; the properties of the background plasma remain constant throughout a simulation.

\subsection{Simulation input, boundary and end conditions}\label{sec:input}

All simulations use a common set of certain input parameters.  The electron number density is set to $n = 1 \times 10^{11}$~cm$^{-3}$, a relatively high value for the coronal density, but one which is necessarily high in order for the deka-keV electrons to be stopped in the corona and which is chosen to correspond to recent analyses of thick target coronal sources \citep[e.g.,][]{2008ApJ...673..576X,2011ApJ...730L..22K,2013ApJ...766...75J}.
For the Coulomb logarithm a typical coronal value of $\ln{\Lambda}=20$ is used. The plasma temperature is assumed uniform
along the $z$ direction, at a value of either $0$~MK, $10$~MK, $20$~MK or $30$~MK. The initial spatial distribution of injected
flux (``acceleration region size'') is assumed to be a Gaussian centred at $z=0$ (which is the position of the coronal loop apex) with an input standard deviation of $d=10\arcsec$, corresponding to a FWHM$=2\sqrt{2\ln 2} \, d=23\arcsec.5$.  The initial pitch angle distribution is taken to be either {\it completely beamed} (that is, half the distribution has $\mu=1$ and the other half $\mu=-1$) or {\it isotropic}. The injected electron energy flux distribution $F_0(E)$ has the form of a power law with spectral index $\delta=4$ or $\delta=7$, up to a maximum energy of $50$ keV, above which the energy-integrated electron flux is negligibly small.

The upper boundary of the $z$ domain is set at a value sufficiently large that no electrons ever {\it spatially} leave the region of computation. For the runs that use the cold target energy loss formula, electrons lose energy monotonically.  Hence an electron is removed from the simulation once its  energy is below $1$~keV. However, for the warm target simulations, electrons of very low energy can still gain energy through Coulomb collisions with more energetic neighbours, as the ensemble evolves to a thermal (Maxwellian) distribution.  Thus electrons in the warm target runs are {\it never} removed; for such runs the particle number is conserved and the electron distribution asymptotically approaches the Maxwellian distribution $F(E)\sim E \exp (-E/k_B T)$.  For this distribution, the flux-averaged energy is

\begin{equation}\label{average-energy}
{\overline E} = \frac {\int^{\infty}_{0} E \, F(E) \, dE}  {\int^{\infty}_{0} F(E) \, dE} = 2 \, k_{B} T \,\,\, .
\end{equation}
Therefore a simulation is terminated when the average energy of the distribution is $2 k_B T$ and the pitch angle distribution becomes approximately isotropic, conditions that approximate the essential features of a Maxwellian. Note that $\overline E$ is {\it not} the average kinetic energy of the three-dimensional phase space distribution $f(v,\mu,z)$ (which is $\langle{mv^{2}}/{2}\rangle=\frac{3}{2}k_{B}T$). After each distance step $\Delta s$, the values of the electron distribution function $F(E,\mu,z)$ are saved into an array. These arrays represent the distribution functions resulting from the continuous injection of electrons with the source function given by Equation (\ref{eq:source}).

\subsection{Gaussian fitting and the determination of the source length FWHM}

The arrays generated by each simulation are energy-binned to give $F(z,\mu)$ in increasing energy bins from $1$~keV to $30$~keV.
The longitudinal extent of the source could be identified as the standard deviation $std(E)=\sqrt{var(E)}$ of the $F(z,\mu)$ spatial distribution in each energy bin, calculated from the second spatial moment
of $F(z,\mu)$.  However, in part because the injected flux
distribution is {\it assumed} to be Gaussian, the forms of $F(z,\mu)$ generally also closely resemble Gaussian forms,
excluding relatively low-intensity components at high $|z|$.  Therefore, just as in Section~\ref{cold_theory}, a Gaussian distribution is instead fitted to each $F(z,\mu)$ distribution and thus used to determine the associated standard
deviations $std_{G}(E)$ and corresponding FWHM$=2\sqrt{2 \ln 2} \, std_{G}(E)$ in each energy bin. In this way, the extent of the source is characterised through the {\it shape} of its core spatial form, rather than through a {\it moment} of the entire distribution. Again, as in Section~\ref{cold_theory}, FWHM$(E)=L_{0}+\alpha E^{2}$ (Equation~(\ref{L})) was fitted to the FWHM versus electron energy results, and values of $\alpha$ and $L_0$ found.

For a cold plasma with an initially beamed pitch angle distribution and no collisional pitch angle scattering, it is expected that $L_{0}=L_{init}=2\sqrt{2 \ln 2} \, d$, the Gaussian FWHM of the input distribution, has a value of $\alpha$ equal
to that found numerically from the fits to $\delta=4$ and $\delta=7$ curves in Figure~\ref{fex_good}. However,
the presence of a finite plasma temperature $T$, an initially broad pitch angle distribution, and/or collisional pitch angle scattering will all change the values of $L_{0}$ and $\alpha$ obtained. The inferred values of the acceleration region density $n$ depend on the value of $\alpha$ ($\alpha \propto 1/n$).  The values of other parameters inferred from $n$ and the acceleration region length $L_0$ -- see Section~\ref{ref:chap2_con} -- are thus dependent upon both the assumed electron distribution and the properties of the target plasma. The results will be used to find, for instance, if the inappropriate use of a one-dimensional cold target assumption changes the inferred number density by a factor larger than the observational uncertainty, and thus determine if a correction should be applied to the results.
\begin{figure}[h]
\vspace{-20pt}
\centering
\includegraphics[width=75mm]{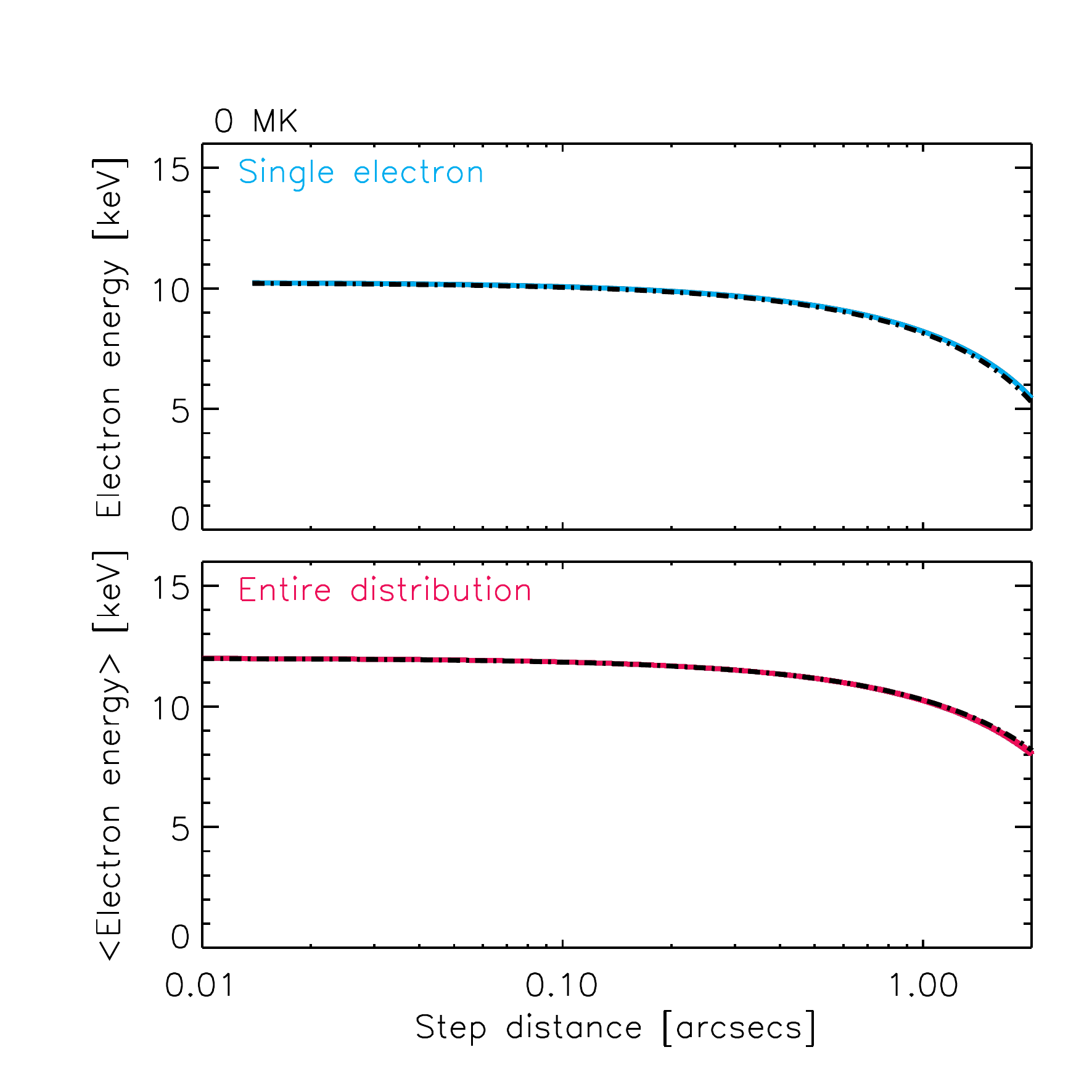}
\includegraphics[width=75mm]{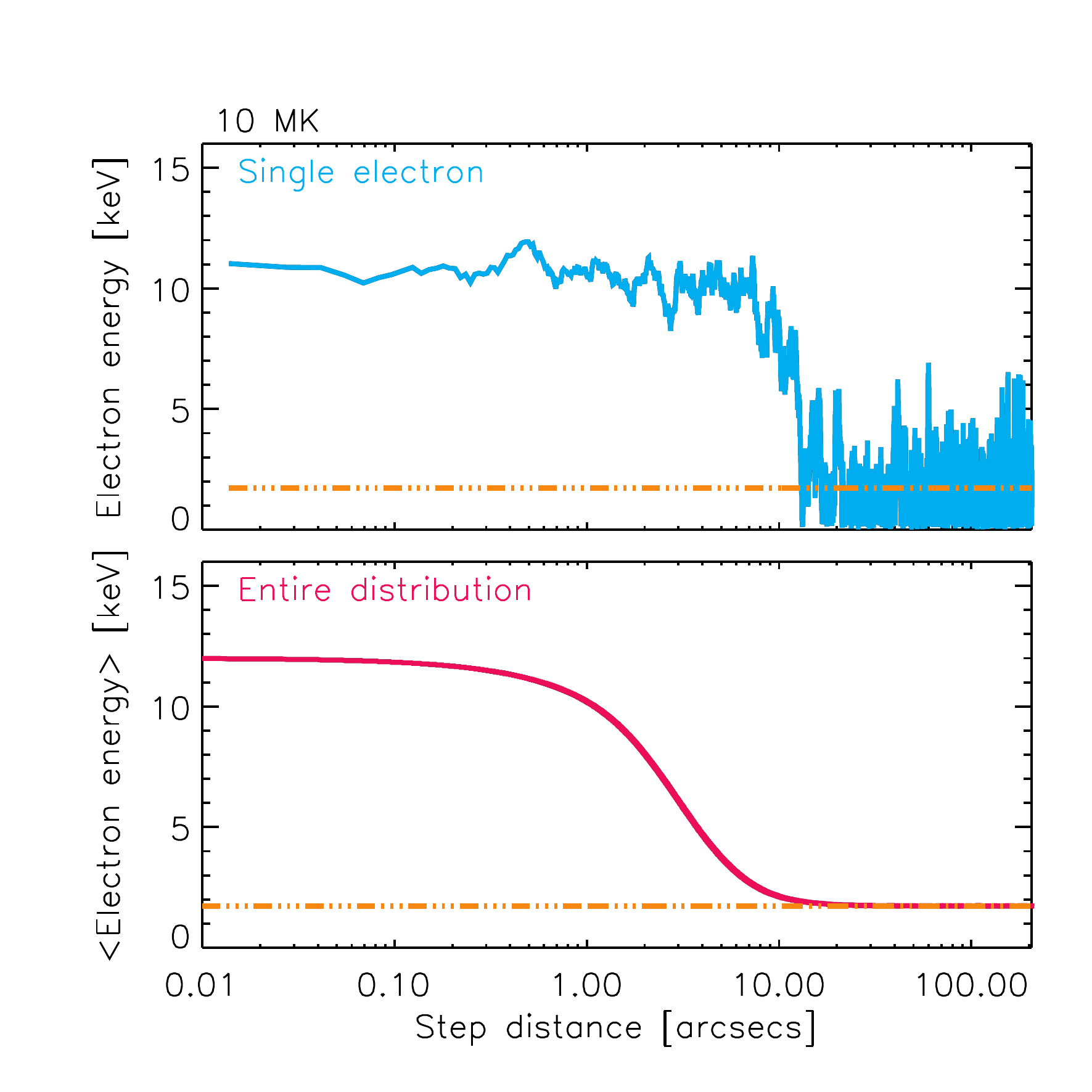}
\includegraphics[width=75mm]{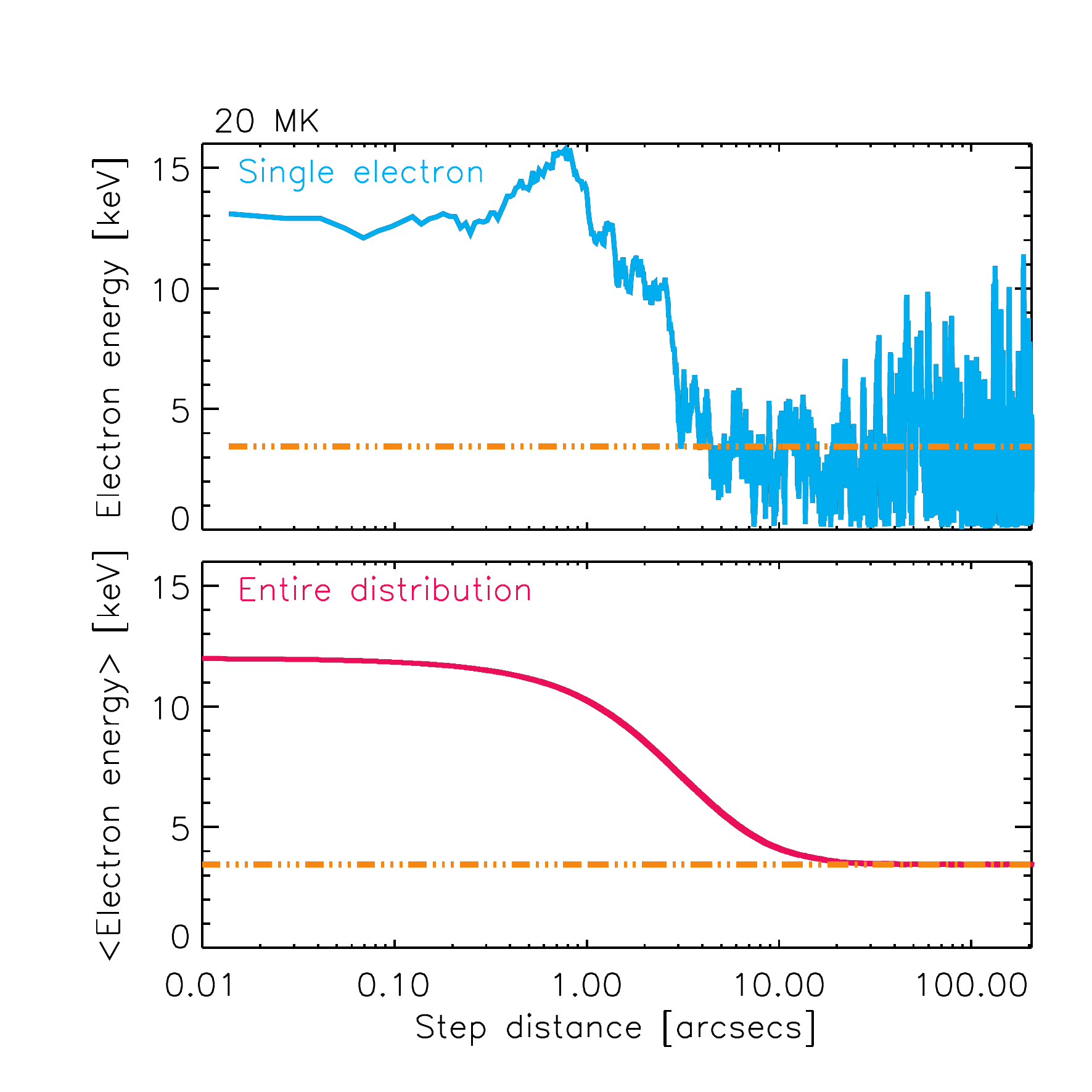}
\includegraphics[width=75mm]{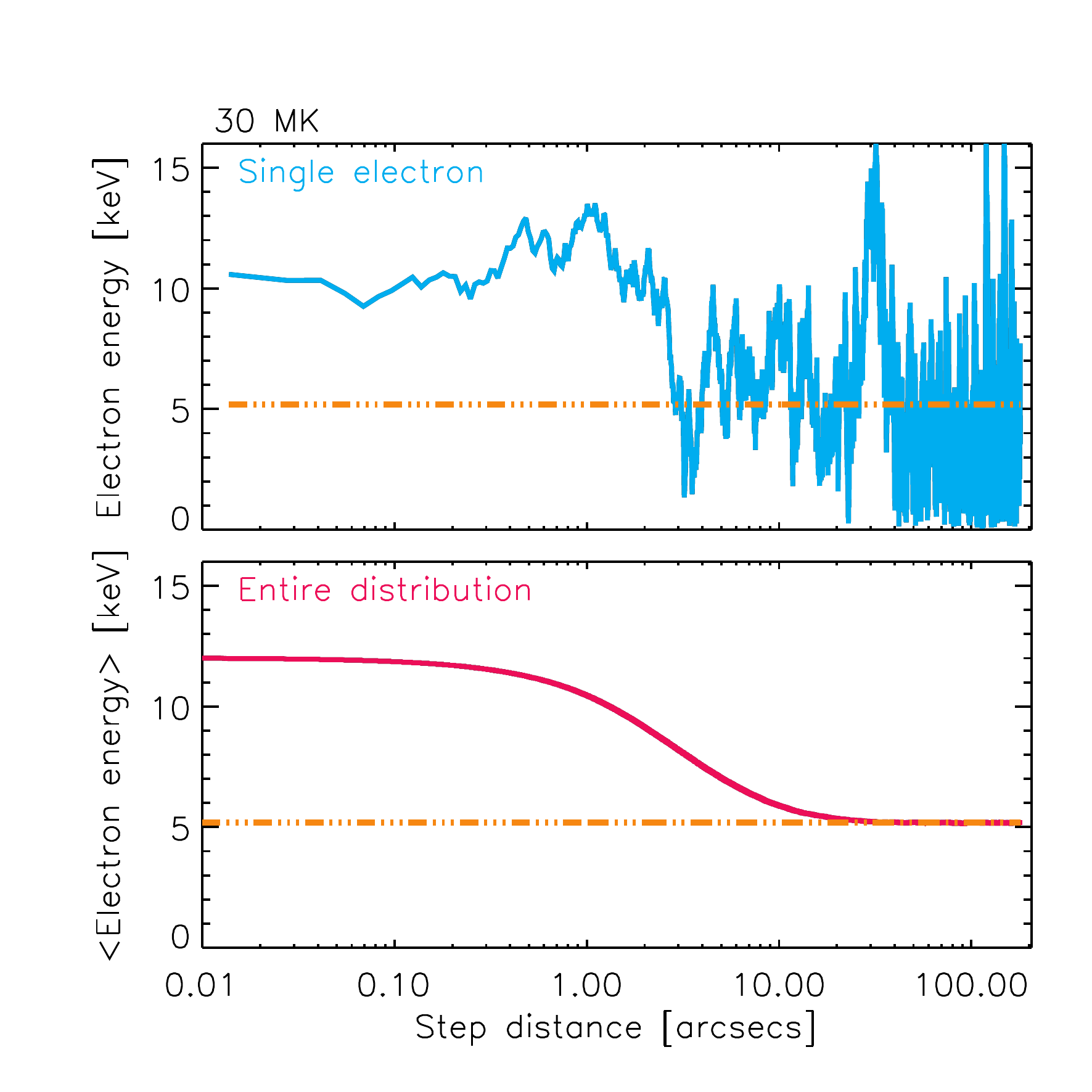}
\caption[The energy $E$ of a single electron and $<E>$ of the entire distribution for $T=0,10,20,30$ MK simulations plotted as a function of the overall distance $\sum \Delta s$ travelled.]{{\it Top panels}: $E$ of a single electron for $T=0,10,20,30$ MK simulations as a function of the overall distance $\sum \Delta s$ travelled. For the chosen $\Delta s = 10^5$~cm $\simeq 10^{-3}$ arc seconds, the change in energy over a single step is small. The randomness in the $T$=10, 20 and 30~MK cases is due to thermal fluctuations that increase with $T$; the error associated is difficult to estimate for a single particle. {\it Bottom panels}: $\langle E \rangle$ of the entire distribution versus $\sum \Delta s$ travelled, for the parameters given in Section \ref{sec:input}.  In contrast to the results for a single particle, these show smooth curves, with only small fluctuations for the $T$=10, 20 and 30~MK cases. Black dashed dot curve (0 MK): analytical cold target solution and orange dashed dot curves (10, 20 and 30 MK): final average energy of the $F(E)$ distribution.}
\label{fig:check_errors}
\end{figure}

\subsubsection{Simulation accuracy and limiting cases}

In general, consideration of the errors associated with stochastic simulations are a complex problem and beyond the scope of this thesis. However,  convergence of the simulation results against limiting analytical solutions can be checked. In the various plots shown in Figure~\ref{fig:check_errors} the energy of a single electron versus the overall step distance travelled (top) and the average energy of the entire distribution against the distance travelled (bottom) are plotted.  This was done for $\delta=7$, and for $T$=0, 10~MK, 20~MK and 30MK. For the cold ($T=0$) case, the error in the energy of a single electron is very small; the stochastic terms in the difference equations~(\ref{eq:sto_x}) through~(\ref{eq:sto_mu}) are negligible and individual electron energies (and hence the average energy of the entire distribution) follow the analytical results very well. However, for a finite temperature target, the stochastic part of the difference equations plays a significant role, the dominance of which increases with $T$. Hence the energy of a single electron fluctuates significantly, especially at low energies.  However, due to ensemble averaging, even for finite target temperatures the average energy of the {\it distribution} exhibits a relatively smooth transition from the starting average energy of the distribution to the final average value of the distribution $F(E)$.
%\begin{figure*}
\begin{sidewaysfigure*}
\centering
\includegraphics[width=21.5cm]{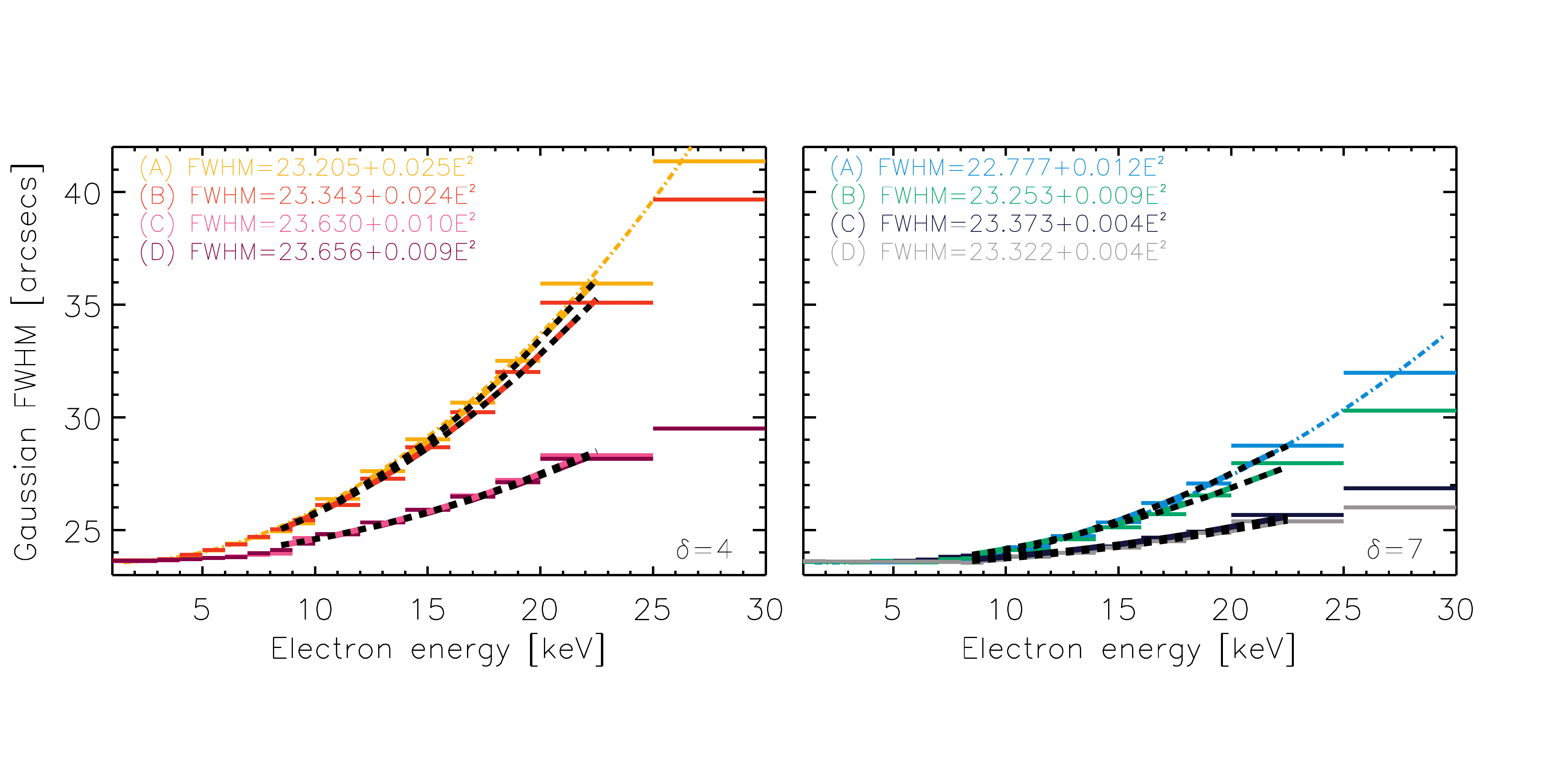}
\caption[Gaussian FWHM versus electron energy $E$ for all cold plasma simulation runs.]{Gaussian FWHM versus electron energy $E$ for all cold plasma simulation runs with $n=1\times10^{11}$ cm$^{-3}$ and $\delta=4$ (left plot) and $\delta=7$ (right plot). The cases shown are for: (A) beamed, no pitch angle scattering (orange, blue), (B) beamed, with pitch angle scattering (red, green), (C) isotropic, no pitch angle scattering (pink, navy) and (D) isotropic, with pitch angle scattering (purple, grey). An equation of the form $L_{0}+\alpha E^{2}$ was fitted to each curve and the values thus found for $L_{0}$ and $\alpha$ are shown on each plot. The fits use energies in the range $\sim 8-25$ keV, matching the energy range fitted to {\em RHESSI} observations. The dashed-dot lines represent the Gaussian FWHM curves fitted to the results of Equation~(\ref{eq:fexgauss}), as in the bottom panels of Figure \ref{fex_good}. As expected, these match well with scenario (A).}
\label{fig:cold_runs}
\end{sidewaysfigure*}
\clearpage
%\end{figure*}

\subsection{Numerical results}

\subsubsection{Cold plasma with collisional pitch angle scattering}

Firstly, the case of a cold target is considered, with different pitch angle injection and scattering scenarios. Eight simulations were performed, corresponding to two spectral indices ($\delta=4$ and $\delta=7$) and:

\begin{itemize}
\item {(A) an injected bi-directional beamed distribution of electrons ($\mu=-1,1$) without collisional pitch angle scattering,}
\item {(B) an injected bi-directional beamed distribution of electrons ($\mu=-1,1$) undergoing collisional pitch angle scattering,}
\item {(C) an initially isotropic pitch angle distribution of electrons without collisional pitch angle scattering, and}
\item {(D) an initially isotropic pitch angle distribution of electrons undergoing collisional pitch angle scattering.}
\end{itemize}

Figure~\ref{fig:cold_runs} shows the Gaussian spatial FWHM plotted against electron energy $E$ for cases (A), (B), (C) and (D), together with fits using Equation (\ref{L}) between $\sim8-25$ keV. This energy range is chosen to match with the energy ranges often used for such observations by {\em RHESSI}. The corresponding values of $\alpha$ and $L_{0}$ for each scenario are shown on Figure~\ref{fig:cold_runs}, and there are two general statements that can be made regarding the results. Firstly, the broader the initial pitch angle distribution, the smaller the source length at a given energy and secondly, the presence of collisional pitch angle scattering acts to slightly decrease the source length at a given electron energy.  Both effects occur because electrons with $|\mu| < 1$ move a correspondingly smaller distance along the magnetic field. The latter effect is greater at higher electron energies but overall the change is rather small (Figure~\ref{fig:cold_runs}).

The case of an initially isotropic distribution, with or without pitch angle scattering, produces the flattest (lowest value of $\alpha$) results for each $\delta$. For example, compared with the initially beamed, scatter-free cases for $\delta=4,7$, the isotropic, scatter-free $\alpha$'s are lower by factors of $\sim2.6$ and $\sim3.5$, respectively.

Since the coefficient $\alpha$ (Equation~(\ref{eq: sd2})) in a one-dimensional cold target formulation is inversely proportional to the ambient density $n$, {\it the reduced penetration distance associated with the presence of an initially broad pitch angle distribution and/or collisional scattering will lead to an overestimate of $n$ if the results are interpreted using the one-dimensional cold target result},  with the exact reduction factor dependent upon the properties of the initial electron distribution and background plasma.

\begin{figure*}
\centering
\includegraphics[width=17cm]{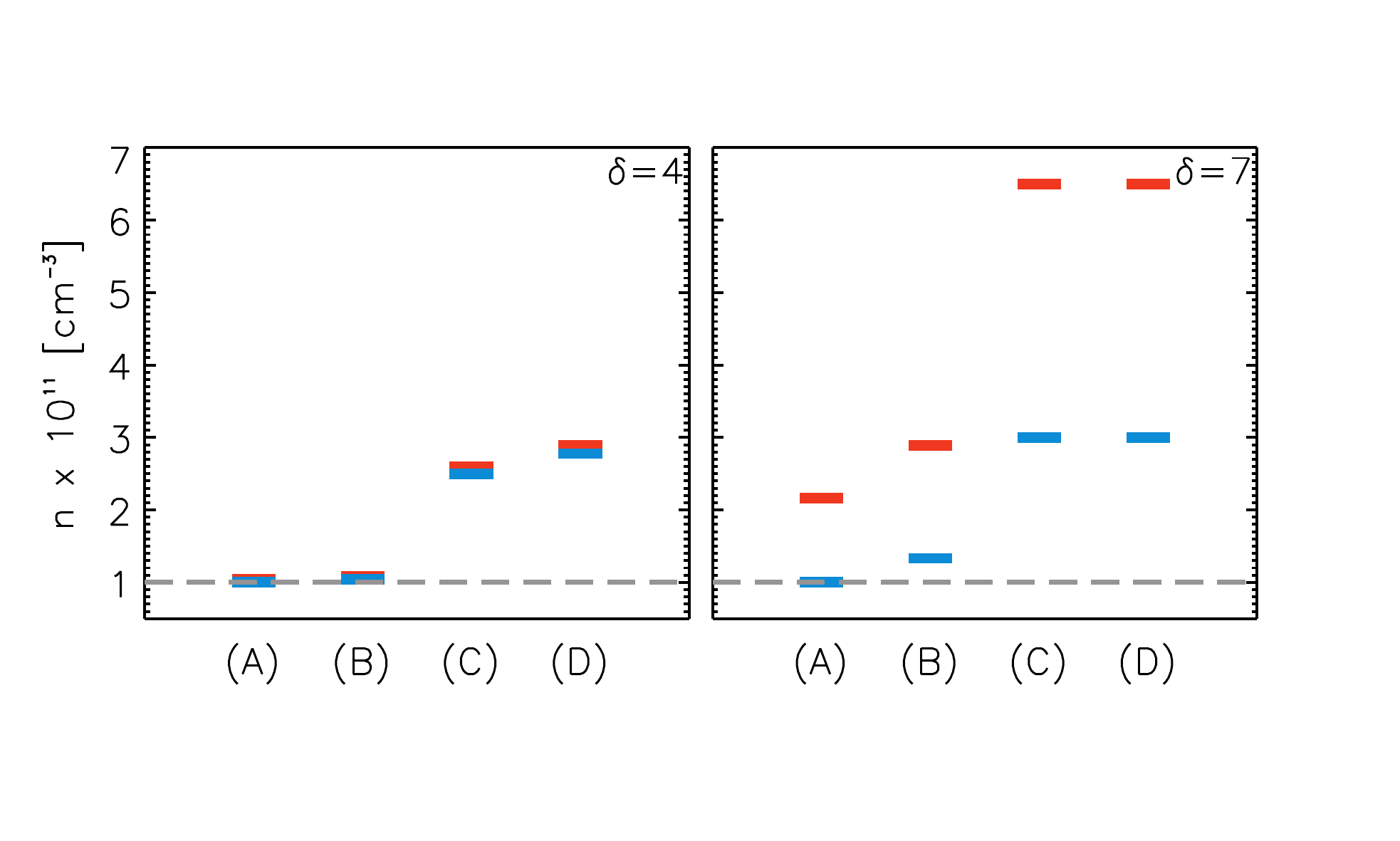}
\caption[For each cold target simulation scenario -- (A), (B), (C) and (D) -- the value of the coefficient $\alpha$ calculated by fitting each curve in Figure \ref{fig:cold_runs} is used to infer a number density $n$ using two different one-dimensional cold target approaches.]{For each cold target simulation scenario -- (A), (B), (C) and (D) -- the value of the coefficient $\alpha$ calculated by fitting each curve in Figure \ref{fig:cold_runs} is used to infer a number density $n$ using two different one-dimensional cold target approaches: (1) point injection $\alpha=1/2Kn$ (red) and (2) an extended Gaussian input that is initially beamed with no pitch angle scattering (blue). The actual number density of $1\times10^{11}$ cm$^{-3}$ is given by the grey dashed line and the inferred value of $n$ is $\sim$ equal or greater than the actual value.}
\label{fig:find_num_6}
\end{figure*}

Values of $n$ were inferred for each of the cases (A), (B), (C) and (D), using two different interpretive approaches:

\begin{enumerate}
\item {$\alpha_{1}=1/2Kn$, i.e., simple one-dimensional propagation within a cold target, giving $\alpha_{1}=0.026$ arcsecond keV$^{-2}$ for $n=1\times10^{11}$ cm$^{-3}$}
\item {$\alpha_{2}$, found using an extended Gaussian input for an initially beamed distribution with no pitch angle scattering, i.e., Equation~(\ref{eq:fexgauss}) and scenario (A).  From the lower right panel of Figure~\ref{fex_good}, for $n=1\times10^{11}$ cm$^{-3}$, $\alpha_{2}=0.026$ arcsecond~keV$^{-2}$ for $\delta=4$ and $\alpha_{2}=0.012$ arcsecond~keV$^{-2}$ for $\delta=7$.}
\end{enumerate}

In Figure \ref{fig:find_num_6}, the actual number density of the region $n=1\times10^{11}$ cm$^{-3}$ is shown by the dashed grey line and the values of $n$ inferred from approaches (1) and (2) are shown by the red and blue points, respectively. The inferred number density can be up to six times too large, with the largest effect being for steep spectra (the $\delta=7$ case) and isotropic injection (cases (C) and (D)).

\subsubsection{Hot plasma and collisional pitch angle scattering}

In this section it is studied how the effect of a finite-temperature target (in the presence of collisional pitch angle scattering) changes the electron transport through the plasma and hence the extent of the source with energy. Six further simulations were considered corresponding to three target temperatures (10~MK, 20~MK and 30~MK), and pitch angle scenario (B), an injected beamed electron distribution including pitch angle scattering, for both $\delta=4$ and $\delta=7$.

\begin{figure*}
\centering
\includegraphics[width=15.5cm]{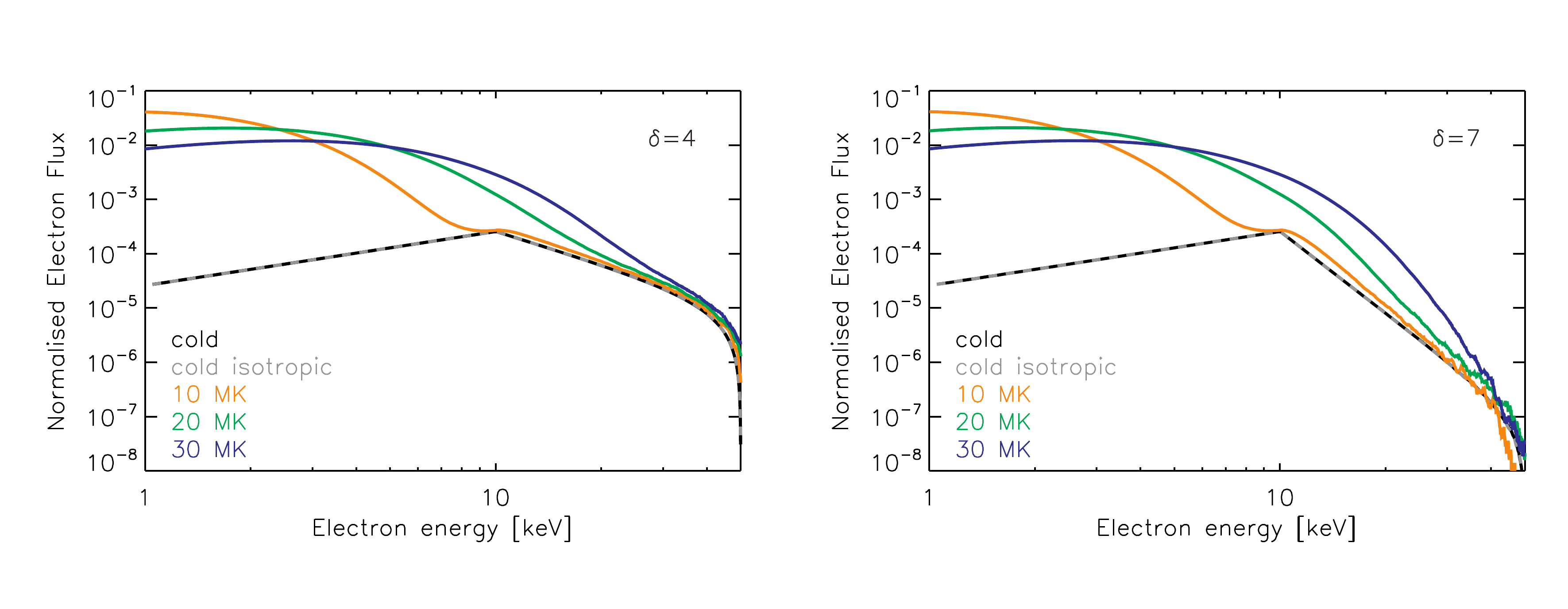}
\includegraphics[width=15.5cm]{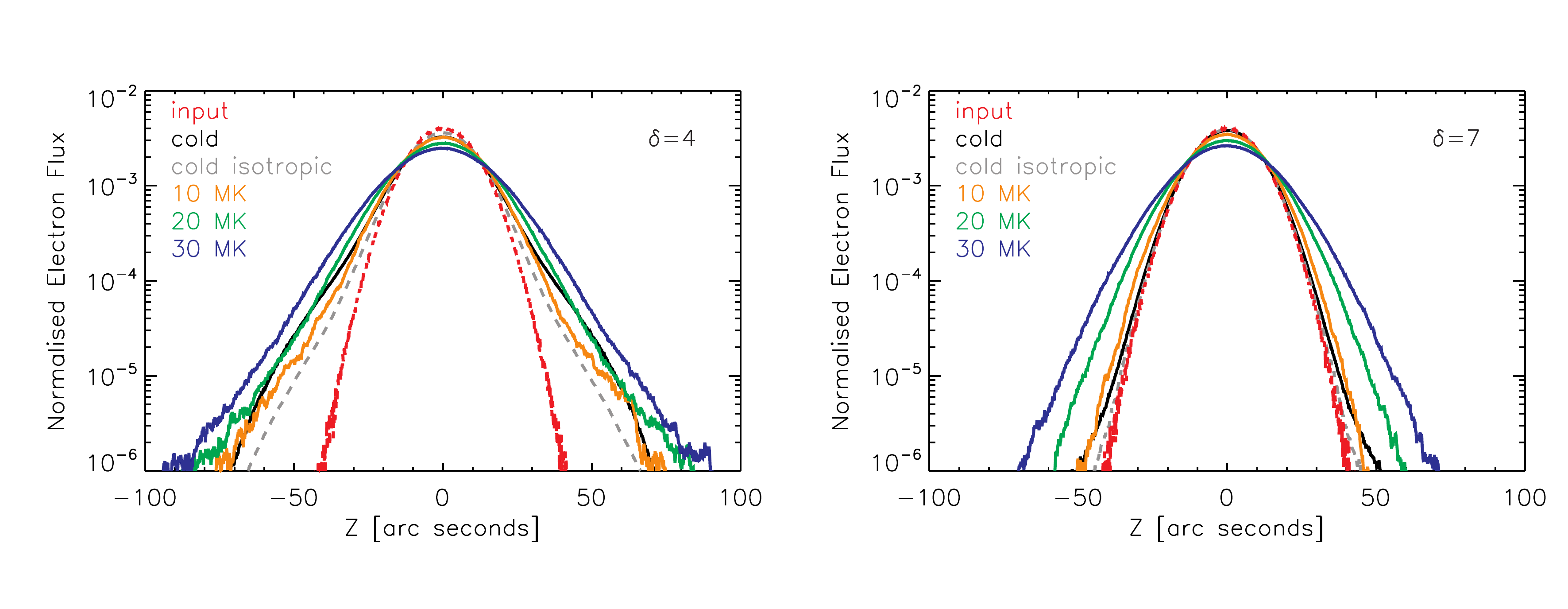}
\caption[Plots of the spatially-integrated spectra and energy-integrated spatial distributions for both cold and hot plasma simulation runs.]{{\it Top panels}: spatially-integrated spectra; {\it bottom panels}: energy-integrated spatial distributions for the following scenarios: (1) cold plasma, initially beamed distribution with pitch angle scattering (black); (2) cold plasma, initially isotropic distribution with pitch angle scattering (grey); and warm target cases with (3) $T$=10~MK (orange), (4) $T$=20~MK (green) and (5) $T$=30~MK (blue), with pitch angle scattering.  Results are shown for both $\delta=4$ (left) and $\delta=7$ (right). The red dashed lines in the bottom panels indicate the the source function $S(E,\mu=1,z)$.}
\label{fig:spectra_x}
\end{figure*}

Figure~\ref{fig:spectra_x} shows both the spatially-integrated spectra and the spectrally-integrated spatial distributions for five different simulations: one-dimensional (beamed) cold target (black), cold target with isotropic injection (grey), and beamed injection in three warm target cases: $T$=10~MK (orange), 20~MK (green) and 30~MK (blue). Figure~\ref{fig:spectra_x} shows only the spatially and spectrally integrated evolutions of the injected electron distribution and does not include the background cold or Maxwellian distribution. The total spatially-integrated spectra are plotted in the top row of panels, for $\delta=4$ (left) and $\delta=7$ (right); the spatial distribution of the spectrally-integrated flux is plotted in the bottom row of panels, again for $\delta=4$ (left) and $\delta=7$ (right).

Not surprisingly, higher temperature targets tend to make the overall electron spectrum more thermal in form. The lower the temperature of the background Maxwellian plasma, the greater the distinction between the thermal part of the distribution at lower energies and the nonthermal power-law component at higher energies. Also, the inclusion of thermal effects tends to broaden the spatial distribution of the electron distribution, with the effect being more pronounced at higher temperatures.  The spatial spread for a given input distribution is larger for a smaller spectral index because of the larger fraction of higher energy electrons in such flat distributions.  It was also found (not shown) that, not surprisingly, the initially beamed distribution (case (B)) shows greater spreads in $z$ than for the same six runs performed for the isotropic injection case (case (D)), see Figure~\ref{fig:spectra_x}.

\begin{sidewaysfigure*}
\centering
\includegraphics[width=18.5cm]{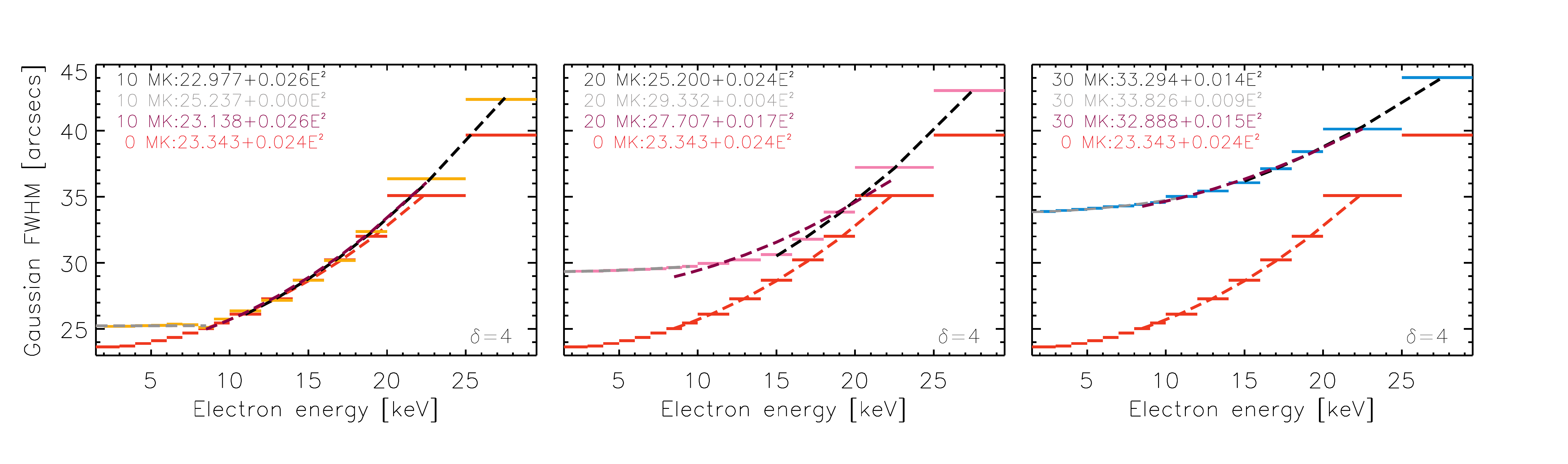}
\includegraphics[width=18.5cm]{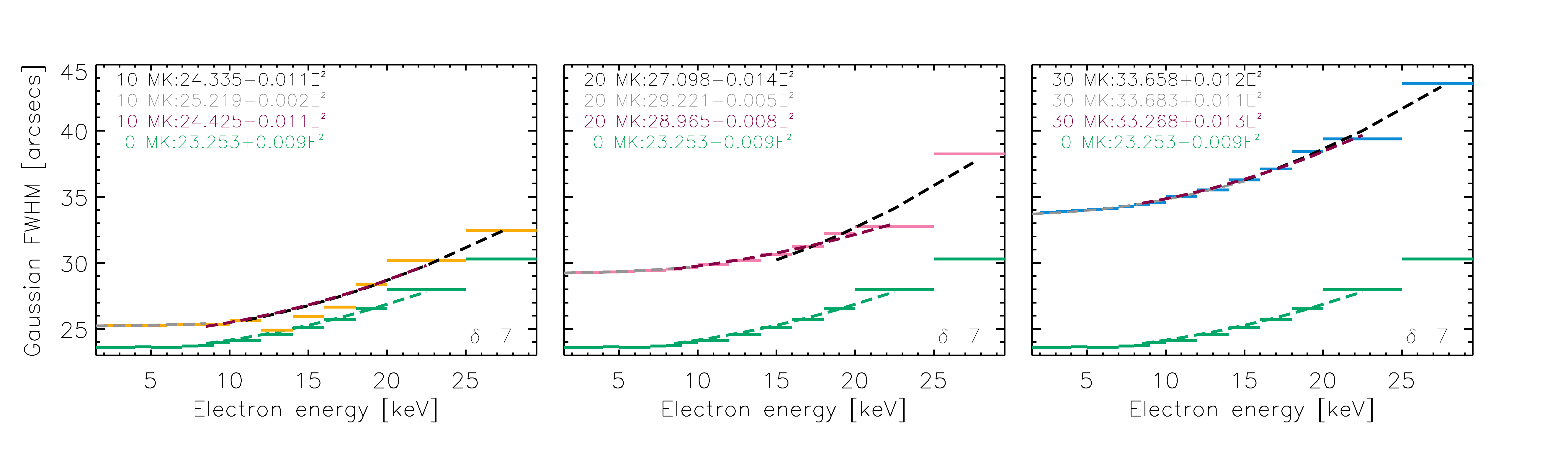}
\caption[Plots of FWHM versus electron energy for the finite temperature plasma simulation runs.]{Plots of FWHM versus electron energy for $\delta=4$ (top) and $\delta=7$ (bottom), and $T=$ 10 MK, 20 MK and 30 MK (left to right) and $n=1\times10^{11}$ cm$^{-3}$. Fits of the form FWHM=$L_0+\alpha E^{2}$ are shown on each plot. The red and green dashed curves show the corresponding results for the  beamed, cold plasma case with pitch angle scattering (scenario (B)). The purple dashed lines show the best fit in the energy range 8-25~keV, the range used in the fit to {\em RHESSI} observations. Also shown, are the two-component fits, one component representing the thermal diffusion at lower energies (grey dashed curve) and another component representing collisional friction that dominates at higher energies (black dashed curve).}
\label{hot_errors}
\end{sidewaysfigure*}

Figure~\ref{hot_errors} shows the results of the Gaussian fits to the computed spatial distributions for all six warm target scenarios, together with the corresponding results for the cold target case.  Compared to the cold target case, the addition of thermal effects results in changes that affect the inferred values of both $n\propto 1/\alpha$ {\it and} $L_{0}$. Firstly, it is obvious from all panels in Figure~\ref{hot_errors} that the value of the $y$-axis intercept $L_0$ (the inferred acceleration region length) increases with temperature; it was found that the magnitude of this increase depends somewhat on the number density $n$ and is relatively independent of the power-law index $\delta$. This effect is purely due to the thermal diffusive nature of the electron transport, both energetically and spatially, at low energies. This result suggests that the temperature of the background plasma must be accounted for, when estimating $L_0$ from such observations. The determination of the actual acceleration region length from the inferred length is discussed further in Section \ref{landn}.

Just as before, curves of the form of Equation (\ref{L}) are fitted between $\sim8-25$~keV, for a better comparison with observations when using imaging algorithms such as VisFwdFit and {\it uv$\_$smooth}.  These are shown by the purple dashed lines and the values of $L_0$ and $\alpha$ from the purple fits are shown on each panel of Figure~\ref{hot_errors}.  However, the presence of a finite background temperature causes the lower energies of the distribution, in particular, to be dominated by thermal diffusion and hence analysis of the curves in Figure~\ref{hot_errors} shows that overall, the FWHM over the entire plotted energy range is not so well-fitted by a single curve of the form FWHM$(E)=L_0+\alpha E^{2}$. This can be clearly seen for the 20 MK, $\delta=4$ curve. Therefore, two other FWHM$(E)=L_0+\alpha E^{2}$ curves are fitted to the results; one component representing the lower energy values that are controlled mainly by thermal diffusion (grey curve) and another component representing higher energies mainly controlled by collisional friction, since the FWHM values should return to match those of a cold target case when $E>>k_{B}T$. The $L_0$ and $\alpha$ values found from the grey and black curves are also shown on each panel of Figure~\ref{hot_errors}.

To illustrate, for the $T$=10~MK case, the FWHM values match those of the cold case (red or green dashed line) after $\sim10$~keV, for both the $\delta=4$ and $\delta=7$ cases. This is because the temperature diffusion is limited to energies below $\sim8$~keV (grey curve); Figure~\ref{hot_errors} clearly shows this transition. Therefore for the 10 MK case, the $8-25$~keV fits (purple) match that of the higher energy black fits and cold cases reasonably well for both $\delta=4$ and $\delta=7$. By $T$=20~MK, the energy range between $8-25$ keV is not so well fitted by a curve of the form of Equation (\ref{L}) and occurs because the trend of the FWHM moves from being dominated by the effects of thermal diffusion to being dominated by the effects of collisional friction at approximately $15$ keV, right in the middle of the range used for the fit. This is clear for the $\delta=4$ case but harder to see for $\delta=7$ case due to the smaller values of $\alpha$. The $\alpha$ values of the friction-dominated fits (black curves) are only approximately the same as for the cold plasma case after $\sim17$ keV. Also the diffusion at 20~MK noticeably influences the length values at all energies plotted, with the FWHM values above $\sim$17~keV lying above those for the cold case. By $T$=30~MK, the entire plotted energy range and the fitted energy range between $8-25$ keV is mainly controlled by thermal diffusion and the $\alpha$ values for both the $\delta=4$ and $\delta =7$ cases are similar. All plotted FWHM values are much larger than that of equivalent cold cases, over $10\arcsec$ at $1$ keV. For the $8-25$ keV fits, the $\delta=4$ value is smaller than that of the equivalent cold case, and the $\delta=7$ value is slightly larger.

%\begin{figure*}
\clearpage

\subsubsection{Inferring the acceleration region length $L_0$ and density $n$}\label{landn}

The thermal diffusion-component (grey dashed) curves in Figure~\ref{hot_errors} use Equation~(\ref{L}) to fit the FWHM values at lower energies, and hence give $L_0$, the inferred length of the acceleration region. For a given temperature, the values of $L_0$ found for both $\delta=4$ and $\delta=7$ are approximately the same, with an average value of $25\arcsec$ for $T$=10~MK, $29\arcsec$ for $T$=20~MK and $34\arcsec$ for $T$=30~MK.

\begin{figure*}
\centering
\includegraphics[width=16.5cm]{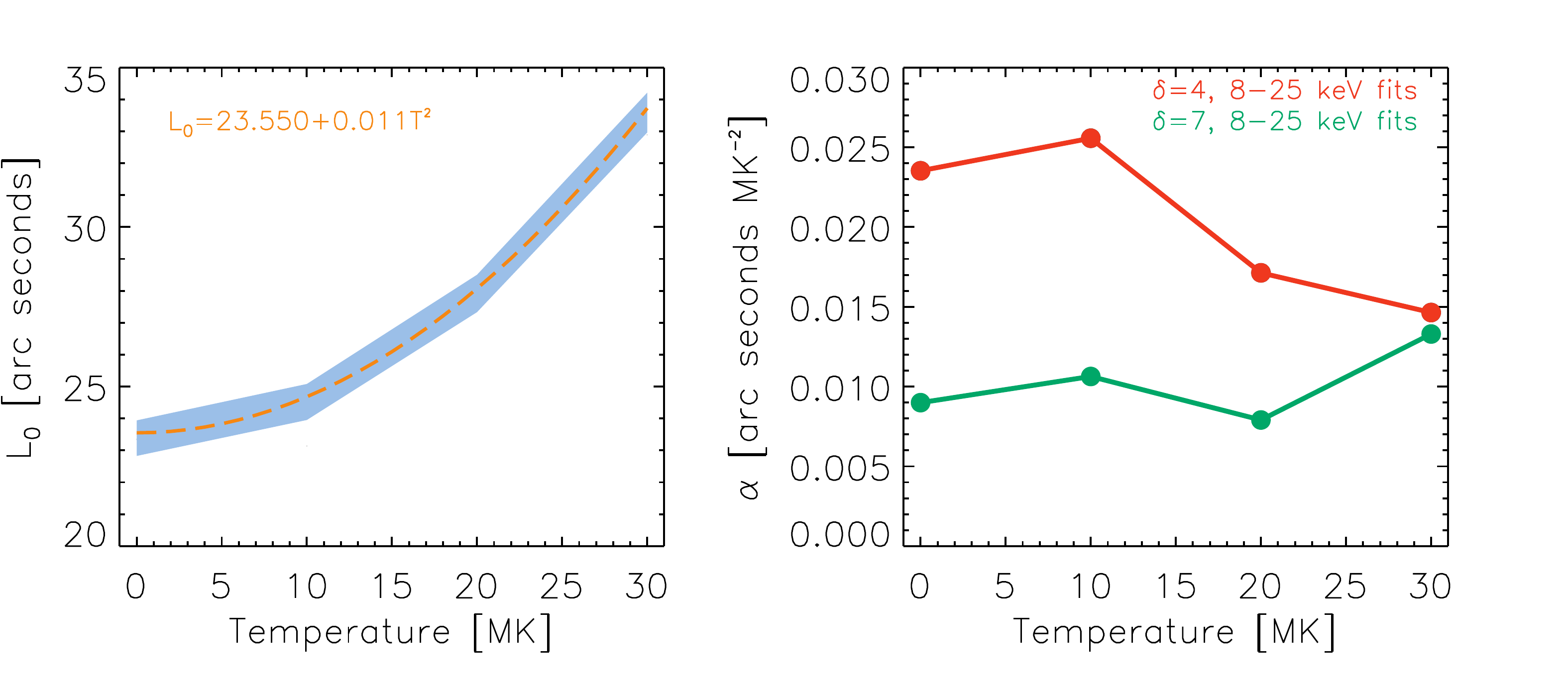}
\caption[Inferred acceleration region length $L_0$ and quadratic fit parameter $\alpha$ versus plasma temperature.]{{\it Left panel}: $L_0$ versus $T$. The blue band represents the area containing the $L_{0}$ values for both the $\delta=4$ and $\delta=7$ low-energy and 8-25 keV fits from Figure \ref{hot_errors} (grey and purple respectively), plotted against temperature $T$. For each of these curves, a function of the form $L_0=L_{0}(T=0)+\xi T^{2}$ is fitted, and average values of $\langle L_{0}(T=0)\rangle=23\arcsec.5$ and $\langle\xi(n=1 \times 10^{11}$~cm$^{-3})\rangle=0.011$ [arcsec MK$^{-2}$] are found, with $L_0=\langle L_{0}(T=0)\rangle+\langle \xi\rangle T^{2}$ represented by the orange dashed line.  {\it Right panel:} $\alpha$ from the 8-25 keV fits (Figure~\ref{hot_errors}) versus $T$, for $\delta=4$ (red) and $\delta=7$ (green).}
\label{hot_runs_2}
\end{figure*}

For the reasons discussed above, it is more likely that the $L_0$ values obtained from the $\sim8-25$~keV fits will be most reliable from observation. Averaged over the two values of $\delta$, these give values of $24\arcsec$ for $T$=10~MK, $28\arcsec$ for $T$=20~MK, and $33\arcsec$ for $T$=30~MK.  These values are only slightly smaller than the values found from the grey dashed fits at lower energies. However, if viable, as low an energy as possible should be used to find the inferred value of $L_0$. Figure \ref{hot_runs_2} (left) plots the values of $L_0$ found for the thermal diffusion-dominated (grey curve) and $8-25$ keV fits against $T$. Each is fitted with a curve of the form,
\begin{equation}
L_0(T,n)=L_{0}(T=0)+\xi(n) T^{2}=23\arcsec.5+\xi(n) T^{2} \,\,\, .
\label{eq:LT}
\end{equation}
By fitting Equation (\ref{eq:LT}) to each, $\xi$ is found for both ``global'' and thermal diffusion-dominated fits, and an average value of $\langle\xi(n=1\times10^{11})\rangle=0.011$ arcsecond MK$^{-2}$ is calculated empirically from the four fits.
To summarize, if the size $L_0(T=0)$ and number density $n$ of the region have been inferred from a cold target analysis, and $n$ is close to $n=1\times10^{11}$ cm$^{-3}$ (as it must be for a viable thick target coronal source to appear), then {\it the actual extent of the acceleration region is less than would be inferred using a cold target formula.}  Quantitatively, the actual size of the acceleration region $L_0$ can be approximated by the expression
\begin{equation}\label{T-correction}
%L_0=L_0(T)-0.011 \, T^{2} \,\,\, ,
L_0=L_0(T=0)-0.011 \, T^{2} \,\,\, ,
\end{equation}
where $L_0(T=0)$ is the value deduced from a fit using the cold target formula to an observation.

The right panels in Figure~\ref{hot_runs_2} also show how $\alpha$ from the $8-25$ keV fits changes with $T$ for both $\delta=4$ and $\delta=7$. For $\delta=4$, $\alpha$ decreases between $T$=10~MK and $T$=30~MK.  This is expected, since for higher temperatures, particle diffusion is controlling the shape of the curve and the $\delta=4$ cold target case has a relatively high $\alpha$ value. However this is not the case for $\delta=7$, where between 10-30 MK $\alpha$ increases with $T$.
\\\\
From the plots in Figure \ref{hot_runs_2}, the values of $\alpha$ for the fits between 8-25 keV can be used to infer a number
density from observations. Two cold target approaches are used: (1) $\alpha={1}/2Kn$, and (2) an extended source Gaussian input as found from Equation (\ref{eq:fexgauss}). Also, using the results from the cold-plasma cases (2) can be expanded to account for the initial beaming of the distribution so that a range of $n$ can be found.  Finally, (3) can be used, which is the same as (2) but accounts for pitch angle scattering.

\begin{figure*}
\centering
\includegraphics[width=17cm]{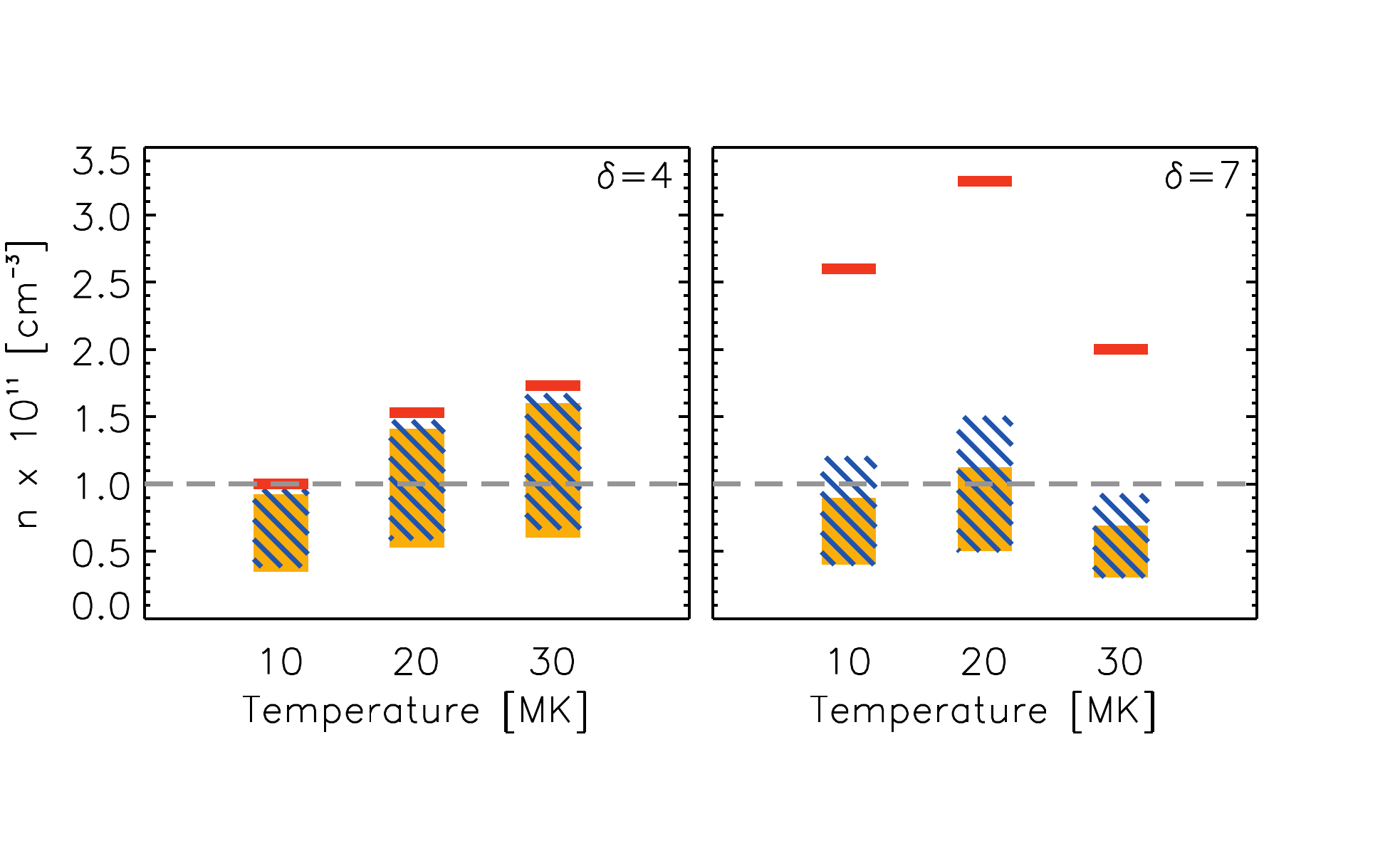}
\caption[Cold plasma fits are applied to the different hot plasma simulation curves to determine an inferred density that can be compared with the actual density of the region.]{Cold plasma fits are applied to the different hot plasma simulation curves to determine an inferred density that can be compared with the actual density of the region: (1) (red lines): $\alpha_{1}={1}/2Kn$. (2) (blue hashed areas): an extended Gaussian injection model with no pitch angle scattering that is initially either beamed (Equation~(\ref{eq:fexgauss})) or isotropic (found from the cold plasma simulation -- see Figure~\ref{fig:cold_runs}). (3) (orange regions): as for (2), but with collisional pitch angle scattering included.  For both (2) and (3), the spread in $n$ occurs due to different electron pitch angle distributions from completely beamed to isotropic. The highest inferred values for $n$ are for a completely beamed distribution.}
\label{new_simn}
\end{figure*}

The inferred values of $n$ for $T=10, 20, 30$~MK and for $\delta=4,7$ are shown in Figure~\ref{new_simn}. For $\delta=4$ the largest inferred value  is $\sim1.7$ times larger than the actual density and the smallest is around three times smaller; for $\delta=7$, the largest value is $\sim3.3$ times larger and the smallest value is again about 3 times smaller. In general, (1) (cold target, point injection, red lines) produces the largest differences, which is not surprising since the input was an extended Gaussian, rather than point-injection, source. However, even this simple analytical case, that accounts very poorly for the true physical properties of the electron distribution, only increases the number density by a factor of about $3$ (for a beamed finite temperature case). (2) and (3) (extended injection models, without and with collisional scattering in the target, respectively) that do not account for the finite temperature of the plasma provide an inferred value for $n$ that is quite close to the true value of $n$, with the biggest uncertainty due to the unknown degree of beaming of the injected distribution.

\section{Discussion and conclusions}\label{ref:chap2_con}
The aim of Chapter \ref{ref:Chapter2} was to understand how the presence of different injected pitch angle distributions, plus the effects of collisional pitch angle scattering and of a finite target temperature change electron transport through a plasma and hence the spatial properties of compact hard X-ray sources in solar flares.

The simulations show three main results:

\begin{enumerate}

\item{Collisional pitch angle scattering alone does not dramatically change the behaviour of source length with electron energy.}

\item{Beaming of the initial electron pitch angle distribution does produce a significant change in the variation of the length of the X-ray source with energy; distributions that are initially beamed produce a larger variation of length with energy, a consequence of the fact that the collisional stopping distance is now projected onto the direction defined by the guiding magnetic field.  The difference in the coefficient $\alpha$ can be up to a factor of $6$ if a beamed approximation is used for a distribution that is in fact completely isotropic. The uncertainty in the initial angular distribution of the injected electrons produces the largest uncertainty in the inferred number density $n$.}

\item{The finite temperature of the target atmosphere leads to thermal diffusion, both in energy and space, and an increase of the inferred acceleration region length. The FWHM versus energy consists of two competing components, one due to thermal diffusion that is dominant at lower energies, and another due to advection that is dominant at higher energies.  Which component predominates depends on a complicated way on the temperature of the region, on the density $n$, and even on the spectral index $\delta$. Therefore the use of a cold target approximation with a single fitted curve to infer properties of the acceleration region should always be used with caution. The results show that applying a cold model to a warm plasma changes the inferred acceleration length $L_0$ by several arc seconds (see Equation~(\ref{T-correction})) and the inferred number density by up to a factor of $3$ (in either direction), depending mainly on the initial beaming of the electron distribution (see Figure~\ref{new_simn}).}
\end{enumerate}

The influence of the effects studied in this chapter also influence the determination of other quantities, such as the acceleration region filling factor $f$ (the fraction of the apparent source volume in which acceleration occurs) and the specific acceleration rate (the fraction of the ambient electron population that is accelerated per unit time). The filling factor $f$ is defined by

\begin{equation}\label{filling-factor}
f = \frac{EM}{n^2 V} \,\,\, ,
\end{equation}
where $V = (\pi W^2/4) L_0$ is the volume of the acceleration region, determined from the inferred value of $L_0$ and the observed lateral extent $W$ of the (cylindrical) acceleration volume, and the emission measure $EM$ is determined from, for example, fits to the spatially-integrated soft X-ray spectrum of the flare.  The effects studied in this chapter show that in general, application of a one-dimensional cold target formula leads to erroneously high inferred values for both the acceleration region length $L_0$ (see Figure~\ref{hot_errors}) and density $n$ (see Figure~\ref{new_simn}).  Use of such erroneously high values of $L_0$ and $n$  leads to an overestimate of the denominator in Equation~(\ref{filling-factor}) and so an underestimate of the filling factor $f$.

In a study of 24 coronal thick target events, using the one-dimensional cold target result~(\ref{cold_theory}) to estimate $L_0$ and $n$, \citet{2013ApJ...766...28G} found filling factors $f$ that were generally somewhat less than unity. The results of this chapter therefore lend support to a value of $f$ being even closer to unity than previously thought. Indeed, given that $f$ cannot exceed unity, this may place constraints on the allowable values of $n$ and $L_0$.  And, since the inferred values of $n$ depend significantly on the pitch angle distribution of the injected electrons, this could conceivably be used to constrain the form of the injected pitch angle distribution.  In particular, broad injected distributions lead to relatively small values of the coefficient $\alpha$ (see Figure~\ref{fig:cold_runs}) and hence to inferred densities that are higher than the actual target density (Figure~\ref{new_simn}).  Correcting for such an effect in the interpretation of a particular event could imply an actual target density that was too small to be compatible with the observationally-inferred emission measure, thus ruling out the hypothesis of a broad injected distribution of accelerated electrons.
The inference of the acceleration region length $L_0$, lateral extent $W$, and density $n$ also gives the number of electrons available for acceleration:

\begin{equation}\label{number_of_particles}
{\cal N} = n \, V = n \, \left ( \frac{\pi W^2}{4} \right ) \, L_0 \,\,\, .
\end{equation}
This, combined with the inference of $d{\cal N}(E_0)/dt$, the rate of electron acceleration beyond energy $E_0$ (obtained rather straightforwardly from spatially-integrated hard X-ray data) provides the value of the {\it specific acceleration rate} (electrons~s$^{-1}$ per ambient electron)

\begin{equation}\label{acceleration-rate}
\eta(E_0) = \frac{1}{\cal{N}} \, \frac{d {\cal N}(E_0)}{dt} \,\,\, .
\end{equation}
Overestimating the value of the acceleration region volume and density through the use of an over-simplistic one-dimensional
cold target model thus causes an overestimate of ${\cal N}$ and, since $d{\cal N}(E_0)/dt$ is fixed, this causes an underestimate
of $\eta(E_0)$.  In their multi-event study, \citet{2013ApJ...766...28G} found typical values for $\eta(E_0=20~{\rm keV})$ were of the order $10^{-2}$~s$^{-1}$ and they compared these values with those predicted from different acceleration models: large scale electric field acceleration (super-Dreicer) \cite[e.g.][]{1993SoPh..146..127L,2008AIPC.1039....3E} and stochastic acceleration \cite[e.g.][]{1996ApJ...461..445M,2012ApJ...754..103B}, both of which could be made to account for such values. The application of the physically realistic source models considered herein will increase $\eta$ even further, and place more profound constraints on the electron acceleration mechanism.

\chapter{The temporal and spatial evolution of solar flare coronal X-ray sources}
\label{ref:Chapter3}

\normalsize{\it This is work can be found in the publication \cite{2013ApJ...766...75J}}

\section{Introduction to the chapter}\label{ref:intro}
Using simulations, Chapter \ref{ref:Chapter2} studied the variation of coronal X-ray source lengths with electron energy. This study was motivated by recent {\it RHESSI} observations of dense coronal X-ray sources and the information that could be inferred from the length increases with X-ray energy. In their study \cite{2008ApJ...673..576X} found that the coronal X-ray loop width, the direction perpendicular to the guiding field, of each event also increased with X-ray energy. \cite{2011ApJ...730L..22K} examined the width changes of one coronal X-ray source and found that the loop width increased proportionally with X-ray energy. However, unlike increases in coronal loop length, changes in width are more difficult to explain since the electrons are bound to the guiding field and collisional cross field transport should be negligible. \cite{2011ApJ...730L..22K} and \cite{2011A&A...535A..18B} inferred that the width increase could be due to the magnetic diffusion of field lines perpendicular to the direction of the field, which is caused by the presence of magnetic turbulence within the loop. The interesting spatial trends shown by these observations and the information inferred indicate the usefulness of observing coronal X-ray loop spatial properties. Chapter \ref{ref:Chapter3} now goes on to further explore this, by examining how the spatial properties: length, width and position of such coronal X-ray sources change in time before, during and after the impulsive phase of the flare.

\subsection{Past studies of coronal loop spatial properties}
Past observations concentrated on studying changes in coronal loop positions, due to the difficultly in trying to quantitatively infer source size changes with {\it RHESSI} and other instruments, as was discussed in Chapter \ref{ref:Chapter1}.
\cite{1996ApJ...459..330F} used the Soft X-ray Telescope (SXT) on-board {\it Yohkoh} to observe the changing locations of post flare loops, which was interpreted as the decrease in height that open field lines undergo after they have reconnected to form closed loops. This study looked at two long duration events near the solar limb and found the presence of loop shrinkage that matched the shrinkage predicted by a simple model of the reconnecting field, but overall the entire flare loop system grew with time.
\cite{2003ApJ...596L.251S}, \cite{2004ApJ...612..546S}, \cite{2006A&A...446..675V} and \cite{2009ApJ...706.1438J} all noted a decrease in the altitude of coronal loop top sources
during the impulsive phase of the flare, until the peak X-ray emission and an increase in altitude after the impulsive phase.
\cite{2003ApJ...596L.251S} also found evidence for an above the loop top source and interpreted the situation as the formation
of a reconnection current sheet between the loop top source and the higher coronal source.
\cite{2006A&A...446..675V} interpreted the decrease as a collapsing magnetic trap \citep{1997ApJ...485..859S,2004A&A...419.1159K}.
The contraction and expansion of the loop source has also been observed in other wavelengths of EUV \citep{2009ApJ...696..121L,2009ApJ...706.1438J}
and radio \citep{2005ApJ...629L.137L,2010ApJ...724..171R}.  \cite{2010ApJ...724..171R} found both changes in the radio loop span and height with time.
More recently, \cite{2012ApJ...749...85G} looked for evidence of collapsing fields using Solar Dynamics Observatory ({\it SDO/AIA} and {\it HMI}) observations.
The loops rose slowly and then moved into a collapse phase during the impulsive phase of the flare, where the loop tops contracted. Lower loops contracted earlier than higher loops and the loop contraction was interpreted as a reduction of magnetic energy as the system relaxed to a state of lower energy, that is, relaxation theory \citep{1974PhRvL..33.1139T}.

In this chapter, three flares with dense coronal X-ray loops are studied, in order to find how the emission lengths, widths and positions change with time at three energy ranges
between 10-25 keV. Using spectroscopy, it is also found how plasma parameters such as emission measure and plasma temperature vary with time, for each flare.
Using a combination of imaging and spectroscopy parameters, the X-ray loop corpulence, volume, plasma number density, thermal pressure
and thermal energy density are inferred during the time evolution of the flare. This chapter will also propose some explanations describing the trends and the processes occurring within the coronal loops.

\section{Chosen events with coronal X-ray emission}
\begin{table}\scriptsize
\caption{Table showing the main parameters of Flares 1, 2 and 3.}\label{table1}
\begin{center}
\begin{tabular}{|l|l|l|l|l|l|}
\hline
 & GOES Class & Date & Obs. time & Peak time (10 keV) & Footpoints (30-40 keV) \\
\hline
Flare 1 &  M3.0 & 23-August-2005 & 14:22:00-14:40:00 & 14:30:00 & 14:36:00 onwards\\
\hline
Flare 2 &  M4.1 & 14/15-April-2002 &23:58:00-00:20:00 & 00:12:00 & 00:05:00 onwards\\
\hline
Flare 3 & M2.6 & 21-May-2004 & 23:42:00-23:58:00 & 23:50:00 & 23:42:00 onwards\\
\hline
\end{tabular}
\end{center}
\end{table}
The three events studied are: 23rd August 2005 from 14:22:00 (Flare 1),
14th/15th April 2002 from 23:58:00 (Flare 2) and 21st May 2004 from 23:40:00 (Flare 3).
All three flares share similar characteristics: {\it GOES\,} M-class flares with similar lightcurves, strong coronal X-ray loop top emission
and only relatively weak footpoint emission. Since the aim of the study is to examine how the properties of coronal X-ray sources
vary with time, these events were chosen as they show a clear coronal X-ray source throughout the rise, peak and decay stages
of X-ray emission and their spatial properties have been previously studied in energy by \cite{2008ApJ...673..576X} and \cite{2011ApJ...730L..22K}.
The coronal X-ray emission during each flare appears clearly, for study, up to $\sim 30$ keV and appears as a simple loop-like shape
connecting weak 30-40 keV HXR footpoints, during certain time intervals. The main parameters of each flare: {\it GOES} class, date,
observation time, peak time at 10 keV and the time of footpoint appearance are given
in Table~\ref{table1}. As discussed in Chapter \ref{ref:Chapter2}, length variations of each of these coronal loops with X-ray energy were studied by \cite{2008ApJ...673..576X}.
Length and width changes with X-ray energy for Flare 2 were also studied by \cite{2011ApJ...730L..22K}. It should be noted that Flare 1
and Flare 3 show similar results as Flare 2 in \cite{2011ApJ...730L..22K}, where both the length and width increase with
energy as $\sim\epsilon^{2}$ and $\sim\epsilon$ respectively but this chapter will concentrate on the size, position and spectral parameter
changes with time.
\subsection{Lightcurves for each event}
The lightcurves for each event are shown in Figure \ref{fig:is_paras} (top row). Flares 1 and 3 have similar lightcurves;
a simple shape with one peak. The lightcurves of four energy bands between 10-40 keV for Flare 1 are shown in Figure \ref{fig:is_paras} (top row, left plot). The study of this event begins
at 14:22:00. At this time, X-ray emission from the 10-20 keV energy bands are slowly rising and reach a peak at $\sim$14:30:00.
After this point, there is a gradual decrease in X-ray emission until $\sim$ 14:40:00, and continues to decrease until 14:50:00, where {\it RHESSI} enters into a night phase and can no longer view the flare. In the 20-40 keV band there are a series of peaks between the observation range of 14:22:00 and 14:40:00.
The lightcurve for Flare 3 is shown
in Figure \ref{fig:is_paras} (top, right plot). From the start of the study at 23:40:00, the X-ray emission from four energy bands between 14-40 keV rises and peaks around 23:50:00. After this peak, the X-ray emission from each energy band decreases. The lightcurves for Flare 2 are more complex and are shown
in Figure \ref{fig:is_paras} (top row, middle plot) for four energy bands from 10-40 keV. During the observational time there are two main peaks in X-ray emission at $\sim$00:03:00 and $\sim$00:12:00, possibly more peaks, which are most prominent in the 20-40 keV energy range.

\begin{figure*}
\vspace{-10pt}
\textbf{Flare 1}\\
\includegraphics[width=15.8cm]{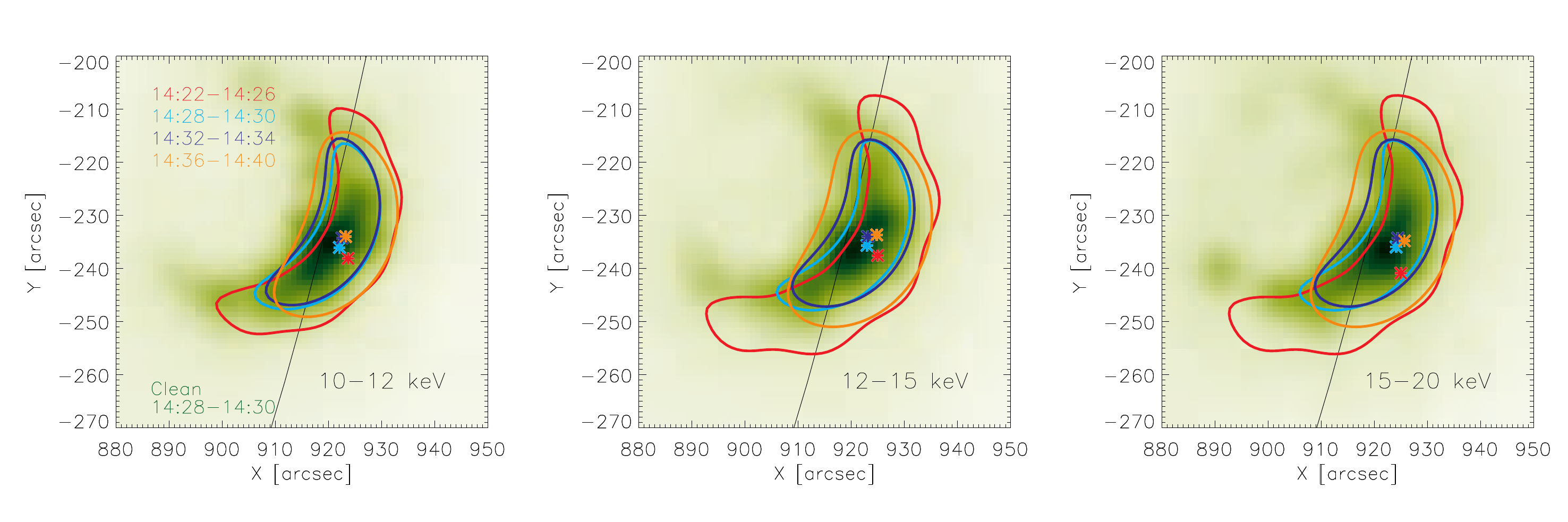}
\vspace{-10pt}
\textbf{Flare 2}\\
\includegraphics[width=15.8cm]{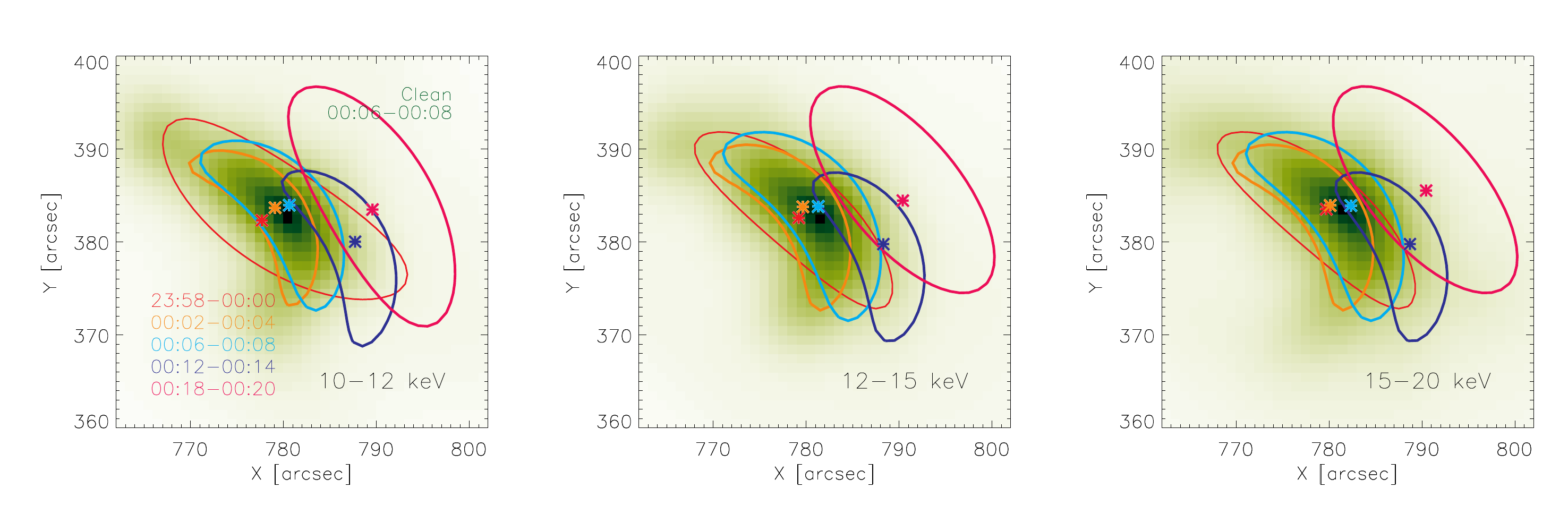}
\vspace{-10pt}
\textbf{Flare 3}\\
\includegraphics[width=15.8cm]{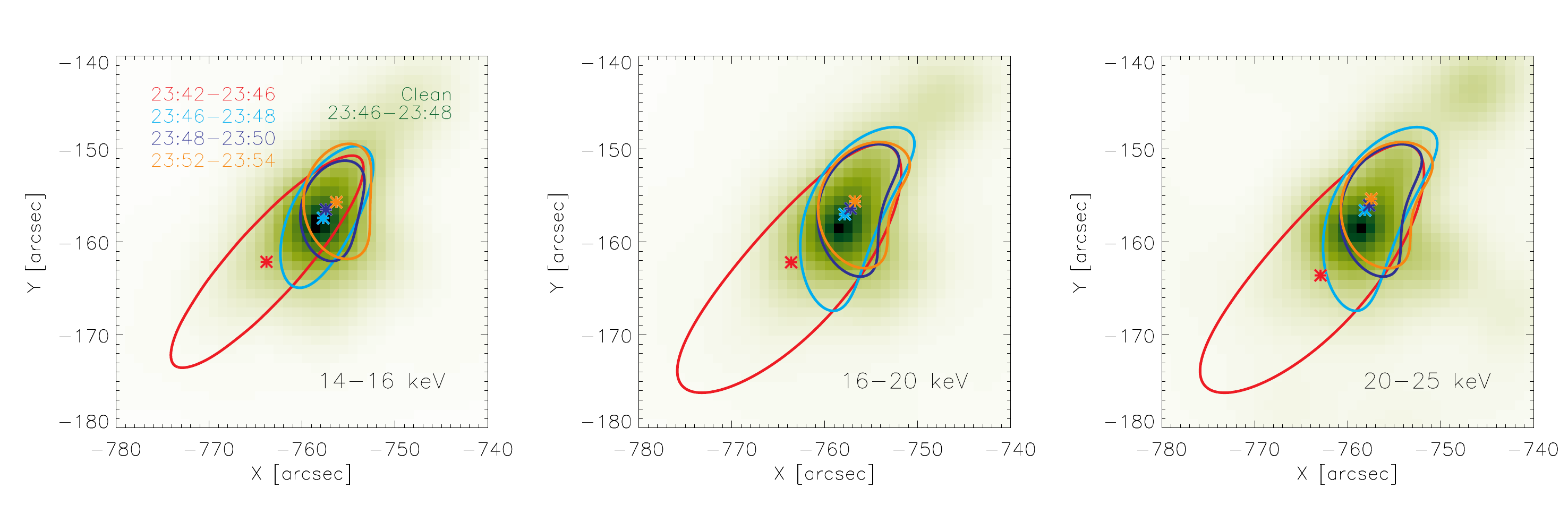}

\caption[CLEAN images and Vis FwdFit contours for Flares 1,2 and 3, at different observational energies and times.]{CLEAN background image (green) at one time and over-plotted Vis FwdFit contours (50\% maximum intensity)
at four times and energies of 10-12 keV (left), 12-15 keV (middle) and 15-20 keV (right) for Flare 1 (top) and Flare 2 (middle) and 14-16 keV (left), 16-20 keV (middle) and 20-25 keV (right) for Flare 3 (bottom). The asterisks denote the loop position for each time. The cyan Vis FwdFit contour matches the time of the CLEAN image.}
\label{images_flare}
\end{figure*}

\subsection{Imaging of each event}
Each event was studied in X-rays with {\it RHESSI} using the imaging algorithms of CLEAN \citep{1974A&AS...15..417H, 2002SoPh..210...61H},
Pixon \citep{1993PASP..105..630P,1996ApJ...466..585M} and Vis FwdFit \citep{2002SoPh..210...61H,2007SoPh..240..241S} (see Chapter \ref{ref:Chapter1}, Section \ref{intro_rhessi}).
Firstly, each event was studied using only CLEAN and Pixon. These imaging algorithms were used to confirm the loop shape of each coronal source and find the energy ranges over which an X-ray  coronal source was present in each flare. Confident that the
chosen events only had a simple and similar loop shape given by CLEAN and Pixon, each event was then studied using Vis FwdFit by fitting a curved elliptical Gaussian to each loop. It is important that the coronal source has a simple, singular loop-like shape so that Vis FwdFit can effectively fit a curved elliptical Gaussian to the X-ray visibilities and give reliable and realistic estimates with errors for the source parameters. Amongst other parameters, Vis FwdFit provides the following spatial parameters: loop length FWHM (full width half maximum), loop width FWHM and the $(x,y)$ centroid position of the loop, which are the mean coordinates of the loop shape. For this study of radial position, the $(x,y)$ position of the loop top is required, not the mean position of the loop shape itself. However, Vis FwdFit creates the elliptical Gaussian shape by placing a set of circular Gaussian sources along the length of the loop. Therefore, the coordinates for the coronal loop top position were simply obtained by extracting the coordinates of the central circular Gaussian.
\begin{figure*}
\centering
\includegraphics[width=7.15cm]{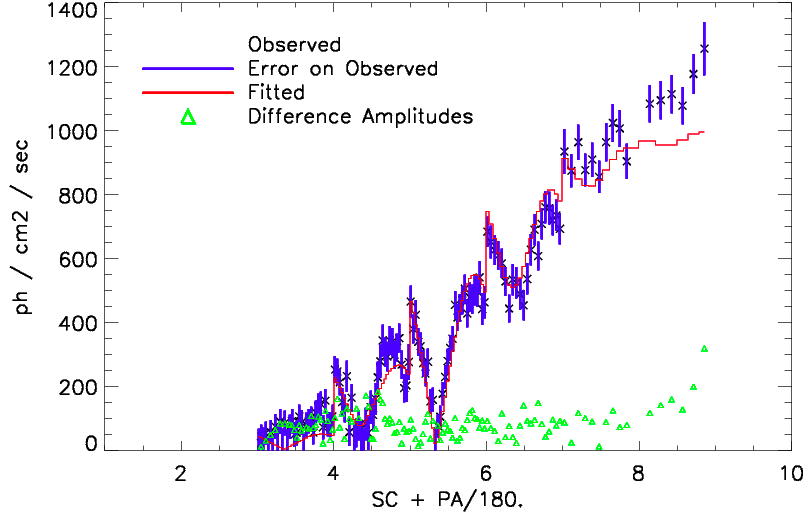}
%\hspace{-10pt}
\hspace{50pt}
\includegraphics[width=7.15cm]{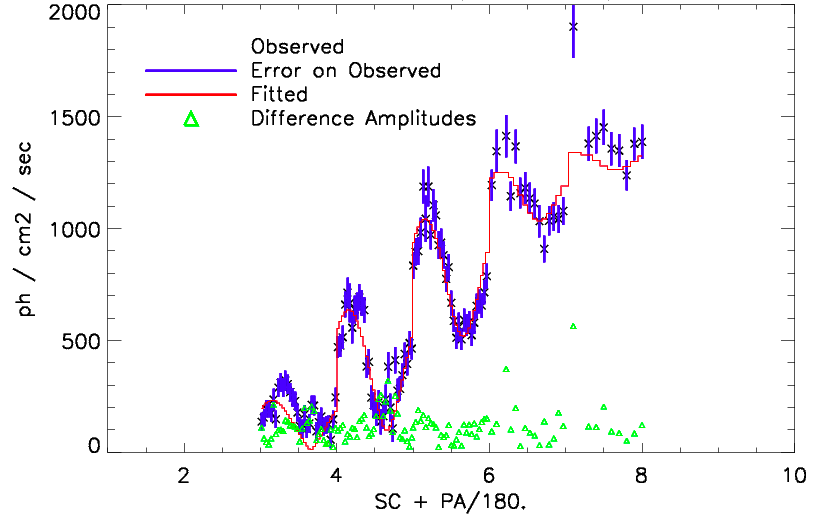}
%\hspace{-10pt}
\includegraphics[width=7.15cm]{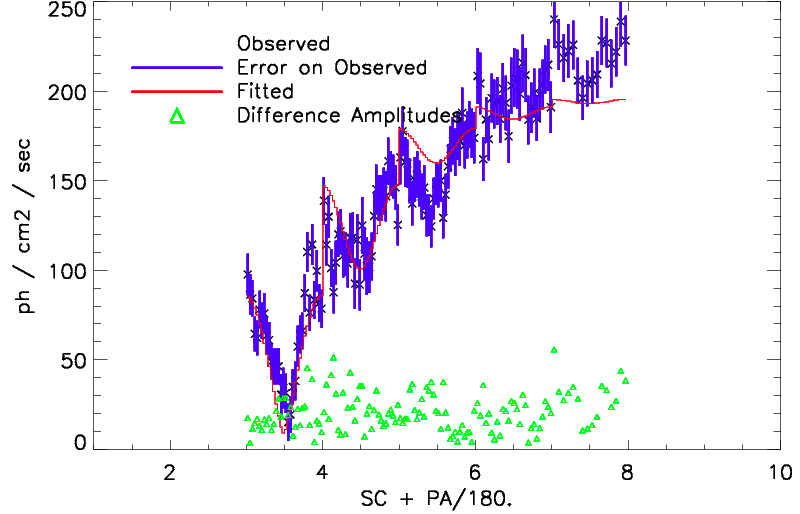}
\caption[The observed {\it RHESSI\,} visibility amplitudes plus the error bars at one chosen time bin for Flare 1, Flare 2 and Flare 3.]{The observed {\it RHESSI\,} visibility amplitudes (black stars) versus RMC (the minor x-axis gives the position angle of the RMC/180$^{\circ}$) plus error bars (blue) at one chosen time bin for Flare 1 (top), Flare 2 (bottom left) and Flare 3 (bottom right). The fitted elliptical Gaussian model is shown by the red line. The difference between the observed and fitted amplitudes is shown in the bottom plots by the green triangles.}
\label{visfits}
\end{figure*}
This is important as a loop that is very curved and approaching the shape of a ring will pull the shape centroid towards the ends of the loop; this could mask small changes in loop position with time and/or energy. This was especially significant for the large looped shape of the 23rd August 2005 event (Flare 1). In order to study changes in time at a specific energy range, the coronal sources of Flares 1, 2 and 3 were imaged over two minute intervals where possible and four minute intervals where the count rates were lower, during the rise and decay phases of the X-ray emission.
The exact time bins used for each flare are shown in Figures \ref{fig:is_paras} and \ref{fig:com_paras}.
The energy ranges of 10-12 keV, 12-15 keV and 15-20 keV were chosen for Flare 1 and Flare 2  and the energy
ranges of 14-16 keV, 16-20 keV and 20-25 keV were chosen for Flare 3, allowing the spatial parameters of the loops to be studied in both energy and time. The energy ranges from 10-20 keV or 14-25 keV were chosen since the X-ray loop appears clearly over these energy ranges allowing Vis FwdFit to be used reliably. An example of how well the loop model fits the observed {\it RHESSI\,} visibility amplitudes of each rotating modulation collimator (RMC) is shown in Figure \ref{visfits}. Images for Flares 1, 2 and 3 are shown
in Figure \ref{images_flare} respectively. Figure \ref{images_flare} shows that Flare 1
is a limb event while both Flares 2 and 3 are disk events. The main parameters of each flare are given in Table~\ref{table1}. Figure \ref{images_flare} plots background CLEAN images of the coronal source for each flare at one chosen time interval
and over-plots Vis FwdFit contours for selected time intervals, one corresponding to the same time interval as the CLEAN image.
Comparing the shape and size of the Vis FwdFit contour with the CLEAN background image at the selected time interval
for each flare shows good agreement between both algorithms. Qualitatively, Vis FwdFit also agrees well with the CLEAN and Pixon images at other times not shown in Figure \ref{images_flare}. For each algorithm, Flare 1 used detectors 3 to 8 due to its relative large size while Flares 2 and 3 used detectors 3 to 7. It should be noted that sources sizes observed with the CLEAN algorithm can be manually changed using a parameter known as the \textit{clean$\_$beam$\_$width$\_$factor} parameter. This parameter can be thought of as representing the instrument point spread function (PSF) which is convolved with the image reconstruction from CLEAN (see Chapter \ref{ref:Chapter1}, Section \ref{rhessi_imaging} and \cite{DennisPernak2009}). However since CLEAN and Pixon were only used to confirm the qualitative shape of the looped sources, the fact that the source sizes observed with CLEAN can be manually changed was of no initial concern and a \textit{clean$\_$beam$\_$width$\_$factor}$=$1.8 was chosen for each flare, since this was suggested by the observations of recent HXR chromospheric sources \citep*{2010ApJ...717..250K}. However, an additional study of source width was performed by plotting the intensity profiles of the CLEAN, Pixon and Vis FwdFit images, along a line through the centre of the loop top, midway and perpendicular to the line connecting the weak footpoints. This is shown in Figure \ref{cuts} (top left). A measure of loop width was then found by: (1) the standard deviation ($std$) of the profile distribution and (2) the $std$ of a Gaussian fit to the profile distribution, providing the loop width FWHM via FWHM$=2\sqrt{2\ln2}\;std$. Figure \ref{cuts} shows good agreement between the widths at each time and those found from Vis FwdFit. The standard deviation of the profile in particular shows for Flare 1, a \textit{clean$\_$beam$\_$width$\_$factor}$\sim$3.0 matches better with the results of Vis FwdFit than a \textit{clean$\_$beam$\_$width$\_$factor}$\sim$1.8. From Figure \ref{images_flare}, it should be noted that the results for Flare 3
are probably less reliable than those of Flare 1 and Flare 2. From the position of the footpoints in the CLEAN image,
it appears as through the southern `loop leg' is tucked underneath the observer's line of sight.
This means that it is harder for Vis FwdFit with a loop to fit it with a correctly shaped loop and usually fits it with a loop
that is slightly too large or departs from a loop-shape. Hence, events analysed for study with Vis FwdFit must be chosen with caution.
\begin{figure*}
\vspace{-20pt}
\hspace{-35pt}
\centering
\includegraphics[width=16.5cm]{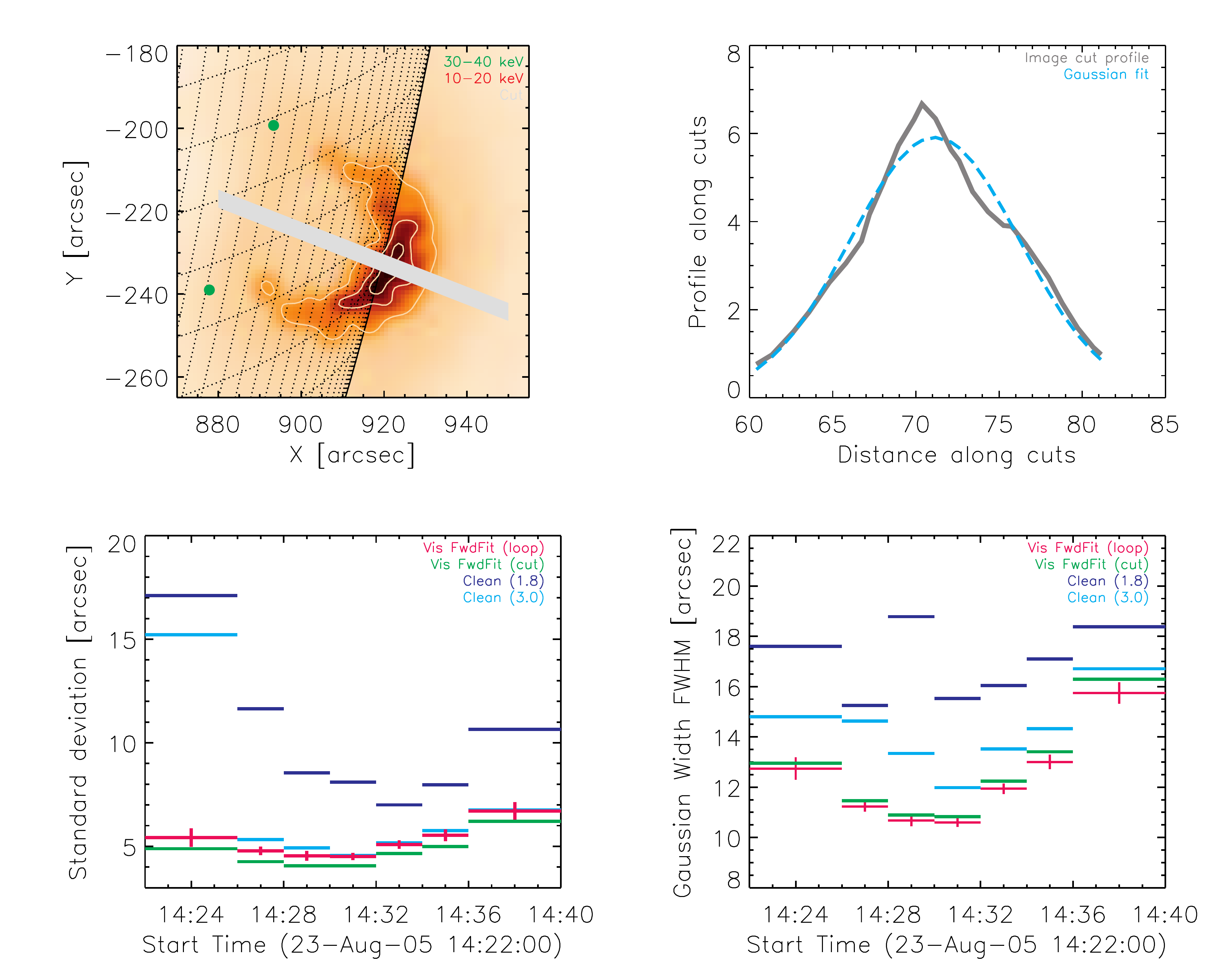}
\caption[For Flare 1, a comparison of the standard deviation of a chosen intensity profile along a line through the loop top perpendicular to line midpoint joining the footpoints, found from the second moment of the distribution, for both CLEAN and Vis FwdFit algorithms.]{{\it Top left:} Flare 1 CLEAN image at one time and energy range of 10-20 keV. The positions of 30-40 keV footpoints are shown by green dots. Loop top intensity profiles are found along the grey band, perpendicular to line midpoint joining the footpoints. {\it Top right:} total intensity profile (grey) and a Gaussian fit to the profile (light blue). {\it Bottom left:} comparison of the standard deviation of the intensity profile, found from the second moment of the distribution, for CLEAN intensity profiles, with either a \textit{clean$\_$sigma$\_$beam$\_$width}$=1.8$ (dark blue) or $3.0$ (light blue) and the Vis FwdFit intensity profile (green). The values given by the Vis FwdFit algorithm are also plotted (pink). {\it Bottom right:} as bottom left, but for the Gaussian FWHM, as found from the fits to the image profiles or from the Vis Fwd Fit algorithm.}
\label{cuts}
\end{figure*}
As was discussed in Chapter \ref{ref:Chapter2}, flares of this type only show relatively weak HXR emission from the chromosphere. It was found that Flare 1 has one weak but clear southern footpoint in the 30-40 keV range. This appears at $\sim$14:36:00, after the peak emission time from 10-20 keV and 14 minutes after the start of the observation start time, corresponding to a bump in the 20-40 keV band shown in the lightcurve for this event. Flare 3 has two very weak HXR footpoints in the 30-40 keV band during the entire observational period. The lightcurve for this event shows that the 25-40 keV band follows the trend of the lower energy bands, all roughly peaking at 23:50:00. Flare 2 is more complex than Flares 1 and 3 due to its multiple lightcurve peaks but it also has two weak footpoints in the 30-40 keV energy range. These appear at $\sim$00:05:00. Figure \ref{fig:is_paras} shows the imaging parameters for Flare 1 (column 1), 2 (column 2) and 3 (column 3):
loop width FWHM (row 2), loop length FWHM (row 3) and loop-top radial position (row 4) for each imaging
energy band. Above these plots, the lightcurve is plotted for each of the imaging energy bands to allow
comparison with changes in the spatial parameters. The dashed lines drawn in each plot represent
the time interval over which the peak X-ray emission occurs for the three energy bands.
For each flare, in general, peaks in the lightcurve represent changes of width, length and source position
parameters with time and this will be discussed in Section \ref{ref:results}
\subsection{Spectroscopy of each event}
\begin{figure*}
\vspace{-20pt}
\includegraphics[scale=.31]{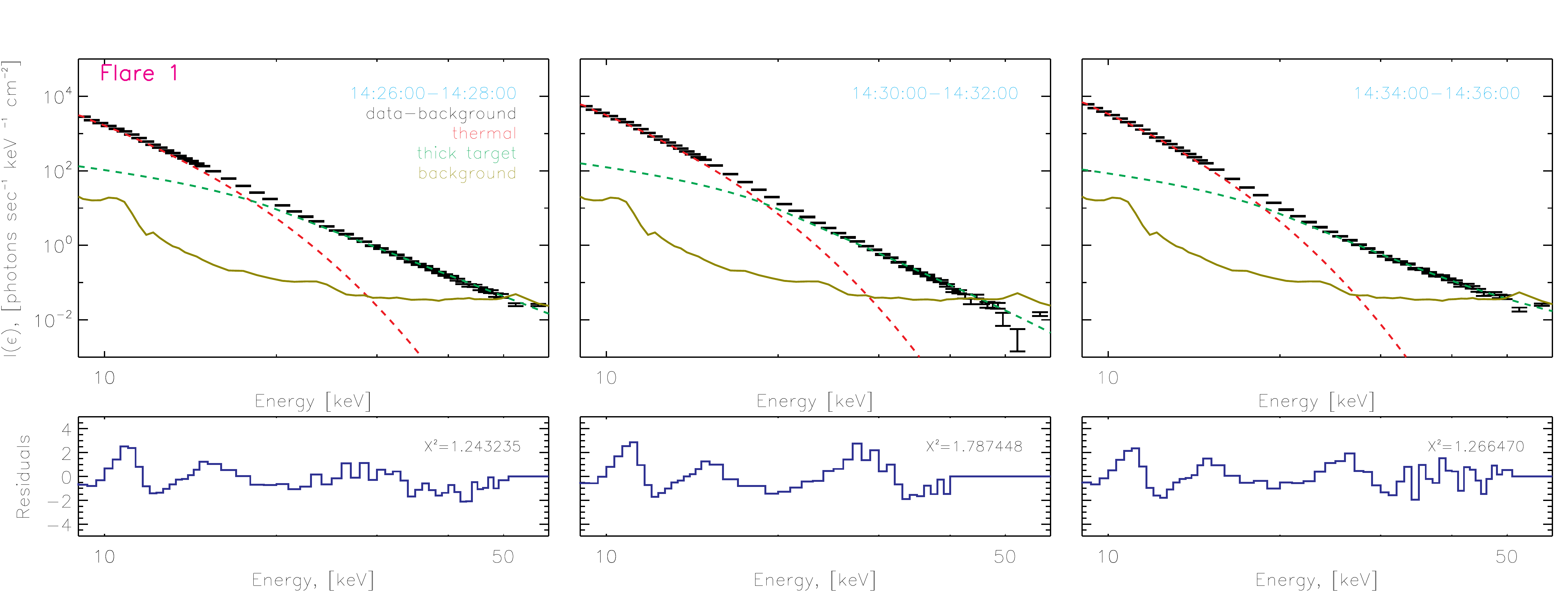}
\includegraphics[scale=.31]{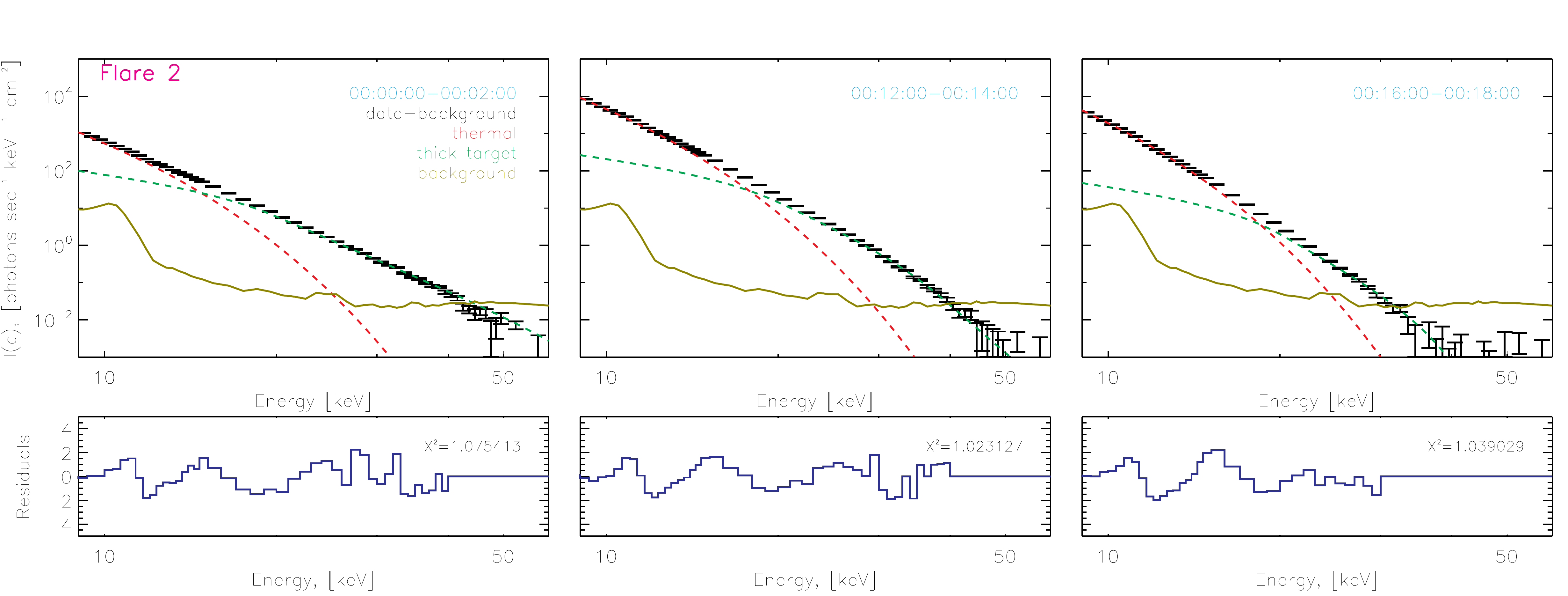}
\includegraphics[scale=.31]{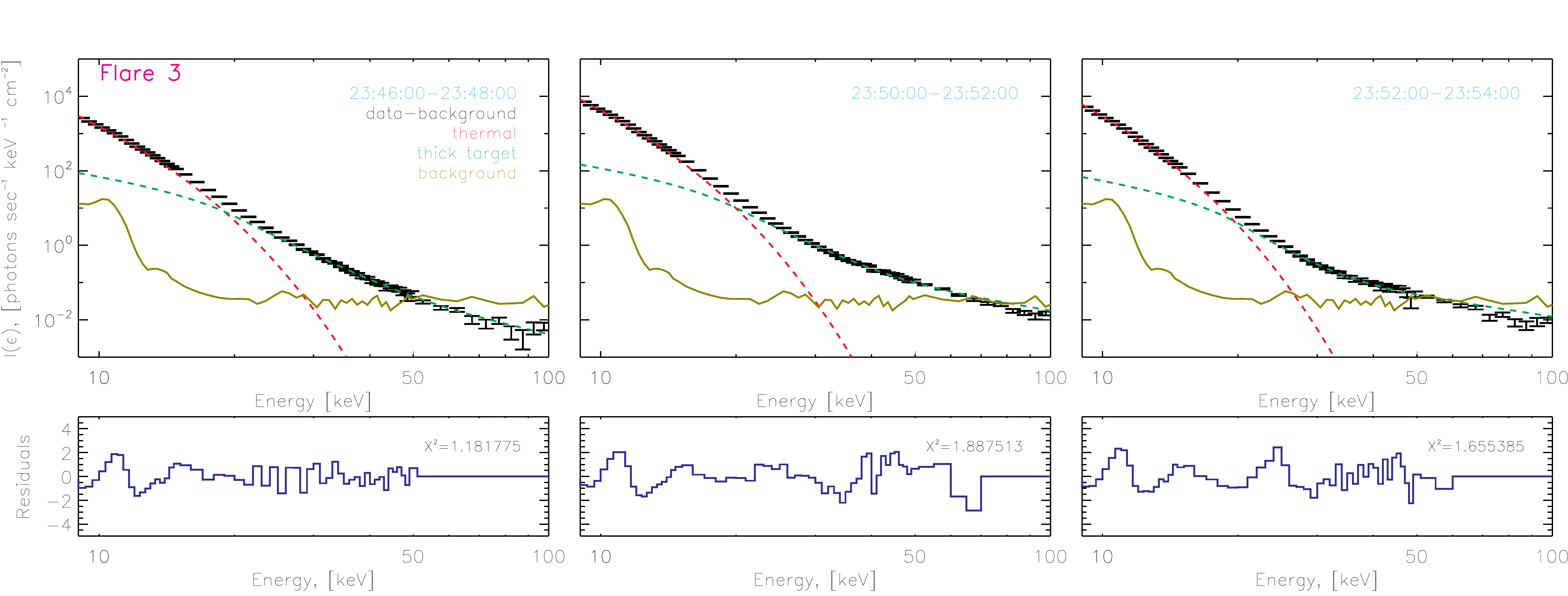}
\caption[Spectra for Flares 1, 2 and 3, at three chosen imaging time bins (during the X-ray rise, peak and decay stages).]{Spectra for Flares 1 ({\it top}), 2 ({\it middle}) and 3 ({\it bottom}), at three chosen imaging time bins (during the X-ray rise, peak and decay stages). Black = data - background, Olive = background, red = thermal fit, green = thick target bremsstrahlung fit. The residuals are plotted below each plot.}
\label{fig:spectra}
\end{figure*}

For each of the imaging time intervals for Flares 1, 2 and 3, energy spectra
were created. The spectra of each flare at each time interval were fitted with a thermal component (continuum only), a non-thermal
component corresponding to thick target bremsstrahlung and two Gaussian line functions for the Fe and Ni lines at 6.7 keV and 8.1 keV. The total fit function used is v$\_$th(continuum)$+$thick2$+$line$+$line. The thermal fits provide information about coronal loop plasma: the emission measure, $EM$, and
the plasma temperature, $T$, that can be used with the imaging parameters to infer the thermal plasma number density, pressure and energy density.
Spectra for Flares 1, 2 and 3 are shown in Figure \ref{fig:spectra} at three selected time bins corresponding to a rise, peak
and decay stage of X-ray emission for each flare. The last two rows of Figure \ref{fig:is_paras} plot how the thermal fit parameters $EM$ and $T$ vary with time for each flare.

\section{Spatial and spectral changes with time}\label{ref:results}
\subsection{Emission measure and plasma temperature}
For all three events, the emission measure $EM$ rises throughout the observation times,
either slowing or decreasing slightly during the last few minutes for each event.
For each event, the plasma temperature $T$ peaks before the peak in X-ray emission
and then slowly decreases after this point, as noted by \cite{1978ApJ...220.1137A}. The plasma temperature decreases much
slower than if it were decreasing by thermal conduction only, suggesting prolonged energy release at the later decay stages of the flare.

\subsection{Loop width}
For all three flares, there is a general pattern for emitting loop width $W$ changes with time.
Before a peak in X-ray emission, the source width $W$ for each flare tends to decrease
and after a peak in X-ray emission, $W$ for each flare increases.
It is also notable, that the rates of both width expansion and contraction (width change in time $dW/dt$) are
approximately the same, as can be seen in Figure \ref{fig:dvwl}.

{\it Flare 1-} For all energies plotted, $W$ decreases until the peak X-ray emission at $\sim$14:30:00-14:32:00 UT
and then increases after this point. The largest change occurs for the highest energy of 15-20 keV.
This falls from $14.5''$ to $11.5''$ at 14:28:00-14:30:00 and then increases until 14:40:00
where the width peaks at $\sim 20''$, producing a final larger loop width than seen at the beginning of the observational time.

{\it Flare 2-} For the energies of 10-12 keV and 12-15 keV, $W$ decreases before the first peak in X-ray emission
at $\sim$00:03:00. After this point $W$ increases before dropping at 00:12:00-00:13:00 where there is another X-ray peak
in the lightcurve. After this peak, $W$ continues to grow again. Before 00:03:00, the width of the 10-12 keV source
falls from around $\sim$10$''$ to $\sim5''$ and after 00:13:00 rises from $\sim$6$''$ to $\sim$13$''$. The 15-20 keV
source also shows this pattern except there is a larger peak at 00:05:00 and then a more pronounced
decrease in $W$ until 00:12:00-00:13:00 UT.

{\it Flare 3-} Again, the change in loop width $W$ with time follows a similar pattern as Flares 1 and 2. For all three energies
ranges considered, $W$ decreases from 23:42:00 to the peak in X-ray emission at 23:50:00.
$W$ then increases after this time. The 14-16 keV and 16-20 keV sources $\sim$ fall
from 7-8$''$ at 23:42:00 to $\sim$5$''$ at 23:50:00 and then increase up to 8-9$''$ at 23:58:00.
The 20-25 keV source falls from $11''$ to $5''$. Note the missing data at the fourth and sixth time bins for the 20-25 keV energy range, as Vis FwdFit was unable to fit successfully at these times, for this energy range.

\subsection{Loop length}

As with loop width, each flare shows general pattern for emitting loop length $L$ changes with time. Before a peak in X-ray emission,
the source length for each flare tends to decrease and after a peak in X-ray emission, the source length either
increases slightly or remains approximately constant within the errors.

{\it Flare 1 - } For all energies plotted, there is a rapid decrease in $L$ until the peak X-ray emission
at $\sim$14:30:00-14:32:00. After the peak, $L$ remains approximately constant (within the error).
The smallest decrease in loop length before the X-ray peak emission occurs for the 10-12 keV source which falls
from $54''$ to $38''$. The decrease in $L$ grows with energy and the highest drop in $L$ occurs
for the highest energies plotted at 15-20 keV, which fall from $74''$ to $38''$ before the peak in X-ray emission.

{\it Flare 2 - } The pattern for length changes with time are similar to that of the width changes. For the 10-12 keV source,
$L$ before the 00:03:00 peak in the lightcurve, falls from $\sim$26$''$ to $\sim$20$''$, rises
to $\sim$24$''$ at 00:11:00, falls to $\sim$20$''$ at 00:15:00 and then increases to $21''$ at 00:17:00. Due to the multiple peaks, $L$ tends to increase after a peak in the lightcurve.

{\it Flare 3 - } For all energies, $L$ drops rapidly between 23:42:00 and 23:50:00 from the range of $25''$ to 30$''$
to $\sim$10$''$ for all energies plotted. Again as with Flare 1, after the X-ray peak the length of the loop remains
approximately steady until the final plotted time of 23:58:00 for all energies.

\subsection{Loop radial position}

For all three flares, peaks in each X-ray lightcurve tend to denote times where the trends in loop radial position $R$ change.

{\it Flare 1- }Since this is a limb event,
the radial position $R$ can show whether the source is moving away or towards the limb. Before the X-ray emission peak
at 14:30:00-14:32:00, the source moves towards the limb, falling a distance of $\sim2''$ at 10-12 keV, 12-15 keV and 15-20 keV.
Then after the peak, the source moves away from the limb. Plotting the actual source positions, shows that the
entire loop structure moves in a U-shape during the time interval of 14:22:00-14:40:00. This was also seen for a number of flares in \cite{2008ApJ...686L..37S}. From the
results of Flare 1, the changes in $R$ with time are comparable to the width changes and smaller than the length changes. The largest change in position is only $2-3''$ while the width decreases by $3''$ and increases by $6''$.
The length shows the largest change with a decrease of $\sim17''$ or more before the peak in X-ray emission.

{\it Flare 2 - }Overall, the radial positions $R$ of the 10-12 keV, 12-15 keV and 15-20 keV sources increase with time.
At points of peak X-ray emission, that is, at $\sim$00:03:00 and $\sim$00:12:00, there are changes in the gradient.
The slope steepens between $\sim$00:03:00 and 00:12:00. Since this is a disk event, it is difficult to say whether there is a change
in source altitude at these peaks (as for Flare 1).  At 10-12 keV, between 23:58:00
and 00:20:00 the source radial direction changes by 9.5$''$. Therefore the overall change in position is larger than the individual changes in loop width and loop length for Flare 2.

{\it Flare 3 - }For all energies, the radial distance $R$ falls with increasing time. There does not seem to be any significant
difference in the radial distance trend after the X-ray emission peak at 23:50:00, apart from the steadier decrease in radial distance
after this time. As with Flare 2, Flare 3 is a disk source and hence it is difficult to determine if there is any altitude change. For all three energies, the radial distance decreases by $\sim$8$''$ between 23:40:00 and 23:58:00.

\begin{figure*}
\vspace{-35pt}
\includegraphics[width=5cm]{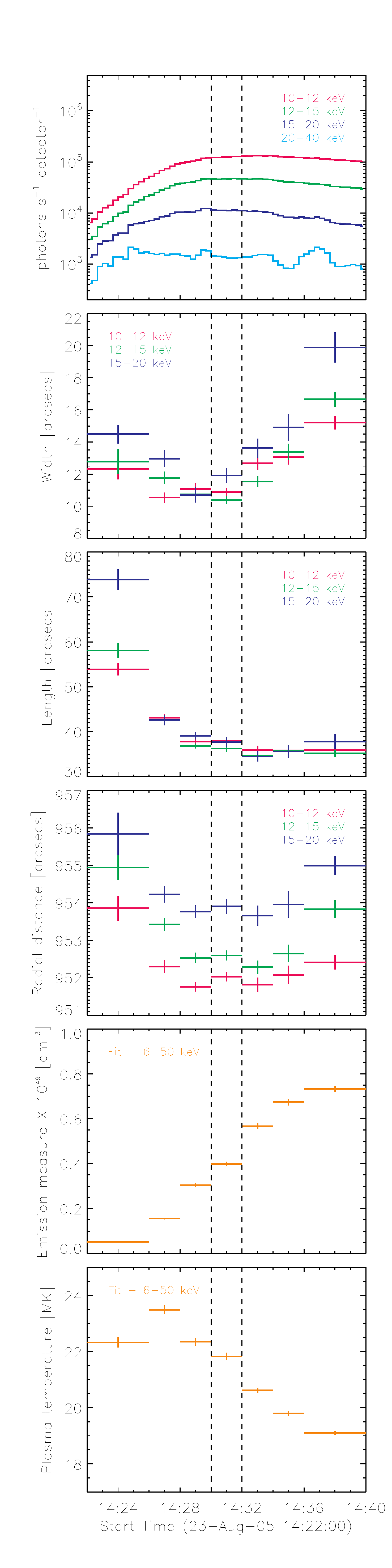}
\includegraphics[width=5cm]{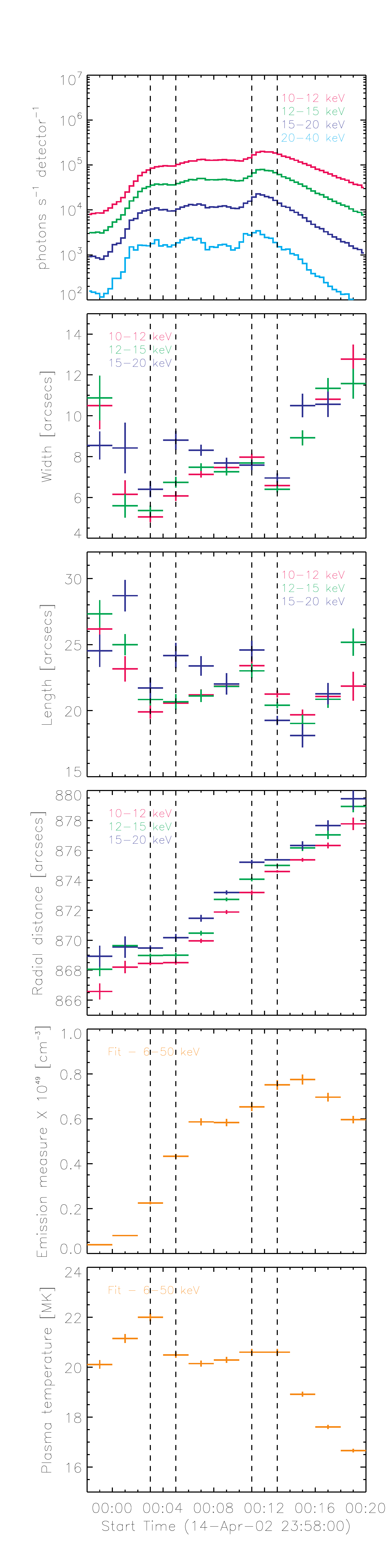}
\includegraphics[width=5cm]{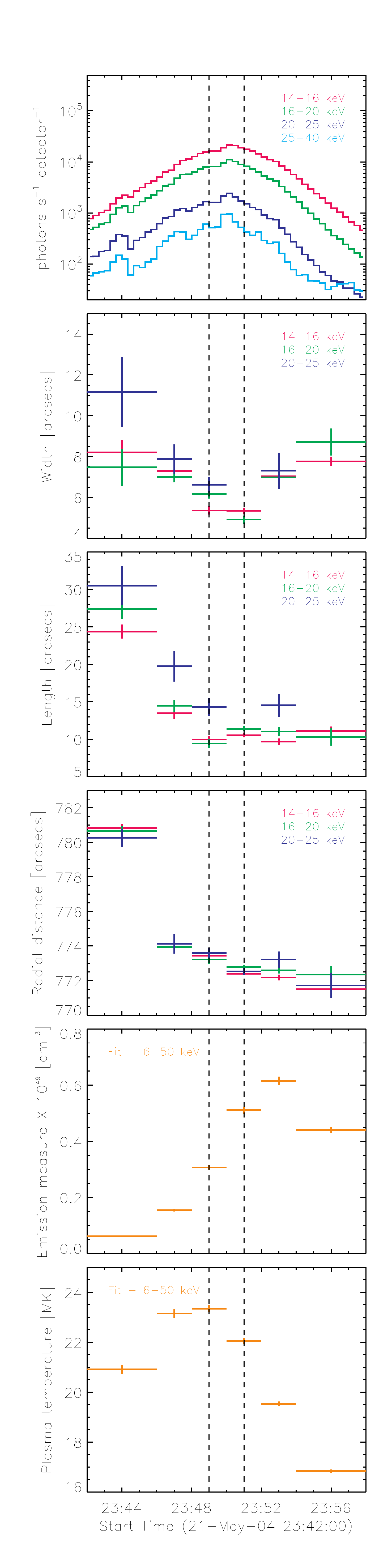}
\caption{{\it Left:} 23-Aug-2005,
{\it middle:} 14-Apr-2002 and {\it right:} 21-May-2004. row 1: lightcurves, row 2: width, row 3: length, row 4: radial position, row 5: emission measure and row 6: plasma temperature, vs. time.
Dashed lines: peak X-ray emission.}
\label{fig:is_paras}
\end{figure*}

\section{Corpulence, volume and other inferred parameters}

\subsection{Loop corpulence}
In the following sections of this chapter, the word `corpulence' will be used to define the shape of a loop. Here corpulence $\mathcal{C}$ will be defined as the ratio of the loop width $W$ to loop length $L$
\begin{equation}\label{corpulence}
\mathcal{C}=\frac{W}{L}.
\end{equation}
This means that for a given loop length FWHM $L$, the loop corpulence $\mathcal{C}$ will increase with increasing loop width FWHM $W$. Therefore a loop has a very high corpulence if $\mathcal{C}>1$ and a low corpulence if $\mathcal{C}<<1$. For a corpulence of 1, the length and width of the loop are equal. Loop corpulence for each event is plotted in the second row of Figure \ref{fig:com_paras} for Flares 1, 2 and 3. In general, the loop corpulence increases with time. This is particularly noticeable for Flare 1, where the loop corpulence increase throughout the observation time for each of the three energy bands. For example, in the 12-15 keV band $\mathcal{C}$ increases from $\sim0.2$ to $\sim0.45$. For Flare 2, the changing corpulence with time is more complex, as expected, and it follows the same trend of both the length and width parameters that are very similar for Flare 2. However, overall $\mathcal{C}$ is larger at the final observational time than at the start time. The corpulence for Flare 3 is similar to that of Flare 1. Overall, it increases throughout the observational time and is larger at the end time than at the start time, with only a small dip at the time of peak X-ray emission. For the 12-15 keV band, $\mathcal{C}$ increases from $\sim0.3$ to $\sim0.8$.

\subsection{Volume, number density, thermal pressure and energy density}
\begin{figure*}
\vspace{-30pt}
\includegraphics[width=5cm]{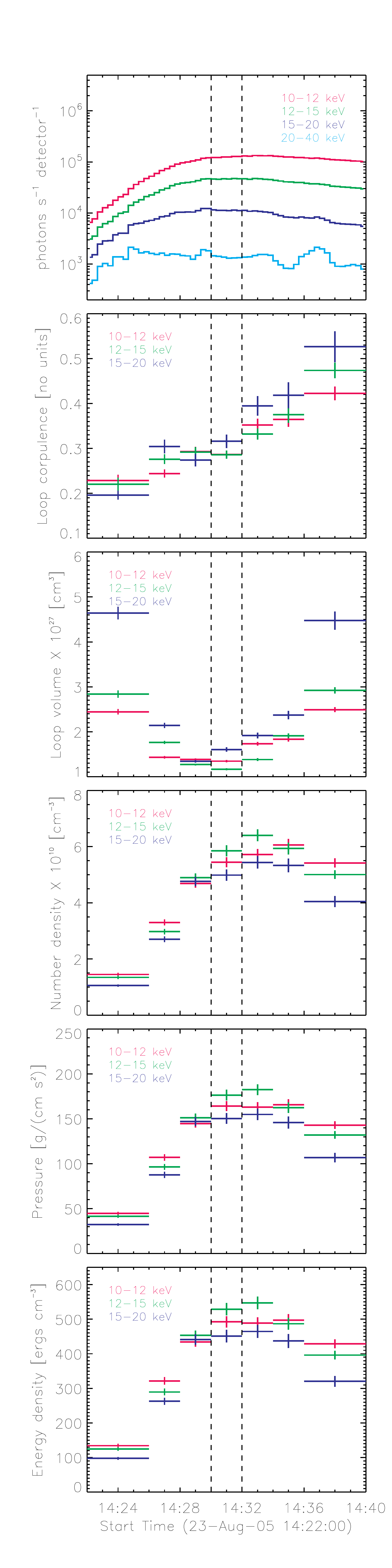}
\includegraphics[width=5cm]{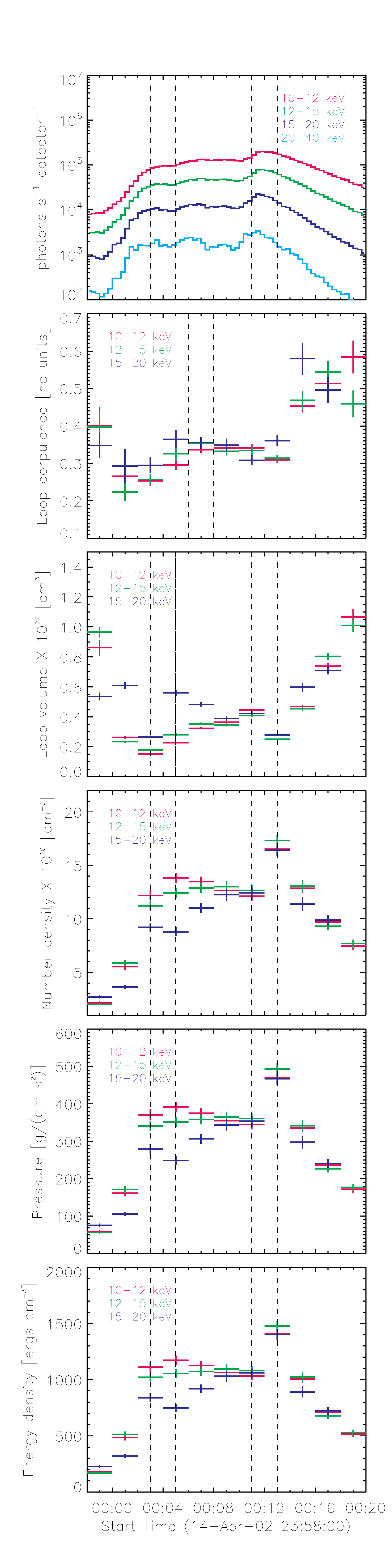}
\includegraphics[width=5cm]{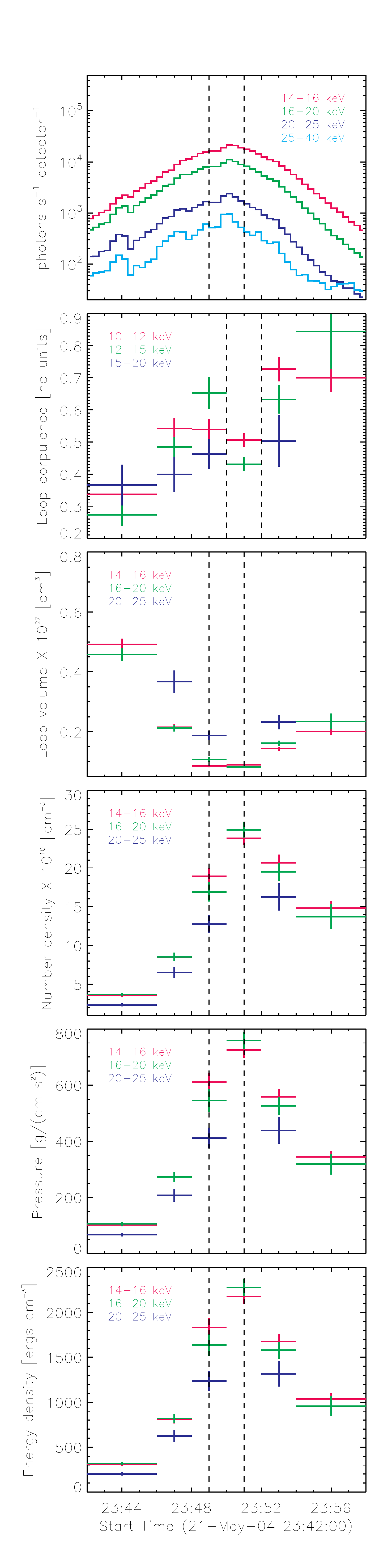}
\caption{{\it Left:} 23-Aug-2005,
{\it middle:} 14-Apr-2002 and {\it right:} 21-May-2004. row 1: lightcurves, row 2: corpulence, row 3: volume,
row 4: number density, row 5: thermal pressure and row 6: thermal energy density, versus time.}
\label{fig:com_paras}
\end{figure*}

From the loop width FWHM $W$ and loop length FWHM $L$, the general changes in the emitting loop volume, $V$, can be inferred
for each flare over time, at each energy band. It is assumed the volume of the loop is given by
$V=\pi W^{2}L/4$ that is, assuming a cylindrical loop. The changes in emitting loop volume with time for Flares 1, 2 and 3 are plotted
in Figure \ref{fig:com_paras} (second row). In general, for all three events, the source volume decreases before a peak
in X-ray emission and increases after a peak in X-ray emission. The changes in plasma number density, thermal pressure and thermal energy density can all be calculated using combinations of loop volume, emission measure
and plasma temperature. The plasma number density, $n$, can be obtained via
$n=\sqrt{EM/V}$, the pressure, $P$, from $P=nk_{B}T$, where $k_{B}$ is the Boltzmann constant
and finally the energy density, $U=3nk_{B}T$. The variation of these quantities with time for Flares 1, 2 and 3
are shown in Figure \ref{fig:com_paras}  (third, fourth and fifth rows respectively). It should be noted that it is assumed that the entire loop volume is emitting, that is the filling factor $f=1$. This means that the calculated values for number density, thermal pressure and energy density are a lower limit, and will increase if $f\le1$.

{\it Flare 1 - }As expected from the width and length results,
the loop volume falls between 14:22:00 and the peak in the X-ray lightcurve at $\sim$14:30:00 and then rises
after this time for all three energies. For all three energy bands, the decrease and increase in loop volume occurs at roughly the same rate. The largest decrease is for the highest 15-20 keV band, falling from $\sim4.7\times10^{27}$ cm$^{3}$ to $\sim1.4\times10^{27}$ cm$^{3}$ at 14:29:00 and then rising after this time to $\sim4.7\times10^{27}$ cm$^{3}$ at the last observational time. The 10-12 keV band falls from $\sim2.4\times10^{27}$ cm$^{3}$ to $\sim1.3\times10^{27}$ cm$^{3}$ at 14:31:00 and then rises back to $\sim2.5\times10^{27}$ cm$^{3}$ at the final observational time. The number density, thermal pressure and energy density for all three energy bands tend to follow
the same pattern, rising to a peak at some time after the peak X-ray emission and then slowly decreasing. For the 10-12 keV band, the number density rises from $1.5\times10^{10}$ cm$^{-3}$ at 14:22:00
to $6.5\times10^{10}$ cm$^{-3}$ at 14:35:00. It then falls to $\sim5.6\times10^{10}$ cm$^{-3}$ at 14:38:00. The 12-15 keV and 15-20 keV bands follow similar patterns, peaking at $\sim$14:33:00.
The pressure rises from $\sim40$ g/[cm s$^{2}$] at 14:22:00
and reaches $170$ g/[cm s$^{2}$]  at 14:31:00, the time where the X-ray emission peaks. After this time the pressure
remains approximately constant at $\sim160-170$ g/[cm s$^{2}$] until the last observation time where it falls to $\sim140$ g/[cm s$^{2}$]. The 12-15 keV and 15-20 keV energy bands peak at $\sim$14:33:00.
The thermal energy density of the plasma is just the thermal plasma pressure multiplied by 3 and hence it follows the same pattern as plasma pressure throughout the observed duration of the flare.
The energy density peaks at $500$ ergs cm$^{-3}$ in the 12-15 keV band at $\sim$14:33:00, after the peak X-ray emission.

{\it Flare 2 - }For the 10-12 keV and 12-15 keV sources, the loop volume falls until the first peak at 00:03:00
and then increases until it reaches 00:12:00, drops at 00:13:00 and then increases again after this time.
As for Flare 1, the number density, thermal pressure and energy
density all follow the same pattern for each energy band, rising to a time at or just after the peak in X-ray emission and then decreasing after this point. The highest number density, thermal pressure and energy density for each energy band occurs at 00:12:00 to 00:14:00, just after the peak in X-rays at 00:11:00-00:13:00. The number density peaks at around $18\times10^{10}$ cm$^{-3}$ in the 12-15 keV band, while the thermal pressure and energy density peak at $\sim500$ g/[cm s$^{2}$] and $1500$ ergs cm$^{-3}$ respectively at this time and energy.

{\it Flare 3 - }The loop volume again falls before the peak in X-ray emission at 23:50:00
and then rises again after this time, for all three energies.
For this flare again, the number density, thermal pressure and energy density all peak just after the peak in X-ray emission. At 23:50:00-23:52:00 the number density peaks at $23\times10^{10}$ cm$^{-3}$, the thermal pressure peaks at $750$ g/[cm s$^{2}$] and the energy density at $\sim2300$ ergs cm$^{-3}$, in the 16-20 keV band.

\section{Summary and discussion}

Using visibility forward fitting, a dedicated study was performed for the first time of changing spatial and spectral properties
of three coronal X-ray loops with time during the flare. All X-ray loops exhibited similar changes in both their spatial and spectral properties and hence the results indicate that a common process is
occurring for all three events; the emitting flaring loop volume is decreasing before the peak in X-rays and increases after the peak in X-rays is reached.
Before the peak X-ray emission, the emitting lengths and widths of each coronal loop decreased with time, indicating that the X-ray emitting region of the loop volume
was contracting; there was a reduction in loop width and length, as the X-ray emission from the region grew. After the X-ray peak, the loop width increased at approximately the same rate as during the contraction stage. For Flares 1 and 3, with one peak in their lightcurves, once the minimum X-ray loop length was reached during the X-ray peak it remained approximately constant, at least within the errors of the results. It was found that a property defined as the loop corpulence (equal to the loop width divided by loop length), in general, increases with time during the observation time.

Similar to previous studies \citep{1978ApJ...220.1137A,1980ApJ...242.1243S,1984ApJ...285..835G,1999ApJ...514..472M},
spectroscopy for each event showed that the plasma temperature initially grew but began to decrease before the peaks in X-ray emission and emission measure. The emission measure for each flare generally grew with dips observed during the final observational times of Flares 2 and 3. At the same time, the number density, thermal pressure and energy density of the plasma
also increased as the X-ray emission grew.  The plasma temperature decreased much slower than by thermal conduction only, even during the X-ray decay phase, suggesting at later stages additional energy
release is required \citep[e.g.][]{2011A&A...531A..57K} to explain the longer lasting X-ray loop emission.
For the limb event (Flare 1), there is a decrease in loop altitude before the peak in X-ray emission of roughly $2''\simeq1.4$ Mm, which is comparable in magnitude to the decrease in loop width but overall, the largest changes occurred for the X-ray loop volume, not the X-ray loop position. Decreases in loop altitude before the peak in X-ray emission have been well noted before and are often referred to as coronal implosion or loop contraction. Some of these observations were briefly discussed in the introduction of this chapter (Section \ref{ref:intro}).

\cite{2004ApJ...612..546S} also studied the loop position changes of Flare 2. From their observations, they did conclude
that the loop was indeed decreasing in altitude before the first peak in the lightcurve at 00:04:00.
They concluded that the loop decreased by $\sim2''$ over a 4 minute period, which is consistent with the radial distance results for Flare 2. However, no altitude decrease before the larger peak in X-ray emission at 00:12:00 was noted, where again there is a decrease in loop length and loop width.

\begin{figure}
\centering
\includegraphics[width=7cm]{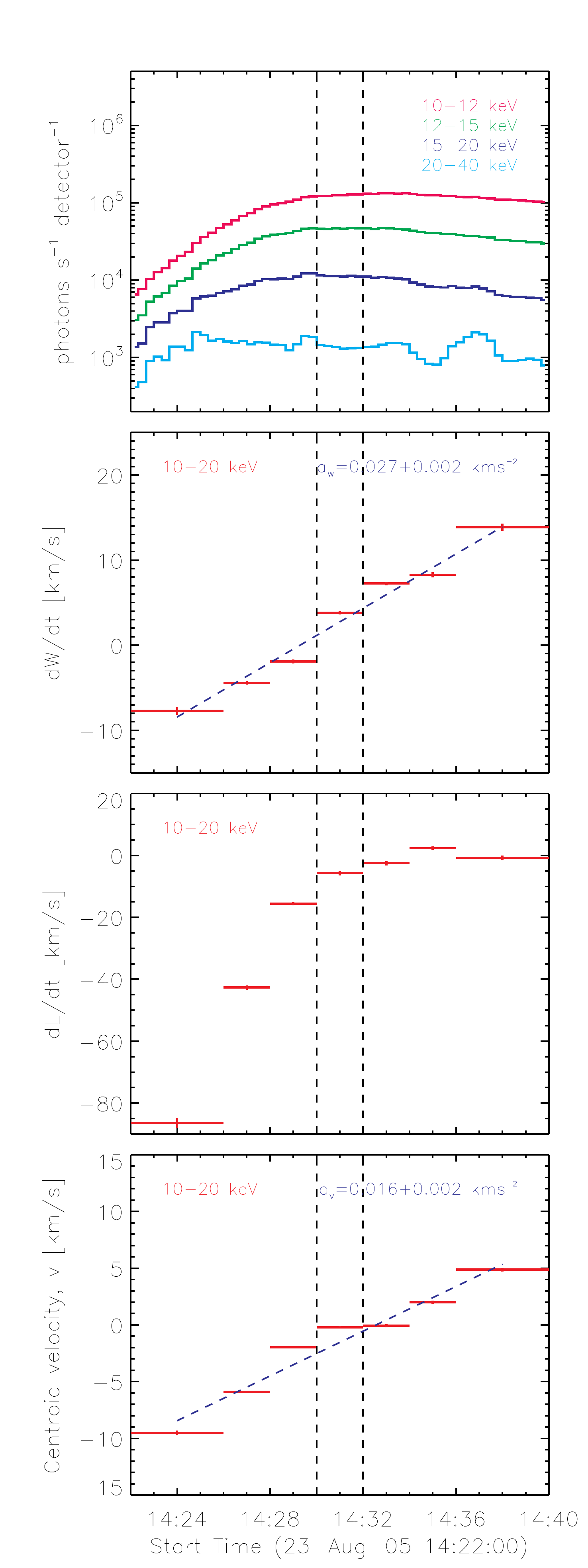}
\caption[For Flare 1: lightcurve (row 1), $dW/dt$ (row 2), $dL/dt$ (row 3) and $dr/dt=v$ (row 4).]{For Flare 1 (23-Aug-2005), lightcurve (row 1), $dW/dt$ (row 2), $dL/dt$ (row 3) and $dR/dt=v$ (row 4). Both $dW/dt$ and $v$ can be
fitted using a straight line indicating constant accelerations $a_{W}$ and $a_{v}$ and hence constant forces.}
\label{fig:dvwl}
\end{figure}

It has been suggested that a decrease in loop altitude may be an indication of collapsing magnetic trap acceleration/heating \citep{1966PASJ...18...57T,1997ApJ...485..859S,2002A&ARv..10..313P,2004A&A...419.1159K,2006A&A...446..675V,2012A&A...546A..85G}.
Although this may be the case for other events, it is not believed that the collapsing magnetic trap is the prime solution
to these observations, firstly due to overall larger loop volume changes over relatively small position changes.
\cite{2006A&A...446..675V} observed the coronal loop of a {\it GOES} X-class flare and found downward velocities as large
as $\sim14$ km/s in 10-15 keV band and $\sim29$ km/s in the 15-20 keV band prior to the peak in X-ray emission.
This is not what is observed for these events. For Flare 1, there is an average downward velocity in the 10-20 keV
band of $\sim4$ km/s, which is generally comparable to the decrease in loop width during this period.
Average changes in loop radial position for Flares 2 and 3 are also $\sim4$ km/s. More convincingly, in a collapsing magnetic
trap model, simple compressive heating during the contraction stage would imply that
$NT\propto1/A$ \citep{1978ApJ...223.1058M,1981ApJ...244..653E}, where $N$ is the number of particles in the region,
$T$ is the plasma temperature and $A$ is the cross-sectional area of the region, given by $\pi W^{2}$, where $W$ is the loop width.
Figure \ref{fig:NTA} plots $\log{NT}$ against $\log{A}$ for each flare and shows this not to be the case.
Straight lines fits to both the contraction and expansion phases show
that the gradients are either greater or less than $-1$.

\begin{figure*}
\centering
\includegraphics[width=15.5cm]{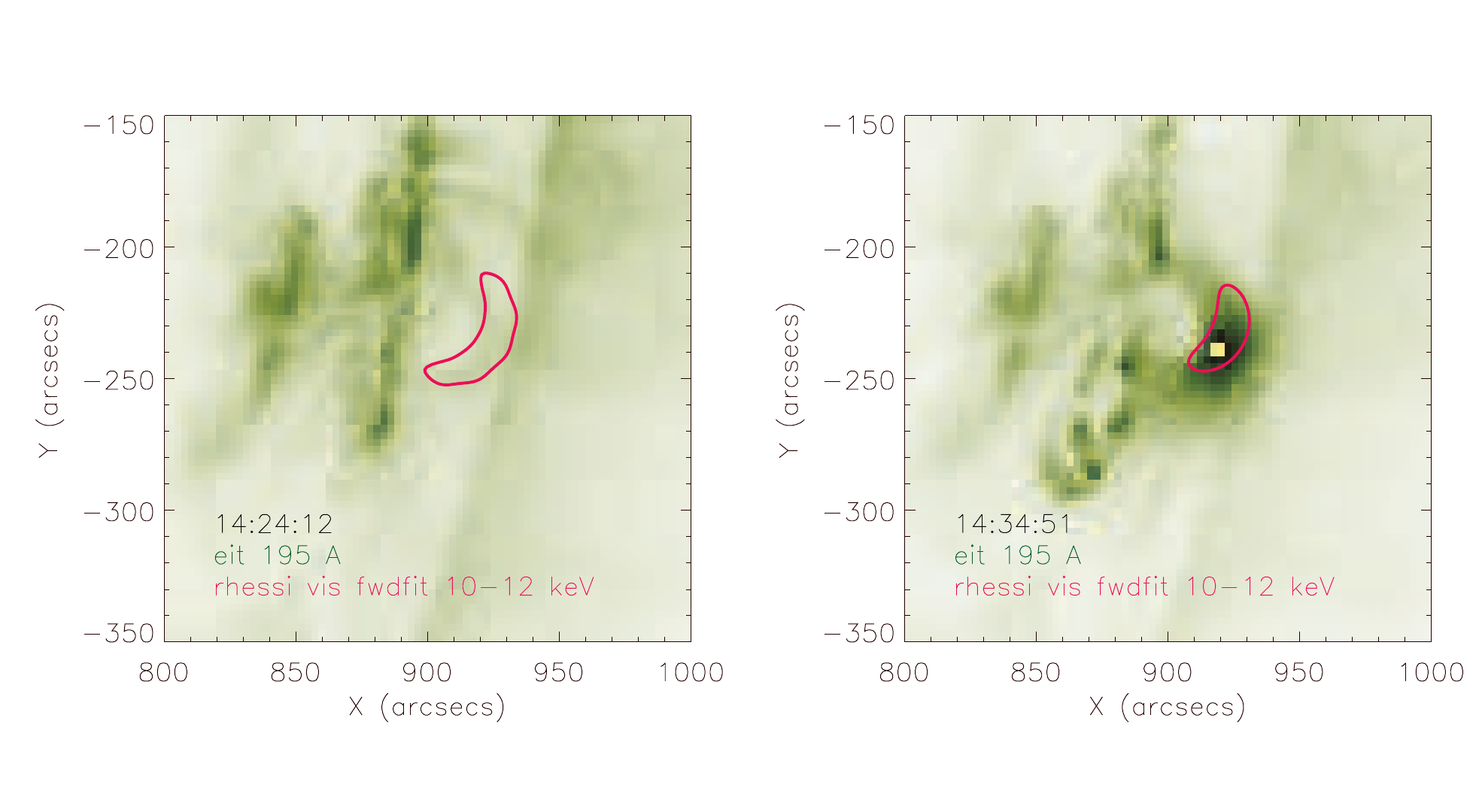}
\caption[{\it SOHO} EIT $195\AA$ images for Flare 1 at the times of 14:21:12 and 14:34:51, corresponding
to the times of rise and peak in X-ray emission.]{{\it SOHO} EIT $195$~{\AA} images for Flare 1 at the times of 14:21:12 and 14:34:51, corresponding
to the times of rise and peak in X-ray emission. {\it RHESSI} 10-12 keV X-ray contours are over plotted in pink.
Note the lack of bright EUV emission at the rise stage of this flare.}
\label{fig:fig_eit}
\end{figure*}

Observations by \cite{2011ApJ...730L..22K} showed how the loop width of the 14/15th April 2002 event (Flare 2) increased with
energy during two times corresponding to the first rise and first decay stages of the observations and suggested that the presence of magnetic
turbulence (the diffusion of field lines) was the cause of the energetic width increases. Hence, the suggestion that magnetic
turbulence in the region can account for the energetic changes in loop width, may also be able to account for the extra
energy in the loop, since both \cite{2011ApJ...730L..22K} and \cite{2011A&A...535A..18B} found the energy density of
magnetic fluctuations to be significant and could be comparable to that of the flaring plasma and higher than the energy density of non-thermal particles. Flares 1 and 3 also show similar length and width increases with energy at a single time range, also suggesting the presence of magnetic turbulence.

Many observations \citep{2005ApJ...629L.137L,2009ApJ...696..121L,2009ApJ...706.1438J,2010ApJ...724..171R,2012ApJ...749...85G} of loop height and length changes have been explained in terms of a reduction in magnetic
pressure. Usually the reduction in magnetic pressure is referred to as Taylor relaxation but this only refers to a special case where the resulting field is linear force free \citep{1974PhRvL..33.1139T}. The reduction in magnetic pressure could also account for the reduction of loop width
or cross-sectional area, as shown in simulations by \cite{2007A&A...472..957J}, and hence the observed trends of number density and pressure.
\cite{2009ApJ...696..121L} studied coronal implosion of one coronal source. They explained the reduction in height of a coronal source
before the peak in X-ray emission in terms of Taylor relaxation. They also suggested that this type of event can only occur
if the coronal loop is already filled with hot, dense plasma before the onset of a new event, that is from a previous event in that region.
The fact there is no EUV emission during the rise phase of Flare 1 seems to correlate with the observations
and suggestions of \cite{2009ApJ...696..121L}. Figure \ref{fig:fig_eit} shows {\it SOHO} (Solar and Heliospheric Observatory)
EIT (Extreme Ultraviolet Imaging Telescope) 195~{\AA} images at the times of 14:21:12 and 14:34:51 for Flare 1.
X-ray emission contours at 10-12 keV for 14:22:00-14:26:00 and 14:34:00-14:36:00 are also over plotted.
From Figure \ref{fig:fig_eit}, during the rise phase there is no bright 195~{\AA} EUV emission emanating from the loop,
only 10-20 keV X-ray emission. After the peak in X-ray emission, EUV emission can be observed from the loop.
There is an overall increase in the number of particles, $N$, within the loop region throughout the
duration of all three events. Due to low coronal densities, chromospheric evaporation probably accounts for this increasing $N$, initially driven by thermal conduction and possibly at later times
by electrons reaching the chromosphere, where there is weak footpoint emission and EUV emission from the loop.

It should be noted that a study of the 23rd July 2002 flare by \cite{2010ApJ...725L.161C} calculated temporal volume changes using the CLEAN algorithm and assuming an elliptical geometry. Overall this flare shows a general trend consistent with the results; an overall decrease in volume before the peak in X-ray emission and an overall increase in volume after the peak in X-ray emission. This flare also shows a peak in plasma temperature before the peak in X-ray emission and a high number density (and hence thermal pressure) after the X-ray emission first peaks.
\begin{figure*}
\centering
\includegraphics[width=7.5cm]{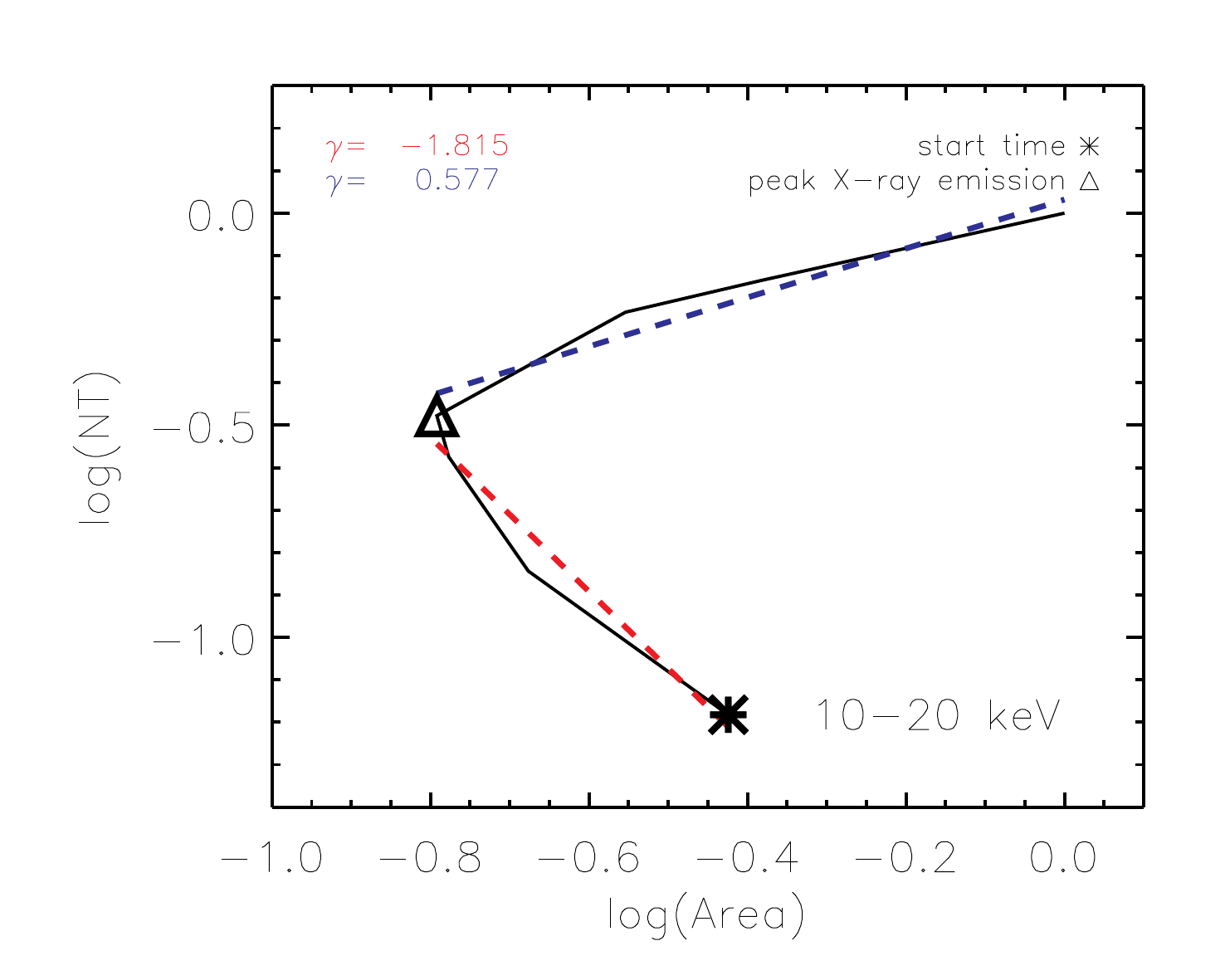}
\hspace{50pt}
\includegraphics[width=7.5cm]{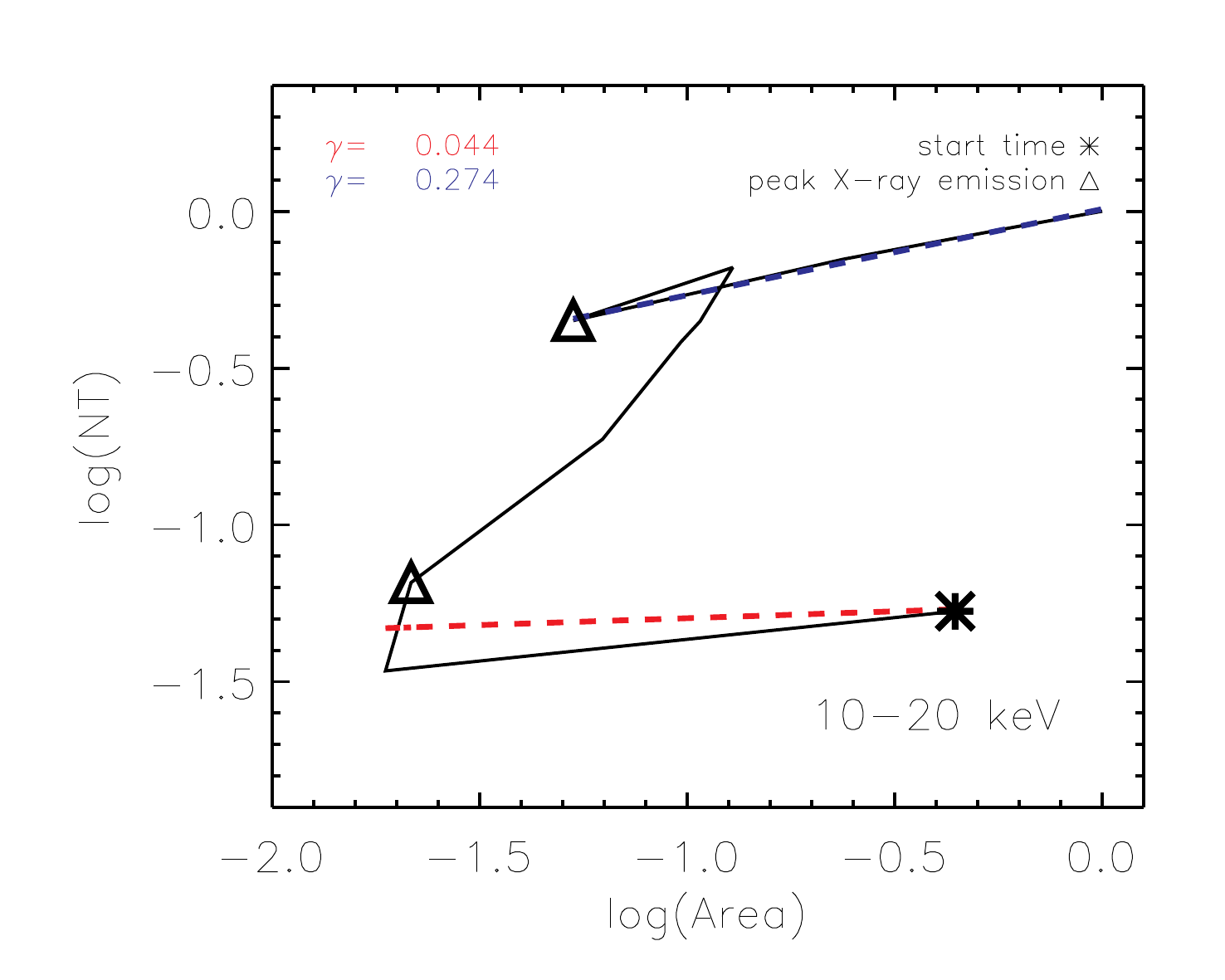}
\includegraphics[width=7.5cm]{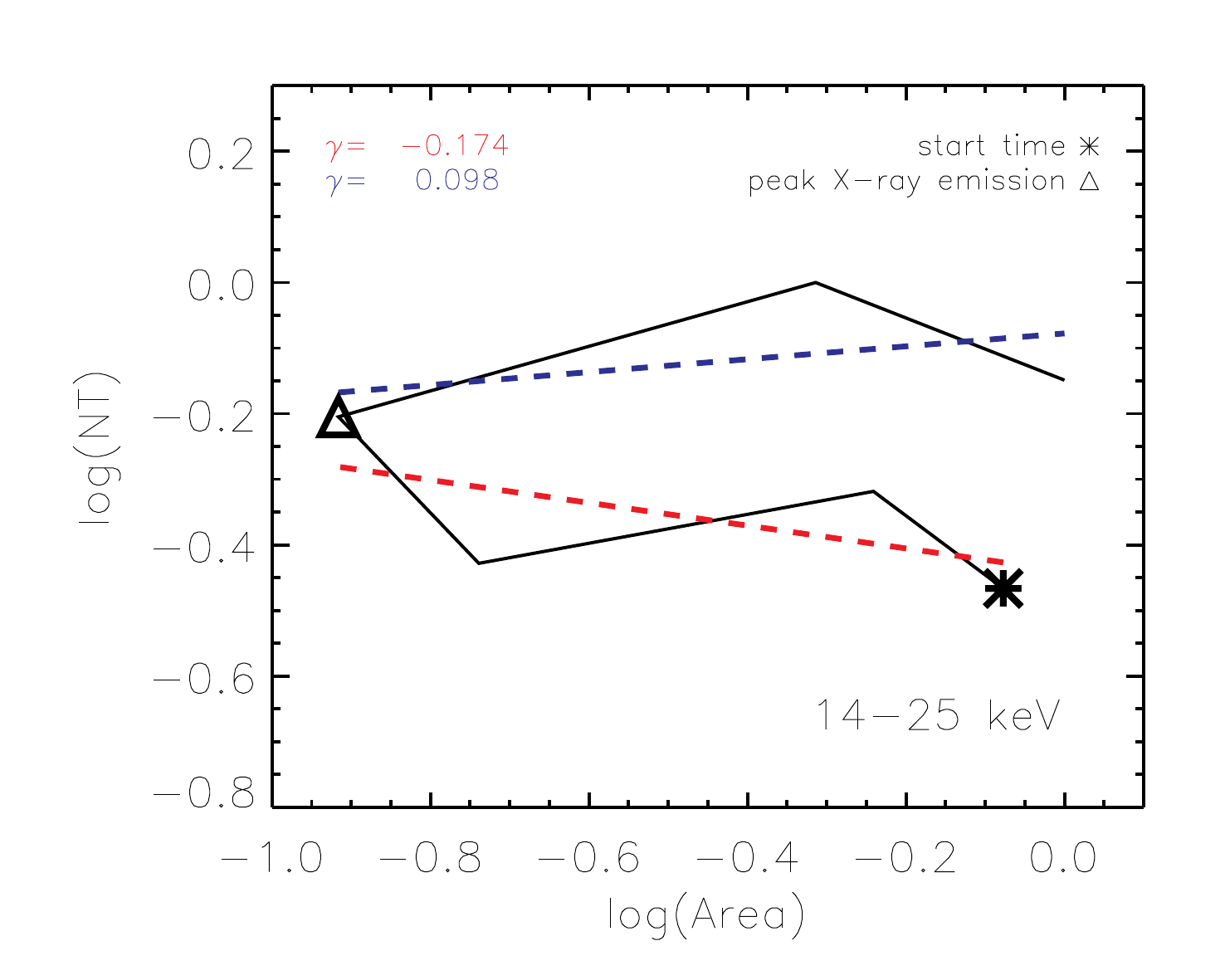}
\caption[Plots of $\log{NT}$ against $\log{1/A}$ for Flares 1, 2 and 3.]{Plots of $\log{NT}$ against $\log{1/A}$ for Flares 1 (top), 2 (bottom left) and 3 (bottom right).  The star represents the start time
of each event while the triangle represents the peaks in X-ray emission for each event. The red and blue
dashed lines represent straight line fits during the compressive (red) and expansive (blue) phases.
For Flare 2, only the first compressive and final expansion phase have been fitted. For simple compressive
heating, it is expected the gradient of each line would be $\gamma=-1$.}
\label{fig:NTA}
\end{figure*}

\subsection{Three temporal phases and suggested explanations for the observations}

\begin{figure*}
%\epsscale{.32}
\includegraphics[width=5.0cm]{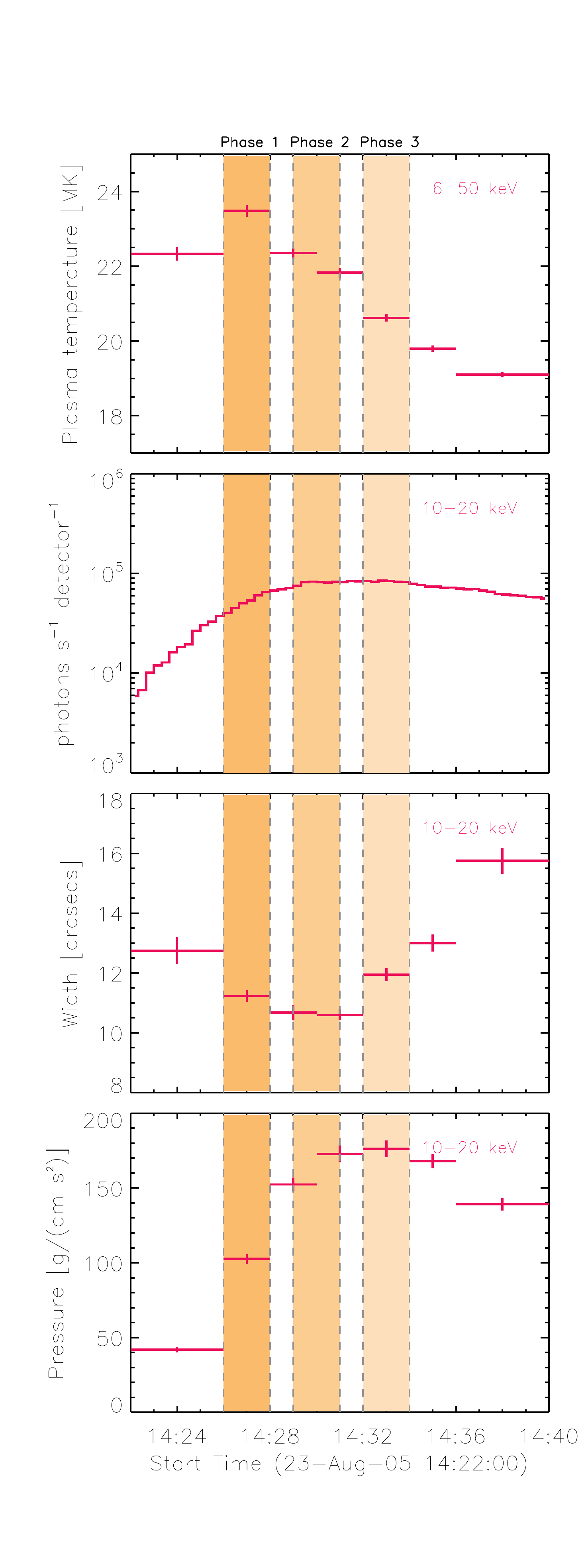}
\includegraphics[width=5.0cm]{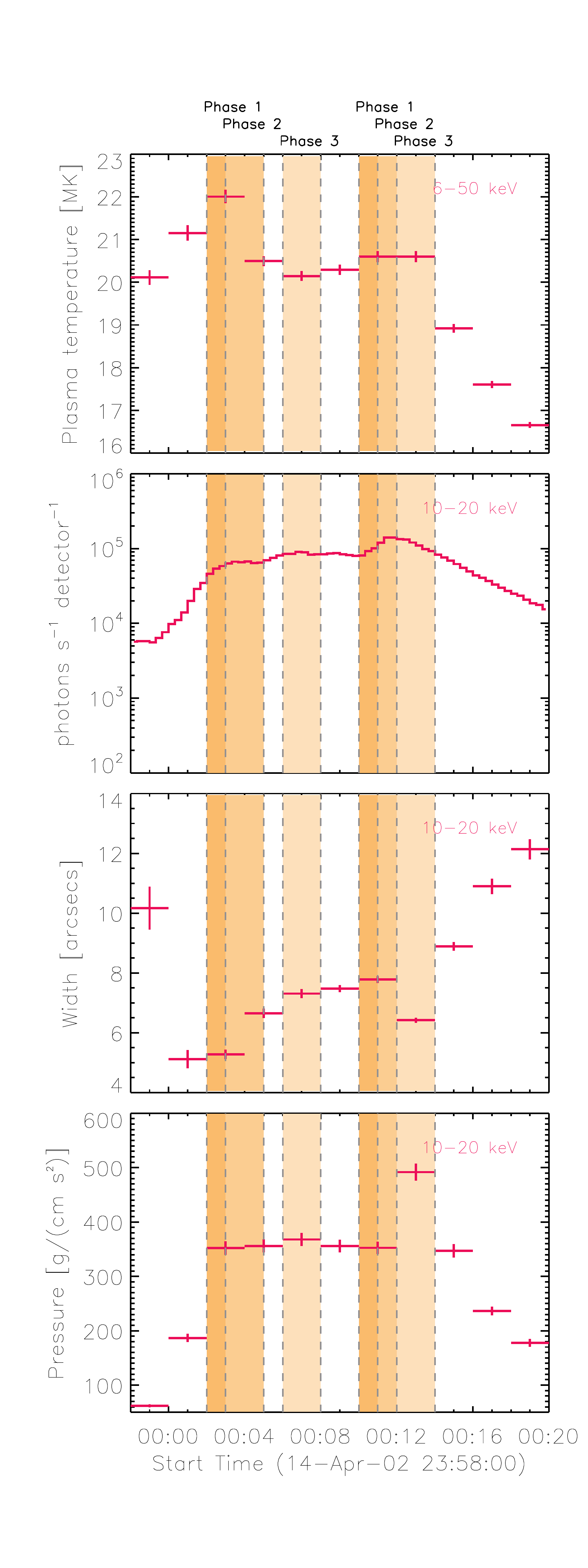}
\includegraphics[width=5.0cm]{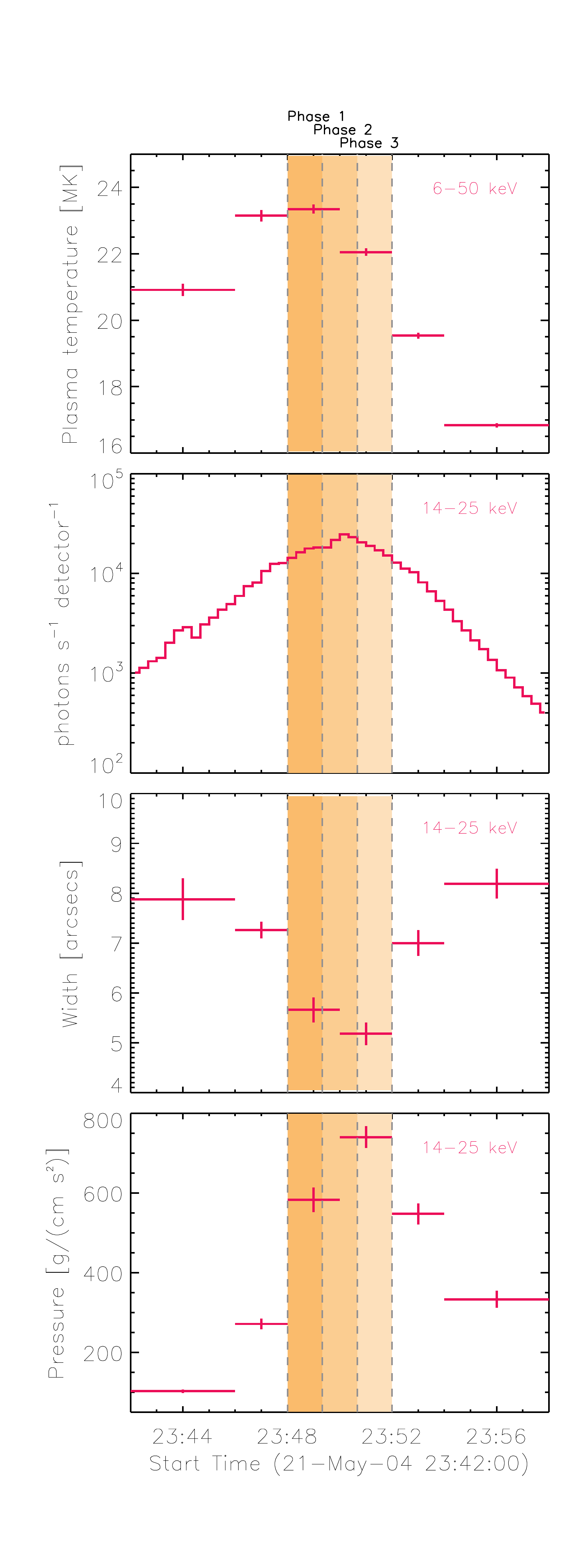}
\caption[Observations of plasma temperature, X-ray emission, loop width and thermal pressure are replotted together for Flares 1,
2 and 3 at one energy band of 10-20 keV (14-25 keV for Flare 3). ]{Observations of plasma temperature, X-ray emission, loop width and thermal pressure are replotted together for Flares 1 (left),
2 (middle) and 3 (right) at one energy band of 10-20 keV (14-25 keV for Flare 3). The orange bars represent three phases: 1. a peak in plasma temperature, 2. a peak in X-ray emission, generally coinciding with the smallest loop width and 3. a peak in thermal pressure.}
\label{fig:modphases}
\end{figure*}

For each flare, Figure \ref{fig:modphases} replots the plasma temperatures, X-ray emissions, emitting loop widths and thermal pressures, but now at a single energy band of 10-20 keV (14-25 keV for Flare 3). From Figures \ref{fig:is_paras} and \ref{fig:com_paras} and now more clearly in Figure \ref{fig:modphases}, it can be observed collectively, that the observations of each flare display three distinct phases and each of these phases will form the basis of the suggested explanation. Each phase is represented by a shaded orange bar. During Phase 1, there is a peak in plasma temperature and during Phase 2, a peak in X-ray emission. At Phase 2 the smallest emitting loop width, length and hence volume occurs. Finally, during Phase 3, the thermal pressure of the region peaks. For Flare 1 (left column), each phase is well separated and can be clearly seen. The pattern can also be seen for Flares 2 (middle column) and 3 (right column) but each phase is not as clearly defined as for Flare 1. During Flare 3, each phase occurs over much shorter time intervals and therefore each of the phases overlap slightly. During Flare 2, there are multiple peaks, which along with the shorter timescales for each process, makes each individual phase harder to see. However, the overall pattern is observed for all three flares.
Figure \ref{fig:modphases} shows that each phase can only be easily seen for slower events. The quicker the event, the harder it is to distinguish between each of the three phases as each phase overlaps in time. Many flares may exhibit a similar pattern but each phase may occur at timescales too fast to be currently observed.

\begin{figure*}
\centering
\includegraphics[width=15.5cm]{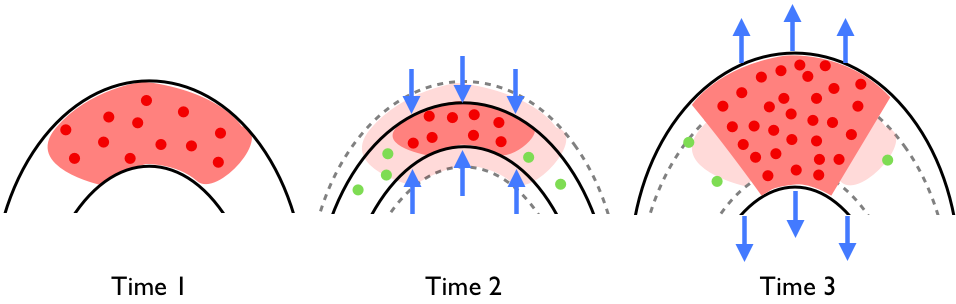}
\caption[Simple cartoon showing the suggested coronal loop evolution with time. ]{Simple cartoon showing the suggested coronal loop evolution with time. At time 1, the coronal region (pink) has a number density $n$ (red dots) and a temperature $T$. At time 2, number density of the region increases due to the cross-sectional area of the loop decreasing and the expansion of material from lower atmospheric layers. At time 3, the loop expands from thermal pressure due to chromospheric evaporation.}
\label{fig:chapt3_dcfig}
\end{figure*}

In order to understand the observations, it is crucial to understand the decreasing X-ray widths before the X-ray peak. Length variations along the field can always be explained in terms of changing number density, but in order to explain the width variations, where electrons are tied to the magnetic field, the most plausible explanation is the movement or the diffusion of magnetic field lines. The energetic loop width increases at a given time for each flare has already been noted, suggesting the presence of turbulence in the region, but overall in time the width at a given energy band shrinks until the X-ray peak is reached, implying the presence of both turbulence and the shrinking of the cross-sectional area of the field at this stage of the flare. Therefore, it is suggested the plasma within the emitting X-ray loop region is tied to the magnetic field lines and hence the contraction of the emitting loop cross-section or width during the rise phase is ultimately due to the contraction of the cross-sectional area of field lines that thread the region or possibly the expansion of field lines above the region. It is also sensible to assume that the loop region in X-rays actually consists of multiple coronal loops that cannot be resolved in X-rays using {\it RHESSI}. Although the region of the loop emitting X-rays can only be observed, it is assumed the entire loop region is contracting since the plasma is tied to the field lines. Hence from these observations, the reason for this cross-sectional width shrinking is unknown, but it can be speculated for further study that it may be due to a reduction in magnetic pressure within the X-ray loop region, as often suggested to describe height decreases (coronal implosion or loop contraction) discussed earlier \citep{2005ApJ...629L.137L,2009ApJ...696..121L,2009ApJ...706.1438J,2010ApJ...724..171R,2012ApJ...749...85G}. The magnetic pressure decreases as the magnetic field relaxes from a non-potential state. Although it is reasonable to assume that during a solar eruptive event such as a flare, the field will reside in a non-potential state, X-ray observations of these events possibly provide evidence for non-potentiality, again through the inference of magnetic turbulence and hence a non-parallel field component within the loop region. The non-potential state and decreasing magnetic pressure could be due to reconnection above the region or maybe even along the loop itself \citep{2004ApJ...608..540V,2011ApJ...729..101G,2012SoPh..277..299G,2013SoPh..284..489G}. One such model is caused by a kink instability as shown in recent simulations by \cite{2011ApJ...729..101G,2012SoPh..277..299G,2013SoPh..284..489G}. In this model, the energy of twisted loops is released by reconnection inside the loop and transferred to plasma heating and particle accelerations.

{\it Phase 1:} During phase 1, the process causing the contraction of loop width/cross-section is also probably responsible for the comparable decrease in loop altitude for Flare 1, and at least the change in loop position observed for Flares 2 and 3 sitting on the solar disk. Plots of $dW/dt$ (width contraction/expansion) and $dR/dt=v$ (centroid velocity) for Flare 1 plotted in Figure \ref{fig:dvwl} show that each parameter can be fitted using a straight line during both the contraction and expansion phases. $dL/dt$ (length contraction/expansion) is also plotted in Figure \ref{fig:dvwl} but its trend cannot be described by a straight line fit, unlike $dW/dt$ and $dR/dt$.
%the suggesting the changes in length are caused by a different process, itself a consequence of the process producing the decreasing loop width. It is also sensible to assume that this process is responsible for the increased plasma temperature and the acceleration of electrons.
The increasing temperature of the region means that energy will be thermally conducted towards the lower levels of the solar atmosphere causing gentle chromospheric evaporation of the denser coronal and chromospheric layers below. The observations show the plasma temperature peaking relatively early before the peak in X-ray emission and then slowly decreasing; slower than by thermal conduction, implying that energy is still being supplied to the loop plasma. This is most likely via the conversion of magnetic energy. Chromospheric evaporation drives plasma into the region producing the increasing number density and hence thermal pressure, along with the shrinking width at this phase. The increasing number density is responsible for the rapid non-linear decreasing X-ray loop length, since electrons accelerated within the region will travel shorter distances before interacting.

{\it Phase 2:} During Phase 2, after the peak in plasma temperature, the loop width stops shrinking. The number density and thermal pressure within the loop are still increasing due to chromospheric evaporation and the length of the emitting region also reaches its lowest point. The loop width may stop shrinking because the process causing the shrinking ceases or it may be due to the balancing of forces within the region. For example, if the reduction of loop width is due to the reduction of magnetic pressure in the loop then at this phase, the growing thermal pressure may finally be high enough to balance the reduction in $B$ pressure.

{\it Phase 3:} During the final phase, the thermal pressure continues to rise within the loop due to the increasing number density from chromospheric evaporation. It is believed the growing thermal pressure in the region is now responsible for the expanding loop width. This expansion, in turn, eventually halts the increasing number density and thermal pressure at a time after the peak in X-ray emission. After Phase 3, the loop width continues to increase and slow decreases in both number density and thermal pressure are observed. The X-ray emission continues to decrease during and after Phase 3 and the emitting loop length during this period remains approximately constant, equal to the minimum loop length in Phase 2, even with a decreasing number density. It is sensible to assume that the acceleration mechanism in the loop is slowing during this time. However, Flare 2 is an exception to this trend with multiple events/X-ray peaks in the lightcurve.

This observational study shows the usefulness of measuring changes in the spatial properties of coronal X-ray sources and combining such observations with the parameters deduced from spectral analysis, in order to deduce how the properties of the flaring corona change during the flare. It is hoped that future observations of such flares will have complementary EUV data from other solar missions such as SDO/AIA. 
\chapter{Solar flare X-ray albedo and the positions and sizes of hard X-ray (HXR) footpoints}
\label{ref:Chapter5}

\normalsize{\it This work can be found in the publication \cite{KontarJeffrey2010} and also \cite{2011A&A...536A..93J}.}
\\\\
\section{Introduction}
Chapters \ref{ref:Chapter5} and \ref{ref:Chapter6} examine solar flare X-ray albedo; an effect that can change the observed properties of HXR sources and hence the interpretation of the observations from instruments such as {\it RHESSI}. A brief discussion of this was given in Chapter \ref{ref:Chapter1}, Section \ref{intro_albedo1}. Here in Chapter \ref{ref:Chapter5}, it is discussed quantitatively for the first time how an albedo component will change the positions and sizes of observed HXR sources and hence the interpretation of the results. The solar atmosphere above HXR sources is optically thin and the X-rays emitted as bremsstrahlung, for example, are directly related to the emitting target electrons. However, depending on the anisotropy of the HXR source and hence the anisotropy of the target electron distribution in the chromosphere, a certain proportion of the X-rays are emitted downwards, towards the denser layers of the solar atmosphere, namely the photosphere. Here they can interact with
free or bound electrons and can be back-scattered towards the observer. This phenomenon was first discussed by \cite{Tomblin1972} and \cite{Santangeloetal1973} and the X-rays back-scattered and emerging from the dense photosphere, ``the albedo patch", are known as the albedo X-rays. The albedo X-rays are viewed alongside those X-rays directly emitted from the HXR source, which are often called the primary X-rays; together as a single observed HXR footpoint.
An isotropic HXR source produces the minimum albedo and even its
flux can account for up to 40\% of the detected flux in the peak albedo energy range between 20 and 50 keV \citep{BaiRamaty1978,ZhangHuang2004,Kontaretal2006,Kasparovaetal2007}. Therefore, all X-ray sources at the solar disk should be viewed as a combination of both the primary and backscattered albedo fluxes. As discussed in Chapter \ref{ref:Chapter1}, Section \ref{intro_albedo1}, accounting for the albedo effect is important for all X-ray solar observations, which can only view disk sources as a combination of the direct X-ray flux and the backscattered X-ray flux. The backscattered component often taints the primary HXR source properties, changing the observed angular, energy, spatial and polarization distributions. At the same time, the usefulness of albedo as diagnostic of electron directivity will be discussed in Chapter \ref{ref:Chapter6}.
Past studies concentrated on the observations of the era; looking at changes in X-ray spectra or the total integrated polarization of the HXR source. Albedo changes the shape of the spatially integrated X-ray spectrum, which is flattened at lower energies up to around 20-30 keV and at higher energies above around 70 keV, the spectrum is steeper than expected from a primary X-ray spectrum alone. Albedo can even produce artificial spectral features in observed spectra \citep{2008SoPh..252..139K}. \cite{Kontaretal2006} developed and implemented an albedo correction for spectral X-ray {\it RHESSI} analysis using a Green's function approximation by \cite{MagdziarzZdziarski1995}. This was discussed in Chapter \ref{ref:Chapter1}. For a more in-depth introduction on the topic of X-ray albedo, an excellent review can be found in \cite{2011SSRv..159..301K}.
Due to the height of a HXR source in the chromosphere ($\sim2\arcsec$), the reflected albedo X-rays come from a rather large area in the photosphere, the albedo patch. The surface brightness of the albedo patch at the solar surface is therefore rather low \citep{BaiRamaty1978} since the flux is spread over this large area. This fact explains the difficulty in imaging the albedo patch \citep{SchmahlHurford2002}. However, imaging algorithms such as VisFwdFit, discussed in Chapter \ref{ref:Chapter1}, have allowed the positions and sizes of HXR footpoints to be found (\cite{2008A&A...489L..57K}, \citet*{2010ApJ...717..250K}), using the spatially integrated moments of the X-ray distribution. In \citet*{2010ApJ...717..250K}, changes in radial height and vertical and horizontal spatial extents with energy, of a HXR footpoint situated at the solar limb, were found. This work is discussed in Chapter \ref{ref:Chapter1}, Section \ref{fp_loop}, demonstrating how the careful use of forward fitting algorithms and the spatially integrated moments of the X-ray distribution can be used to infer the positions, sizes and even shapes of HXR footpoints. It must be ensured that the albedo component, which should be present as part of every HXR footpoint source, is properly accounted for, before the conditions within the chromosphere or the properties of the radiating electron distribution are deduced from these spatial changes.  Even though the albedo patch is difficult to image directly, the use of imaging algorithms such as Vis FwdFit may actually help to `see' the presence of an albedo component. Since the moments of the distribution are integrated over the full area of the source, algorithms such as Vis FwdFit can better account for the low intensity and diffusiveness of the albedo component, than other imaging methods currently available.

\section{The modelling of X-ray transport in the photosphere}
\begin{figure}
\includegraphics[width=15cm]{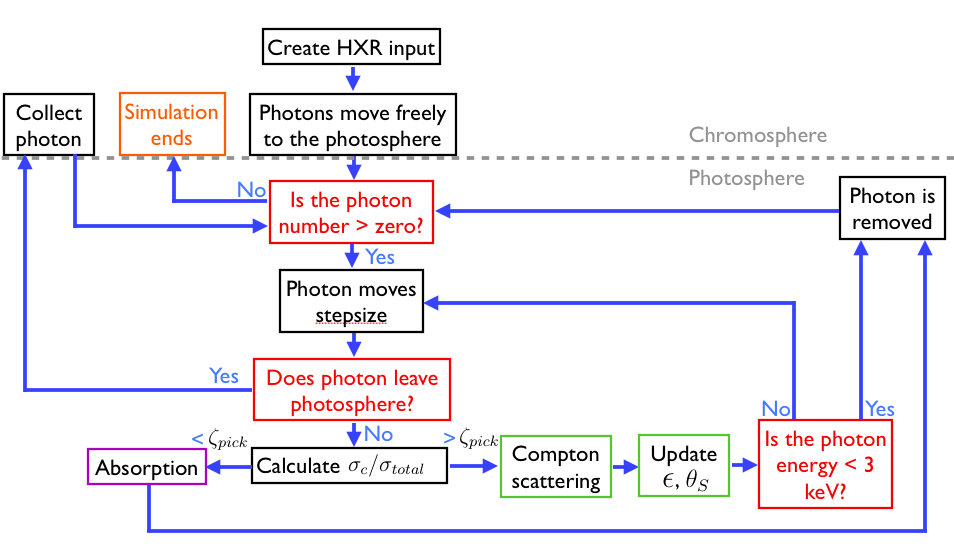}
\centering
\caption[A flow chart showing the main steps involved in the Monte Carlo photon transport simulations in the photosphere.]{A flow chart showing the main steps involved in the Monte Carlo photon transport simulations in the photosphere and the creation of the HXR distribution in the chromosphere.}
\label{fig:mcs_flow_chart}
\end{figure}

In order to study the backscattered X-ray flux and albedo effect, Monte Carlo (MC) simulations are used to model photon transport in the photosphere, starting with a hundred million photons per run from a chosen HXR source created in the chromosphere. Each of the main steps involved in the MC photon transport code are described in the following sections and can be seen graphically as a flow chart in Figure \ref{fig:mcs_flow_chart}.
\subsection{The modelling of a hard X-ray footpoint source}
An unpolarized HXR source is modelled in space as
a two-dimensional circular Gaussian distribution in the plane
parallel to the solar surface
\begin{equation}
I(x,y)\propto\exp\left(-\frac{x^{2}}{2 var_{x}}-\frac{y^{2}}{2 var_{y}}\right),
\label{eq:I_xy}
\end{equation}
with width (standard deviation) $=\sqrt{var_{x}}=\sqrt{var_{y}}$, at a chosen height $h=1$ Mm ($\sim1\arcsec.4$) in the chromosphere, above a layer defined as the photosphere (see Section \ref{ref:photo}). It is assumed that the HXR source has zero extent in $z$, the direction perpendicular to the solar surface. Although, the results of \citet*{2010ApJ...717..250K} showed that HXR sources at lower energies sit at a higher point in the chromosphere and have a greater extent in $z$, modelling $z$ with a finite extent would unnecessarily complicate the initial results and hence is neglected for clarity. For the HXR energy range, a source at $1$ Mm above the photosphere is chosen to match with recent X-ray observations
\citep{Aschwandenetal2002,2008A&A...489L..57K,Pratoetal2009,SaintHilaireetal2010,MrozekKowalczuk2010,2010ApJ...717..250K,BattagliaKontar2011}.
The X-ray energy spectrum in the chromosphere is simply input as a power law,
\begin{equation}
I(\epsilon)\sim\epsilon^{-\gamma}
\end{equation}
with a spectral
index of $\gamma$, for X-ray energies $\epsilon$ between 3 keV and 300 keV, an energy range typical of HXR footpoints observed by {\it RHESSI}. Typical HXR footpoint spectral index values are also used: $\gamma=2,3,4$ \citep[e.g.,][]{1991ApJ...379..381M, Kasparovaetal2007}

\subsection{X-ray transport and interaction in the photosphere}\label{ref:photo}
In this model, it is assumed that X-rays move freely without interaction until they reach the photosphere and hence their positions on the solar surface are simply calculated from their initial $(x,y)$ positions and emission angles $(\theta,\phi)$.
The photosphere is defined as a layer with a hydrogen number density of $1.16\times10^{17}$ cm$^{-3}$ \citep{Vernazzaetal1981}.  An
X-ray interaction with the photospheric medium can either be by Compton scattering or by
photoelectric absorption. For X-rays with energies less than $\sim10$ keV, photoelectric
absorption is the more probable process while Compton scattering dominates above $\sim10$ keV. This can be seen in Figure~\ref{fig:cross sections}. Within the photosphere, each X-ray photon moves a
step-size $ss$ before an interaction and this is calculated using $ss=-l\ln \zeta_{step}$ (see Appendix \ref{ref:App2} for more information), where $l$ is the
photon mean free path and $\zeta_{step}\in[0,1]$ is drawn from a uniform random distribution. The photon mean free path is calculated by
$l=1/n_{H}\sigma_{total}$, where $\sigma_{total}=\sigma_{c}+\sigma_{a}$, the addition of
the Compton scattering cross section $\sigma_{c}$ and
photoelectric absorption cross section $\sigma_{a}$. For a number density of $1.16\times10^{17}$ cm$^{-3}$, the photon mean free path, $l$, is of the order 100 km. Hence, the earlier assumption that photons travel freely until they reach the photosphere is valid since even a high chromospheric density of $\sim1\times10^{15}$ cm$^{-3}$ would give a mean free path $l\sim1000$ km $=1$ Mm, the same size as the chosen source height. When a photon is absorbed, it is simply removed from the simulations.  For each
photon, one of the two processes is chosen by calculating the ratio of $\sigma_{c}/\sigma_{total}$.
Another random number $\zeta_{pick}$ is then sampled from a uniform distribution between 0 and 1.
If the ratio is greater than $\zeta_{pick}$ then the photon is Compton scattered and if the ratio
is less than $\zeta_{pick}$, then the photon is absorbed.
These simulations also differ from previous work as the curvature of the Sun is included and a photon exits the photosphere when it satisfies the condition
$z>z_{\bigodot}=\sqrt{R_{\bigodot}^{2}-x^{2}-y^{2}}-R_{\bigodot}$, where $R_{\bigodot}$ is the
radius of the Sun which is taken to be $6.96\times10^{10}$ cm $\sim960''$. The extent of the albedo patch is limited by properly modelling the curvature of the Sun. The photons are allowed to scatter
multiple times until they exit the photosphere or are removed by
absorption. Photons are also removed when their energy falls below 3 keV since X-rays below this energy cannot be observed by {\it RHESSI} and a photon cannot gain energy from an interaction with an electron in photospheric conditions. Most photons will leave the photosphere during their first scatter, with subsequent scatterings
producing less and less photons. This is shown in terms of Green's functions in \cite{Kontaretal2006}.
Photons that exit the photosphere with $\cos\theta > 0$ are collected into selected angular
or energy bins corresponding to HXR sources sitting at any chosen heliocentric
angle on/above the solar disk.
\begin{figure}
\includegraphics[width=15cm]{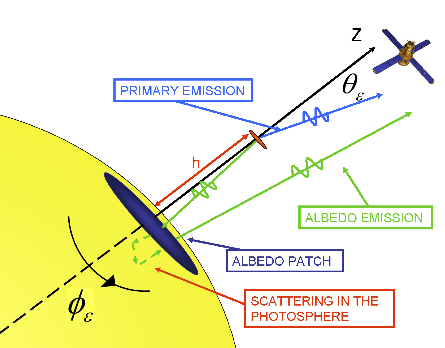}
\centering
\caption[Cartoon showing how X-rays emitted in the chromosphere via the Coulomb interaction can travel to the photosphere, Compton scatter, head out into interplanetary space and then be detected alongside X-rays directly emitted from the chromosphere.]{Cartoon showing how X-rays emitted in the chromosphere via the Coulomb interaction can travel to the photosphere, Compton scatter, head out into interplanetary space and then be detected alongside X-rays directly emitted from the chromosphere. The polar coordinates of the emitted X-rays $\theta_{\epsilon}$ and $\phi_{\epsilon}$ are also shown.}
\label{fig:ac2}
\end{figure}
\subsection{Photoelectric absorption}
\begin{figure}
\centering
\includegraphics[width=12cm]{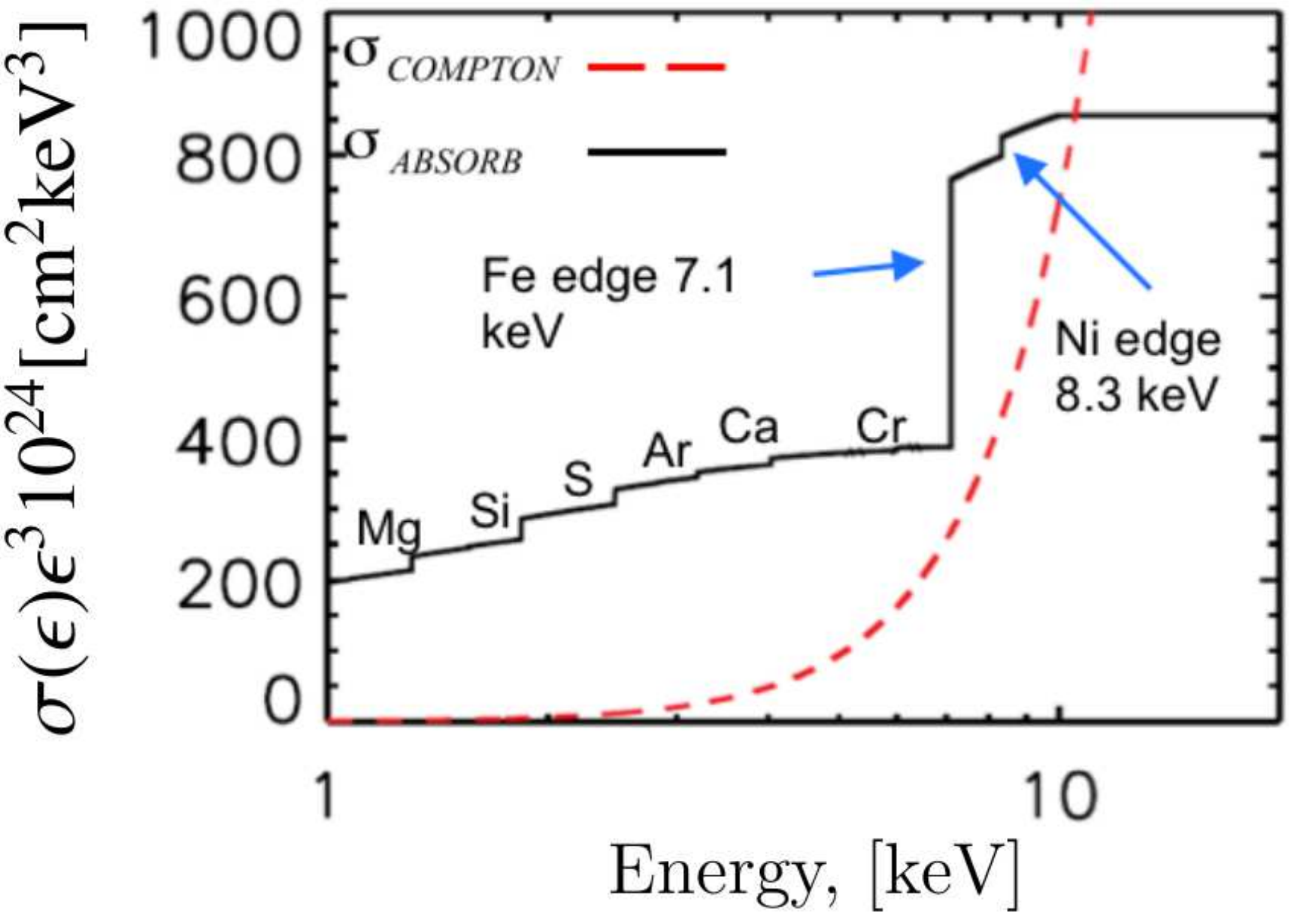}
\caption[Absorption $\sigma_{a}$ and Compton
$\sigma_{c}$ cross sections plotted at low energies below 10 keV.]{Absorption  $\sigma_{a}$ (black solid) and Compton
$\sigma_{c}$ (red dashed) cross sections plotted at low energies below 10 keV and
multiplied by $10^{24}\epsilon^{3}$ ($\epsilon$ in keV) for clarity. The absorption cross section
is calculated using photospheric element abundances by \cite{Asplundetal2009}.}
\label{fig:cross sections}
\end{figure}

For photons with energies below $\sim $10 keV, photoelectric absorption is the most probable photon
interaction in the photosphere. The process of absorption is heavily dependent upon the abundance of chemical
elements within the photosphere. Absorption was therefore
modelled using the latest known solar photospheric abundances taken from \cite{Asplundetal2009}.
Absorption cross section codes for the most important elements of H, He, C, N, O, Ne, Na, Mg, Al,
Si, S, Cl, Ar, Ca, Cr, Fe and Ni were adapted from \cite{Balucinska-ChurchMcCammon1992}. For energies
higher than $10$ keV, the absorption
cross section was approximated by $\sigma_a(\epsilon_{0})\propto\epsilon_{0}^{-3}$.  A comparison of the Compton scattering and absorption
cross sections is shown in Figure \ref{fig:cross sections} (with $\sigma_{a}$ and $\sigma_{c}$ multiplied by $10^{24}
\epsilon^{3}$ ($\epsilon$ in keV) for comparison with \cite{MorrisonMcCammon1983}). Any differences between
Figure \ref{fig:cross sections} and \cite{MorrisonMcCammon1983} are
due to the newer element abundances \citep{Asplundetal2009} and updated absorption cross section codes
(in particular Helium) being used in these simulations.
\subsection{Compton scattering}\label{ref:cs}
Similar to previous MC simulations \citep{BaiRamaty1978, MagdziarzZdziarski1995}, Compton scattering is modelled using the Klein-Nishina \citep{KleinNishina1929} differential scattering cross section for unpolarized X-ray radiation. This is valid
at all energies of interest and is given by,
\begin{equation}
\frac{d\sigma_{c}(\theta_{S},\epsilon)}{d\Omega}=\frac{1}{2}r_{0}^{2}\left(\left(\frac{\epsilon}{\epsilon_{0}}\right)^{3}+
\frac{\epsilon_{0}}{\epsilon}-\left(\frac{\epsilon}{\epsilon_{0}}\right)^{2}\sin^{2}\theta_{S}\right).
\label{eq:sigma_c}
\end{equation}
where $\epsilon_{0}$ is the initial photon energy, $\epsilon$ is the new photon energy, $\theta_{S}$ is the angle between the initial and after scattering photon directions and $r_{0} = 2.82\times10^{-13}$ cm is the classical electron radius. In MC simulations, when a Compton scattering occurs, the properties of the outgoing photon: energy $\epsilon$ and polar
scattering angle $\theta_{S}$, need to be updated.
Since MC simulations operate by drawing numbers randomly from a given distribution, this means
that $\theta_{S}$ can be easily found by matching each value of
$\theta_{S}\in[0^{\circ},180^{\circ}]$ with a random number $\zeta_{\theta}\in[0,1]$ drawn from a uniform distribution, for every value of
$\epsilon$ using,
\begin{equation}
\zeta_{\theta}=\frac{2\pi}{\sigma_{c}}\int_{0}^{\theta_{S}=\pi}\frac{d\sigma_{c}(\theta_{S},\epsilon)}{d\Omega}\sin\theta_{S}\;\;d\theta_{S}.
\end{equation}
New values of $\theta_{S}$ are simply
drawn at each scattering using the photon energy before a scattering and a random, uniform number $\zeta_{\theta}$ between
$0$ and $1$. Once the new scattering angles $\theta_{S}$ are obtained then the new photon energy $\epsilon$ can
be easily found using,
\begin{equation}
\epsilon=\frac{\epsilon_{0}}{1+\frac{\epsilon_{0}}{mc^{2}}(1-\cos\theta_{S})}.
\end{equation}

In the simulations, the Klein-Nishina cross section is
multiplied by $Z_{photo}=1.18$ to take account of elements higher than hydrogen that are present
within the photosphere. $Z_{photo}$ indicates the average atomic number and the
number of electrons per hydrogen atom in the photosphere, and is given by
\begin{equation}
Z_{photo}=\frac{\sum\limits _{Z}Z10^{A_{Z}}}{10^{A_{H}}}
\end{equation}
where $A_{Z}$ is the $\log_{10}$ abundance of an element with atomic number Z relative to
hydrogen while $A_{H}=12$ is the $\log_{10}$ abundance of hydrogen \citep{Asplundetal2009}.
In this case, where the X-ray distribution is completely isotropic and unpolarized, the azimuthal scattering angle in the plane perpendicular to the incoming direction of the photon during a scattering $\phi_{S}$ can just be drawn randomly from a uniform distribution between 0 and 2$\pi$.  This is not the case for a polarized X-ray distribution, which will be described in Chapter \ref{ref:Chapter6}. In the photosphere, it is very unlikely that the opposite case where the photon gains energy during an interaction with an electron, named inverse Compton scattering, will occur. For this to occur, a photon must interact with an electron with a kinetic energy larger than that of the photon energy, and this is unlikely in photospheric conditions.

\section{The position and sizes of backscattered and observed hard X-ray sources}

The escaping photons are accumulated to create the brightness distribution $I(x,y)$ over a given energy and solid angle. The total primary or reflected flux is then just an integral over the corresponding area $\int I(x,y)dxdy$ which is the zeroth moment of the brightness distribution. In these simulations, the source positions and sizes of each component
(primary and albedo) and the total source (primary plus albedo) can be found using the first and second moments of the distribution; the first and second normalised moments which are the mean and variance of the distribution respectively.

\subsection{The moments of the hard X-ray distribution}
Using solar disk centred coordinates, the centroid position $(\bar{x},\bar{y})$ of both the albedo source alone and total observed source can be found by the mean,
\begin{equation}
\bar{x}=\frac{\int_{0}^{\infty}xI(x,y)dxdy}{\int_{0}^{\infty}I(x,y)dxdy}\;\;\;,\;\;\;
\bar{y}=\frac{\int_{0}^{\infty}yI(x,y)dxdy}{\int_{0}^{\infty}I(x,y)dxdy},
\end{equation}
and the spatial extent in each direction $(x,y)$ by the variance $(var_{x},var_{y})$
\begin{equation}
var_{x}=\frac{\int_{0}^{\infty}(x-\bar{x})^{2}I(x,y)dxdy}{\int_{0}^{\infty}I(x,y)dxdy}\;\;\;,\;\;\;
var_{y}=\frac{\int_{0}^{\infty}(y-\bar{y})^{2}I(x,y)dxdy}{\int_{0}^{\infty}I(x,y)dxdy}.
\end{equation}
Although the primary distribution is initially
Gaussian, the albedo (and hence total observed) distribution will have a complex shape that is no longer Gaussian. To quantify the sizes we use `Gaussian' Full Width Half Maximum (FWHM) defined as FWHM$_{x,y}=2\sqrt{2\ln2var_{x,y}}$, using the square of the variance and not the actual FWHM of the complex distribution. This allows a simple comparison with {\it RHESSI} measurements \citep{2008A&A...489L..57K, DennisPernak2009, Pratoetal2009}. All FWHM values in this chapter and also Chapter \ref{ref:Chapter6} are measured in this way.

\subsection{Resulting brightness distributions}
\begin{sidewaysfigure*}[ht]
\centering
\includegraphics[width=22cm]{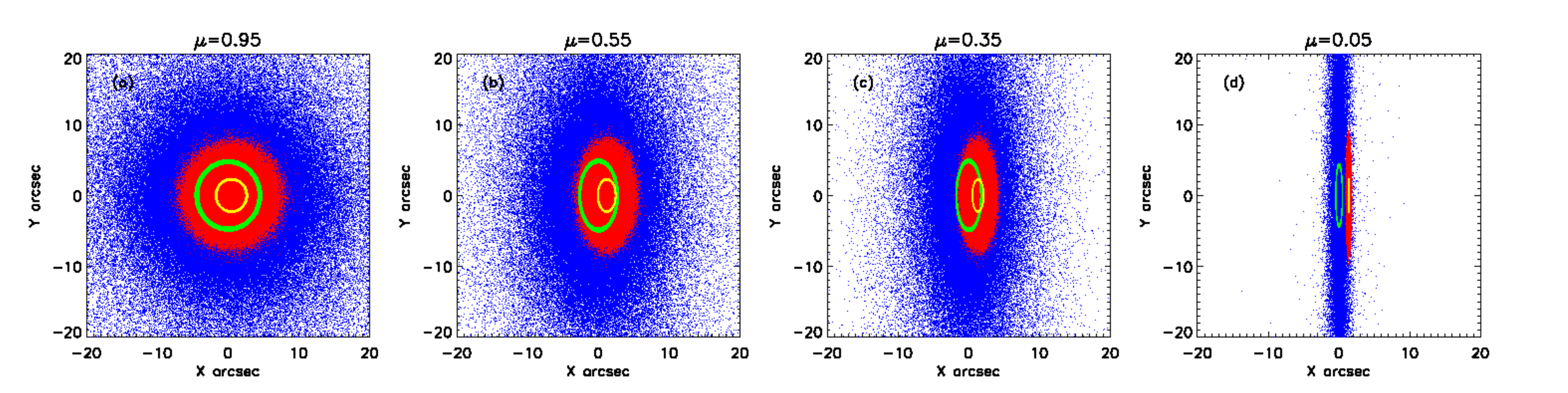}
\caption[The X-ray scatter distributions of the primary photons and the Compton back-scattered photons.]{The X-ray scatter distributions of the primary photons (red dots) and the Compton back-scattered photons (blue dots) for a primary source at $h=1$ Mm with width $d=\sqrt{var_{x,y}}=1.5h$ (giving a FWHM$_{x,y}\sim4\arcsec .9$) between 20 and 50 keV for four viewing angles given by $\mu$. The yellow and green ellipses show the FWHM sizes for the primary and combined sources respectively.}
\label{fig:paper_plots}
\end{sidewaysfigure*}

\begin{figure}
\centering
\includegraphics[width=16.5cm]{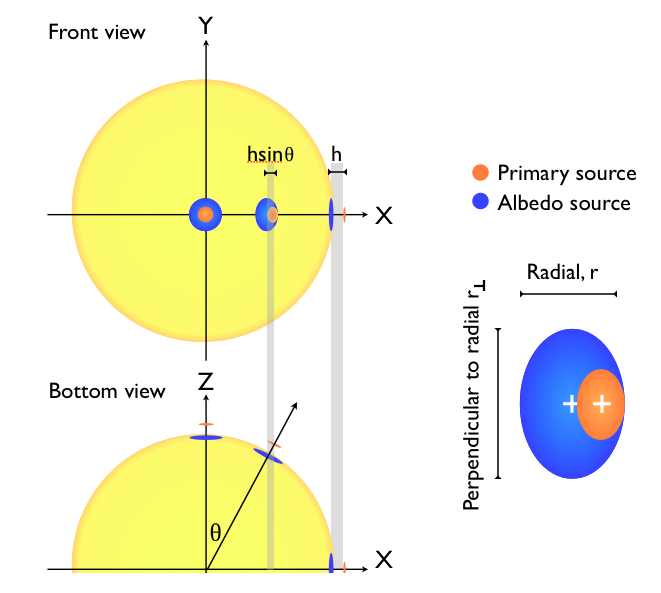}
\caption[Diagram showing a HXR primary source at three different heliocentric angles $\theta$ above the solar disk and the corresponding albedo patch at a shifted location of $h\sin\theta$.]{Diagram showing a HXR primary source (orange) at three different heliocentric angles $\theta$ above the solar disk and the corresponding albedo patch (blue) at a shifted location of $h\sin\theta$, where $\theta=0^{\circ}$ ($\mu=1$) is located at the solar centre and $\theta=90^{\circ}$ ($\mu=0$) is located at the edge of the Sun, the solar limb. The radial $r$ and perpendicular to radial $r_{\perp}$ directions are also shown.}
\label{fig:albedo_pic_fb}
\end{figure}

%\begin{figure}
%\centering
%\includegraphics[width=16.5cm]{paper_plots.pdf}
%\caption{The X-ray scatter distributions of the primary photons (red dots) and the Compton back-scattered photons (blue dots) for a primary source at $h=1$ Mm with width $d=\sqrt{var_{x,y}}=1.5h$ (giving a FWHM$_{x,y}\sim4\arcsec .9$) between 20 and 50 keV for four viewing angles given by $\mu$. The yellow and green ellipses show the FWHM sizes for the primary and combined sources respectively.}
%\label{fig:paper_plots}
%\end{figure}
Figure \ref{fig:paper_plots} shows the primary and escaping photon brightness distributions for a completely isotropic HXR footpoint located at the disk centre. Similar to previous results \citep{BaiRamaty1978}, for a chosen compact primary source of size $d=1.5h$ (FWHM$_{x,y}\sim4\arcsec .9$), the back-scattered albedo photons are reflected from an area much larger than the primary source. The reflected photons change the spatial distribution of the observed photons and produce a halo around the primary source.
\clearpage
\clearpage
The scattered X-ray flux depends on the cosine of the heliocentric angle of the source ($\mu=\cos(\theta)$) or equivalently on the position of the source at the solar disk, $\mu=\sqrt{1-(x^{2}+y^{2})/R_{\bigodot}^{2}}$. A circular X-ray source located above the centre of the disk will produce a circular albedo patch, as can be seen in the first plot of Figure \ref{fig:paper_plots}. Naturally, the location of the HXR source and albedo patch will coincide at the disk centre, so albedo will not change the source position. However, the albedo will make the source larger than the input primary size. The albedo contribution becomes asymmetric if the source is located away from the disk centre at a given heliocentric angle $\theta$ (Figure \ref{fig:paper_plots} b-d). A diagram depicting this scenario for different locations on the solar disk is shown in Figure \ref{fig:albedo_pic_fb}.

Due to the spherical symmetry of the Sun, there are two distinct directions: radial along the line connecting the centre of the Sun and the X-ray source $r$, and perpendicular to the radial $r_{\perp}$. This is shown in Figure \ref{fig:albedo_pic_fb}. There is no change in centroid position in the $r_{\perp}$ direction for a spherically symmetric primary source. Qualitatively, in the $r$ direction, the albedo component causes a centroid shift towards the disk centre that rises from $0$ at $\mu=1$ (solar centre), peaks at a position less than $\mu=1$ (this will be discussed further in the following sections) and reduces to $0$ at $\mu=0$ (solar limb). This pattern emerges since the centroid position of the albedo component is located at a position $h\sin\theta$ disk-ward of the primary centroid position, where $\theta$ is the heliocentric angle of the source (see Figure \ref{fig:albedo_pic_fb}). However, the intensity of the albedo component falls as we move closer to the limb and hence the position of the total source peaks at a heliocentric angle where the combined contribution of $h\sin\theta$ and the albedo intensity is greatest.
Figure \ref{fig:paper_plots} also shows how the source size varies in the $r_{\perp}$ direction, with the FWHM of the total source generally decreasing at lower $\mu$, since the albedo intensity falls as $\mu\rightarrow0$. In the radial direction, the FWHM of the total and primary sources decrease close to linear due to a simple projection effect. To study the albedo component only, source sizes are only examined in the $r_{\perp}$ direction and the source positions are examined in the radial direction $r$. Similar to the spatially integrated albedo \citep{Kontaretal2006}, the shift in centroid position and the growth of the source size are also energy and $\mu$ dependent. In the following section, the position and source size changes are studied for various: (a) spectral index of the HXR primary source, (b) HXR primary source size, and (c) X-ray directivity (the ratio of downward to upward emitted photons), separately. All the results are shown in Figure \ref{fig:albedo_2}.

\subsection{Changes due to hard X-ray spectral index}
In order to study changes due to spectral index only, a completely isotropic HXR primary source is input into the simulation. An isotropic source should produce the minimum albedo contribution and hence the smallest changes in source size and position. Using an isotropic HXR primary source with a FWHM$\sim4\arcsec.9$ and three spectral indices of $\gamma=2,3,4$, Figure \ref{fig:albedo_2} a-d shows that the albedo contribution from a smaller spectral index $\gamma$ produces the largest shift in position and a larger total source size. The lowest modelled spectral index of $\gamma=2$ produces the greatest shift of $0\arcsec.5$ at $\mu=0.5-0.6$ and $\sim30$ keV. This spectral index also produces the largest source size, compared with the other spectral indices of $\gamma=3,4$ modelled, and has a resulting FWHM$\sim9\arcsec.5$ at $\mu=1$.

\subsection{Changes due to hard X-ray primary source size}
For the same reasons in the previous section, an isotropic source is input into the simulations to study changes due to an input primary HXR source size. For an isotropic HXR primary source with fixed spectral index of $\gamma=3$, three primary source sizes are tested: FWHM$\sim0\arcsec$ (point), FWHM$\sim4\arcsec.9$ and FWHM$\sim14\arcsec.6$. As expected, it is found all primary source sizes produce the same shift in centroid position. The maximum shift in position occurs at $\mu=0.5-0.6$ and $\sim30$ keV for all sources modelled. These results can be seen in Figure \ref{fig:albedo_2} e-h.  Although the FWHM of the total source grows with increasing primary size, it is observed that the relative size of the total to the primary source is smaller for a larger primary source. This indicates that a larger primary source should have a smaller relative size increase due to albedo since the brightness distribution of a large primary source is less influenced by the reflected photons but nevertheless the source will look larger than the true primary size of a source between $10-100$ keV at all heliocentric angles. More importantly, even a primary point source will be seen as a source of finite size and an initial point source produces a total source with a FWHM peaking around $7\arcsec$ (Figure \ref{fig:albedo_2} f, h). There is even an increase of $\sim5\arcsec$ at the solar limb between $20-50$ keV where the albedo contribution is smallest.

\subsection{Changes due to hard X-ray anisotropy}
\begin{sidewaysfigure*}[ht]
\centering
%  \begin{minipage}[c]{0.60\textwidth}
\includegraphics[width=16.5cm]{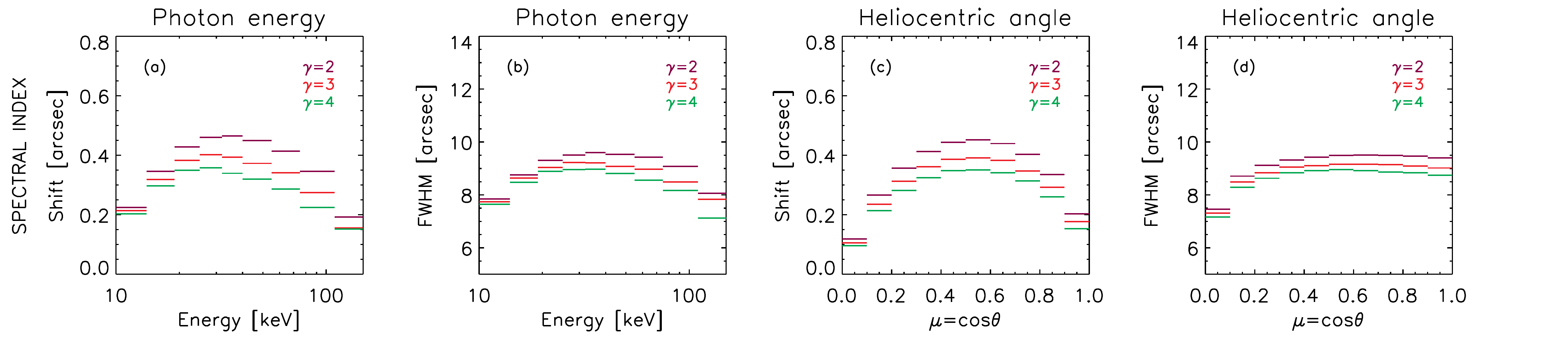}
\includegraphics[width=16.5cm]{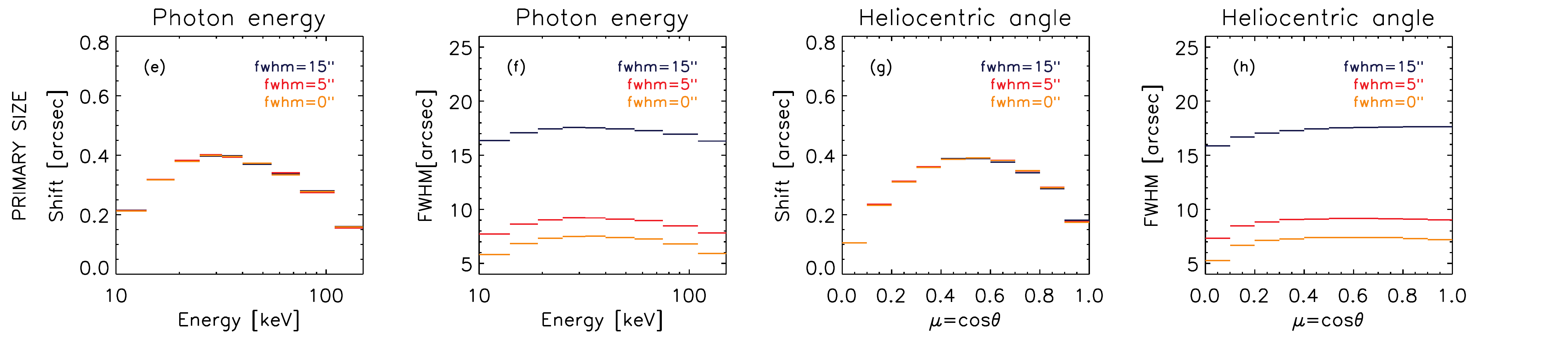}
\includegraphics[width=16.5cm]{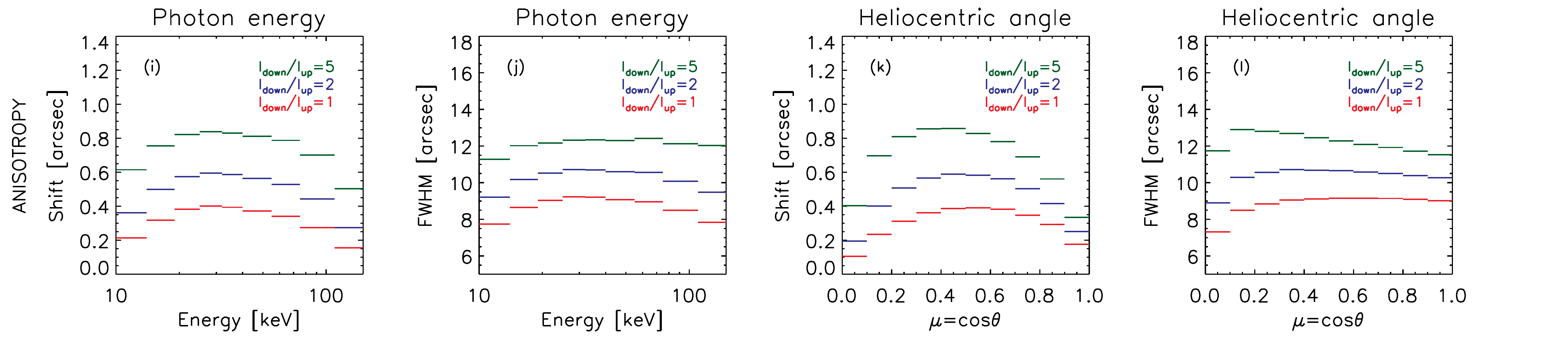}
 % \end{minipage}\hfill
 % \begin{minipage}[c]{0.30\textwidth}
\caption[Plots of the source position shift in the radial direction and source size FWHM in the perpendicular to radial direction due to albedo, against X-ray energy and heliocentric angle.]{Spectral index dependency (panels a-d)): the source position shift is in the radial direction due to albedo (source $\mu= 0.55$) and source size FWHM is in the perpendicular to radial direction for various spectral indices $\gamma$ for an isotropic source with FWHM$\sim4\arcsec.9$: green-$\gamma= 4$; red-$\gamma= 3$; purple-$\gamma=2$. Primary source size dependency (panels e-h)): isotropic primary source with $\gamma= 3$: orange-point source, red-FWHM$\sim4\arcsec.9$, blue-FWHM$\sim14\arcsec.6$. Anisotropy dependency (ratio of downward to upward directed fluxes) (panels i-l)): simulations are for the primary source with FWHM$\sim4\arcsec.9$ and spectral index $\gamma= 3$: red-anisotropy $= 1$ (isotropic), blue-anisotropy$=2$, green-anisotropy$= 5$. Graphs as a function of energy are for $\mu = 0.55$ and the graphs as a function of $\mu$ are for energies between 20 and 50 keV.}
\label{fig:albedo_2}
% \end{minipage}
\end{sidewaysfigure*}

Using a HXR primary source with a FWHM$\sim4\arcsec.9$ and $\gamma=3$, three photon anisotropies are modelled using the ratio of downward to upward flux of: (1) $I_{down}/I_{up}=1$ (isotropic), (2) $I_{down}/I_{up}=2$ and (3) $I_{down}/I_{up}=5$. The shift in centroid position is larger for a higher initial downward anisotropy for all $\mu$ and energies, shown in Figure \ref{fig:albedo_2} i, k. All shifts follow the general trend and tend towards zero at the centre ($\mu=1$) and the limb ($\mu=0$). A directivity of $5$ produces a peak difference of $0\arcsec.9$ and even an isotropic source produces a peak difference of $0\arcsec.4$. The shift in source position peaks near $\mu=0.4-0.6$ and $\sim$30 keV for a downward anisotropy of $2$ and an isotropic source, but the shift peaks at a lower $\mu=0.4-0.5$ for a downward directivity of $5$. The stronger downward beaming of the primary source also leads to larger apparent source sizes for all $\mu$ and energies (Figure \ref{fig:albedo_2} j, l). It should be observed that the total FWHM produced for a directivity of $5$ peaks at $\mu=0.15$ (Figure \ref{fig:albedo_2} 2p) giving an apparent FWHM$\sim13\arcsec$. Since the fraction of reflected photons reduces with $\mu$ the FWHM in perpendicular direction can be expected to slowly decrease from disk centre to limb, but the FWHM actually increases, peaks at $\mu\sim0.15$ and only then starts to decrease. This effect is due to the angular dependence of the Compton cross section. This is because the azimuthal-independent (that is, assuming that the scattering is isotropic in azimuthal angle) Compton cross section is anisotropic and peaks at angles less than $90^{\circ}$ (see Figure \ref{cs_graphs} in Chapter \ref{ref:Chapter1}), which allows a larger number of photons to scatter into an observer direction for flares close to the limb. It is this anisotropy in the scattering of the photons that causes the FWHM to peak at an angle smaller than $\mu=1.0$. The observation of this effect is particularly clear in the case of high downward directivity (Figure \ref{fig:albedo_2} l). It should be noted that if the anisotropy of the photon distribution is created from a given radiating electron distribution, then as the directivity of the electron distribution increases,
\clearpage
\clearpage
a higher proportion of the X-ray distribution will be concentrated over a narrower solid angle of emission, instead of a higher proportion of X-rays emitted downward at all angles, as was assumed in this simple simulation. The photon directivity is properly modelled from a chosen electron distribution in Chapter \ref{ref:Chapter6}.

\section{Discussion and conclusions}

The results of the simulations show that albedo can substantially affect the precise position and source size measurements of X-ray sources. Therefore, the effect of albedo should always be considered when the sizes or positions of X-ray sources are analysed. The only exception is occulted flares or possibly limb flares. However, this assumption should be used with caution particularly if the anisotropy of the source is high. The albedo displacement of source position is radially directed towards the disk centre and depends on the anisotropy of X-ray radiation, the X-ray source size and the spectral index of the primary source. Similar to total reflected flux, the displacement of HXR source position is energy dependent. The largest displacement can be observed in the range between $30-50$ keV at $\mu\sim0.5$ (heliocentric angle $\sim60^{\circ}$). The shift in centroid position in this energy range is $0\arcsec.1-0\arcsec.5$ for an isotropic (minimum albedo) source $1\arcsec.4$ above the photosphere and this can be up to $\sim0\arcsec.9$ for a downward beaming with factor of $5$. Because of the albedo, X-ray source sizes will be energy dependent, larger in the perpendicular to radial direction, and elliptical even for a spherically symmetric primary source. In the perpendicular to radial direction, the largest growth in source size occurs for sources close to the solar disk centre, in the energy range between $30-50$ keV, where albedo is the strongest. Thus, an isotropic primary source with FWHM$\sim4\arcsec.9$ at $1\arcsec.4$ above the photosphere will have an apparent FWHM size of $\sim9\arcsec$ in the energy range $20-50$ keV for sources in a wide range of heliocentric angles from $0^{\circ}$ to $80^{\circ}$.

The simulations demonstrate that X-ray sources will have a minimum size. An isotropic point source at $1$ Mm above the photosphere will be measured by {\it RHESSI} as a source with a FWHM size of $\sim7\arcsec$ across. This result can explain larger X-ray footpoint sizes than EUV or optical ones e.g. \cite{Kasparovaetal2005}. \cite{DennisPernak2009} reported that the average semi-minor axis of 18 double source flares is about $4\arcsec$, while a few of the X-ray source sizes were found to be consistent with line sources along the flare ribbons.  %While the quantitative comparison with {\it RHESSI} observations requires additional work, it should be noted that zero sizes are either the artefacts of the algorithms used or are caused by the very low source heights, since the size of the albedo patch will decrease with the height of the HXR source above the chromosphere.

The energy dependent character of albedo predicts that the source size as measured by {\it RHESSI} should grow with energy from $10$ keV up to $\sim30$ keV. Considering a large primary source of $14\arcsec.6$ across, for example a flaring loop, it is found that the source will grow up to $\sim18\arcsec$ at $\sim30$ keV. The X-ray anisotropy results show that spatial changes due to albedo have a great diagnostic potential for finding the anisotropy of the radiating electron distribution in the chromosphere. The source size changes due to albedo were not applied to the coronal X-ray sources studied in Chapters \ref{ref:Chapter2} and \ref{ref:Chapter3} because coronal X-ray sources sit at much greater heights ($\sim\ge15\arcsec$) than the chromospheric X-ray sources studied here ($1$ Mm) and hence, changes in X-ray source size and position will be smaller at coronal heights due to albedo. However, future work will fully investigate the albedo contribution when studying coronal X-ray source sizes and positions. 
\chapter{Solar flare X-ray albedo and spatially resolved polarization of hard X-ray (HXR) footpoints}
\label{ref:Chapter6}

\normalsize{\it This work can be found in the publications \cite{2011A&A...536A..93J} and \cite{KontarJeffrey2010}.}

\section{Introduction}

As discussed in Chapter \ref{ref:Chapter1}, a major insight regarding the angular properties of HXR footpoints comes directly from the X-ray polarization. The anisotropy and polarization of an X-ray distribution produced by bremsstrahlung will increase with the
anisotropy of the electron distribution \citep[e.g.][]{1959RvMP...31..920K, GlucksternHull1953, ElwertHaug1970,Brown1972, Haug1972, LeachPetrosian1983},
and hence in theory HXR polarization allows the anisotropy of the
emitting electron distribution in the chromosphere to be inferred. However, compared with the other HXR observables: energy, spatial
location, source size and time of arrival for example, polarization measurements through the years
\citep[see][as a review]{2011SSRv..159..301K} have been fraught with difficulties and the measurements
often met with skepticism. Nonetheless many missions have reported measurements
of HXR polarization from solar flares \citep{Tindoetal1970, Tindoetal1972, Nakadaetal1974,
Tindoetal1976, Lemenetal1982, McConnelletal2004, Boggsetal2006, 2006SoPh..239..149S, McConnelletal2007}.

Compton scattering is polarization dependent; the total integrated polarization of an HXR source will be
altered by its albedo component. However, the albedo
component will not only change the total polarization of the observed HXR footpoint; there will be spatial variations in polarization across the extent of the albedo source and hence the observed HXR footpoint.
Chapter \ref{ref:Chapter5} discussed the importance of compensating for the Compton backscattered albedo component when interpreting the positions and sizes of HXR sources. In Chapter \ref{ref:Chapter6}, the usefulness of albedo polarization as a valuable diagnostic tool will be discussed.
Spatially resolved polarization measurements across a HXR source caused by albedo, have the advantage over spatially integrated measurements since both the magnitude and the direction of polarization will change with X-ray directivity; allowing maps of the albedo and primary components to be created. Understanding how these two parameters change with the photon anisotropy is essential and provides a new method of investigating the entire photon anisotropy from a single HXR source.
However, although there are a
number of simulations for the spatially integrated polarization signal in flares
\citep{ElwertHaug1970,Haug1972,Zharkovaetal1995,Emslieetal2008,Zharkovaetal2010} and how an albedo component changes the spatially integrated polarization \citep[e.g.][]{Henoux1975,LangerPetrosian1977,BaiRamaty1978}, until now spatially resolved polarization had not been investigated. The work shown in this chapter and published in \cite{2011A&A...536A..93J} is the only known prediction of the spatially resolved hard X-ray
polarization due to albedo.

For the first time in solar physics, spatially resolved polarization across HXR sources at various locations
on the solar disk, taking into account the influence of albedo, is computed, for various emitting electron
populations. The simulations \citep{2011A&A...536A..93J} will also predict the angular resolution and preferred energy range required for such future polarization observations. Finally, the usefulness of future observations such as these will be discussed. Chapter \ref{ref:Chapter6} will also briefly examine how changing the maximum electron energy available during bremsstrahlung can alter spatially integrated polarization measurements with photon energy, possibly providing a new method for finding the maximum electron cutoff energy.

\section{Defining the polarization of an X-ray distribution}
The polarization state of incoherent radiation can be completely described using four Stokes
parameters \citep{Stokes1852,Chandrasekhar1960}. The Stokes pseudovector consists of these four
parameters and takes the form of
\begin{equation}
S= \left[
\begin{matrix}
I \\
Q\\
U\\
V
\end{matrix}
\right]=[I\:Q\:U\:V]^{T}.
\end{equation}
A pseudovector is a vector-like object that is invariant under an inversion of its coordinate axes and is often called an axial vector. The first Stokes parameter $I$ is the normalised
total intensity of the photon beam and hence always equal to 1, while $Q/I$, $U/I$ and $V/I$ will have values between $-1$ and
$1$. The second and third normalised Stokes parameters are used to define linear polarization with $1$ or $-1$
indicating that the beam or photon packet is completely polarized with the sign providing the
direction of polarization. The fourth parameter is used to describe circular polarization. However,
bremsstrahlung emission in the solar corona or chromosphere only produces radiation that is
linearly polarized. In order to produce circularly polarised radiation via bremsstrahlung, the
spins of the radiating electrons need to be aligned and the magnetic field in the
corona or chromosphere is not strong enough for this alignment. Also, Compton scattering cannot produce circularly polarized emission, that is linear photons can not become circularly polarized during a Compton scattering. This means that only the
first three Stokes parameters are required and the fourth can be set to zero throughout the
simulations. Generally in X-ray and gamma ray astronomy the polarization of radiation is measured
using the degree of polarization ($DOP$) and the polarization angle $\Psi$, which is the preferred
direction of the electric field. These are defined using the Stokes parameters as,
\begin{equation}
DOP=\frac{\sqrt{Q^{2}+U^2}}{I},
\label{eq:DOP}
\end{equation}
and
\begin{equation}\label{eq:Psi}
\Psi=\frac{1}{2}\arctan\left(\frac{-U}{-Q}\right),
\end{equation}
where the angle $\Psi$ is chosen to lie within the quadrant between $[-180^{\circ},180^{\circ}]$,
so that $\arctan\left(\frac{+0}{+0}\right)=+0^{\circ},\; \arctan\left(\frac{+0}{-0}\right)=+180^{\circ},\; \arctan\left(\frac{-0}{+0}\right)=-0^{\circ}$ and
$\arctan\left(\frac{-0}{-0}\right)=-180^{\circ}$. The negatives introduced into Equation (\ref{eq:Psi}) ensure that a
negative $Q$ gives $0$ and a positive $Q$ gives $90^{\circ}$. Hence with this definition, when
$\Psi=0^{\circ}$, the observed radiation is polarized parallel to the radial direction at the solar
disk and when $\Psi=90^{\circ}$, the radiation is polarized perpendicular to the radial direction.
The opposite definition of radiation polarized parallel to the radial direction having $\Psi=90^{\circ}$ is equally valid as long as the definition is consistent.
\begin{figure}
\includegraphics[width=15.5cm]{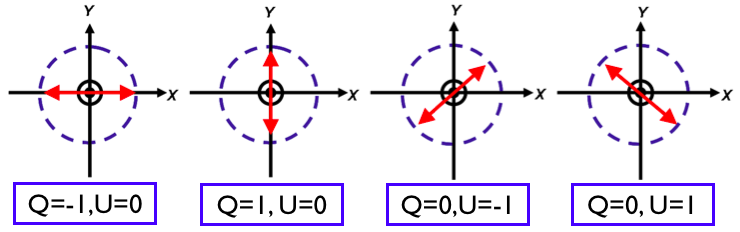}
\caption[Diagram showing the preferred direction of the electric field for a photon travelling out of the page,  for each of the possible values of the linear Stokes parameters $Q$ and $U$.]{Diagram showing the preferred direction of the electric field for a photon travelling out of the page,  for each of the possible values of the linear Stokes parameters $Q$ and $U$. A completely unpolarised source has $Q=0$ and $U=0$ and no preferred direction of the electric field.}
\label{fig:stokes}
\end{figure}
The Stokes parameters are also frame dependent and hence have to be updated by the
use of rotation matrices \citep[e.g.,][]{Hovenier1983} when moving between different coordinate frames. In the simulations, the Stokes pseudovector will be initially defined in the source frame and must be rotated to the scattering frame during each Compton scattering. The pseudovector must then be rotated back to the source frame before the results are examined. These rotations will be properly described in Section \ref{ref:upps}. The $DOP$ remains unchanged by a rotation but the polarization angle $\Psi$ is measured with respect to the new frame.

For simplicity, it is assumed that the flare loop and the dominant direction of the electrons lie parallel with the local solar vertical. This means that both the spatially integrated and spatially resolved bremsstrahlung polarization direction $\Psi$ only ever equal $0^{\circ}$ or $90^{\circ}$, since $U$ is always close to zero \citep{BaiRamaty1978} (see Equation \ref{eq:Psi}). Here, it should be noted that the HXR source is always assumed to be much smaller than the solar disk. However, Compton scattering in the photosphere can
produce non-zero values of $U$. This means that the spatially resolved Compton scattered $\Psi$ can have values other than $0^{\circ}$ or $90^{\circ}$ in the solar disk frame, which is the frame of the HXR source and can therefore provide us with additional information regarding the anisotropy of the electron distribution. The spatially integrated albedo Stokes parameters again sum to produce $\Psi$ values of either $0^{\circ}$ or $90^{\circ}$. Therefore this means the spatially integrated polarization angle $\Psi$ never provides us with any information regarding the anisotropy of the electron distribution. However, \cite{Emslieetal2008} found that the spatially integrated direction of the
polarization angle is related to the dominant direction of electrons or equivalently, the tilt of the guiding field or loop with respect to the local solar vertical; with values of spatially integrated $\Psi$ other than $0^{\circ}$ or $90^{\circ}$ revealing that the loop does not lie parallel to the local solar vertical.

\section{HXR footpoint bremsstrahlung polarization}

\subsection{The radiating electron distribution}
In order to create X-ray distributions in the chromosphere with varying anisotropy and polarization, for input into the simulations, an X-ray emitting electron distribution in the chromosphere is chosen to have the following form,
\begin{equation}
F(E,\beta)\propto E^{-\delta_{T}}\exp\left({-\frac{(1+\cos\beta)^{2}}{\Delta\nu^{2}}}\right).
\label{eq:F_E_beta}
\end{equation}
$F(E,\beta)$ is the electron flux distribution [electrons cm$^{-2}$ s$^{-1}$ keV$^{-1}$] in the chromosphere, $E$ is the electron energy and $\beta\in[0^{\circ},180^{\circ}]$ is the pitch-angle of the emitting electrons
velocity to the local magnetic field with $\beta=0^{\circ}$ directed away from the Sun along the local solar vertical. The energy dependence follows a power law as shown by observations and is produced by an injected electron distribution of $\delta=\delta_{T}+2$ \citep[see][as recent
reviews and Chapter \ref{ref:Chapter1}]{2011SSRv..159..107H,2011SSRv..159..301K}. The electron angular distribution is modelled as a Gaussian distribution; allowing the angular anisotropy of the electron distribution to be easily controlled by a single parameter $\Delta\nu$. The smaller the value of $\Delta\nu$, the greater the proportion of the electron distribution, and hence the resulting bremsstrahlung X-ray emission directed towards the photosphere.

\subsection{The emitted primary X-ray photon distribution}

\begin{figure*}
\centering
\includegraphics[width=14cm]{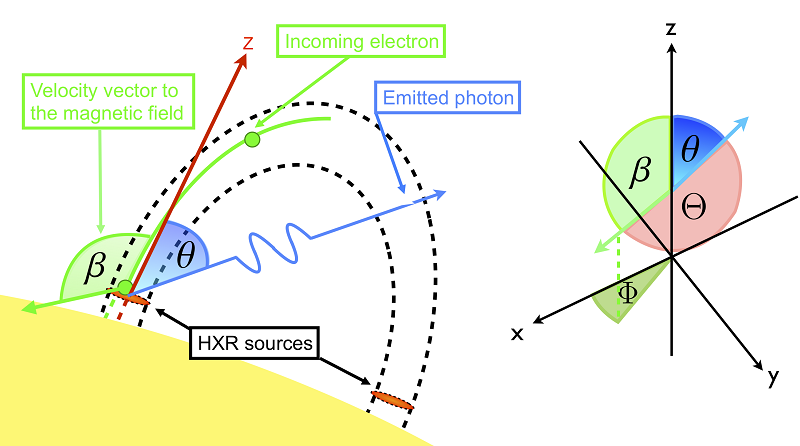}
\caption[A cartoon of a typical solar flare scenario where an electron in the chromosphere, transported along the guiding field from the corona interacts by Coulomb collisions producing a HXR photon. ]{{\it Left:} A cartoon of a typical solar flare scenario where an electron in the chromosphere, transported along the guiding field from the corona interacts by Coulomb collisions producing an HXR photon. {\it Right:} Diagram indicating all the angles defined in Equation \ref{eq:cosTheta}.}
\label{fig:epfig3}
\end{figure*}

The intensity of an X-ray photon distribution $I(\epsilon,\theta)$ [photons s$^{-1}$ cm$^{-2}$ keV$^{-1}$] produced by bremsstrahlung for a
chosen electron distribution $F(E,\beta)$ is given by
\begin{eqnarray}
I(\epsilon,\theta)\propto\int^{\infty}_{E=\epsilon}\int^{2\pi}_{\Phi=0}\int^{\pi}_{\beta=0}F(E,\beta)
\sigma(E,\epsilon,\Theta) \sin\beta d\beta d\Phi dE
\label{eq:I}
\end{eqnarray}
where $\epsilon$ is the X-ray energy and $\sigma(E,\epsilon,\Theta)$ is the total (averaged over all polarization
states) angle-dependent bremsstrahlung cross-section
\citep{ElwertHaug1970,BaiRamaty1978,Massoneetal2004}. $\theta\in[0^{\circ},180^{\circ}]$ is the
photon polar emission angle measured from the local solar vertical with $\theta=0^{\circ}$ directed
away from the Sun, as for $\beta$.
$\Phi\in[0^{\circ},360^{\circ}]$ is the corresponding electron azimuthal angle measured in the plane perpendicular to the local solar vertical and
$\Theta(\beta,\Phi,\theta)$ is the angle between the plane of emission (at angle $\beta$)
and the plane of observation (at angle $\theta$). Figure \ref{fig:epfig3} depicts this scenario pictorially showing the angles $\beta$, $\Phi$, $\theta$ and $\Theta$.
The X-ray emission angle
is described by $\mu=\cos\theta$, where $\mu$ from $0$ to $1$ corresponds to emission away from the
Sun, and $\mu$ from $-1$ to $0$ corresponds to emission towards the solar surface. The X-ray emission
angle $\mu=\cos\theta$ is related to the electron pitch-angle $\beta$ by:
\begin{equation}
\cos\Theta=\cos\theta\cos\beta+\sin\theta\sin\beta\cos\Phi.
\label{eq:cosTheta}
\end{equation}
Viewing the outward emission from $\mu=0$ to $\mu=1$ corresponds to observing the HXR source at a selected
heliocentric angle on the solar disk, that is $\mu=0$ corresponds to $90^{\circ}$ and is equivalent to viewing a HXR footpoint source sitting at the solar limb.

Just as with the X-ray intensity, $I$, the linear Stokes parameters $Q$ and $U$ can be calculated in a similar manner:
\begin{equation}
Q(\epsilon,\theta)\propto\int^{\infty}_{E=\epsilon}\int^{2\pi}_{\Phi=0}\int^{\pi}_{\beta=0}F(E,\beta)\sigma_{Q}(E,\epsilon,\Theta)
 \sin\beta d\beta d\Phi dE,
\label{eq:Q}
\end{equation}
\begin{equation}
U(\epsilon,\theta)\propto\int^{\infty}_{E=\epsilon}\int^{2\pi}_{\Phi=0}\int^{\pi}_{\beta=0}F(E,\beta)\sigma_{U}(E,\epsilon,\Theta)
\sin\beta d\beta d\Phi dE ,
\label{eq:U}
\end{equation}
\\
with the only difference being the use of either $\sigma(E,\epsilon,\Theta)$,
$\sigma_{Q}(E,\epsilon,\Theta)$ or $\sigma_{U}(E,\epsilon,\Theta)$.
$\sigma(E,\epsilon,\Theta)$, $\sigma_{Q}(E,\epsilon,\Theta)$ and $\sigma_{U}(E,\epsilon,\Theta)$ are the polarization
dependent cross-sections for bremsstrahlung taken from \cite{GlucksternHull1953} and also following
the form used in \cite{Haug1972} and \cite{Emslieetal2008}. They are given by
\begin{equation}
\sigma(E,\epsilon,\Theta)=\sigma_{\perp}(E,\epsilon,\Theta)+\sigma_{\parallel}(E,\epsilon,\Theta),
\end{equation}
\begin{equation}
\sigma_{Q}(E,\epsilon,\Theta)=(\sigma_{\perp}(E,\epsilon,\Theta)-\sigma_{\parallel}(E,\epsilon,\Theta))\cos2\Theta ,
\end{equation}
and
\begin{equation}
\sigma_{U}(E,\epsilon,\Theta)=(\sigma_{\perp}(E,\epsilon,\Theta)-\sigma_{\parallel}(E,\epsilon,\Theta))\sin2\Theta,
\end{equation}
where $\sigma_{\perp}(E,\epsilon,\Theta)$ and $\sigma_{\parallel}(E,\epsilon,\Theta)$ are the perpendicular and parallel components of the
bremsstrahlung cross-section respectively. $Q(\epsilon,\theta)$ and $U(\epsilon,\theta)$ are normalised between [-1,1] by dividing
through by $I(\epsilon,\theta)$.

\section{Photon transport in the photosphere and changes in hard X-ray polarization}

\subsection{Monte Carlo simulation inputs}
\begin{figure}
\includegraphics[width=15cm]{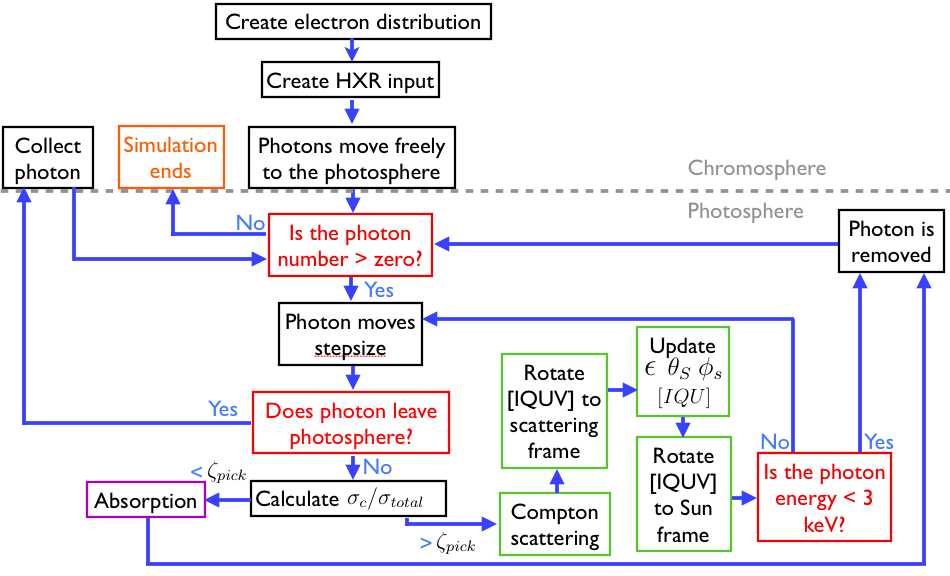}
\centering
\caption{An updated version of the steps in the MC simulations including polarization and the creation of a HXR distribution via a chosen electron distribution in the chromosphere.}
\label{fig:mcs_flow_chart_pol}
\end{figure}

In order to study changes in HXR polarization due to a given target-averaged electron angular distribution, the energy, angular and polarization properties of the input
photon distributions for the MC code are determined via the chosen input electron distributions
given by Equation \ref{eq:F_E_beta}. In all simulation runs, a mean electron spectrum of spectral index $\delta_{T}=2$ is used. This means that the injected electron distribution has a typical solar flare value of $\delta=4$. In order to test how both the total integrated and spatially resolved HXR polarization change with electron directivity, three values of $\Delta\nu$ are used to describe various pitch-angle distributions of electrons in the chromosphere: $\Delta\nu=4.0$ , $\Delta\nu=0.5$ or $\Delta\nu=0.1$. A $\Delta\nu=4.0$ electron distribution produces an approximately isotropic, unpolarised
photon distribution, while the $\Delta\nu=0.5$ and $\Delta\nu=0.1$ electron distributions produce photon distributions with progressively greater beaming towards the photosphere.
In the MC simulations, distributions of energy $\epsilon$, angle $\theta$ and polarization $Q$ and $U$
are numerically created from $I(\epsilon,\theta)$, $Q(\epsilon,\theta)$ and $U(\epsilon,\theta)$ (Equations \ref{eq:I}, \ref{eq:Q} and \ref{eq:U}). Each photon input azimuthal angle $\phi$ is simply drawn from a uniform, random distribution between $0$ and $2\pi$. The spatial properties are determined via Equation \ref{eq:I_xy}, as described in Chapter \ref{ref:Chapter5}, again at a height $h=1$ Mm. To achieve the best statistics feasible, $10^8$ photons are used in every MC code run.
Once the HXR source is created from a given target electron distribution, the MC code runs as described in Chapter \ref{ref:Chapter5} with the only differences due to the inclusion of polarization, which will be described in the following sections. An updated flow chart showing the main steps of the MC photon transport code, including polarization, is shown in Figure \ref{fig:mcs_flow_chart_pol}.

\subsection{Photoelectric absorption and hard X-ray polarization}
Chapter \ref{ref:Chapter5} describes how photoelectric absorption is the dominant interaction process in the photosphere below $\sim $10 keV. The probability of photoelectric absorption
is assumed to be independent of polarization \citep{Poutanenetal1996}.
Only the angular distribution of the ejected electron is dependent upon the photon polarization,
which is not modelled in these simulations.

\subsection{Compton scattering and hard X-ray polarization}\label{ref:upps}
As described in Chapter \ref{ref:Chapter5}, above $\sim$10 keV, Compton scattering is the dominant interaction process in the photosphere. The polarization
dependent differential Compton scattering cross-section is given by \cite{KleinNishina1929}, but
the form used in these simulations is from \cite{McMaster1961} and \cite{BaiRamaty1978}
\begin{equation}
\frac{d\sigma_{c}}{d\Omega}=\frac{1}{2}r_{0}^{2}\left(\frac{\epsilon}{\epsilon_{0}}\right)^{2}{\bigg\lgroup}\frac{\epsilon}{\epsilon_{0}}+
\frac{\epsilon_{0}}{\epsilon}-\sin^{2}\theta_{S}{\bigg\lgroup}1-Q\cos2\phi_{S}
 -U\sin2\phi_{S}{\bigg\rgroup}{\bigg\rgroup},
\end{equation}
where $r_{0}=2.82\times10^{-13}$cm is the classical electron radius, $\epsilon_{0}$ is the energy
of the incoming photon, $\epsilon$ is the energy of the outgoing photon, $\theta_{S}$ is the polar
scattering angle, $\phi_{S}$ is the azimuthal scattering angle and $Q$ and $U$ are the linear
Stokes parameters respectively \citep{McMaster1961, BaiRamaty1978}. The maximum change in $DOP$
occurs when $\theta_{S}=90^{\circ}$ and no change in $DOP$ occurs for a backscattering at
$180^{\circ}$. The azimuthal scattering angle $\phi_{S}$ also has a non uniform dependency on the incoming polarization state. If the HXR photon distribution is completely isotropic and hence unpolarized, then the Klein-Nishina cross section returns to the unpolarized form used in Chapter \ref{ref:Chapter5} Equation \ref{eq:sigma_c}, since the linear Stokes parameters $Q$ and $U$ are zero.

In the MC simulations, when a Compton scattering occurs, the properties of the outgoing photon: energy $\epsilon$, polar
scattering angle $\theta_{S}$, azimuthal angle $\phi_{S}$ and Stokes parameters $Q$ and $U$, need to be updated. New polar scattering angles $\theta_{S}$
for each photon can be found by integrating the polarization dependent Klein-Nishina differential cross section over $\phi_{S}$ to produce a differential cross-section that is only dependent on $\epsilon$ and $\theta_{S}$, that is the unpolarized form given by Equation \ref{eq:sigma_c}. The new energy $\epsilon$ and scattering angle $\theta_{S}$ can then be found by the same method shown in Chapter \ref{ref:Chapter5} Section \ref{ref:cs} while the new $\phi_{S}$ and Stokes parameters $Q$ and $U$ are found by the method described below.

\subsection{Updating photon polarization states}\label{ref:uppps}
If the photon distribution is completely isotropic and unpolarised then the azimuthal scattering
angle $\phi_{S}$ can just be sampled from a uniform distribution between $0$ and $2\pi$, but this
is not true for the more general polarization dependent case. The probability density function of
obtaining a value of $\phi_{S}$ between $\phi_{S}$ and $\phi_{S}+d\phi_{S}$ can be described by:
\begin{equation*}
P(\phi_{S})=\frac{1}{2\pi}\frac{d\sigma_{c}(\epsilon,\theta_{S},\phi_{S})/d\Omega}{d\sigma_{c}(\epsilon,\theta_{S})/d\Omega}=
\end{equation*}
\begin{equation}
=\frac{1}{2\pi}\frac{\frac{\epsilon_{0}}{\epsilon}+\frac{\epsilon}{\epsilon_{0}}-
\sin^{2}\theta_{S}(1-Q\cos2\phi_{S}-U\sin2\phi_{S})}{\frac{\epsilon_{0}}{\epsilon}+\frac{\epsilon}{\epsilon_{0}}-\sin^{2}\theta_{S}}
\end{equation}
with the maximum value of this function given by:
\begin{equation}
P_{max}(\phi_S)=\frac{1}{2\pi}\frac{\frac{\epsilon_{0}}{\epsilon}+\frac{\epsilon}{\epsilon_{0}}-
\sin^{2}\theta_{S}\left(1-\sqrt{Q^{2}+U^{2}}\right)}{(\frac{\epsilon_{0}}{\epsilon}+
\frac{\epsilon}{\epsilon_{0}}-\sin^{2}\theta_{S})}.
\end{equation}
Firstly, a value of $\phi_{S}$ is sampled between $0$ and $2\pi$. The condition that
$P(\phi_S)<P_{max}(\phi_S)$ is then used to accept a value of $\phi_{S}$ and provides a method for
sampling values of $\phi_{S}$ for each photon, using the new values $\theta_S$ and $\epsilon$
already calculated for each photon. This method is repeated until the condition is
satisfied for each photon and each photon is provided with an azimuthal scattering angle
\citep{Salvatetal2008}.

Due to Compton scattering, the Stokes parameters have to be updated using the scattering matrix T
\citep{McMaster1961, BaiRamaty1978}				\begin{equation} 		T(\epsilon,\theta_{S})= 		
\left( 		\begin{array}{ccc}
  		\frac{\epsilon}{\epsilon_{0}}+\frac{\epsilon_{0}}{\epsilon}+\sin^{2}\theta_{s}&
  \sin^{2}\theta_{s}  &0   \\
  		\sin^{2}\theta_{s}&\cos^{2}\theta_{s}+1   & 0  \\
  		0& 0  &2\cos\theta_{s}
		\end{array} 		\right).
\label{eq:T_matrix}
		\end{equation} Before a scattering, the Stokes parameters have to be rotated by $\phi_S$ so
that they are defined relative to the plane of scattering, using the rotation matrix $M(\phi_{S})$ given by:
		
		\begin{equation} 		M(\phi_{S})= 		\left( 		\begin{array}{ccc} 		1&0&0 \\
  		0 &\cos2\phi_{S}  &\sin2\phi_{S} \\
  		0 &-\sin2\phi_{S}  &\cos2\phi_{S} \\
		\end{array} 		\right). 		\end{equation}
		
 After a scattering, the Stokes
parameters have to be rotated again so that they are defined relative to the starting position of
the source, and are rotated by the rotation matrix
$M(-\Xi)$.
		
		\begin{equation} 		M(-\Xi)= 		\left( 		\begin{array}{ccc} 		1&0&0 \\
  		0 &\cos-2\Xi  &\sin-2\Xi \\
  		0 &-\sin-2\Xi  &\cos-2\Xi \\
		\end{array} 		\right), 		\end{equation}
\begin{figure}
\centering
\includegraphics[width=90mm]{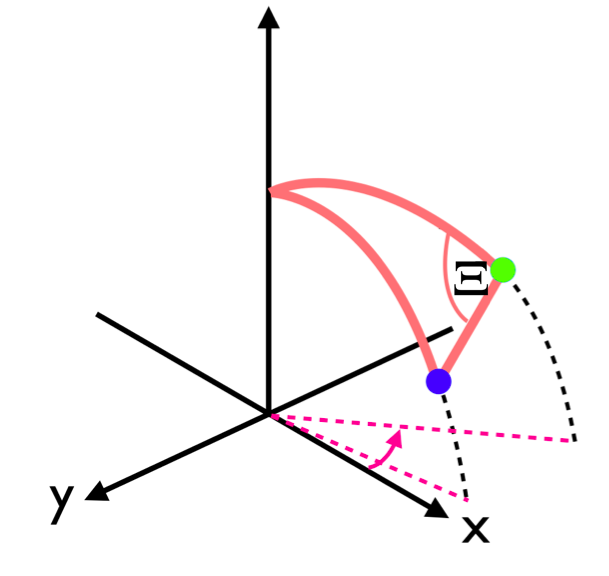}
\caption[The position of the photon before scattering and after scattering
and the angle $\Xi$ that determines the final rotation of the Stokes parameters back into the
frame of the source from the scattering frame.]{The position of the photon before scattering (blue) and after scattering (green)
and the angle $\Xi$ that determines the final rotation of the Stokes parameters back into the
frame of the source from the scattering frame.}
\label{fig:rotation}
\end{figure}
where
\begin{equation}
\cos\Xi=\pm\frac{w'-\cos\theta_{s}w}{\sqrt{1-\cos^{2}\theta_{s}}\sqrt{1-w^2}}.
\label{eq:cosGamma}
\end{equation}
$w$ and $w'$ are the current and previous $z$ direction cosines respectively \citep{Hovenier1983}.
The $\pm$ in equation (\ref{eq:cosGamma}) is present due to the negative sign being used when
$\pi\leq\phi_{s}\leq2\pi$ and the positive sign for $0\leq\phi_{S}<\pi$. The angle $\Xi$ is
shown in Figure \ref{fig:rotation}. $\Xi$ is the angle between the scattering plane and the
normal plane in the frame of the source. Therefore during a Compton scattering the order of the
rotations on the Stokes pseudovector $[I Q U]^{T}$ is $M(\phi_{S})T(\epsilon,\theta_{S})M(-\Xi)$.

\section{Integrated distribution of hard X-ray polarization}

\subsection{Hard X-ray polarization and electron directivity}

\begin{figure*}
\centering
\includegraphics[width=17cm]{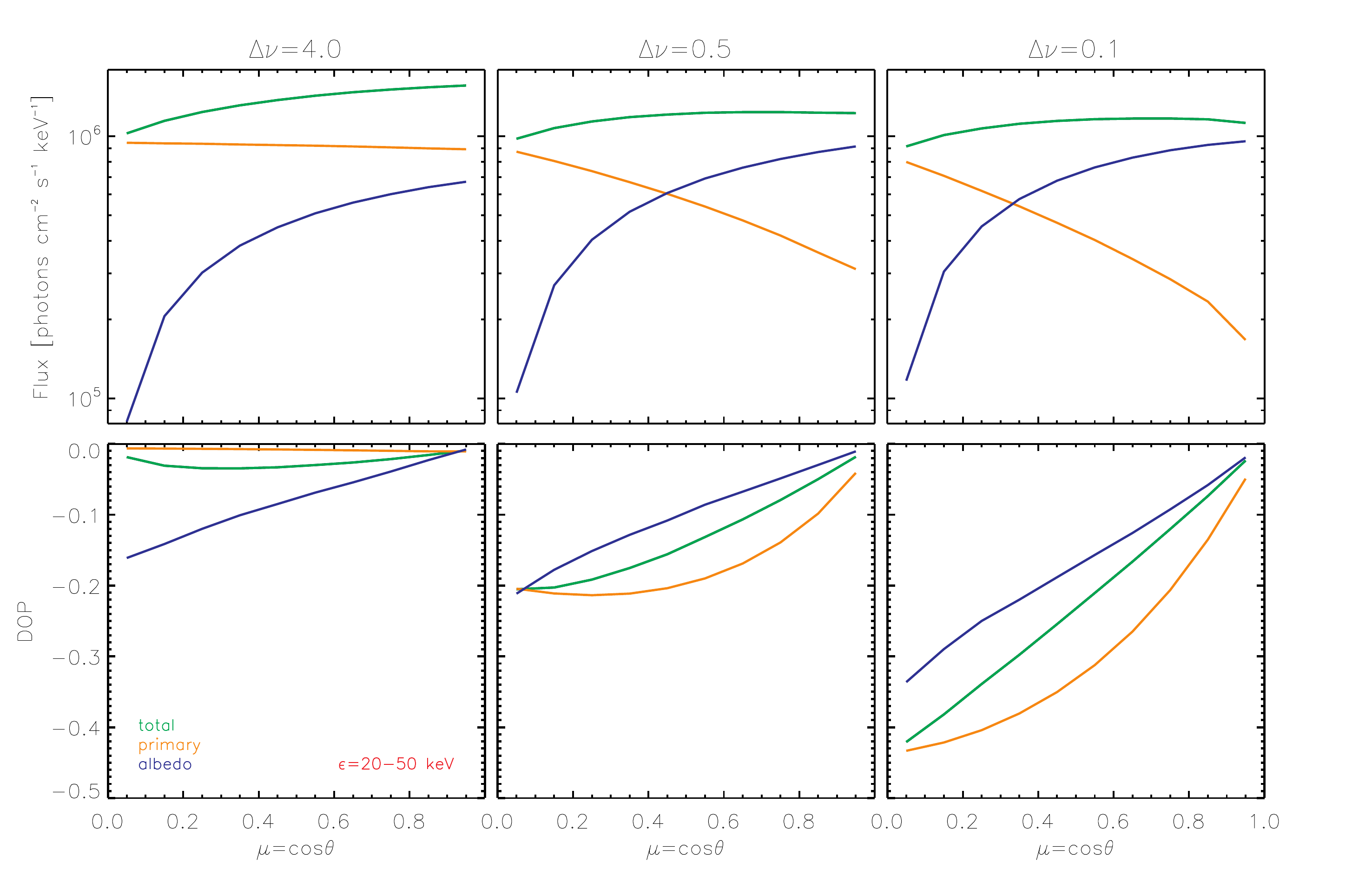}
\caption[Plots of the photon flux and spatially integrated $DOP$ against heliocentric angle for the upward primary,
albedo and total components, for each MC simulation input.]{Photon flux (top row) and spatially integrated $DOP$ (bottom row) for the upward primary (orange),
albedo (blue) and total (green) components for both the $\Delta\nu= 4.0$
(1st column), the $\Delta\nu= 0.5$ (2nd column) and the $\Delta\nu=0.1$ (last column) created
photon distributions respectively, at the peak albedo energies of 20-50 keV, for different
locations on the solar disk from the centre ($\mu=1.0$) to the limb ($\mu=0.0$). Note here that a
negative $DOP$ denotes that the direction of polarization $\Psi$ across the source is parallel to the
radial direction. The $DOP$ for $\Delta\nu=4.0$ photon distribution (near isotropic
distribution) shows nearly the same result as \cite{BaiRamaty1978}, as expected.}
\label{fig:Fig1}
\end{figure*}

Figure \ref{fig:Fig1} shows the resulting total integrated flux and degree of polarization for each simulation run using the $\Delta\nu=4.0$, $\Delta\nu=0.5$ and $\Delta\nu=0.1$ electron pitch-angle distributions.
Each HXR component: primary towards the observer only (orange), backscattered albedo (blue) and also the total observed HXR emission (green) are plotted against emission angle or equivalently heliocentric angle $\mu=\cos\theta$. Each are shown between
20-50 keV, where the HXR albedo emission peaks. As expected, even though the
$\Delta\nu=0.1$ distribution has a greater downward beaming and hence a smaller proportion of its emission is directed towards the observer than the $\Delta\nu=0.5$ distribution, within the 20-50 keV range the difference in anisotropy between the two primary distributions cannot be clearly seen. However, there is a clear difference in the $DOP$ of both distributions with disk location over 20-50 keV (second row Figure \ref{fig:Fig1}). The total integrated polarization angle $\Psi$ is not shown due to the fact that it remains constant with emission angle between the energy range of 20-50 keV at $\Psi=0^{\circ}$ and hence provides no information concerning the directivity of the electron distribution in the chromosphere. In Figure \ref{fig:Fig1} $\Psi=0^{\circ}$ is indicated by the negative value of spatially integrated $DOP$.

\subsection{Hard X-ray polarization and the high energy cutoff in the electron distribution}
Spatially integrated polarization is dependent on the highest energy in the electron distribution, called the high cutoff energy
 \citep{Heristchi1987}. When calculating spatially integrated polarization,
equation (\ref{eq:DOP}) reduces to $DOP=Q$ and Equation (\ref{eq:Psi}) reduces to $\Psi=\frac{1}{2}\arctan\left(\frac{-0}{-Q}\right)$
as $U$ sums to zero for a single measurement across the entire source. A negative $DOP$ indicates that the polarization angle
is parallel to the radial direction ($\Psi=0^{\circ}$), while a positive $DOP$ indicates that the polarization angle is
perpendicular to the radial direction ($\Psi=90^{\circ}$).

\begin{figure}[h]
\vspace{-20pt}
\centering
\includegraphics[width=14.55cm]{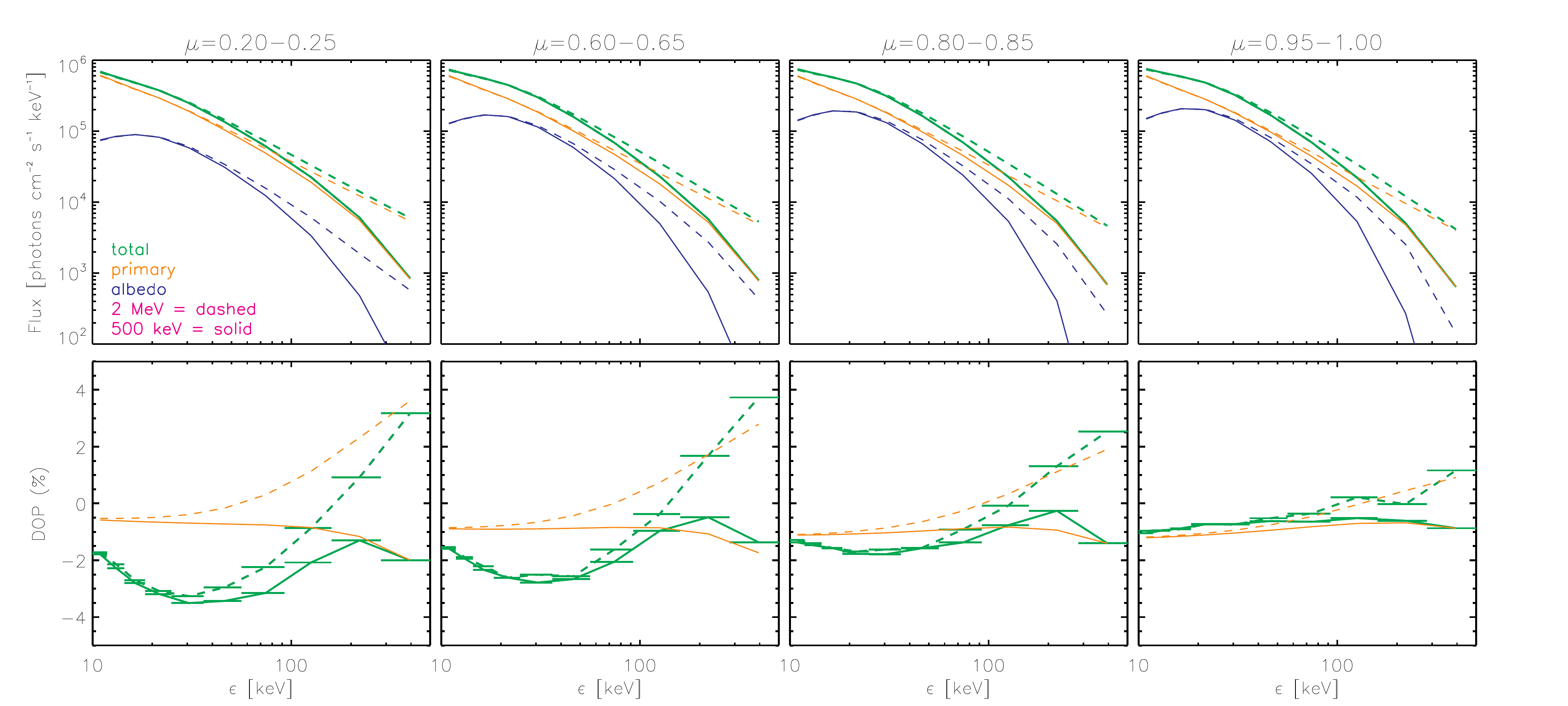}
\includegraphics[width=14.55cm]{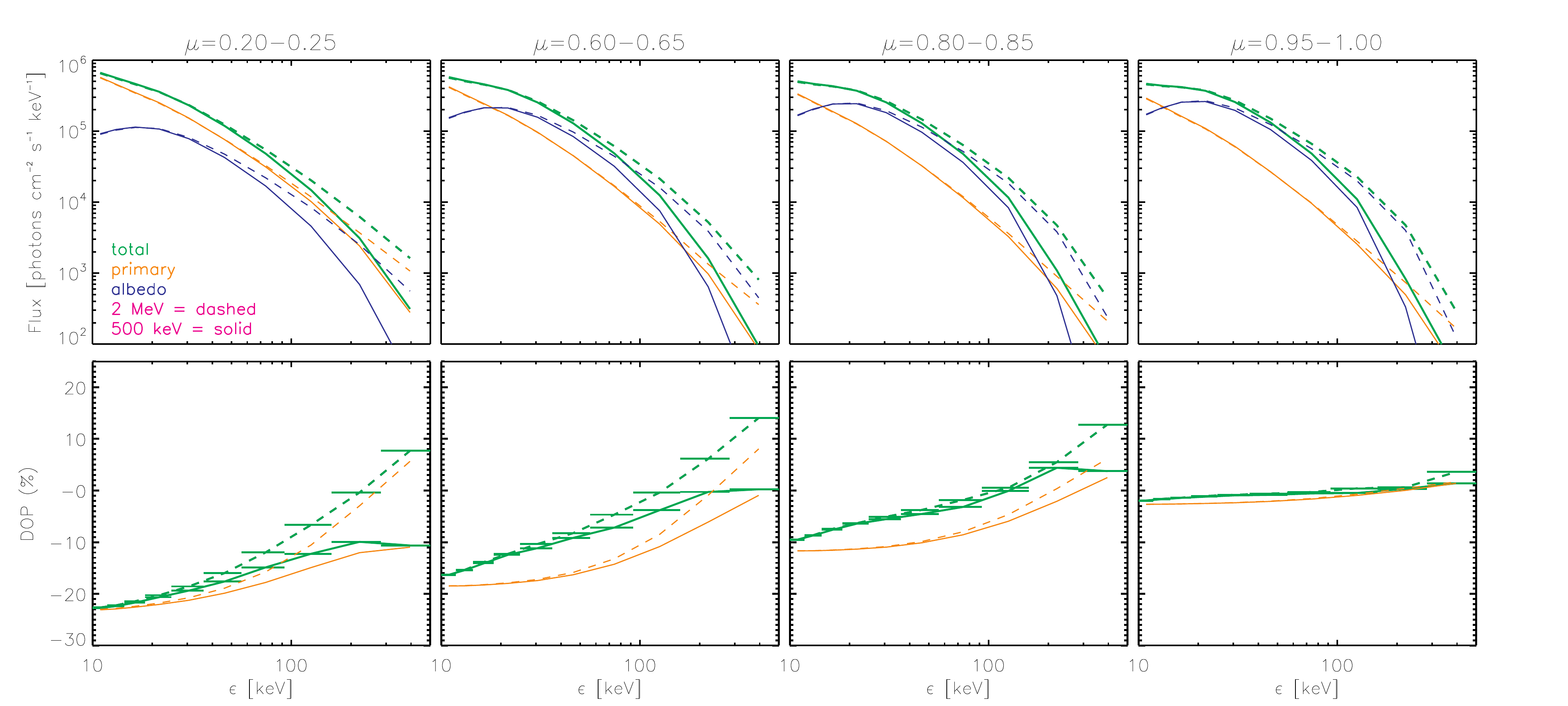}
\includegraphics[width=14.55cm]{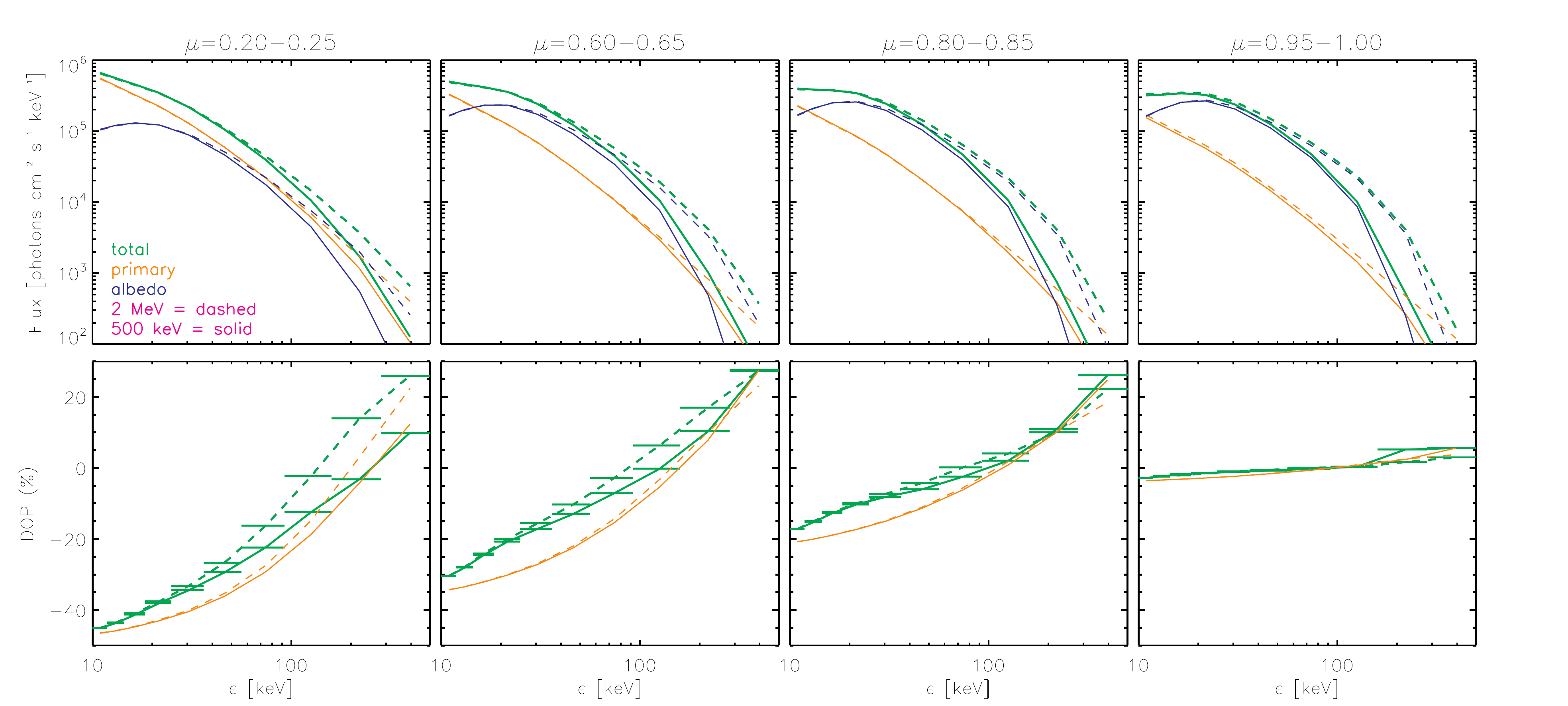}
\caption[Plots of photon flux and spatially integrated $DOP$ for the upward primary,
albedo and total components against X-ray energy.]{Flux and spatially integrated $DOP$ for total source (green), primary (orange) and albedo (blue) vs. energy $\epsilon$ using $E_{cutoff} = 500$ keV (solid) and $E_{cutoff} = 2$ MeV (dashed) for $\Delta\nu=4$ (top), $\Delta\nu=0.5$ (middle) and $\Delta\nu=0.1$ (bottom).}
\label{fig:energy_dnu_all}
\end{figure}

For the three electron distributions of $\Delta\nu=4.0$, $\Delta\nu=0.5$ and
$\Delta\nu=0.1$, simulations were run with two high cutoff electron energies of $E_{cutoff}=500$
keV and $E_{cutoff}=2$ MeV. Figure \ref{fig:energy_dnu_all} plots the flux and spatially integrated polarization across the total source (green)
and the primary source only (orange) against photon energy $\epsilon$ at four disk locations
$\mu\in[0.20-0.25],[0.60-0.65],[0.80-0.85],[0.95-1.00]$ for $\Delta\nu=4.0$, $\Delta\nu=0.5$ and
$\Delta\nu=0.1$ (top to bottom) respectively. The important property to observe here is not the
magnitude but the sign of the $DOP$ or whether $\Psi=0^{\circ}$ or $\Psi=90^{\circ}$.

Using an electron distribution with $\Delta\nu=4.0$ (Figure \ref{fig:energy_dnu_all} rows 1 and 2) and a cutoff energy of $E_{cutoff}=500$ keV
produces a photon distribution with a negative $DOP$ at all photon energies and disk locations, while using the same distribution
with a cutoff energy of $E_{cutoff}=2$ MeV creates a photon distribution where the $DOP$ changes from negative to positive at
$\sim100-200$ keV at all disk locations.  During bremsstahlung, in order to conserve energy, electrons with higher energies will scatter through larger angles.
When a photon is scattered through a large angle, its polarization is more likely to be directed perpendicular to the plane of emission
($\Psi=90^{\circ}$) rather than parallel to the plane of emission ($\Psi=0^{\circ}$). Therefore, a change in the direction of polarization
(from $\Psi=0^{\circ}$ to $\Psi=90^{\circ}$) indicates the presence of higher energies in the electron distribution, greater than $\sim1$ MeV.

As the beaming of the electron distribution increases ($\Delta\nu=0.5$ and $\Delta\nu=0.1$ distributions), the above statement
does not hold and it becomes more likely that both electron distributions with cutoff energies of either $E_{cutoff}=500$ keV and $E_{cutoff}=2$ MeV
will produce photons with $\Psi=90^{\circ}$. For the $\Delta\nu=0.5$ distribution (Figure
\ref{fig:energy_dnu_all} rows 3 and 4), as the source moves towards the solar centre, the
photon distribution created by the $E_{cutoff}=500$ keV also produces photons at higher energies (again $\sim100-200$ keV)
with $\Psi=90^{\circ}$. For the very beamed $\Delta\nu=0.1$ distribution (Figure \ref{fig:energy_dnu_all} rows 5 and 6), both
the $E_{cutoff}=500$ keV and $E_{cutoff}=2$ MeV distributions produce high ($\sim100-200$ keV) energy photons with $\Psi=90^{\circ}$
at all disk locations.
Therefore using the direction of the spatially integrated polarization angle $\Psi$ as an indicator for high energies in the electron
distribution becomes less and less useful as the beaming of the photon/electron distribution
increases. The increased beaming causes lower and lower energy photons to scatter at larger angles, especially at locations closer to the solar centre,
hence producing photons with $\Psi=90^{\circ}$ in the $E_{cutoff}=500$ keV distributions.
However this method may be useful when the anisotropy of the HXR distribution in the chromosphere is close to isotropic as has been suggested by recent observations \citep{KontarBrown2006,2013SoPh..284..405D}.

\section{Spatial distribution of hard X-ray polarization}

\subsection{Single Compton scatter for an isotropic unpolarised source}

In order to demonstrate the spatial variation in polarization due to Compton scattering, the
easiest example to consider is the albedo patch created by an initially isotropic, unpolarised
point source at a height $h$ above the photosphere (see Chapter \ref{ref:Chapter5} and \cite{KontarJeffrey2010}). For this example, the variation in
polarization across the source can be described analytically by \citep{McMaster1961} if no energy losses are assumed,
\begin{equation}
DOP=\frac{1-\cos^{2}\theta_{S}}{1+\cos^{2}\theta_{S}},
\label{eq:DOP_1scat}
\end{equation}
where
\begin{equation}
\cos\theta_{S}=\cos\theta\cos(\pi-\theta_{i})+\sin\theta\sin(\pi-\theta_{i})\cos\phi.
\label{eq:cosTheta_S}
\end{equation}
$\theta_{S}$ is the scattering angle, $\theta_{i}$ is the emission angle measured from the local
solar vertical, $\theta$ is the heliocentric angle on the solar disk and $\phi$ is the azimuthal angle
measured in the solar disk plane. In Equation (\ref{eq:DOP_1scat}), the scattering angle
$\theta_{S}$ determines the $DOP$. $\theta_{S}$ is related to distance $r$ from the centre of the
albedo patch by Equation (\ref{eq:cosTheta_S}) though $r=h\tan(\pi-\theta_{i})$, and hence the $DOP$ at any point across the albedo patch located at any heliocentric angle on the solar disk can be
easily calculated. Note that for this simple example, Equation (\ref{eq:DOP_1scat}) assumes that
the energy difference between the incoming and scattered photon is negligible,
though this is only the case for low HXR energies of $\sim10$ keV. The $DOP$ for higher energies must be calculated using the $T$ scattering matrix, which is used in the simulations (Equation (\ref{eq:T_matrix}) and discussed in Section \ref{ref:uppps}) \citep{McMaster1961} .

\begin{figure}
\centering
\includegraphics[width=14cm]{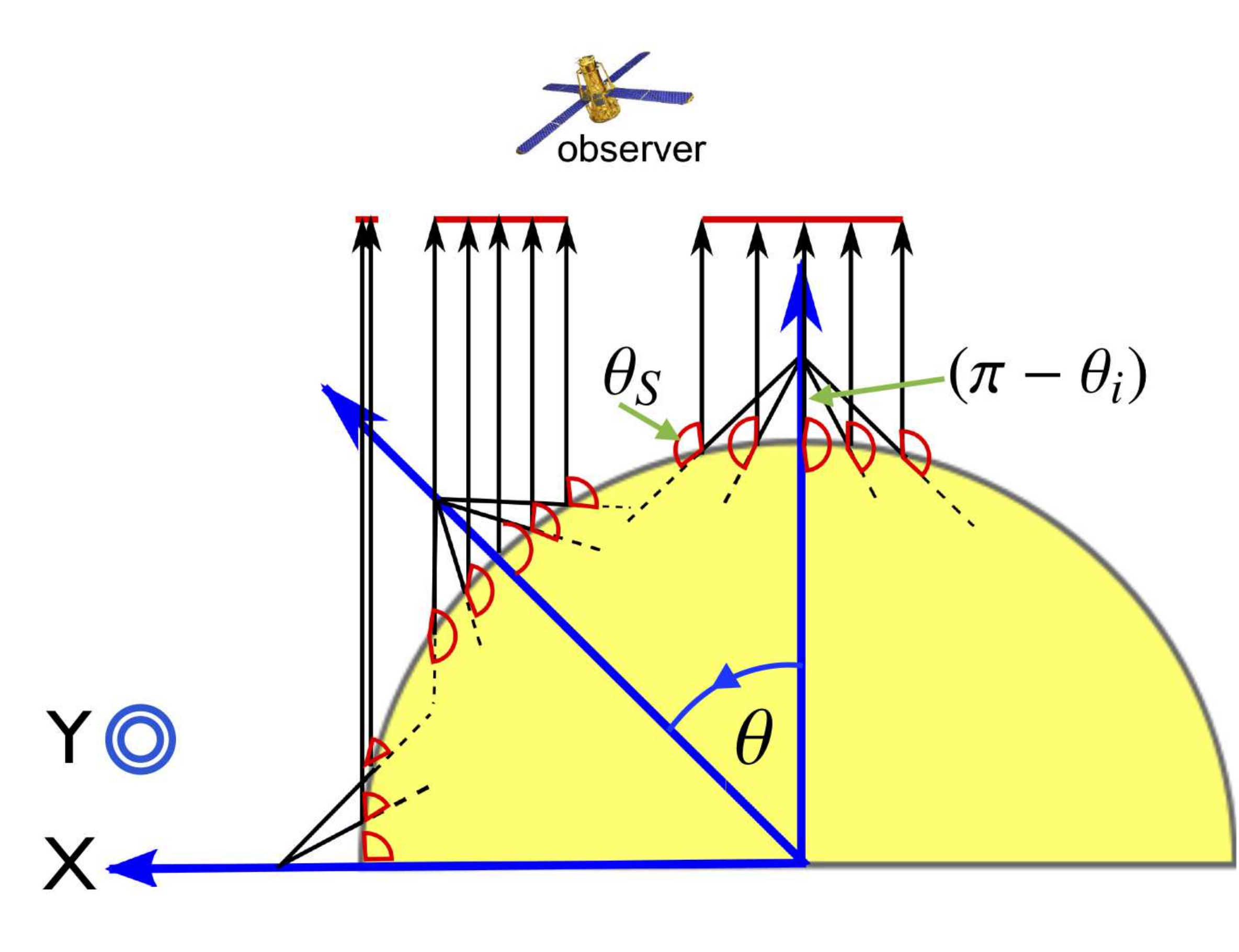}
\caption[Diagram of a single Compton scattering in the photosphere
for three heliocentric angles of $0^{\circ}$, $45^{\circ}$ and $90^{\circ}$.]{Diagram of a single Compton scattering in the photosphere
for three heliocentric angles of $0^{\circ}$, $45^{\circ}$ and $90^{\circ}$. For a single scatter,
the DOP tends to $100\%$ as the scattering angle approaches $90^{\circ}$, producing a variation in
polarization across the extent of the source.}
\label{fig:cartoon1s}
\end{figure}

\begin{figure*}
\centering
\includegraphics[width=17cm]{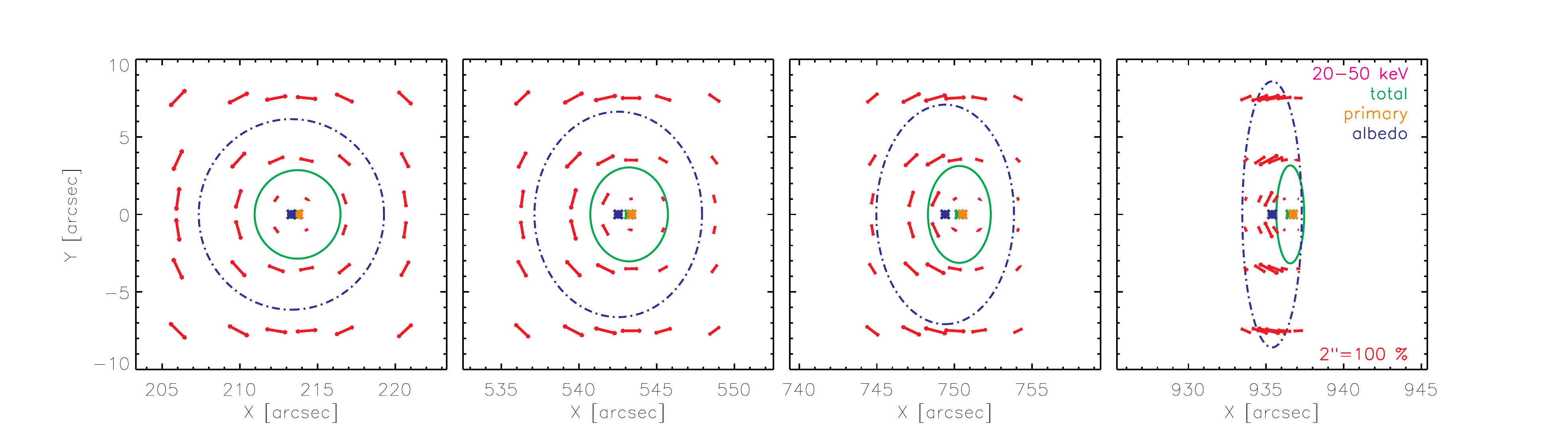}
\caption[Albedo polarization maps for an isotropic, unpolarised point source sitting above the
photosphere at four different locations after a single Compton scatter in the photosphere.]
{Albedo polarization maps for an isotropic, unpolarised point source sitting above the
photosphere
at four different radial locations of $X=214'', 543'', 750'', 936''$ (left to right) at $Y=0''$ (corresponding
$\mu=0.97,0.82,0.62,0.22$) after a single Compton scatter in the photosphere. All results are shown
at the peak albedo energies of 20-50 keV. The length of each red arrow indicates the $DOP$ and the
direction of each arrow depicts the polarization angle $\Psi$ within the chosen plotting bin. The
solar radial direction (or $X$ axis for this case) is defined as the $\Psi = 0^{\circ}$ position. An
arrow length of $2''$ corresponds to a maximum $DOP$ of $100 \%$. The green and blue ellipses give
the FWHM of the total and albedo sources respectively, while the green, blue and orange asterisks
give the centroid position of the total, albedo and primary sources.}
\label{fig:mapsss}
%\end{figure*}
%\begin{figure*}
\centering
\includegraphics[width=17cm]{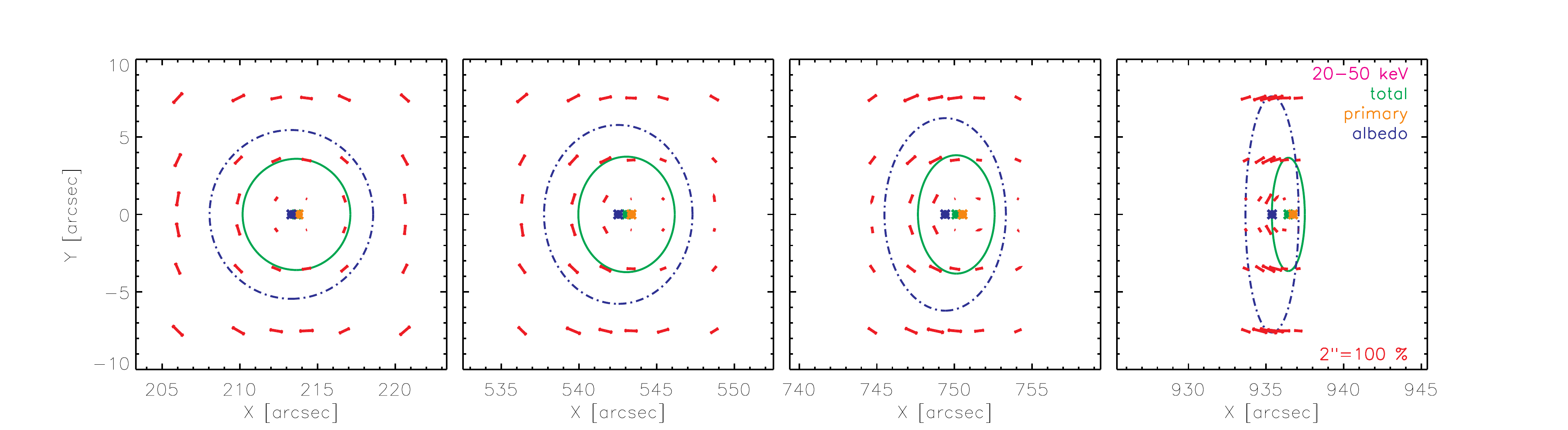}
\caption[Albedo polarization maps as in Figure \ref{fig:mapsss}, but for the case of multiple
Compton scatterings
in the photosphere.]{Albedo polarization maps as in Figure \ref{fig:mapsss}, but for the case of multiple
Compton scatterings
in the photosphere. Multiple scatterings have acted to decrease the $DOP$ at all points across each
source but the polarization angle at each point on the map remains the same.}
\label{fig:mapsms}
\end{figure*}

Figure \ref{fig:cartoon1s} shows a cartoon of this single scattering from an isotropic point source
for three different heliocentric angles of $0^{\circ}$, $45^{\circ}$ and $90^{\circ}$.  For a
source located exactly above the solar centre ($0^{\circ}$) at a height $h$, the resulting variation across the
photospheric albedo patch is radially symmetric. As the radial distance $r$ from the source centre
increases, the scattering angle $\theta_{S}$ of any observed radiation will decrease from $180^{\circ}$ towards $90^{\circ}$,
causing the $DOP$ to grow from $0\%$ to $\sim 100\%$.
Radiation scattered in the photosphere at a location directly below the HXR source, which is a $180^{\circ}$ backscatter,  and emitted
towards the observer will experience no change in its $DOP$. This
statement is true for HXR sources at any heliocentric angle $\theta_{i}$ but the
projection effects at angles $\theta_{i}>0^{\circ}$ will create an asymmetry in the polarization pattern along the radial direction,
whereas the polarization pattern in the perpendicular to radial direction always remains
symmetrical. The described pattern can be seen in Figure \ref{fig:mapsss} which shows the
polarization maps for a single Compton scattering at four disk locations
ranging from the solar centre to the limb, along $Y=0''$ at $X=214'' , 543'' , 750'', 936''$.
These locations are equivalent to $\mu=0.97,0.82,0.62,0.22$ ($\theta_{i}=14^{\circ},35^{\circ},52^{\circ},77^{\circ}$) and denote the approximate positions of the total observed HXR source,
which are shifted from the primary HXR position due to the albedo component \citep{KontarJeffrey2010}, which was discussed in Chapter \ref{ref:Chapter5}.
The polarization across the HXR source at any location on the solar disk can always be measured with respect to the radial line connecting the solar
disk centre and the centre of the source. Therefore due to the symmetry of the problem, source
locations at $Y=0''$ considered in this thesis can straightforwardly be applied to any solar disk
location.  In Figure \ref{fig:mapsss}, the dotted blue ellipse denotes the FWHM of the diffuse albedo
component, the solid green ellipse denotes the FWHM of the total observed source and the orange,
blue and green asterisks indicate the centroid positions of primary, albedo and total observed
sources respectively. The polarization angle $\Psi$ follows a steady pattern across all
sources. $\Psi$ is always at an angle tangential to the line connecting the desired position and
the location of the source centroid position. Hence at disk centre locations the $\Psi$ pattern is also symmetrical
across the source.

Figure \ref{fig:mapsms} shows the same albedo
polarization maps as in Figure \ref{fig:mapsss} but for multiple Compton scatterings.
The overall pattern for the $DOP$ and $\Psi$ are preserved but due to multiple scatterings the overall $DOP$ across
all locations over the albedo patch has decreased. A single scattering $DOP$ of $\sim100\%$ near the
edge of the source has been reduced to $\sim50\%$ by multiple scatterings. All other simulations
shown below use multiple Compton scattering in the photosphere.

The albedo pattern from a primary source at a greater
height than 1 Mm (say from a coronal source) should produce the same albedo polarization
pattern but over a much greater area in the photosphere. The albedo patch for such a source should therefore be very large with a very low intensity and hence should not alter the properties of coronal sources to the same extent as chromospheric sources. The polarization pattern is always plotted at the peak albedo energies 20-50 keV since this is where the greatest change in source polarization should occur.

\subsection{Anisotropic source at a height of h=1~Mm (1''.4) and size of 5''}

For a chosen chromospheric HXR source height of $h =1$ Mm $(1\arcsec.4)$ and a primary source size of
FWHM$\sim5''$, simulations were performed for all three photon distributions created by the
$\Delta\nu= 4.0$, $\Delta\nu= 0.5$ and $\Delta\nu= 0.1$ electron distributions. This is the same height that was used in Chapter \ref{ref:Chapter5} and again was chosen to match chromospheric HXR source height measurements
\citep{2008A&A...489L..57K,Pratoetal2009,SaintHilaireetal2010,MrozekKowalczuk2010,2010ApJ...717..250K,BattagliaKontar2011}.
Again, all the results shown here are for the energy range of 20-50 keV, where albedo emission peaks,
producing the largest distortion to the primary component but the best opportunity for the
detection of the albedo component.

Figures \ref{fig:maps_dnu4}, \ref{fig:maps_dnu05} and \ref{fig:maps_dnu01} each plot the resulting polarisation maps for four HXR sources
(resulting from the primary and albedo components) created by the
$\Delta\nu= 4.0$, $\Delta\nu=0.5$ and $\Delta\nu=0.1$ electron distributions respectively. As with Figures \ref{fig:mapsss} and \ref{fig:mapsms},
each figure plots four HXR sources positioned at $\mu\sim0.97,0.82,0.62,0.22$.
In Figures \ref{fig:maps_dnu4}, \ref{fig:maps_dnu05} and \ref{fig:maps_dnu01}, the dotted ellipses denote the FWHM of the total source (green),
the primary source (orange) and the albedo source (blue) and
the correspondingly coloured asterisks denote the $(x,y)$ centroid position of the total source and
the primary and albedo components respectively. Figures \ref{fig:slicesX_dnu4}, \ref{fig:slicesX_dnu05} and \ref{fig:slicesX_dnu01} plot intensity,
$I$ (top row), $DOP$ (middle row) and $\Psi$ (bottom row) along the radial
direction $X$ centred at $Y= 0''$ (across a of bin width$= 2''$) for each of the maps in Figures \ref{fig:maps_dnu4}, \ref{fig:maps_dnu05} and \ref{fig:maps_dnu01}. Figures
\ref{fig:slicesY_dnu4},  \ref{fig:slicesY_dnu05} and \ref{fig:slicesY_dnu01} plot the $DOP$ (top row) and $\Psi$ (bottom row) along the perpendicular to radial direction $Y$
centred at X$=213.6'' ,543.9'' ,750.1'' ,936.6''$ (again across a bin width$=2''$ ) for each of the maps in Figures \ref{fig:maps_dnu4}, \ref{fig:maps_dnu05} and \ref{fig:maps_dnu01}.

\subsubsection{Quasi-isotropic distribution of electrons with $\Delta\nu=4.0$}

\begin{figure}[h]
\vspace{-30pt}
\includegraphics[width=17cm]{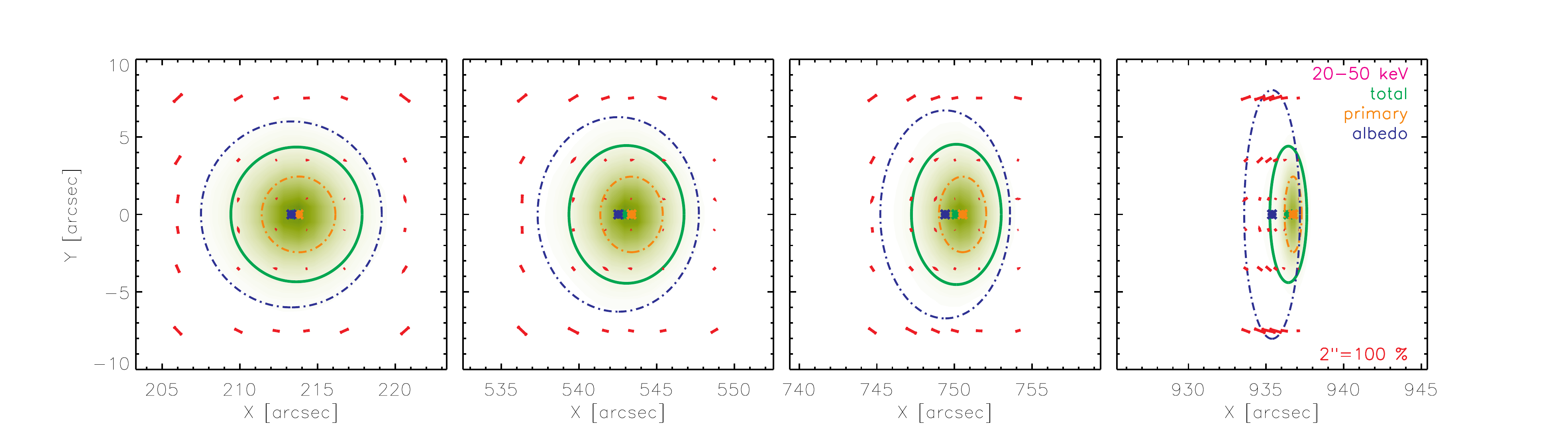}
\caption[Total X-ray brightness and polarization maps for $\Delta\nu=4.0$ electron distribution.]{Total X-ray brightness and polarization maps for $\Delta\nu=4.0$. The total sources sit at 4 disk locations of $X=213.6'', 542.9'', 750.1'', 936.6''$ at $Y=0''$(corresponding to $\mu\sim0.97,0.82,0.62,0.22$). Green, blue and orange ellipses/dots give the FWHM and centroid positions of the total, albedo and primary sources.}
\label{fig:maps_dnu4}
\includegraphics[width=17cm]{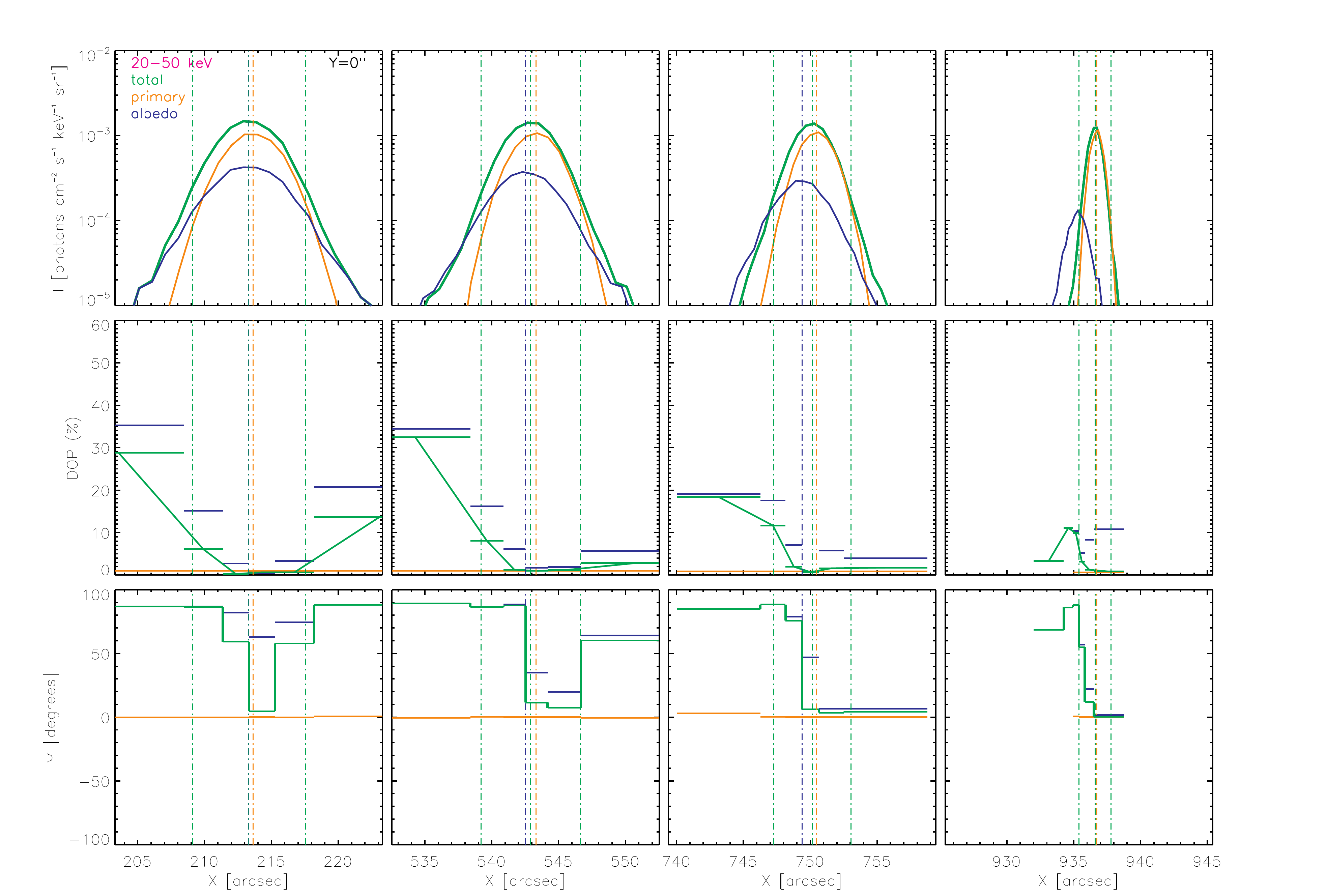}
\caption[$I$, $DOP$ and $\Psi$ radial slices along $X$ at $Y=0''$ for the sources in Figure \ref{fig:maps_dnu4} for the
$\Delta\nu=4.0$ electron distribution.]{$I$ and $DOP$ radial slices along $X$ at $Y=0''$ for the sources in Figure \ref{fig:maps_dnu4}
($\Delta\nu=4.0$). Colours as in Figure~\ref{fig:maps_dnu4} and dash-dot lines denote the centroid positions and the FWHMs of the total observed source, primary and albedo sources.}
\label{fig:slicesX_dnu4}
\end{figure}

\begin{figure*}
\includegraphics[width=17cm]{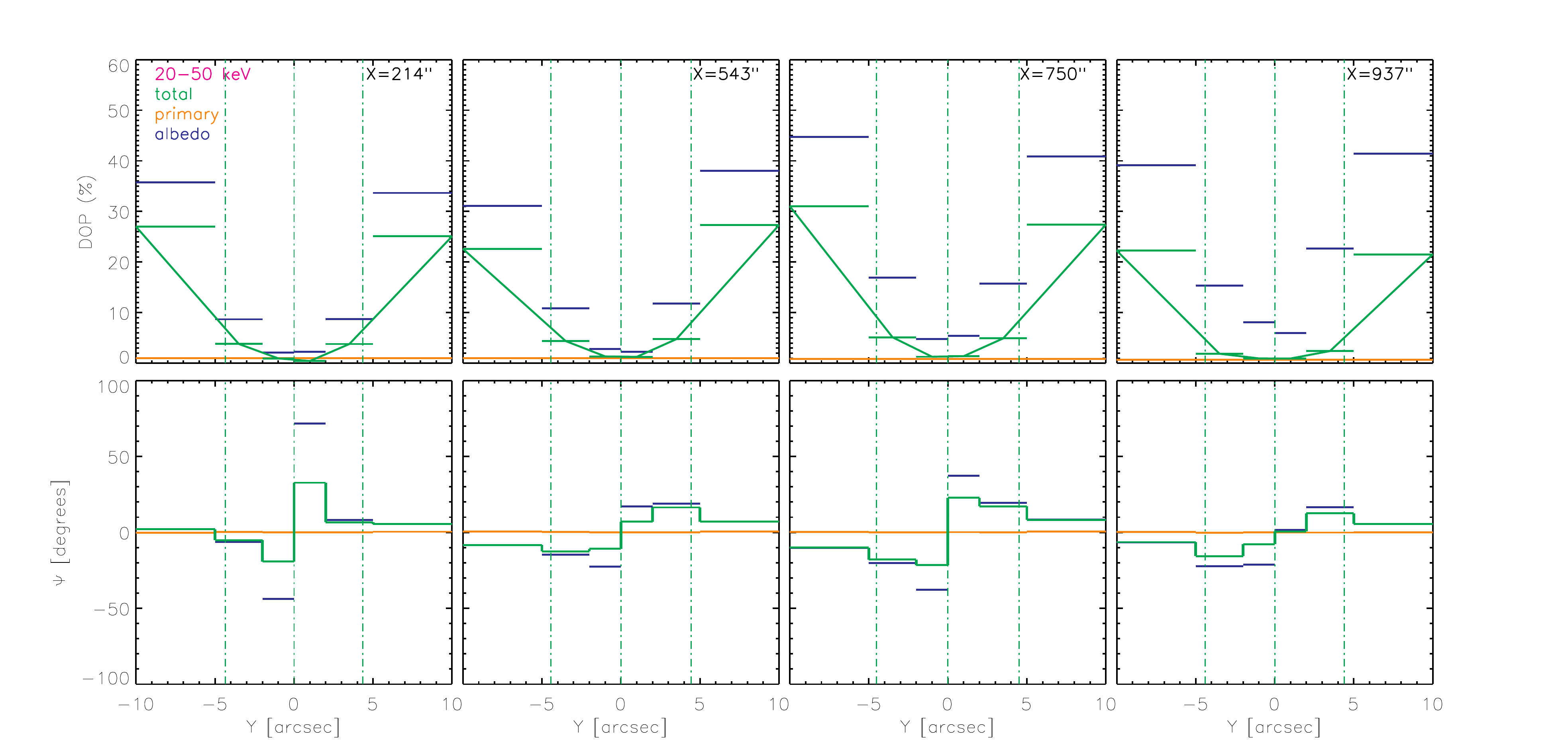}
\caption[Perpendicular to radial slices through each of the sources shown in Figure
\ref{fig:maps_dnu4} for the $\Delta\nu=4.0$ electron distribution.]{Perpendicular to radial slices through each of the sources shown in Figure
\ref{fig:maps_dnu4} for the $\Delta\nu=4.0$ distribution.
Each of the perpendicular to radial slices are taken along $Y$  at $X=213.6'', 542.9'', 750.1'', 936.6''$ for the $DOP$ and polarization angle $\Psi$. The lines and
colours are as in Figures \ref{fig:maps_dnu4}, \ref{fig:slicesX_dnu4}. The $DOP$ and the magnitude
of the polarization angle $\Psi$ remain symmetrical along the source perpendicular to radial direction.}
\label{fig:slicesY_dnu4}
\end{figure*}

The HXR photon distribution produced by the $\Delta\nu=4.0$ electron distribution is approximately
unpolarised and isotropic. Therefore, both the spatially integrated and spatially resolved polarization measurements across the primary source at all locations on the solar disk produce
$DOP\sim 0\%$ and $\Psi=0^{\circ}$ (radial) at $20-50$ keV.
The albedo component produces asymmetrical $DOP$ and $\Psi$ variations along the source radial direction (Figure \ref{fig:slicesX_dnu4})
while along the source perpendicular to radial direction (Figure \ref{fig:slicesY_dnu4}),
variations in albedo $DOP$ and $\Psi$ are approximately symmetrical, since the centroid positions of the primary and albedo components always coincide in the perpendicular to radial direction (see Chapter \ref{ref:Chapter5}).
The simulated HXR sources plotted in Figure \ref{fig:maps_dnu4} have a finite source size of $\sim5''$.
Compared with a point source (Figure \ref{fig:mapsms}), this produces two main differences:
i) photons leave the source from different positions above the photosphere and
ii) for certain distributions and disk locations, the primary polarization will dominate over the extent of the primary source.
The first property reduces the $DOP$ at all points across the source, compared with the albedo patch created by a point source. Due to the second property, the polarization variation caused by albedo may be slightly masked by the primary component within the source FWHM, especially for cases where the primary component is dominant, that is, isotropic or near isotropic distributions.

For the quasi-isotropic $\Delta\nu=4.0$ distribution, the primary component is the dominant component at
all four disk locations, with the albedo contribution falling as the source location moves towards
the limb (Figure \ref{fig:slicesX_dnu4}- first row). Hence, the primary component dominates within the
FWHM of the total source, while the albedo component dominates after this boundary.
The second and third rows of Figure \ref{fig:slicesX_dnu4} demonstrate common radial trends in $DOP$
and $\Psi$, not only across the extent of each individual source but also between sources at different disk locations.
For a quasi-isotropic distribution, at a particular disk location (other than the disk centre), the highest $DOP$ along the radial direction is observed at the disk-centre-side of the source (where the albedo dominates). This falls to approximately zero within the FWHM of the total source (where the primary dominates) and then increases again towards the limb-side of the source (where the albedo again dominates), but always remains lower than the $DOP$ at the disk-centre-side. Comparing the four disk locations, the $DOP$
at all points along the radial direction of a single source decreases as the source location nears the limb.
In the radial direction, spatially resolved $DOP$ can achieve values as high as $\sim30\%$ at disk centre locations.
The albedo component produces the distinctive $\Psi$ variation shown in Figure \ref{fig:maps_dnu4}.
Along the radial direction, $\Psi=90^{\circ}$ at the disk-centre-side of the source, falls to zero within the FWHM extent and then rises again at the limb-side of the source.
Along the perpendicular to radial direction of a single source (Figure \ref{fig:slicesY_dnu4}), $DOP$ (first row) and the magnitude of $\Psi$ (second row) are symmetrical due to no projection effects and the centres of the primary and albedo components always coinciding. As with the radial direction, in the perpendicular to radial direction, spatially resolved $DOP$ can achieve values as high as $\sim30\%$.
While the magnitude of $\Psi$ across the
source is symmetrical at each disk location in the perpendicular to radial direction,
$\Psi$ itself behaves as an odd function along Y, with a $180^{\circ}$ rotational symmetry about the source centre, increasing from the radial at the upper source edge to $\left|{\Psi>0^{\circ}}\right|$ and then back to the radial direction at the lower source edge.

\subsubsection{Beamed electron distributions $\Delta\nu=0.5$ and $\Delta\nu=0.1$}

\begin{figure}[h]
\vspace{-30pt}
\includegraphics[width=17cm]{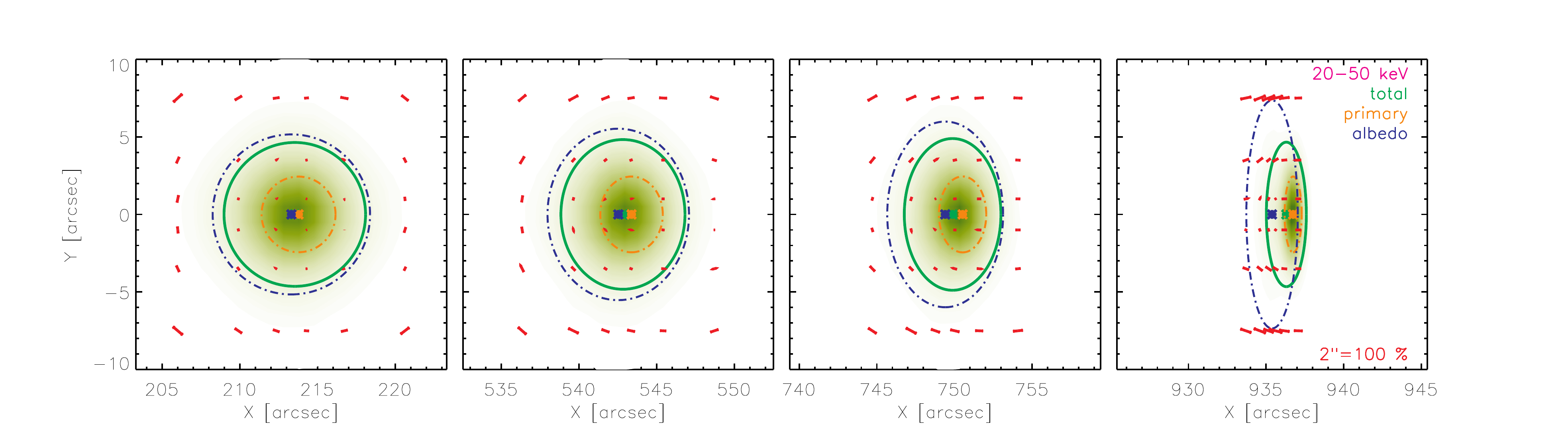}
\caption[Total X-ray brightness and polarization maps for the photon distribution created by the $\Delta\nu=0.5$ electron distribution.]{Total X-ray brightness and polarization maps for the photon distribution created by the $\Delta\nu=0.5$ electron distribution for a $5''$ primary source. Colours and symbols as in Figure~\ref{fig:maps_dnu4}.}
%%Total X-ray brightness and polarization maps for the photon distribution created by the $\Delta\nu=0.5$
%electron distribution for a $5''$ primary
%source. The resulting total sources sit at four disk locations of $X=213.1'', 542.6'', 750.0'', 936.5''$ at $Y=0''$ (corresponding to $\mu\sim0.97,0.82,0.62,0.22$). The green, blue and orange ellipses give the FWHM of the
%total, albedo and primary sources respectively.  Green, blue and orange asterisks give the centroid
%position of the total, albedo and primary sources. For disk locations closer to the solar centre,
%the albedo component is the dominant component over the primary component.}
\label{fig:maps_dnu05}
%\end{figure*}

%\begin{figure*}
\includegraphics[width=17cm]{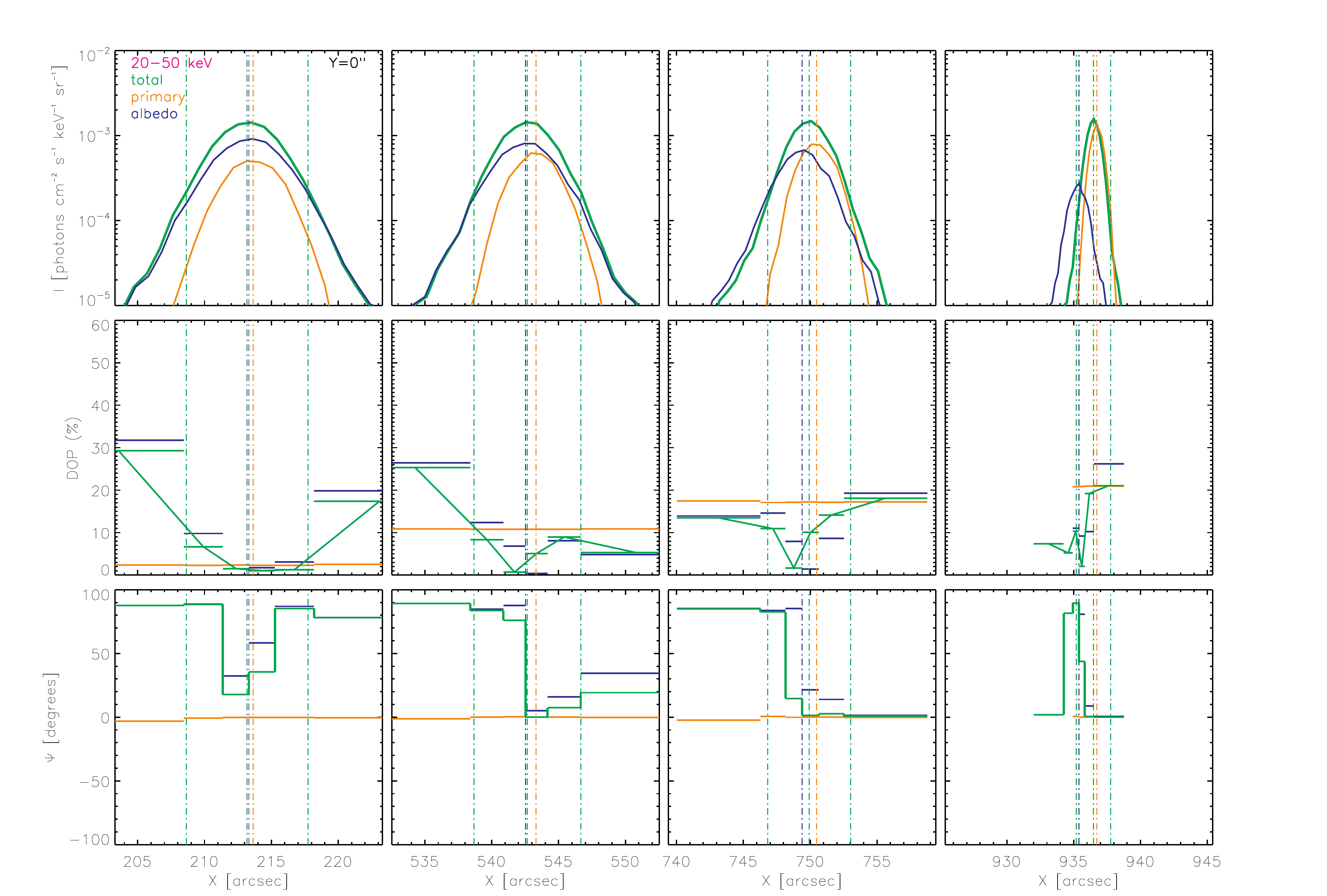}
\caption[Radial slices (along $X$) through $Y=0''$ for the intensity, $I$, the $DOP$ and $\Psi$
for each of the sources in Figure \ref{fig:maps_dnu05} for the $\Delta\nu=0.5$ electron distribution.]{Radial slices (along $X$) through $Y=0''$ for the intensity, $I$, the $DOP$ and $\Psi$
for each of the sources in Figure \ref{fig:maps_dnu05} for the $\Delta\nu=0.5$ distribution. Again, the source locations, colours and symbols are the same as in Figures \ref{fig:maps_dnu4} and \ref{fig:slicesX_dnu4} for $\Delta\nu=4.0$.}%orange=primary, blue=albedo and green=total. The green dash-dot
%lines denote the centroid position and the FWHM of the total observed source while the orange and
%blue lines denote the centroid positions of the primary and albedo sources respectively.}
\label{fig:slicesX_dnu05}
\end{figure}

\begin{figure*}
\includegraphics[width=17cm]{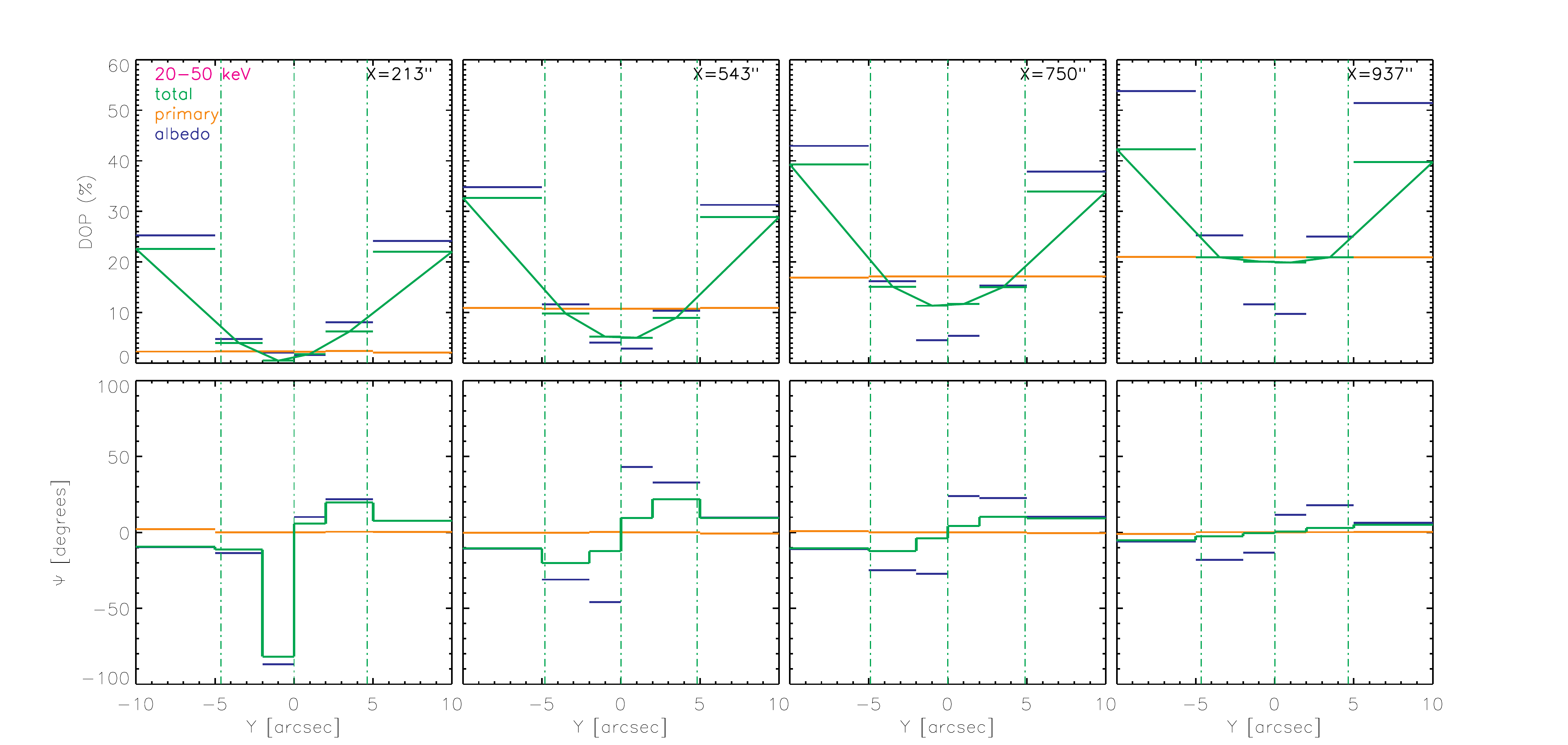}
\caption[Perpendicular to radial slices through each of the sources shown in Figure
\ref{fig:maps_dnu05} for the $\Delta\nu=0.5$ electron distribution.]{Perpendicular to radial slices through each of the sources shown in Figure
\ref{fig:maps_dnu05} for the $\Delta\nu=0.5$ distribution.
Each of the perpendicular to radial slices are taken along $Y$  at $X=213.1'', 542.6'', 750.0'', 936.5''$ for the $DOP$ and polarization angle $\Psi$. The lines and
colours are as in Figures \ref{fig:maps_dnu05}, \ref{fig:slicesX_dnu05}.}
\label{fig:slicesY_dnu05}
\end{figure*}

\begin{figure}[h]
\vspace{-30pt}
\includegraphics[width=17cm]{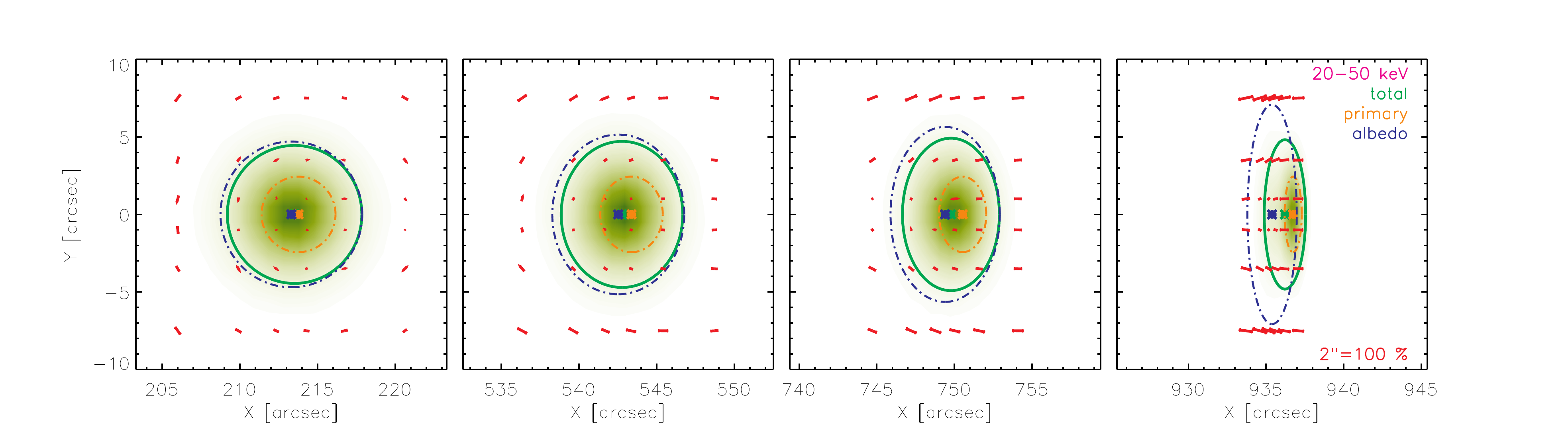}
\caption[Total X-ray brightness and polarization maps for the photon distribution created by the $\Delta\nu=0.1$ electron distribution.]{Total X-ray brightness and polarization maps for the photon distribution created by the $\Delta\nu=0.1$ electron distribution for a $5''$ primary source. The source locations, colours and symbols as in Figure~\ref{fig:maps_dnu4}.}
%The resulting total sources sit at four disk locations of $X=213.0'', 542.5'', 749.9'', 936.4''$ at $Y=0''$ (corresponding to $\mu\sim0.97,0.82,0.62,0.22$). The green, blue and orange ellipses give the FWHM of the
%total, albedo and primary sources respectively.  Green, blue and orange asterisks give the centroid
%position of the total, albedo and primary sources. At all disk locations the albedo
%component is the dominant component over the primary component.}
\label{fig:maps_dnu01}
\includegraphics[width=17cm]{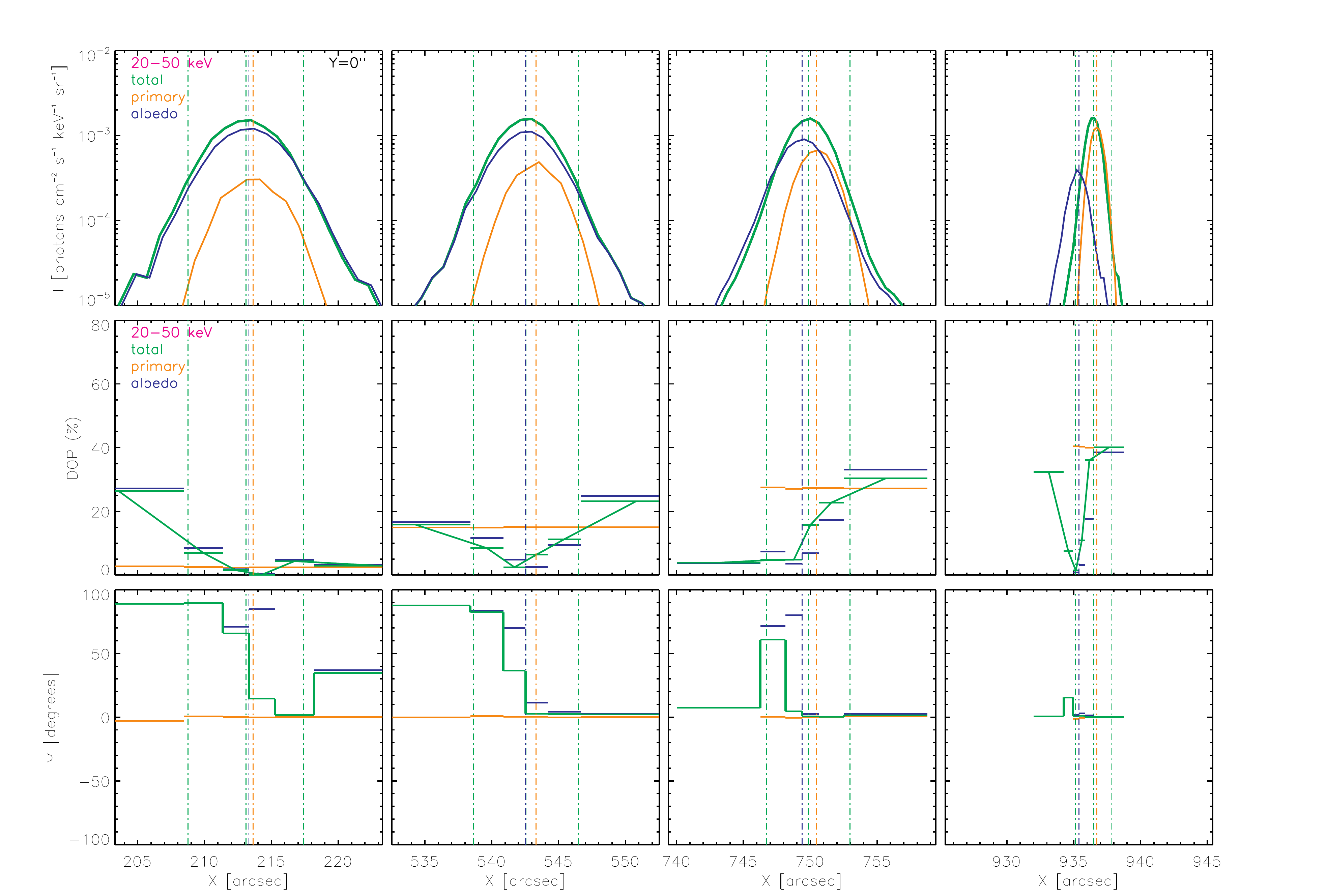}
\caption[Radial slices (along $X$) through $Y=0''$ for the intensity, $I$, the $DOP$ and $\Psi$
for each of the sources in Figure \ref{fig:maps_dnu01} for the $\Delta\nu=0.1$ electron distribution.]{Radial slices (along $X$) through $Y=0''$ for the intensity, $I$, the $DOP$ and $\Psi$
for each of the sources in Figure \ref{fig:maps_dnu01} for the $\Delta\nu=0.1$ distribution. Again, the locations, colours and symbols are the same as in Figures \ref{fig:maps_dnu4} and \ref{fig:slicesX_dnu4} for $\Delta\nu=4.0$.}
%orange=primary, blue=albedo and green=total. The green dash-dot
%lines denote the centroid position and the FWHM of the total observed source while the orange and
%blue lines denote the centroid positions of the primary and albedo sources respectively.}
\label{fig:slicesX_dnu01}
\end{figure}

\begin{figure*}
\includegraphics[width=17cm]{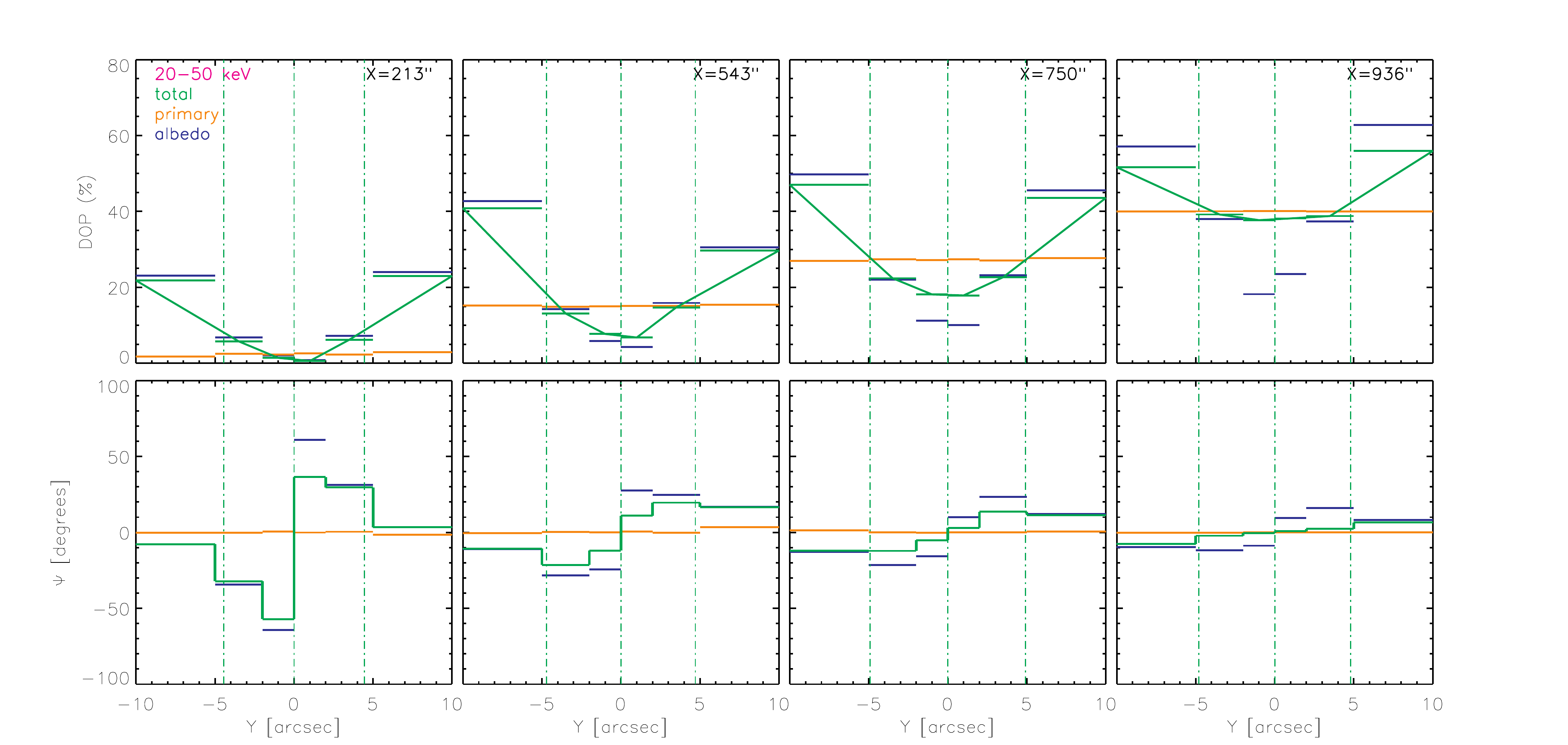}
\caption[Perpendicular to radial slices through each of the sources shown in Figure
\ref{fig:maps_dnu01} for the $\Delta\nu=0.1$ electron distribution.]{Perpendicular to radial slices through each of the sources shown in Figure
\ref{fig:maps_dnu01} for the $\Delta\nu=0.1$ distribution.
Each of the perpendicular to radial slices are taken along $Y$  at $X=213.0'', 542.5'', 749.9'', 936.4''$ for the $DOP$ and polarization angle $\Psi$. The lines and
colours are as in Figures \ref{fig:maps_dnu01}, \ref{fig:slicesX_dnu01}.}
\label{fig:slicesY_dnu01}
\end{figure*}

All plots for the photon distribution created by the $\Delta\nu=0.5$ electron distribution are shown in Figures (\ref{fig:maps_dnu05}-\ref{fig:slicesY_dnu05}), while Figures (\ref{fig:maps_dnu01}-\ref{fig:slicesY_dnu01}) show all plots for the photon distribution created by the $\Delta\nu=0.1$ electron distribution. Comparison of Figures \ref{fig:maps_dnu4}, \ref{fig:maps_dnu05} and \ref{fig:maps_dnu01} demonstrate that increased beaming towards the photosphere produces smaller, more concentrated and intense albedo patches.%This is because a larger fraction of the photon distribution is directed at smaller angles towards the photosphere.

For the $\Delta\nu=0.5$ distribution, Figure \ref{fig:slicesX_dnu05} plots the radial intensity, $I$ (top) (along $X$ at $Y=0''$). The first two disk locations are albedo dominated, the third disk location has approximately equal contributions from the primary and albedo components and only the disk location closest to the limb is primary dominated. The primary $DOP$ can rise as high as $\sim20\%$ at the limb and the primary $\Psi$ is radial at all locations.
Figure \ref{fig:slicesX_dnu01} (top) plots radial intensity slices (along $X$ at $Y=0''$) for the $\Delta\nu=0.1$ distribution. As expected, the first three disk locations are albedo dominated and the primary $DOP$ can reach $\sim40\%$ at the limb. Again, the primary $\Psi$ is radial at all disk locations.

As with the quasi-isotropic $\Delta\nu=4.0$ distribution, common trends can be observed across individual sources at particular disk locations and between disk locations for both the $\Delta\nu=0.5$ and $\Delta\nu=0.1$ distributions. More importantly for observations and anisotropy deduction purposes, trends between each of the three simulated distributions ($\Delta\nu=4.0$, $\Delta\nu=0.5$ and $\Delta\nu=0.1$) can be observed, along the radial and perpendicular to radial directions, at any chosen disk location. The most notable trends are observed in the radial (X) direction, and it is these trends that may help to deduce the anisotropy of the photon distribution for a HXR source sitting at a given disk location. Trends can be observed at all disk locations, but in this example the patterns are most noticeable in the second and third disk locations plotted. In both of these locations, the disk-centre-side $DOP$ falls with increased beaming while the limb-side $DOP$ rises with increased beaming.

Comparing the third disk location (for example) in Figures \ref{fig:slicesX_dnu4}, \ref{fig:slicesX_dnu05} and \ref{fig:slicesX_dnu01} shows how the radial $DOP$ at the limb-side of the source rises with increased beaming from $\sim2\%$ for the $\Delta\nu=4.0$ distribution to $\sim18\%$ for the $\Delta\nu=0.5$ distribution to $\sim30\%$ for the $\Delta\nu=0.1$ distribution. The radial $DOP$ at the disk-centre-side of the source falls with increased beaming from $\sim18\%$ ($\Delta\nu=4.0$) to $\sim14\%$ ($\Delta\nu=0.5$)  to $\sim4\%$ ($\Delta\nu=0.1$). The polarization angle $\Psi$ also produces similar patterns with changing anisotropy. A clear example of this can be observed by comparing the second disk location plotted in Figures \ref{fig:slicesX_dnu4}, \ref{fig:slicesX_dnu05} and \ref{fig:slicesX_dnu01}. Along the radial direction, disk-centre-side $\Psi$ generally stays at $\Psi=90^{\circ}$ for all photon anisotropies, while the outer limb-side $\Psi$ falls significantly with increased beaming, from $\Psi=60^{\circ}$ ($\Delta\nu=4.0$) to $\Psi=20^{\circ}$ ($\Delta\nu=0.5$) to $\Psi=0^{\circ}$ ($\Delta\nu=0.1$).

Therefore, the $DOP$ and $\Psi$ patterns prescribe how spatially resolved polarization measurements could be used to determine the beaming of the photon distribution. It should be noted that a (near) disk-centre source produces a slightly different trend in radial DOP with increasing photon anisotropy. The $DOP$ at the disk-centre-side remains approximately the same for all photon anisotropies while the $DOP$ at the limb-side falls with increased beaming (this is the opposite trend to other disk locations).

Comparing each disk location along the perpendicular to radial direction (Y) in Figures \ref{fig:slicesY_dnu4}, \ref{fig:slicesY_dnu05} and \ref{fig:slicesY_dnu01} shows that greater beaming increases the $DOP$ over the whole extent of the source at any given location (except at the disk centre where the spatially resolved polarization along Y is approximately the same for all three distributions). This spatial increase is most noticeable at limb locations where from the source
centre to the source edge, $DOP$ increases from $\sim0\%$ to $\sim20\%$ ($\Delta\nu=4.0$), from
$\sim20\%$ to $\sim40\%$ ($\Delta\nu=0.5$) and from $\sim40\%$ to $\sim55\%$ ($\Delta\nu=0.1$).

\section{Discussion and conclusions}
The simulation results show that Compton backscattering produces a clear and distinct albedo polarization pattern across a HXR source at the peak albedo energies of 20-50 keV. Trends can be observed for both of the measured polarization parameters, $DOP$ and $\Psi$, and are clear indications of the existence of an albedo component in comparison with the constant, radial polarization of the primary emission. This means that spatially resolved polarization can be used to probe structure within HXR footpoint sources, helping to distinguish between the bremsstrahlung source and the albedo source. More importantly, at a single disk location, spatially resolved $DOP$ and $\Psi$ exhibit clear variations with changing photon directivity, along both the radial and perpendicular to radial directions and can be used to determine the anisotropy of the electron distribution.
Therefore, to take advantage of these properties requires reliable polarization measurements with an angular resolution of $\sim5''-10''$ over the peak albedo energies of $\sim20-50$ keV.

The simulations also suggest that for approximately isotropic HXR sources, spatially integrated polarization angle measurements, $\Psi$, from low to high energies,
with consideration of the changes due to albedo, could help indirectly infer the highest energy in
the electron distribution. For near isotropic sources implied by {\it RHESSI} X-ray observations
\citep{Kasparovaetal2007,KontarBrown2006}, changes in $\Psi$ from the radial ($\Psi=0^{\circ}$) to the perpendicular to radial direction ($\Psi=90^{\circ}$) may help indicate the presence of high energy electrons ($\geq1$ MeV) present in the electron distribution. Changes in spatially integrated $DOP$ measurements, from low to high energies, will also help determine the anisotropy of the photon distribution.

Currently, when observing solar flares, the instrumentation required to produce spatially resolved polarization measurements is not available. It is doubtful whether proposed {\it near} future missions such as the Gamma-Ray Imager/Polarimeter for Solar Flares ({\it GRIPS}) \citep{Shihetal2009} may have the capability to measure polarization over albedo energies, even though it should be able to measure polarization across $12\arcsec.5$. The best polarization measurements are likely to be over the range of 150-650 keV with a $\sim$4$\%$ minimum detectable polarization \citep{Shihetal2009}. At energies greater than $\sim$100 keV, the albedo flux drops off steeply, thus it is unlikely that any albedo component could be detected at these energies and the polarization across the observed HXR source would only be from the bremsstrahlung emission. Therefore, good polarization measurements at 150-650 keV from flares with high fluxes will give a direct measurement of the primary component and $DOP/\Psi$ measurements at these energies may be used to infer the high energy cutoff in the electron distribution, for relatively isotropic distributions. However, it should be noted that most flares have a relatively low flux after $\sim100$ keV since the X-ray flux generally decreases as a power law. Therefore, only the highest X-class flares will probably give enough counts for a reliable polarization measurement. Hence, the importance of understanding the albedo component and how it can be used as a beneficial diagnostic tool is stressed since the most reliable future polarization measurements from flares will probably be over an energy range of 10-50 keV, where the photon flux is high. 
\chapter{Conclusions and final remarks}
\label{ref:Chapter7}

The aim of this thesis was to study changes in the spatial, spectral and polarization properties of solar flare X-ray sources, in order to determine how these properties are related to the energetic and angular properties of an emitting electron distribution, the properties of a flaring coronal or chromospheric plasma, and X-ray interactions in the photosphere such as Compton scattering. The work in this thesis was performed by modelling and using observations with {\it RHESSI}.

In Chapter \ref{ref:Chapter2}, stochastic simulations were used to model electron transport in a dense corona. Different injected electron pitch angle distributions, undergoing collisional pitch angle scattering within a finite temperature plasma were simulated, and this work can also be found in \cite{2014ApJ...787...86J}. This study was partly in response to recent {\it RHESSI} observations of coronal thick target X-ray sources. The thick target model is applicable to the corona for these events because of their high number densities of the order $10^{11}$ cm$^{-3}$. The lengths of such X-ray sources increase quadratically with energy and often a simple one-dimensional cold target collisional transport model is used to estimate the number density and the length of an acceleration region, using a quadratic fit function of the form $FWHM=L_0+\alpha\;E^{2}$ where $\alpha\propto1/n$. The simulations showed how the presence of collisional pitch angle scattering alone does not dramatically change the behaviour of source length with electron energy. However, it was shown that the beaming of the injected electron pitch angle distribution did produce a significant change in the variation of X-ray source length with energy. It was found that injecting a beamed electron distribution produced a larger variation of length with energy, since electrons with velocity completely aligned to the guiding magnetic field, move the greatest distance through a plasma. The uncertainty in the initial angular distribution of the injected electrons produces the largest uncertainty in the inferred number density $n$, which can be up to a factor of $\sim6$ larger. The finite temperature of the target atmosphere leads to thermal diffusion, both in energy and space, and causes an increase of the inferred acceleration region length $L_0$. The results show that the application of a one-dimensional cold target approach to a warm target changes the inferred acceleration length $L_0$ by up to $\sim10\arcsec$ for a $30$ MK plasma, and the equation $L_0=L_0(T=0)-0.011 \, T^{2}$ was found empirically in order to estimate the true length of the acceleration region from observation, for number density values close to $n\sim1\times10^{11}$ [cm$^{-3}$]. It was also found that the FWHM versus energy curve consisted of two competing components, one due to thermal diffusion that is dominant at lower energies, and another due to collisional friction that is dominant at higher energies where $E>>k_{B}T$. The dominant component depends on the temperature of the region, on the density, and the spectral index of the electron distribution, and it was shown that the application of a one-dimensional cold target approach to a warm target produced an inferred number density different by a factor of $\sim3$ (in either direction), again, depending mainly on the initial beaming of the electron distribution.

It is often assumed that the initial accelerated electron distribution and even the target electron distribution in the corona or chromosphere is completely beamed. However, many recent {\it RHESSI observations} are consistent with a lack of anisotropy, including \cite{KontarBrown2006}, \cite{Kasparovaetal2007} and \cite{2013SoPh..284..405D}; with all studies using the presence of a photospheric albedo component to determine the electron distribution isotropy via a single spacecraft stereoscopic method or a centre-to-limb statistical method. Hence simulating possibly more realistic isotropic electron distributions, in general produces a more gradual variation of source length with energy, that is, smaller values of the quadratic fit coefficient $\alpha$. Therefore, depending on the electron distribution spectral index, observed steep behaviours (high values of $\alpha$) may be indicative of other processes at work within the coronal region. For instance, throughout the simulations it was assumed that the length of the acceleration region length $L_0$ was independent of electron energy $E$. However, depending on the acceleration process, this may not be the case. For example, if $L_0$ grows with energy, this may produce a larger value of $\alpha$ than expected and hence the analysis of this effect may indicate the properties of the acceleration mechanism itself. As was discussed in Chapter \ref{ref:Chapter2}, the stochastic simulations are not self-consistent, that is, they do not account for changes in the properties of the background plasma, specifically the temperature increase of the background plasma due to the energy loss of the injected electron distribution. Further improvements could also be made to include spatial variations in temperature and/or density along the loop, as would be expected in a real flaring region.  Also, it should be noted that a recent study by \citet{2014ApJ...780..176K} shows how the presence of {\it non-collisional} pitch angle scattering (for example, involving magnetic field inhomogeneities) results in a different non-quadratic predicted behaviour for the variation of source length with energy. Moreover, the code developed for this work could be rather straightforwardly extended to the study of magnetic diffusion of particles across the guiding field in a warm target \citep[e.g.,][]{2011A&A...535A..18B} and hence to study the variation of source length with energy in this alternative scenario.

In Chapter \ref{ref:Chapter3}, temporal changes of the spatial properties of dense coronal X-ray sources were analysed for the first time, using observations from {\it RHESSI}. The work can also be found in \cite{2013ApJ...766...75J}. Similar to the studies of both coronal and chromospheric X-ray source spatial parameters with energy, this study was made possible by the use of the forward-fitting imaging algorithm {\it Visibility Forward Fitting}. This algorithm can only be used for specific X-ray source shapes where one or two X-ray sources have a simple form that can be well-fitted by either a circular, elliptical or curved elliptical Gaussian shape. Since many X-ray sources have a more complex shape, only three dense coronal X-ray flares were analysed reliably using this method. However, the study found that all three X-ray loops exhibited similar and interesting temporal trends in both their observed spatial properties and inferred physical properties such as temperature, number density and thermal pressure, estimated from a combined spectral and spatial analysis of each X-ray source. Peaks in X-ray emission denoted periods where there were changes in the loop spatial dynamics and most interestingly of all, a study of temperature, volume, number density and thermal pressure showed how the X-ray loops went through three phases denoted by three clear peaks: the first in temperature, the second in X-ray emission and the third in thermal pressure. Before the peak in X-ray emission, the loop length, loop width and radial position decreased. After the X-ray peak the loop width and radial position rose again while the loop length seemed to remain approximately constant - at least for the two flares that only have one clear X-ray peak in their lightcurve. It was found that at the start of the observation time, the X-ray loops are relatively long and thin and then they become smaller, before growing in width at the later decay stages of X-ray emission. Hence in order to describe this changing morphology, a new parameter named {\it corpulence, $\mathcal{C}$} was defined; the ratio of loop width $W$ to loop length $L$, and it was found, in general, that corpulence increased with time. Overall, the volume of the loop decreased before the peak in X-ray emission, but the relationship between temperature and volume did not support simple compressive heating, as in a collapsing magnetic trap model. The most difficult observation to explain is that of the decreasing loop widths before the peak in X-ray emission, since electrons should be unable to move across the guiding field lines threading the corona. This leads to the suggestion that the field lines themselves are being squeezed together during this time. In the discussion of Chapter \ref{ref:Chapter3}, it was tentatively suggested that this could be due to Taylor relaxation, where the release of magnetic energy during a flare may cause the surrounding field lines to shrink or contract in some manner. However it should be stressed that the cause is unknown, and further future investigation of similar coronal X-ray loops is required to help explain this observation. Another possible cause of this contraction may be the sausage-pinch effect, whereby the emitting plasma is pushed together by a generated Lorentz force due to the current flowing in the parallel direction producing an azimuthal magnetic field that compresses the plasma. However, we would expect the parallel field along which electrons are flowing to cancel any sausage-pinch effects. However, the lack of flares with this type of X-ray source morphology, and especially those with {\it RHESSI} observations that can be analysed with algorithms such as Vis FwdFit, since the launch of {\it SDO}/AIA has proved rather frustrating. The observation of the spatial and temporal properties of separate loops within the `blob' region viewed in X-rays may help to better understand the trends in loop width and position. Further, it was suggested that thermal conduction causes chromospheric evaporation, leading to the increasing number density and thermal pressure in the loop. This could cause the loop lengths to decrease, as electrons interact at shorter and shorter distances. Eventually, the increasing thermal pressure in the region could balance the process causing the loop to shrink, and cause the increase in loop width after the peak X-ray emission.

In Chapter \ref{ref:Chapter5}, Compton scattering and photoelectric absorption of X-rays in the photosphere were studied via a Monte Carlo simulation of photon transport, in order to study how X-ray albedo photons, backscattered from the photosphere, can alter the spatial properties of a HXR source (the primary source), namely its size and position. The results of simulations can also be found in the publication \cite{KontarJeffrey2010}. These changes were inspected by simulating HXR chromospheric footpoint sources at different heliocentric angles above the solar disk, at a typical chromospheric footpoint height of $1\arcsec.4$ and using X-ray energies between 3 and 300 keV, matching recent {\it RHESSI} observations, as discussed in Chapter \ref{ref:Chapter5}. The results from all the simulations showed a general trend for both the measured source sizes and the shifts from the true primary source position. The greatest alteration in source size and position occurred between the energies of 20-50 keV, the energy range over which the albedo emission peaks. Generally the largest source increases due to albedo are observed nearer the solar centre since the greatest fraction of albedo X-rays are emitted towards an observer at this heliocentric angle, hence with the smallest source size increases occurring at solar limb. The largest shifts in source position occur at mid-heliocentric angles with the measured position nearing its true primary position both at the solar centre, where the positions of both the primary and albedo sources coincide, and near the solar limb, where the albedo flux is minimal. The resulting albedo component, and hence total observed source from primary HXR sources, with different spectral indices, source sizes and anisotropy were simulated. It was found that the lower the spectral index of the photon distribution, that is, the harder the spectrum, the greater the source size increase and position shift over all disk positions and photon energies, compared with similar HXR sources with larger spectral indices. It is observed that the smaller the true HXR source size, the greater the relative increase due to the albedo X-rays. Therefore, the results show that interestingly even a HXR point source that is isotropic, and hence has the minimum backscattered albedo flux, may be observed as a HXR source as large as $7\arcsec$ across, when located near the solar centre and viewed at a peak albedo energy range of 20-50 keV. The larger the initial HXR source, the smaller the relative increase due to the albedo X-rays, over all energies and heliocentric angles. It was found that the position shifts seem to be generally independent of primary HXR source size if all other factors such as spectral index and beaming remain the same. The contribution from an albedo component depends greatly on the anisotropy of the HXR source. If the HXR source initially throws a greater proportion of photons into the photosphere than towards e.g.,{\it RHESSI}, then the albedo component can be very large compared to the true direct component. Large downward directivities from anisotropic sources can produce shifts as large as $0\arcsec.8$.

The results of Chapter \ref{ref:Chapter5} and \cite{KontarJeffrey2010}, and as discussed in the conclusion of Chapter \ref{ref:Chapter5}, suggest inevitably that all HXR sources in the chromosphere, and {\it even} those with an isotropic pitch angle distribution, should have an albedo component capable of altering their spatial properties, and hence the information deduced from such observations. Hence, the results were also suggestive of the initial HXR emission coming from a volume in the chromosphere smaller than suggested by {\it RHESSI} observations. For example, \citet*{2011ApJ...739...96K} examined both high resolution optical and hard X-ray observations of a flare that occurred on the 6th December 2006. The G-band observations resolved the width of the flare ribbon to be somewhere between $0\arcsec.5$ and $1\arcsec.8$. However, the {\it RHESSI} hard X-ray observations were unresolved and suggested that the HXR source width was even smaller than that of the G-band observations. Even though this flare is located close to the solar limb, the results shown here suggest that its spatial properties should be tainted by an albedo component that would increase the size of the observed HXR source. The simulations shown in Chapter \ref{ref:Chapter5} and \cite{KontarJeffrey2010}, were used to estimate the size of the primary HXR source which should be smaller than that found from observations. However, the size of the primary HXR source could not be easily estimated since the HXR source was unresolved and hence the observed HXR source width of $\sim1\arcsec.1$ {\it with} the presence of an albedo component was an upper limit, and suggestive that the HXR sources were extremely point-like in this flare.

In Chapter \ref{ref:Chapter6}, a new study of solar flare X-ray polarization due to the presence of a photospheric albedo component was presented. For the first time in solar physics, this chapter simulated how an albedo component can produce spatial changes in polarization across a single HXR source. The results of Chapter \ref{ref:Chapter6} are also published in \cite{2011A&A...536A..93J}. The Monte Carlo simulations used in Chapter \ref{ref:Chapter5} were adapted to study spatially resolved polarization due to albedo. In order to achieve this, HXR sources created from three different electron directivities were input into the simulations; from the near isotropic to a highly beamed distribution. The first purpose of the study was to examine how the degree {\it and} direction of polarization change in space, across a single observed HXR footpoint, due to the inevitable presence of a Compton scattered albedo component. The polarization changes in space across an X-ray source were presented as polarization maps and cuts were taken along the radial (line connecting the Sun centre and the centre of the HXR source) and perpendicular to radial lines. Clear spatial changes in both the degree and direction of polarization were found for each case, which importantly are dependent on the beaming of the electron distribution. In general, for most HXR source locations, not located close to the disk centre, it was found for a given heliocentric angle, that the radial $DOP$ at the limb-side of the HXR source rises with increased beaming, while the $DOP$ at the disk-centre-side of the HXR source decreases with increased beaming. Similarly, for a radial measurement of $\Psi$, the angle of polarization along the source, the limb-side $\Psi$ decreases with increased beaming while the disk-side $\Psi$ generally stays at $90^{\circ}$. For example, for the HXR source simulated at a disk location of $\mu\sim0.60$, the radial $DOP$ at the limb-side of the source rises with increased beaming from $\sim2\%$ for the isotropic $\Delta\nu=4.0$ distribution to $\sim18\%$ for the (mildly anisotopic) $\Delta\nu=0.5$ distribution to $\sim30\%$ for the (very anisotropic) $\Delta\nu=0.1$ distribution. The radial $DOP$ at the disk-centre-side of the source falls with increased beaming from $\sim18\%$ ($\Delta\nu=4.0$) to $\sim14\%$ ($\Delta\nu=0.5$) to $\sim4\%$ ($\Delta\nu=0.1$). The polarization angle $\Psi$ for a HXR source at a disk location of $\mu\sim0.8$ shows a similar pattern. Along the radial direction, disk-centre-side $\Psi$ generally stays at $\Psi =90^{\circ}$ for all photon anisotropies, while the outer limb-side $\Psi$ falls significantly with increased beaming, from $\Psi=60^{\circ}$ ($\Delta\nu=4.0$) to $\Psi=20^{\circ}$ ($\Delta\nu=0.5$) to $\Psi=0^{\circ}$ ($\Delta\nu=0.1$).

The second purpose of the simulations was to assess the usefulness of measuring spatially resolvable polarization, across a single HXR source, as a possible future diagnostic tool. The simulations found that not only are these results useful, they also require no manipulation, that is, there is no need to separate the primary and albedo components as in Chapter \ref{ref:Chapter5}; these are now used together as one single measurement to determine the photon, and hence electron anisotropy in the chromosphere. As well as other methods involving the use of albedo, X-ray polarization and specifically as suggested here, {\it spatially resolvable albedo polarization} measurements could provide another method of reliably determining the anisotropy of the electron distribution from an {\it individual flare}. Spatially resolvable polarization measurements are not only useful for finding changes across a single HXR source but, more realistically near-future polarization missions might be able to determine the total source polarization of each HXR footpoint in the chromosphere and that of an X-ray coronal source individually. This could provide a method of mapping the electron pitch angle distribution during the flare from the corona to the chromosphere, with the contribution of photospheric albedo providing additional information about individual changes across a single HXR source. Even a polarization mission with a lower angular resolution than that required to see changes across individual HXR sources from $10-100$ keV as discussed in Chapter \ref{ref:Chapter6} could provide essential electron pitch angle information from the corona and chromosphere separately.

Much of the work within this thesis was based on the excellent imaging spectroscopy methods currently available with {\it RHESSI}. The work in this thesis also looks towards future X-ray missions; instruments with higher angular resolution, possibly with direct X-ray imaging capabilities that can more reliably examine the changing sub-arc second lengths, widths and positions of X-ray sources. It is hoped that the work in this thesis can also encourage the development of future X-ray polarisation instruments that can eventually measure spatially resolvable polarization across a single HXR source. 
\newpage

\appendix
\chapter{Calculating the photon stepsize}\label{ref:App2}

In Chapters \ref{ref:Chapter5} and \ref{ref:Chapter6}, a Monte Carlo simulation is used to model the movement and interaction of photons through the solar photosphere. Before an interaction, each photon must move a certain distance called the step size $ss$, which is a certain fraction of its mean free path $l$. The mean free path of a photon is given by,
\begin{equation}\label{ap2_1}
l=\frac{1}{n\sigma_{T}}=\frac{1}{n_{H}\left(\sigma_{a}+\sigma_{c}\right)}
\end{equation}
where $n$ is the number density of the photosphere and $\sigma_{T}$ is the total cross section, that is, the sum of all attenuation processes, which for this case is the sum of the Compton scattering $\sigma_{c}$ and absorption $\sigma_{a}$ cross sections, as discussed in Chapters \ref{ref:Chapter5} and \ref{ref:Chapter6}.

To calculate the photon step size through the solar photosphere, accounting for the processes of absorption and scattering, the {\it Beer-Lambert law} \citep[cf.][]{2002phat.book.....H} is used. This is given by
\begin{equation}\label{ap2_2}
I(z)=I_{0}(z)\exp(-z/l)=I_{0}(z)\exp(-n_{H}(\sigma_{c}+\sigma{a})z)
\end{equation}
where $I$ is the intensity of light after travelling a distance $z$ through a material (in this case) that absorbs and scatters a proportion of the light, and $I_{0}$ is the starting intensity.

The probability density function (PDF) used to describe this can then be given by,
\begin{equation}\label{ap2_3}
p(ss)=\frac{1}{l}\exp(-ss/l)
\end{equation}
where $z$ has been replaced by the step size $ss$. The {\it Inverse transform Method} is then employed; this maps each value of the PDF to a randomly generated number $\zeta_{step}$ between $0$ and $1$ using the cumulative distribution of the PDF \citep[e.g.,][]{Salvatetal2008}, given by,
\begin{equation}\label{ap2_4}
\zeta_{step}=\int_{0}^{ss}p(ss^{'})dss^{'}
=\int_{0}^{ss}\frac{1}{l}\exp(-ss^{'}/l)dss^{'}=1-\exp(-ss/l)
\end{equation}
Re-arranging Equation (\ref{ap2_4}) then gives,
\begin{equation}\label{ap2_5}
ss=-l\ln(1-\zeta_{step})\equiv-l\ln(\zeta_{step}),
\end{equation}
which is the step size $ss$ used in the Monte Carlo simulation in Chapters \ref{ref:Chapter5} and \ref{ref:Chapter6}.

\bibliographystyle{aa1}
\addcontentsline{toc}{chapter}{Bibliography}
\bibliography{main_4arxiv}

\begin{thebibliography}{195}
\expandafter\ifx\csname natexlab\endcsname\relax\def\natexlab#1{#1}\fi

\bibitem[{{Antiochos} \& {Sturrock}(1978)}]{1978ApJ...220.1137A}
{Antiochos}, S.~K. \& {Sturrock}, P.~A. 1978, \apj, 220, 1137

\bibitem[{{Antonucci} \& {Dennis}(1983)}]{1983SoPh...86...67A}
{Antonucci}, E. \& {Dennis}, B.~R. 1983, \solphys, 86, 67

\bibitem[{{Antonucci} {et~al.}(1982){Antonucci}, {Gabriel}, {Acton},
  {Leibacher}, {Culhane}, {Rapley}, {Doyle}, {Machado}, \&
  {Orwig}}]{1982SoPh...78..107A}
{Antonucci}, E., {Gabriel}, A.~H., {Acton}, L.~W., {et~al.} 1982, \solphys, 78,
  107

\bibitem[{{Aschwanden}(2004)}]{2004psci.book.....A}
{Aschwanden}, M.~J. 2004, {Physics of the Solar Corona. An Introduction}
  ({Praxis Publishing Ltd}), p. 24

\bibitem[{{Aschwanden} {et~al.}(2002){Aschwanden}, {Brown}, \&
  {Kontar}}]{Aschwandenetal2002}
{Aschwanden}, M.~J., {Brown}, J.~C., \& {Kontar}, E.~P. 2002, \solphys, 210,
  383

\bibitem[{{Asplund} {et~al.}(2009){Asplund}, {Grevesse}, {Sauval}, \&
  {Scott}}]{Asplundetal2009}
{Asplund}, M., {Grevesse}, N., {Sauval}, A.~J., \& {Scott}, P. 2009, \araa, 47,
  481

\bibitem[{{Avrett} \& {Loeser}(2008)}]{2008ApJS..175..229A}
{Avrett}, E.~H. \& {Loeser}, R. 2008, \apjs, 175, 229

\bibitem[{{Bai} \& {Ramaty}(1978)}]{BaiRamaty1978}
{Bai}, T. \& {Ramaty}, R. 1978, \apj, 219, 705

\bibitem[{{Balucinska-Church} \&
  {McCammon}(1992)}]{Balucinska-ChurchMcCammon1992}
{Balucinska-Church}, M. \& {McCammon}, D. 1992, \apj, 400, 699

\bibitem[{{Bastian} {et~al.}(2007){Bastian}, {Fleishman}, \&
  {Gary}}]{2007ApJ...666.1256B}
{Bastian}, T.~S., {Fleishman}, G.~D., \& {Gary}, D.~E. 2007, \apj, 666, 1256

\bibitem[{{Battaglia} \& {Kontar}(2011{\natexlab{a}})}]{BattagliaKontar2011}
{Battaglia}, M. \& {Kontar}, E.~P. 2011{\natexlab{a}}, \apj, 735, 42

\bibitem[{{Battaglia} \& {Kontar}(2011{\natexlab{b}})}]{2011A&A...533L...2B}
{Battaglia}, M. \& {Kontar}, E.~P. 2011{\natexlab{b}}, \aap, 533, L2

\bibitem[{{Bian} {et~al.}(2012){Bian}, {Emslie}, \&
  {Kontar}}]{2012ApJ...754..103B}
{Bian}, N., {Emslie}, A.~G., \& {Kontar}, E.~P. 2012, \apj, 754, 103

\bibitem[{{Bian} {et~al.}(2011){Bian}, {Kontar}, \&
  {MacKinnon}}]{2011A&A...535A..18B}
{Bian}, N.~H., {Kontar}, E.~P., \& {MacKinnon}, A.~L. 2011, \aap, 535, A18

\bibitem[{{Boggs} {et~al.}(2006){Boggs}, {Coburn}, \&
  {Kalemci}}]{Boggsetal2006}
{Boggs}, S.~E., {Coburn}, W., \& {Kalemci}, E. 2006, \apj, 638, 1129

\bibitem[{{Brown}(1971)}]{1971SoPh...18..489B}
{Brown}, J.~C. 1971, \solphys, 18, 489

\bibitem[{{Brown}(1972)}]{Brown1972}
{Brown}, J.~C. 1972, \solphys, 26, 441

\bibitem[{{Brown} {et~al.}(2002){Brown}, {Aschwanden}, \&
  {Kontar}}]{2002SoPh..210..373B}
{Brown}, J.~C., {Aschwanden}, M.~J., \& {Kontar}, E.~P. 2002, \solphys, 210,
  373

\bibitem[{{Brown} {et~al.}(2003){Brown}, {Emslie}, \&
  {Kontar}}]{2003ApJ...595L.115B}
{Brown}, J.~C., {Emslie}, A.~G., \& {Kontar}, E.~P. 2003, \apjl, 595, L115

\bibitem[{{Brown} {et~al.}(2010){Brown}, {Mallik}, \&
  {Badnell}}]{2010A&A...515C...1B}
{Brown}, J.~C., {Mallik}, P.~C.~V., \& {Badnell}, N.~R. 2010, \aap, 515, C1+

\bibitem[{{Caspi} \& {Lin}(2010)}]{2010ApJ...725L.161C}
{Caspi}, A. \& {Lin}, R.~P. 2010, \apjl, 725, L161

\bibitem[{{Chandrasekhar}(1960)}]{Chandrasekhar1960}
{Chandrasekhar}, S. 1960, {Radiative transfer} (Dover, New York)

\bibitem[{{Chen} \& {Petrosian}(2013)}]{2013ApJ...777...33C}
{Chen}, Q. \& {Petrosian}, V. 2013, \apj, 777, 33

\bibitem[{{Cheng} {et~al.}(1981){Cheng}, {Feldman}, \&
  {Doschek}}]{1981A&A....97..210C}
{Cheng}, C.-C., {Feldman}, U., \& {Doschek}, G.~A. 1981, \aap, 97, 210

\bibitem[{{Christe} {et~al.}(2013){Christe}, {Shih}, {Rodriguez}, {Cramer},
  {Gregory}, {Gaskin}, {Chavis}, {Smith}, \& {HOPE/HEROES
  Team}}]{2013SPD....44...76C}
{Christe}, S., {Shih}, A.~Y., {Rodriguez}, M., {et~al.} 2013, in AAS/Solar
  Physics Division Meeting, Vol.~44, AAS/Solar Physics Division Meeting, 76

\bibitem[{{Chubb} {et~al.}(1966){Chubb}, {Kreplin}, \&
  {Friedman}}]{1966JGR....71.3611C}
{Chubb}, T.~A., {Kreplin}, R.~W., \& {Friedman}, H. 1966, \jgr, 71, 3611

\bibitem[{{Chupp} \& {Ryan}(2009)}]{2009RAA.....9...11C}
{Chupp}, E.~L. \& {Ryan}, J.~M. 2009, Research in Astronomy and Astrophysics,
  9, 11

\bibitem[{{Cline} {et~al.}(1968){Cline}, {Holt}, \&
  {Hones}}]{1968JGR....73..434C}
{Cline}, T.~L., {Holt}, S.~S., \& {Hones}, Jr., E.~W. 1968, \jgr, 73, 434

\bibitem[{{Cohen} {et~al.}(2010){Cohen}, {Dimits}, {Friedman}, \&
  {Caflisch}}]{2010ITPS...38.2394C}
{Cohen}, B.~I., {Dimits}, A.~M., {Friedman}, A., \& {Caflisch}, R.~E. 2010,
  IEEE Transactions on Plasma Science, 38, 2394

\bibitem[{{Compton}(1923)}]{1923PhRv...21..483C}
{Compton}, A.~H. 1923, Physical Review, 21, 483

\bibitem[{{Culhane} \& {Acton}(1970)}]{1970MNRAS.151..141C}
{Culhane}, J.~L. \& {Acton}, L.~W. 1970, \mnras, 151, 141

\bibitem[{{Dennis}(1985)}]{1985SoPh..100..465D}
{Dennis}, B.~R. 1985, \solphys, 100, 465

\bibitem[{{Dennis} \& {Pernak}(2009)}]{DennisPernak2009}
{Dennis}, B.~R. \& {Pernak}, R.~L. 2009, \apj, 698, 2131

\bibitem[{{Dickson} \& {Kontar}(2013)}]{2013SoPh..284..405D}
{Dickson}, E.~C.~M. \& {Kontar}, E.~P. 2013, \solphys, 284, 405

\bibitem[{{Doschek} {et~al.}(1980){Doschek}, {Feldman}, {Kreplin}, \&
  {Cohen}}]{1980ApJ...239..725D}
{Doschek}, G.~A., {Feldman}, U., {Kreplin}, R.~W., \& {Cohen}, L. 1980, \apj,
  239, 725

\bibitem[{{Duijveman} {et~al.}(1982){Duijveman}, {Hoyng}, \&
  {Machado}}]{1982SoPh...81..137D}
{Duijveman}, A., {Hoyng}, P., \& {Machado}, M.~E. 1982, \solphys, 81, 137

\bibitem[{{Elwert} \& {Haug}(1970)}]{ElwertHaug1970}
{Elwert}, G. \& {Haug}, E. 1970, \solphys, 15, 234

\bibitem[{{Emslie}(1978)}]{1978ApJ...224..241E}
{Emslie}, A.~G. 1978, \apj, 224, 241

\bibitem[{{Emslie}(1981)}]{1981ApJ...244..653E}
{Emslie}, A.~G. 1981, \apj, 244, 653

\bibitem[{{Emslie}(2003)}]{2003ApJ...595L.119E}
{Emslie}, A.~G. 2003, \apjl, 595, L119

\bibitem[{{Emslie} {et~al.}(2008{\natexlab{a}}){Emslie}, {Bradsher}, \&
  {McConnell}}]{Emslieetal2008}
{Emslie}, A.~G., {Bradsher}, H.~L., \& {McConnell}, M.~L. 2008{\natexlab{a}},
  \apj, 674, 570

\bibitem[{{Emslie} \& {Brown}(1980)}]{1980ApJ...237.1015E}
{Emslie}, A.~G. \& {Brown}, J.~C. 1980, \apj, 237, 1015

\bibitem[{{Emslie} {et~al.}(2008{\natexlab{b}}){Emslie}, {Hurford}, {Kontar},
  {Massone}, {Piana}, {Prato}, \& {Xu}}]{2008AIPC.1039....3E}
{Emslie}, A.~G., {Hurford}, G.~J., {Kontar}, E.~P., {et~al.}
  2008{\natexlab{b}}, in American Institute of Physics Conference Series, Vol.
  1039, American Institute of Physics Conference Series, ed. G.~{Li}, Q.~{Hu},
  O.~{Verkhoglyadova}, G.~P. {Zank}, R.~P. {Lin}, \& J.~{Luhmann}, 3--10

\bibitem[{{Emslie} {et~al.}(2003){Emslie}, {Kontar}, {Krucker}, \&
  {Lin}}]{2003ApJ...595L.107E}
{Emslie}, A.~G., {Kontar}, E.~P., {Krucker}, S., \& {Lin}, R.~P. 2003, \apjl,
  595, L107

\bibitem[{{Falewicz} {et~al.}(2011){Falewicz}, {Siarkowski}, \&
  {Rudawy}}]{2011ApJ...733...37F}
{Falewicz}, R., {Siarkowski}, M., \& {Rudawy}, P. 2011, \apj, 733, 37

\bibitem[{{Feldman} {et~al.}(1994){Feldman}, {Seely}, {Doschek}, {Strong},
  {Acton}, {Uchida}, \& {Tsuneta}}]{1994ApJ...424..444F}
{Feldman}, U., {Seely}, J.~F., {Doschek}, G.~A., {et~al.} 1994, \apj, 424, 444

\bibitem[{{Fleishman} {et~al.}(2011){Fleishman}, {Kontar}, {Nita}, \&
  {Gary}}]{2011ApJ...731L..19F}
{Fleishman}, G.~D., {Kontar}, E.~P., {Nita}, G.~M., \& {Gary}, D.~E. 2011,
  \apjl, 731, L19

\bibitem[{{Fletcher} {et~al.}(2011){Fletcher}, {Dennis}, {Hudson}, {Krucker},
  {Phillips}, {Veronig}, {Battaglia}, {Bone}, {Caspi}, {Chen}, {Gallagher},
  {Grigis}, {Ji}, {Liu}, {Milligan}, \& {Temmer}}]{2011SSRv..159...19F}
{Fletcher}, L., {Dennis}, B.~R., {Hudson}, H.~S., {et~al.} 2011, \ssr, 159, 19

\bibitem[{{Forbes} \& {Acton}(1996)}]{1996ApJ...459..330F}
{Forbes}, T.~G. \& {Acton}, L.~W. 1996, \apj, 459, 330

\bibitem[{{Galloway} {et~al.}(2005){Galloway}, {MacKinnon}, {Kontar}, \&
  {Helander}}]{2005A&A...438.1107G}
{Galloway}, R.~K., {MacKinnon}, A.~L., {Kontar}, E.~P., \& {Helander}, P. 2005,
  \aap, 438, 1107

\bibitem[{{Gardiner}(1994)}]{1994hsmp.book.....G}
{Gardiner}, C.~W. 1994, {Handbook of stochastic methods for physics, chemistry
  and the natural sciences} (Springer Series in Synergetics, Berlin: Springer,
  |c1994, 2nd ed.~1985.~Corr.~3rd printing 1994)

\bibitem[{{Gary}(2001)}]{2001SoPh..203...71G}
{Gary}, G.~A. 2001, \solphys, 203, 71

\bibitem[{{Gluckstern} \& {Hull}(1953)}]{GlucksternHull1953}
{Gluckstern}, R.~L. \& {Hull}, M.~H. 1953, Physical Review, 90, 1030

\bibitem[{{Gordovskyy} \& {Browning}(2011)}]{2011ApJ...729..101G}
{Gordovskyy}, M. \& {Browning}, P.~K. 2011, \apj, 729, 101

\bibitem[{{Gordovskyy} \& {Browning}(2012)}]{2012SoPh..277..299G}
{Gordovskyy}, M. \& {Browning}, P.~K. 2012, \solphys, 277, 299

\bibitem[{{Gordovskyy} {et~al.}(2013){Gordovskyy}, {Browning}, {Kontar}, \&
  {Bian}}]{2013SoPh..284..489G}
{Gordovskyy}, M., {Browning}, P.~K., {Kontar}, E.~P., \& {Bian}, N.~H. 2013,
  \solphys, 284, 489

\bibitem[{{Gosain}(2012)}]{2012ApJ...749...85G}
{Gosain}, S. 2012, \apj, 749, 85

\bibitem[{{Grady} {et~al.}(2012){Grady}, {Neukirch}, \&
  {Giuliani}}]{2012A&A...546A..85G}
{Grady}, K.~J., {Neukirch}, T., \& {Giuliani}, P. 2012, \aap, 546, A85

\bibitem[{{Gunkler} {et~al.}(1984){Gunkler}, {Canfield}, {Acton}, \&
  {Kiplinger}}]{1984ApJ...285..835G}
{Gunkler}, T.~A., {Canfield}, R.~C., {Acton}, L.~W., \& {Kiplinger}, A.~L.
  1984, \apj, 285, 835

\bibitem[{{Guo} {et~al.}(2012{\natexlab{a}}){Guo}, {Emslie}, {Kontar},
  {Benvenuto}, {Massone}, \& {Piana}}]{2012A&A...543A..53G}
{Guo}, J., {Emslie}, A.~G., {Kontar}, E.~P., {et~al.} 2012{\natexlab{a}}, \aap,
  543, A53

\bibitem[{{Guo} {et~al.}(2012{\natexlab{b}}){Guo}, {Emslie}, {Massone}, \&
  {Piana}}]{2012ApJ...755...32G}
{Guo}, J., {Emslie}, A.~G., {Massone}, A.~M., \& {Piana}, M.
  2012{\natexlab{b}}, \apj, 755, 32

\bibitem[{{Guo} {et~al.}(2013){Guo}, {Emslie}, \&
  {Piana}}]{2013ApJ...766...28G}
{Guo}, J., {Emslie}, A.~G., \& {Piana}, M. 2013, \apj, 766, 28

\bibitem[{{Harrison} {et~al.}(2013){Harrison}, {Craig}, {Christensen},
  {Hailey}, {Zhang}, {Boggs}, {Stern}, {Cook}, {Forster}, {Giommi},
  {Grefenstette}, {Kim}, {Kitaguchi}, {Koglin}, {Madsen}, {Mao}, {Miyasaka},
  {Mori}, {Perri}, {Pivovaroff}, {Puccetti}, {Rana}, {Westergaard}, {Willis},
  {Zoglauer}, {An}, {Bachetti}, {Barri{\`e}re}, {Bellm}, {Bhalerao},
  {Brejnholt}, {Fuerst}, {Liebe}, {Markwardt}, {Nynka}, {Vogel}, {Walton},
  {Wik}, {Alexander}, {Cominsky}, {Hornschemeier}, {Hornstrup}, {Kaspi},
  {Madejski}, {Matt}, {Molendi}, {Smith}, {Tomsick}, {Ajello}, {Ballantyne},
  {Balokovi{\'c}}, {Barret}, {Bauer}, {Blandford}, {Brandt}, {Brenneman},
  {Chiang}, {Chakrabarty}, {Chenevez}, {Comastri}, {Dufour}, {Elvis}, {Fabian},
  {Farrah}, {Fryer}, {Gotthelf}, {Grindlay}, {Helfand}, {Krivonos}, {Meier},
  {Miller}, {Natalucci}, {Ogle}, {Ofek}, {Ptak}, {Reynolds}, {Rigby},
  {Tagliaferri}, {Thorsett}, {Treister}, \& {Urry}}]{2013ApJ...770..103H}
{Harrison}, F.~A., {Craig}, W.~W., {Christensen}, F.~E., {et~al.} 2013, \apj,
  770, 103

\bibitem[{{Haug}(1972)}]{Haug1972}
{Haug}, E. 1972, \solphys, 25, 425

\bibitem[{{Haug}(1975)}]{1975SoPh...45..453H}
{Haug}, E. 1975, \solphys, 45, 453

\bibitem[{{Haug}(1989)}]{1989A&A...218..330H}
{Haug}, E. 1989, \aap, 218, 330

\bibitem[{{Haug}(1997)}]{1997A&A...326..417H}
{Haug}, E. 1997, \aap, 326, 417

\bibitem[{{Haug}(1998)}]{1998SoPh..178..341H}
{Haug}, E. 1998, \solphys, 178, 341

\bibitem[{{Henoux}(1975)}]{Henoux1975}
{Henoux}, J.~C. 1975, \solphys, 42, 219

\bibitem[{{Heristchi}(1987)}]{Heristchi1987}
{Heristchi}, D. 1987, \apj, 323, 391

\bibitem[{{H{\"o}gbom}(1974)}]{1974A&AS...15..417H}
{H{\"o}gbom}, J.~A. 1974, \aaps, 15, 417

\bibitem[{{Holman} {et~al.}(2011){Holman}, {Aschwanden}, {Aurass}, {Battaglia},
  {Grigis}, {Kontar}, {Liu}, {Saint-Hilaire}, \&
  {Zharkova}}]{2011SSRv..159..107H}
{Holman}, G.~D., {Aschwanden}, M.~J., {Aurass}, H., {et~al.} 2011, \ssr, 159,
  107

\bibitem[{{Houdek} \& {Gough}(2011)}]{2011MNRAS.418.1217H}
{Houdek}, G. \& {Gough}, D.~O. 2011, \mnras, 418, 1217

\bibitem[{{Houghton}(2002)}]{2002phat.book.....H}
{Houghton}, J. 2002, {The Physics of Atmospheres} (The Physics of Atmospheres,
  by John Houghton, pp.~336.~ISBN 0521804566.~Cambridge, UK: Cambridge
  University Press, March 2002.)

\bibitem[{{Hovenier} \& {van der Mee}(1983)}]{Hovenier1983}
{Hovenier}, J.~W. \& {van der Mee}, C.~V.~M. 1983, \aap, 128, 1

\bibitem[{{Hoyng} {et~al.}(1981){Hoyng}, {Duijveman}, {Machado}, {Rust},
  {Svestka}, {Boelee}, {de Jager}, {Frost}, {Lafleur}, {Simnett}, {van Beek},
  \& {Woodgate}}]{1981ApJ...246L.155H}
{Hoyng}, P., {Duijveman}, A., {Machado}, M.~E., {et~al.} 1981, \apjl, 246,
  L155+

\bibitem[{{Hurford} {et~al.}(2002){Hurford}, {Schmahl}, {Schwartz}, {Conway},
  {Aschwanden}, {Csillaghy}, {Dennis}, {Johns-Krull}, {Krucker}, {Lin},
  {McTiernan}, {Metcalf}, {Sato}, \& {Smith}}]{2002SoPh..210...61H}
{Hurford}, G.~J., {Schmahl}, E.~J., {Schwartz}, R.~A., {et~al.} 2002, \solphys,
  210, 61

\bibitem[{{Ishikawa} {et~al.}(2011){Ishikawa}, {Krucker}, {Takahashi}, \&
  {Lin}}]{2011ApJ...737...48I}
{Ishikawa}, S., {Krucker}, S., {Takahashi}, T., \& {Lin}, R.~P. 2011, \apj,
  737, 48

\bibitem[{{Jackson}(1962)}]{1962clel.book.....J}
{Jackson}, J.~D. 1962, {Classical Electrodynamics} (New York, Wiley)

\bibitem[{{Janse} \& {Low}(2007)}]{2007A&A...472..957J}
{Janse}, {\AA}.~M. \& {Low}, B.~C. 2007, \aap, 472, 957

\bibitem[{{Jeffrey} \& {Kontar}(2011)}]{2011A&A...536A..93J}
{Jeffrey}, N.~L.~S. \& {Kontar}, E.~P. 2011, \aap, 536, A93

\bibitem[{{Jeffrey} \& {Kontar}(2013)}]{2013ApJ...766...75J}
{Jeffrey}, N.~L.~S. \& {Kontar}, E.~P. 2013, \apj, 766, 75

\bibitem[{{Jeffrey} {et~al.}(2014){Jeffrey}, {Kontar}, {Bian}, \&
  {Emslie}}]{2014ApJ...787...86J}
{Jeffrey}, N.~L.~S., {Kontar}, E.~P., {Bian}, N.~H., \& {Emslie}, A.~G. 2014,
  \apj, 787, 86

\bibitem[{{Joshi} {et~al.}(2009){Joshi}, {Veronig}, {Cho}, {Bong}, {Somov},
  {Moon}, {Lee}, {Manoharan}, \& {Kim}}]{2009ApJ...706.1438J}
{Joshi}, B., {Veronig}, A., {Cho}, K.-S., {et~al.} 2009, \apj, 706, 1438

\bibitem[{{Karlick{\'y}} \& {Kosugi}(2004)}]{2004A&A...419.1159K}
{Karlick{\'y}}, M. \& {Kosugi}, T. 2004, \aap, 419, 1159

\bibitem[{{Karney}(1986)}]{1986CoPhR...4..183K}
{Karney}, C. 1986, Computer Physics Reports, 4, 183

\bibitem[{{Ka{\v s}parov{\'a}} {et~al.}(2005){Ka{\v s}parov{\'a}},
  {Karlick{\'y}}, {Kontar}, {Schwartz}, \& {Dennis}}]{Kasparovaetal2005}
{Ka{\v s}parov{\'a}}, J., {Karlick{\'y}}, M., {Kontar}, E.~P., {Schwartz},
  R.~A., \& {Dennis}, B.~R. 2005, \solphys, 232, 63

\bibitem[{{Ka{\v s}parov{\'a}} {et~al.}(2007){Ka{\v s}parov{\'a}}, {Kontar}, \&
  {Brown}}]{Kasparovaetal2007}
{Ka{\v s}parov{\'a}}, J., {Kontar}, E.~P., \& {Brown}, J.~C. 2007, \aap, 466,
  705

\bibitem[{{Klein} \& {Nishina}(1929)}]{KleinNishina1929}
{Klein}, O. \& {Nishina}, T. 1929, Zeitschrift fur Physik, 52, 853

\bibitem[{{Koch} \& {Motz}(1959)}]{1959RvMP...31..920K}
{Koch}, H.~W. \& {Motz}, J.~W. 1959, Reviews of Modern Physics, 31, 920

\bibitem[{{Ko{\l}oma{\'n}ski} {et~al.}(2011){Ko{\l}oma{\'n}ski}, {Mrozek}, \&
  {B{\c a}k-St{\c e}{\'s}licka}}]{2011A&A...531A..57K}
{Ko{\l}oma{\'n}ski}, S., {Mrozek}, T., \& {B{\c a}k-St{\c e}{\'s}licka}, U.
  2011, \aap, 531, A57

\bibitem[{{Kontar} {et~al.}(2014){Kontar}, {Bian}, {Emslie}, \&
  {Vilmer}}]{2014ApJ...780..176K}
{Kontar}, E.~P., {Bian}, N.~H., {Emslie}, A.~G., \& {Vilmer}, N. 2014, \apj,
  780, 176

\bibitem[{{Kontar} \& {Brown}(2006)}]{KontarBrown2006}
{Kontar}, E.~P. \& {Brown}, J.~C. 2006, \apjl, 653, L149

\bibitem[{{Kontar} {et~al.}(2011{\natexlab{a}}){Kontar}, {Brown}, {Emslie},
  {Hajdas}, {Holman}, {Hurford}, {Ka{\v s}parov{\'a}}, {Mallik}, {Massone},
  {McConnell}, {Piana}, {Prato}, {Schmahl}, \&
  {Suarez-Garcia}}]{2011SSRv..159..301K}
{Kontar}, E.~P., {Brown}, J.~C., {Emslie}, A.~G., {et~al.} 2011{\natexlab{a}},
  \ssr, 159, 301

\bibitem[{{Kontar} {et~al.}(2008{\natexlab{a}}){Kontar}, {Dickson}, \& {Ka{\v
  s}parov{\'a}}}]{2008SoPh..252..139K}
{Kontar}, E.~P., {Dickson}, E., \& {Ka{\v s}parov{\'a}}, J. 2008{\natexlab{a}},
  \solphys, 252, 139

\bibitem[{{Kontar} {et~al.}(2007){Kontar}, {Emslie}, {Massone}, {Piana},
  {Brown}, \& {Prato}}]{2007ApJ...670..857K}
{Kontar}, E.~P., {Emslie}, A.~G., {Massone}, A.~M., {et~al.} 2007, \apj, 670,
  857

\bibitem[{{Kontar} {et~al.}(2011{\natexlab{b}}){Kontar}, {Hannah}, \&
  {Bian}}]{2011ApJ...730L..22K}
{Kontar}, E.~P., {Hannah}, I.~G., \& {Bian}, N.~H. 2011{\natexlab{b}}, \apjl,
  730, L22

\bibitem[{{Kontar} {et~al.}(2010){Kontar}, {Hannah}, {Jeffrey}, \&
  {Battaglia}}]{2010ApJ...717..250K}
{Kontar}, E.~P., {Hannah}, I.~G., {Jeffrey}, N.~L.~S., \& {Battaglia}, M. 2010,
  \apj, 717, 250

\bibitem[{{Kontar} {et~al.}(2008{\natexlab{b}}){Kontar}, {Hannah}, \&
  {MacKinnon}}]{2008A&A...489L..57K}
{Kontar}, E.~P., {Hannah}, I.~G., \& {MacKinnon}, A.~L. 2008{\natexlab{b}},
  \aap, 489, L57

\bibitem[{{Kontar} \& {Jeffrey}(2010)}]{KontarJeffrey2010}
{Kontar}, E.~P. \& {Jeffrey}, N.~L.~S. 2010, \aap, 513, L2+

\bibitem[{{Kontar} {et~al.}(2006){Kontar}, {MacKinnon}, {Schwartz}, \&
  {Brown}}]{Kontaretal2006}
{Kontar}, E.~P., {MacKinnon}, A.~L., {Schwartz}, R.~A., \& {Brown}, J.~C. 2006,
  \aap, 446, 1157

\bibitem[{{Kontar} {et~al.}(2012){Kontar}, {Ratcliffe}, \&
  {Bian}}]{2012A&A...539A..43K}
{Kontar}, E.~P., {Ratcliffe}, H., \& {Bian}, N.~H. 2012, \aap, 539, A43

\bibitem[{{Korchak}(1967)}]{1967SvA....11..258K}
{Korchak}, A.~A. 1967, Soviet Astronomy, 11, 258

\bibitem[{{Kovalev} \& {Korolev}(1981)}]{1981SvA....25..215K}
{Kovalev}, V.~A. \& {Korolev}, O.~S. 1981, \sovast, 25, 215

\bibitem[{{Kretzschmar}(2011)}]{2011A&A...530A..84K}
{Kretzschmar}, M. 2011, \aap, 530, A84

\bibitem[{{Krucker} \& {Battaglia}(2014)}]{2014ApJ...780..107K}
{Krucker}, S. \& {Battaglia}, M. 2014, \apj, 780, 107

\bibitem[{{Krucker} {et~al.}(2011{\natexlab{a}}){Krucker}, {Christe},
  {Glesener}, {Ishikawa}, {McBride}, {Glaser}, {Turin}, {Lin}, {Gubarev},
  {Ramsey}, {Saito}, {Tanaka}, {Takahashi}, {Watanabe}, {Tanaka}, {Tajima}, \&
  {Masuda}}]{2011SPIE.8147E...4K}
{Krucker}, S., {Christe}, S., {Glesener}, L., {et~al.} 2011{\natexlab{a}}, in
  Society of Photo-Optical Instrumentation Engineers (SPIE) Conference Series,
  Vol. 8147, Society of Photo-Optical Instrumentation Engineers (SPIE)
  Conference Series

\bibitem[{{Krucker} {et~al.}(2011{\natexlab{b}}){Krucker}, {Hudson}, {Jeffrey},
  {Battaglia}, {Kontar}, {Benz}, {Csillaghy}, \& {Lin}}]{2011ApJ...739...96K}
{Krucker}, S., {Hudson}, H.~S., {Jeffrey}, N.~L.~S., {et~al.}
  2011{\natexlab{b}}, \apj, 739, 96

\bibitem[{{Langer} \& {Petrosian}(1977)}]{LangerPetrosian1977}
{Langer}, S.~H. \& {Petrosian}, V. 1977, \apj, 215, 666

\bibitem[{{Leach} \& {Petrosian}(1983)}]{LeachPetrosian1983}
{Leach}, J. \& {Petrosian}, V. 1983, \apj, 269, 715

\bibitem[{{Lee} {et~al.}(2013){Lee}, {Lim}, {Choe}, {Kim}, \&
  {Jang}}]{2013ApJ...769L..11L}
{Lee}, J., {Lim}, D., {Choe}, G.~S., {Kim}, K.-S., \& {Jang}, M. 2013, \apjl,
  769, L11

\bibitem[{{Lemen} {et~al.}(1982){Lemen}, {Chanan}, {Hughes}, {Laser}, {Novick},
  {Rochwarger}, {Sackson}, \& {Tramiel}}]{Lemenetal1982}
{Lemen}, J.~R., {Chanan}, G.~A., {Hughes}, J.~P., {et~al.} 1982, \solphys, 80,
  333

\bibitem[{{Lemons} {et~al.}(2009){Lemons}, {Winske}, {Daughton}, \&
  {Albright}}]{2009JCoPh.228.1391L}
{Lemons}, D.~S., {Winske}, D., {Daughton}, W., \& {Albright}, B. 2009, Journal
  of Computational Physics, 228, 1391

\bibitem[{{Li} \& {Gan}(2005)}]{2005ApJ...629L.137L}
{Li}, Y.~P. \& {Gan}, W.~Q. 2005, \apjl, 629, L137

\bibitem[{{Lifshitz} \& {Pitaevskii}(1981)}]{1981phki.book.....L}
{Lifshitz}, E.~M. \& {Pitaevskii}, L.~P. 1981, {Physical kinetics} (Course of
  theoretical physics, Oxford: Pergamon Press, 1981)

\bibitem[{{Lin} {et~al.}(2002){Lin}, {Dennis}, {Hurford}, {Smith}, {Zehnder},
  {Harvey}, {Curtis}, {Pankow}, {Turin}, {Bester}, {Csillaghy}, {Lewis},
  {Madden}, {van Beek}, {Appleby}, {Raudorf}, {McTiernan}, {Ramaty}, {Schmahl},
  {Schwartz}, {Krucker}, {Abiad}, {Quinn}, {Berg}, {Hashii}, {Sterling},
  {Jackson}, {Pratt}, {Campbell}, {Malone}, {Landis}, {Barrington-Leigh},
  {Slassi-Sennou}, {Cork}, {Clark}, {Amato}, {Orwig}, {Boyle}, {Banks},
  {Shirey}, {Tolbert}, {Zarro}, {Snow}, {Thomsen}, {Henneck}, {McHedlishvili},
  {Ming}, {Fivian}, {Jordan}, {Wanner}, {Crubb}, {Preble}, {Matranga}, {Benz},
  {Hudson}, {Canfield}, {Holman}, {Crannell}, {Kosugi}, {Emslie}, {Vilmer},
  {Brown}, {Johns-Krull}, {Aschwanden}, {Metcalf}, \&
  {Conway}}]{2002SoPh..210....3L}
{Lin}, R.~P., {Dennis}, B.~R., {Hurford}, G.~J., {et~al.} 2002, \solphys, 210,
  3

\bibitem[{{Lin} {et~al.}(1981){Lin}, {Schwartz}, {Pelling}, \&
  {Hurley}}]{1981ApJ...251L.109L}
{Lin}, R.~P., {Schwartz}, R.~A., {Pelling}, R.~M., \& {Hurley}, K.~C. 1981,
  \apjl, 251, L109

\bibitem[{{Litvinenko} \& {Somov}(1993)}]{1993SoPh..146..127L}
{Litvinenko}, Y.~E. \& {Somov}, B.~V. 1993, \solphys, 146, 127

\bibitem[{{Liu} {et~al.}(2009){Liu}, {Wang}, \&
  {Alexander}}]{2009ApJ...696..121L}
{Liu}, R., {Wang}, H., \& {Alexander}, D. 2009, \apj, 696, 121

\bibitem[{{Liu} {et~al.}(2008){Liu}, {Petrosian}, {Dennis}, \&
  {Jiang}}]{2008ApJ...676..704L}
{Liu}, W., {Petrosian}, V., {Dennis}, B.~R., \& {Jiang}, Y.~W. 2008, \apj, 676,
  704

\bibitem[{{Longair}(1981)}]{1981cup..book.....L}
{Longair}, M.~S. 1981, {High energy astrophysics} (Cambridge and New York,
  Cambridge University Press, 1981.~420 p.)

\bibitem[{{MacKinnon} \& {Craig}(1991)}]{1991A&A...251..693M}
{MacKinnon}, A.~L. \& {Craig}, I.~J.~D. 1991, \aap, 251, 693

\bibitem[{{Maetzler} {et~al.}(1978){Maetzler}, {Bai}, {Crannell}, \&
  {Frost}}]{1978ApJ...223.1058M}
{Maetzler}, C., {Bai}, T., {Crannell}, C.~J., \& {Frost}, K.~J. 1978, \apj,
  223, 1058

\bibitem[{{Magdziarz} \& {Zdziarski}(1995)}]{MagdziarzZdziarski1995}
{Magdziarz}, P. \& {Zdziarski}, A.~A. 1995, \mnras, 273, 837

\bibitem[{{Massone} {et~al.}(2009){Massone}, {Emslie}, {Hurford}, {Prato},
  {Kontar}, \& {Piana}}]{2009ApJ...703.2004M}
{Massone}, A.~M., {Emslie}, A.~G., {Hurford}, G.~J., {et~al.} 2009, \apj, 703,
  2004

\bibitem[{{Massone} {et~al.}(2004){Massone}, {Emslie}, {Kontar}, {Piana},
  {Prato}, \& {Brown}}]{Massoneetal2004}
{Massone}, A.~M., {Emslie}, A.~G., {Kontar}, E.~P., {et~al.} 2004, \apj, 613,
  1233

\bibitem[{{Masuda} {et~al.}(1994){Masuda}, {Kosugi}, {Hara}, {Tsuneta}, \&
  {Ogawara}}]{1994Natur.371..495M}
{Masuda}, S., {Kosugi}, T., {Hara}, H., {Tsuneta}, S., \& {Ogawara}, Y. 1994,
  \nat, 371, 495

\bibitem[{{McConnell} {et~al.}(2007){McConnell}, {Ryan}, {Smith}, {Hurford},
  {Fivian}, {Lin}, {Emslie}, \& {Hajdas}}]{McConnelletal2007}
{McConnell}, M.~L., {Ryan}, J.~M., {Smith}, D.~M., {et~al.} 2007, in Bulletin
  of the American Astronomical Society, Vol.~38, American Astronomical Society
  Meeting Abstracts \#210, 211--+

\bibitem[{{McConnell} {et~al.}(2002){McConnell}, {Ryan}, {Smith}, {Lin}, \&
  {Emslie}}]{McConnelletal2002}
{McConnell}, M.~L., {Ryan}, J.~M., {Smith}, D.~M., {Lin}, R.~P., \& {Emslie},
  A.~G. 2002, \solphys, 210, 125

\bibitem[{{McConnell} {et~al.}(2003){McConnell}, {Smith}, {Emslie}, {Hurford},
  {Lin}, \& {Ryan}}]{McConnelletal2003}
{McConnell}, M.~L., {Smith}, D.~M., {Emslie}, A.~G., {et~al.} 2003, in Bulletin
  of the American Astronomical Society, Vol.~35, AAS/High Energy Astrophysics
  Division 7, 616--+

\bibitem[{{McConnell} {et~al.}(2004){McConnell}, {Smith}, {Emslie}, {Hurford},
  {Lin}, \& {Ryan}}]{McConnelletal2004}
{McConnell}, M.~L., {Smith}, D.~M., {Emslie}, A.~G., {et~al.} 2004, Advances in
  Space Research, 34, 462

\bibitem[{{McKenzie} {et~al.}(1980){McKenzie}, {Broussard}, {Landecker},
  {Rugge}, {Young}, {Doschek}, \& {Feldman}}]{1980ApJ...238L..43M}
{McKenzie}, D.~L., {Broussard}, R.~M., {Landecker}, P.~B., {et~al.} 1980,
  \apjl, 238, L43

\bibitem[{{McMaster}(1961)}]{McMaster1961}
{McMaster}, W.~H. 1961, Reviews of Modern Physics, 33, 8

\bibitem[{{McTiernan} {et~al.}(1999){McTiernan}, {Fisher}, \&
  {Li}}]{1999ApJ...514..472M}
{McTiernan}, J.~M., {Fisher}, G.~H., \& {Li}, P. 1999, \apj, 514, 472

\bibitem[{{McTiernan} \& {Petrosian}(1991)}]{1991ApJ...379..381M}
{McTiernan}, J.~M. \& {Petrosian}, V. 1991, \apj, 379, 381

\bibitem[{{Mertz} {et~al.}(1986){Mertz}, {Nakano}, \&
  {Kilner}}]{1986JOSAA...3.2167M}
{Mertz}, L.~N., {Nakano}, G.~H., \& {Kilner}, J.~R. 1986, Journal of the
  Optical Society of America A, 3, 2167

\bibitem[{{Metcalf} {et~al.}(1996){Metcalf}, {Hudson}, {Kosugi}, {Puetter}, \&
  {Pina}}]{1996ApJ...466..585M}
{Metcalf}, T.~R., {Hudson}, H.~S., {Kosugi}, T., {Puetter}, R.~C., \& {Pina},
  R.~K. 1996, \apj, 466, 585

\bibitem[{{Miller} {et~al.}(1996){Miller}, {Larosa}, \&
  {Moore}}]{1996ApJ...461..445M}
{Miller}, J.~A., {Larosa}, T.~N., \& {Moore}, R.~L. 1996, \apj, 461, 445

\bibitem[{{Morrison} \& {McCammon}(1983)}]{MorrisonMcCammon1983}
{Morrison}, R. \& {McCammon}, D. 1983, \apj, 270, 119

\bibitem[{{Mrozek} \& {Kowalczuk}(2010)}]{MrozekKowalczuk2010}
{Mrozek}, T. \& {Kowalczuk}, J. 2010, Central European Astrophysical Bulletin,
  34, 57

\bibitem[{{Mrozek} \& {Tomczak}(2004)}]{2004A&A...415..377M}
{Mrozek}, T. \& {Tomczak}, M. 2004, \aap, 415, 377

\bibitem[{{Nakada} {et~al.}(1974){Nakada}, {Neupert}, \&
  {Thomas}}]{Nakadaetal1974}
{Nakada}, M.~P., {Neupert}, W.~M., \& {Thomas}, R.~J. 1974, \solphys, 37, 429

\bibitem[{{Parnell} \& {De Moortel}(2012)}]{2012RSPTA.370.3217P}
{Parnell}, C.~E. \& {De Moortel}, I. 2012, Royal Society of London
  Philosophical Transactions Series A, 370, 3217

\bibitem[{{Petrosian} \& {Donaghy}(1999)}]{1999ApJ...527..945P}
{Petrosian}, V. \& {Donaghy}, T.~Q. 1999, \apj, 527, 945

\bibitem[{{Phillips} {et~al.}(2006){Phillips}, {Chifor}, \&
  {Dennis}}]{2006ApJ...647.1480P}
{Phillips}, K.~J.~H., {Chifor}, C., \& {Dennis}, B.~R. 2006, \apj, 647, 1480

\bibitem[{{Phillips} \& {Dennis}(2012)}]{2012ApJ...748...52P}
{Phillips}, K.~J.~H. \& {Dennis}, B.~R. 2012, \apj, 748, 52

\bibitem[{{Piana} {et~al.}(2007){Piana}, {Massone}, {Hurford}, {Prato},
  {Emslie}, {Kontar}, \& {Schwartz}}]{2007ApJ...665..846P}
{Piana}, M., {Massone}, A.~M., {Hurford}, G.~J., {et~al.} 2007, \apj, 665, 846

\bibitem[{{Pina} \& {Puetter}(1993)}]{1993PASP..105..630P}
{Pina}, R.~K. \& {Puetter}, R.~C. 1993, \pasp, 105, 630

\bibitem[{{Poutanen} {et~al.}(1996){Poutanen}, {Nagendra}, \&
  {Svensson}}]{Poutanenetal1996}
{Poutanen}, J., {Nagendra}, K.~N., \& {Svensson}, R. 1996, \mnras, 283, 892

\bibitem[{{Prato} {et~al.}(2009){Prato}, {Emslie}, {Kontar}, {Massone}, \&
  {Piana}}]{Pratoetal2009}
{Prato}, M., {Emslie}, A.~G., {Kontar}, E.~P., {Massone}, A.~M., \& {Piana}, M.
  2009, \apj, 706, 917

\bibitem[{{Priest} \& {Forbes}(2000)}]{2000mare.book.....P}
{Priest}, E. \& {Forbes}, T. 2000, {Magnetic Reconnection} (Magnetic
  Reconnection, by Eric Priest and Terry Forbes, pp.~612.~ISBN
  0521481791.~Cambridge, UK: Cambridge University Press, June 2000.)

\bibitem[{{Priest} \& {Forbes}(2002)}]{2002A&ARv..10..313P}
{Priest}, E.~R. \& {Forbes}, T.~G. 2002, \aapr, 10, 313

\bibitem[{{Ramaty} {et~al.}(1979){Ramaty}, {Kozlovsky}, \&
  {Lingenfelter}}]{1979ApJS...40..487R}
{Ramaty}, R., {Kozlovsky}, B., \& {Lingenfelter}, R.~E. 1979, \apjs, 40, 487

\bibitem[{{Raymond} \& {Smith}(1977)}]{1977ApJS...35..419R}
{Raymond}, J.~C. \& {Smith}, B.~W. 1977, \apjs, 35, 419

\bibitem[{{Reznikova} {et~al.}(2010){Reznikova}, {Melnikov}, {Ji}, \&
  {Shibasaki}}]{2010ApJ...724..171R}
{Reznikova}, V.~E., {Melnikov}, V.~F., {Ji}, H., \& {Shibasaki}, K. 2010, \apj,
  724, 171

\bibitem[{{Saint-Hilaire} {et~al.}(2010){Saint-Hilaire}, {Krucker}, \&
  {Lin}}]{SaintHilaireetal2010}
{Saint-Hilaire}, P., {Krucker}, S., \& {Lin}, R.~P. 2010, \apj, 721, 1933

\bibitem[{{Sakao}(1994)}]{1994PhDT.......335S}
{Sakao}, T. 1994, PhD thesis, (University of Tokyo), (1994)

\bibitem[{{Sakao} {et~al.}(1996){Sakao}, {Kosugi}, {Masuda}, {Yaji},
  {Inda-Koide}, \& {Makishima}}]{1996AdSpR..17...67S}
{Sakao}, T., {Kosugi}, T., {Masuda}, S., {et~al.} 1996, Advances in Space
  Research, 17, 67

\bibitem[{{Salvat} {et~al.}(2008){Salvat}, {Fernandez-Varea}, \&
  {Sempau}}]{Salvatetal2008}
{Salvat}, F., {Fernandez-Varea}, J.~M., \& {Sempau}, J. 2008, France: OECD
  Nuclear Energy Agency, Issy-les-Moulineaux

\bibitem[{{Santangelo} {et~al.}(1973){Santangelo}, {Horstman}, \&
  {Horstman-Moretti}}]{Santangeloetal1973}
{Santangelo}, N., {Horstman}, H., \& {Horstman-Moretti}, E. 1973, \solphys, 29,
  143

\bibitem[{{Schmahl} \& {Hurford}(2002)}]{SchmahlHurford2002}
{Schmahl}, E.~J. \& {Hurford}, G.~J. 2002, \solphys, 210, 273

\bibitem[{{Schmahl} {et~al.}(2007){Schmahl}, {Pernak}, {Hurford}, {Lee}, \&
  {Bong}}]{2007SoPh..240..241S}
{Schmahl}, E.~J., {Pernak}, R.~L., {Hurford}, G.~J., {Lee}, J., \& {Bong}, S.
  2007, \solphys, 240, 241

\bibitem[{{Schwartz} {et~al.}(2002){Schwartz}, {Csillaghy}, {Tolbert},
  {Hurford}, {Mc Tiernan}, \& {Zarro}}]{2002SoPh..210..165S}
{Schwartz}, R.~A., {Csillaghy}, A., {Tolbert}, A.~K., {et~al.} 2002, \solphys,
  210, 165

\bibitem[{{Shen} {et~al.}(2008){Shen}, {Zhou}, {Ji}, {Wang}, {Cao}, \&
  {Wang}}]{2008ApJ...686L..37S}
{Shen}, J., {Zhou}, T., {Ji}, H., {et~al.} 2008, \apjl, 686, L37

\bibitem[{{Shih} {et~al.}(2009){Shih}, {Lin}, {Hurford}, {Boggs}, {Zoglauer},
  {Wunderer}, {Sample}, {Turin}, {McBride}, {Smith}, {Tajima}, {Luke}, \&
  {Amman}}]{Shihetal2009}
{Shih}, A.~Y., {Lin}, R.~P., {Hurford}, G.~J., {et~al.} 2009, in AAS/Solar
  Physics Division Meeting, Vol.~40, AAS/Solar Physics Division Meeting,
  18.10--+

\bibitem[{{Siding} \& {Spicer}(1980)}]{1980ApJ...242.1243S}
{Siding}, K.~G. \& {Spicer}, D.~S. 1980, \apj, 242, 1243

\bibitem[{{Smith} {et~al.}(2002){Smith}, {Lin}, {Turin},
  {et~al.}}]{2002SoPh..210...33S}
{Smith}, D.~M., {Lin}, R.~P., {Turin}, P., {et~al.} 2002, \solphys, 210, 33

\bibitem[{{Smith} {et~al.}(2003){Smith}, {Share}, {Murphy}, {Schwartz}, {Shih},
  \& {Lin}}]{2003ApJ...595L..81S}
{Smith}, D.~M., {Share}, G.~H., {Murphy}, R.~J., {et~al.} 2003, \apjl, 595, L81

\bibitem[{{Somov} \& {Kosugi}(1997)}]{1997ApJ...485..859S}
{Somov}, B.~V. \& {Kosugi}, T. 1997, \apj, 485, 859

\bibitem[{{Stix}(2004)}]{2004suin.book.....S}
{Stix}, M. 2004, {The sun : an introduction} (2nd ed., Berlin: Springer, 2004.~
  ISBN: 3540207414)

\bibitem[{{Stokes}(1852)}]{Stokes1852}
{Stokes}, G., G. 1852, Trans. Cambridge Phil. Soc, 9, 399

\bibitem[{{Suarez-Garcia} {et~al.}(2006){Suarez-Garcia}, {Hajdas}, {Wigger},
  {Arzner}, {G{\"u}del}, {Zehnder}, \& {Grigis}}]{2006SoPh..239..149S}
{Suarez-Garcia}, E., {Hajdas}, W., {Wigger}, C., {et~al.} 2006, \solphys, 239,
  149

\bibitem[{{Sui} \& {Holman}(2003)}]{2003ApJ...596L.251S}
{Sui}, L. \& {Holman}, G.~D. 2003, \apjl, 596, L251

\bibitem[{{Sui} {et~al.}(2004){Sui}, {Holman}, \&
  {Dennis}}]{2004ApJ...612..546S}
{Sui}, L., {Holman}, G.~D., \& {Dennis}, B.~R. 2004, \apj, 612, 546

\bibitem[{{Syrovatskii} \& {Shmeleva}(1972)}]{1972SvA....16..273S}
{Syrovatskii}, S.~I. \& {Shmeleva}, O.~P. 1972, \sovast, 16, 273

\bibitem[{{Takakara} \& {Kai}(1966)}]{1966PASJ...18...57T}
{Takakara}, T. \& {Kai}, K. 1966, \pasj, 18, 57

\bibitem[{{Takakura} {et~al.}(1995){Takakura}, {Kosugi}, {Sakao}, {Makishima},
  {Inda-Koide}, \& {Masuda}}]{1995PASJ...47..355T}
{Takakura}, T., {Kosugi}, T., {Sakao}, T., {et~al.} 1995, \pasj, 47, 355

\bibitem[{{Taylor}(1974)}]{1974PhRvL..33.1139T}
{Taylor}, J.~B. 1974, Physical Review Letters, 33, 1139

\bibitem[{{Tindo} {et~al.}(1970){Tindo}, {Ivanov}, {Mandel'Stam}, \&
  {Shuryghin}}]{Tindoetal1970}
{Tindo}, I.~P., {Ivanov}, V.~D., {Mandel'Stam}, S.~L., \& {Shuryghin}, A.~I.
  1970, \solphys, 14, 204

\bibitem[{{Tindo} {et~al.}(1972){Tindo}, {Ivanov}, {Valn{\'{\i}}{\v c}ek}, \&
  {Livshits}}]{Tindoetal1972}
{Tindo}, I.~P., {Ivanov}, V.~D., {Valn{\'{\i}}{\v c}ek}, B., \& {Livshits},
  M.~A. 1972, \solphys, 27, 426

\bibitem[{{Tindo} {et~al.}(1976){Tindo}, {Shurygin}, \&
  {Steffen}}]{Tindoetal1976}
{Tindo}, I.~P., {Shurygin}, A.~I., \& {Steffen}, W. 1976, \solphys, 46, 219

\bibitem[{{Tomblin}(1972)}]{Tomblin1972}
{Tomblin}, F.~F. 1972, \apj, 171, 377

\bibitem[{{Tomczak} \& {Ciborski}(2007)}]{2007A&A...461..315T}
{Tomczak}, M. \& {Ciborski}, T. 2007, \aap, 461, 315

\bibitem[{{Vernazza} {et~al.}(1981){Vernazza}, {Avrett}, \&
  {Loeser}}]{Vernazzaetal1981}
{Vernazza}, J.~E., {Avrett}, E.~H., \& {Loeser}, R. 1981, \apjs, 45, 635

\bibitem[{{Veronig} \& {Brown}(2004)}]{2004ApJ...603L.117V}
{Veronig}, A.~M. \& {Brown}, J.~C. 2004, \apjl, 603, L117

\bibitem[{{Veronig} {et~al.}(2006){Veronig}, {Karlick{\'y}}, {Vr{\v s}nak},
  {Temmer}, {Magdaleni{\'c}}, {Dennis}, {Otruba}, \&
  {P{\"o}tzi}}]{2006A&A...446..675V}
{Veronig}, A.~M., {Karlick{\'y}}, M., {Vr{\v s}nak}, B., {et~al.} 2006, \aap,
  446, 675

\bibitem[{{Vilmer} {et~al.}(2011){Vilmer}, {MacKinnon}, \&
  {Hurford}}]{2011SSRv..159..167V}
{Vilmer}, N., {MacKinnon}, A.~L., \& {Hurford}, G.~J. 2011, \ssr, 159, 167

\bibitem[{{Vlahos} {et~al.}(2004){Vlahos}, {Isliker}, \&
  {Lepreti}}]{2004ApJ...608..540V}
{Vlahos}, L., {Isliker}, H., \& {Lepreti}, F. 2004, \apj, 608, 540

\bibitem[{{Wheatland} \& {Melrose}(1995)}]{1995SoPh..158..283W}
{Wheatland}, M.~S. \& {Melrose}, D.~B. 1995, \solphys, 158, 283

\bibitem[{{Woods} {et~al.}(2004){Woods}, {Eparvier}, {Fontenla}, {Harder},
  {Kopp}, {McClintock}, {Rottman}, {Smiley}, \& {Snow}}]{2004GeoRL..3110802W}
{Woods}, T.~N., {Eparvier}, F.~G., {Fontenla}, J., {et~al.} 2004, \grl, 31,
  10802

\bibitem[{{Woods} {et~al.}(2006){Woods}, {Kopp}, \&
  {Chamberlin}}]{2006JGRA..11110S14W}
{Woods}, T.~N., {Kopp}, G., \& {Chamberlin}, P.~C. 2006, Journal of Geophysical
  Research (Space Physics), 111, 10

\bibitem[{{Xu} {et~al.}(2008){Xu}, {Emslie}, \&
  {Hurford}}]{2008ApJ...673..576X}
{Xu}, Y., {Emslie}, A.~G., \& {Hurford}, G.~J. 2008, \apj, 673, 576

\bibitem[{{Zhang} \& {Huang}(2004)}]{ZhangHuang2004}
{Zhang}, J. \& {Huang}, G.~L. 2004, \solphys, 219, 135

\bibitem[{{Zharkova} {et~al.}(1995){Zharkova}, {Brown}, \&
  {Syniavskii}}]{Zharkovaetal1995}
{Zharkova}, V.~V., {Brown}, J.~C., \& {Syniavskii}, D.~V. 1995, \aap, 304, 284

\bibitem[{{Zharkova} {et~al.}(2010){Zharkova}, {Kuznetsov}, \&
  {Siversky}}]{Zharkovaetal2010}
{Zharkova}, V.~V., {Kuznetsov}, A.~A., \& {Siversky}, T.~V. 2010, \aap, 512,
  A8+

\end{thebibliography}

\end{document}